\documentclass[11pt]{article}
\usepackage[T1]{fontenc}
\usepackage{lmodern}
\usepackage{amsmath,amssymb,amsthm,mathtools,bm}
\usepackage[margin=1in]{geometry}
\usepackage{booktabs,longtable,array,multirow,graphicx,placeins,pdflscape}
\usepackage[authoryear,round]{natbib}
\usepackage{float}
\usepackage[colorlinks=true,linkcolor=blue,citecolor=blue,urlcolor=blue]{hyperref}
\usepackage{microtype}
\usepackage{enumitem}
\setlist{nosep}
\newtheorem{theorem}{Theorem}
\newtheorem{lemma}{Lemma}
\newtheorem{auxlemma}{Lemma}[section]
\newtheorem{proposition}{Proposition}
\newtheorem{corollary}{Corollary}
\theoremstyle{remark}\newtheorem{remark}{Remark}
\newtheorem{example}{Example}
\newcommand{\E}{\mathbb E}
\newcommand{\Prb}{\mathbb P}
\newcommand{\Var}{\operatorname{Var}}
\newcommand{\Cov}{\operatorname{Cov}}
\newcommand{\ind}{\mathbf 1}
\newcommand{\op}{o_{\Prb}}
\newcommand{\Op}{O_{\Prb}}
\newcommand{\se}{\operatorname{SE}}

\newcommand{\tblnote}[1]{\par\smallskip\noindent\begin{minipage}{\linewidth}\TableFont \textit{Note:} #1\end{minipage}}

\usepackage{caption}

\usepackage{siunitx}
\newcommand{\TableFont}{\fontsize{9}{11}\selectfont}
\newsavebox{\UniformMeasure}
\newlength{\UniformWidth}
\DeclareCaptionFont{uniformtable}{\fontsize{9}{11}\selectfont}
\newcommand{\TableEndRule}{\noalign{\nobreak\vskip\aboverulesep
  \nobreak\hrule height\heavyrulewidth\nobreak\vskip\belowrulesep\nobreak}}

\usepackage{algorithm,algpseudocode}
\usepackage[title]{appendix}

\newcommand{\result}[1]{\csname value@#1\endcsname}
\expandafter\def\csname value@FtwoGain\endcsname{3.19}
\expandafter\def\csname value@FtwoMCSE\endcsname{0.57}
\expandafter\def\csname value@FtwoCoverage\endcsname{0.9540}
\expandafter\def\csname value@FtwoSERatio\endcsname{1.018}
\expandafter\def\csname value@FthreeGain\endcsname{8.50}
\expandafter\def\csname value@FthreeMCSE\endcsname{0.82}
\expandafter\def\csname value@FthreeCoverage\endcsname{0.9507}
\expandafter\def\csname value@FthreeSERatio\endcsname{1.003}
\expandafter\def\csname value@FzeroGain\endcsname{-0.41}
\expandafter\def\csname value@FzeroMCSE\endcsname{0.20}
\expandafter\def\csname value@FzeroCoverage\endcsname{0.9520}
\expandafter\def\csname value@FzeroSERatio\endcsname{0.997}
\expandafter\def\csname value@FheavyGain\endcsname{-0.60}
\expandafter\def\csname value@FheavyMCSE\endcsname{0.30}
\expandafter\def\csname value@FheavyCoverage\endcsname{0.9465}
\expandafter\def\csname value@FheavySERatio\endcsname{0.986}
\expandafter\def\csname value@FoverlapGain\endcsname{-0.90}
\expandafter\def\csname value@FoverlapMCSE\endcsname{0.32}
\expandafter\def\csname value@FoverlapCoverage\endcsname{0.9485}
\expandafter\def\csname value@FoverlapSERatio\endcsname{0.987}
\expandafter\def\csname value@FquadSmallGain\endcsname{0.87}
\expandafter\def\csname value@FquadSmallMCSE\endcsname{0.60}
\expandafter\def\csname value@FquadSmallCoverage\endcsname{0.9455}
\expandafter\def\csname value@FquadSmallSERatio\endcsname{0.986}
\expandafter\def\csname value@FweakGain\endcsname{-0.24}
\expandafter\def\csname value@FweakMCSE\endcsname{0.24}
\expandafter\def\csname value@FweakCoverage\endcsname{0.9413}
\expandafter\def\csname value@FweakSERatio\endcsname{0.991}
\expandafter\def\csname value@CquadSmallGain\endcsname{2.66}
\expandafter\def\csname value@CquadSmallMCSE\endcsname{0.56}
\expandafter\def\csname value@CquadSmallCoverage\endcsname{0.9523}
\expandafter\def\csname value@CquadSmallSERatio\endcsname{1.013}
\expandafter\def\csname value@CquadLargeGain\endcsname{2.79}
\expandafter\def\csname value@CquadLargeMCSE\endcsname{0.53}
\expandafter\def\csname value@CquadLargeCoverage\endcsname{0.9497}
\expandafter\def\csname value@CquadLargeSERatio\endcsname{0.999}
\expandafter\def\csname value@CgridSmallGain\endcsname{4.86}
\expandafter\def\csname value@CgridSmallMCSE\endcsname{0.53}
\expandafter\def\csname value@CgridSmallCoverage\endcsname{0.9481}
\expandafter\def\csname value@CgridSmallSERatio\endcsname{0.996}
\expandafter\def\csname value@CgridLargeGain\endcsname{4.87}
\expandafter\def\csname value@CgridLargeMCSE\endcsname{0.49}
\expandafter\def\csname value@CgridLargeCoverage\endcsname{0.9524}
\expandafter\def\csname value@CgridLargeSERatio\endcsname{1.007}
\expandafter\def\csname value@CresourceGain\endcsname{12.53}
\expandafter\def\csname value@CresourceMCSE\endcsname{0.68}
\expandafter\def\csname value@CresourceCoverage\endcsname{0.9536}
\expandafter\def\csname value@CresourceSERatio\endcsname{1.005}
\expandafter\def\csname value@CstrongAnchorGain\endcsname{7.44}
\expandafter\def\csname value@CstrongAnchorMCSE\endcsname{0.59}
\expandafter\def\csname value@CstrongAnchorCoverage\endcsname{0.9480}
\expandafter\def\csname value@CstrongAnchorSERatio\endcsname{0.988}
\expandafter\def\csname value@LzeroGain\endcsname{-1.14}
\expandafter\def\csname value@LzeroMCSE\endcsname{0.42}
\expandafter\def\csname value@LzeroCoverage\endcsname{0.9485}
\expandafter\def\csname value@LzeroSERatio\endcsname{1.008}
\expandafter\def\csname value@LmoderateGain\endcsname{4.99}
\expandafter\def\csname value@LmoderateMCSE\endcsname{0.99}
\expandafter\def\csname value@LmoderateCoverage\endcsname{0.9505}
\expandafter\def\csname value@LmoderateSERatio\endcsname{1.008}
\expandafter\def\csname value@LstrongSmallGain\endcsname{12.40}
\expandafter\def\csname value@LstrongSmallMCSE\endcsname{1.41}
\expandafter\def\csname value@LstrongSmallCoverage\endcsname{0.9495}
\expandafter\def\csname value@LstrongSmallSERatio\endcsname{1.008}
\expandafter\def\csname value@LeffectLowGain\endcsname{13.34}
\expandafter\def\csname value@LeffectLowMCSE\endcsname{1.38}
\expandafter\def\csname value@LeffectLowCoverage\endcsname{0.9475}
\expandafter\def\csname value@LeffectLowSERatio\endcsname{1.009}
\expandafter\def\csname value@LeffectHighGain\endcsname{14.56}
\expandafter\def\csname value@LeffectHighMCSE\endcsname{1.39}
\expandafter\def\csname value@LeffectHighCoverage\endcsname{0.9460}
\expandafter\def\csname value@LeffectHighSERatio\endcsname{0.989}
\expandafter\def\csname value@LheavySmallGain\endcsname{6.36}
\expandafter\def\csname value@LheavySmallMCSE\endcsname{1.40}
\expandafter\def\csname value@LheavySmallCoverage\endcsname{0.9435}
\expandafter\def\csname value@LheavySmallSERatio\endcsname{0.997}
\expandafter\def\csname value@LheavyLargeGain\endcsname{11.24}
\expandafter\def\csname value@LheavyLargeMCSE\endcsname{1.28}
\expandafter\def\csname value@LheavyLargeCoverage\endcsname{0.9510}
\expandafter\def\csname value@LheavyLargeSERatio\endcsname{0.990}
\expandafter\def\csname value@LCPmin\endcsname{0.9435}
\expandafter\def\csname value@LCPmax\endcsname{0.9585}
\expandafter\def\csname value@LSEmin\endcsname{0.988}
\expandafter\def\csname value@LSEmax\endcsname{1.034}
\expandafter\def\csname value@RCPmin\endcsname{0.9413}
\expandafter\def\csname value@RCPmax\endcsname{0.9585}
\expandafter\def\csname value@RSEmin\endcsname{0.986}
\expandafter\def\csname value@RSEmax\endcsname{1.019}
\expandafter\def\csname value@CCPmin\endcsname{0.9447}
\expandafter\def\csname value@CCPmax\endcsname{0.9536}
\expandafter\def\csname value@CSEmin\endcsname{0.988}
\expandafter\def\csname value@CSEmax\endcsname{1.013}
\expandafter\def\csname value@RotEstimate\endcsname{0.479474}
\expandafter\def\csname value@RotSE\endcsname{0.011034}
\expandafter\def\csname value@RotCILow\endcsname{0.457847}
\expandafter\def\csname value@RotCIHigh\endcsname{0.501101}
\expandafter\def\csname value@RotSup\endcsname{0.481482}
\expandafter\def\csname value@RotSupSE\endcsname{0.011451}
\expandafter\def\csname value@RotFittedGain\endcsname{7.15}
\expandafter\def\csname value@RotMaskGain\endcsname{13.83}
\expandafter\def\csname value@RotMaskMCSE\endcsname{2.19}
\expandafter\def\csname value@RotMaskLow\endcsname{9.54}
\expandafter\def\csname value@RotMaskHigh\endcsname{18.12}
\expandafter\def\csname value@RotBootGain\endcsname{4.55}
\expandafter\def\csname value@RotBootMCSE\endcsname{1.67}
\expandafter\def\csname value@RotBootLow\endcsname{1.27}
\expandafter\def\csname value@RotBootHigh\endcsname{7.83}
\expandafter\def\csname value@RotMaskMean\endcsname{0.45968}
\expandafter\def\csname value@RotMaskZ\endcsname{2.57}
\expandafter\def\csname value@RotMaskFractionBelow\endcsname{0.994}
\expandafter\def\csname value@RotBootSD\endcsname{0.011021}
\expandafter\def\csname value@RotMeanIFSE\endcsname{0.010979}
\expandafter\def\csname value@RotBootSERatio\endcsname{0.996}
\expandafter\def\csname value@RotCensorHR\endcsname{1.817}
\expandafter\def\csname value@RotCensorLow\endcsname{1.652}
\expandafter\def\csname value@RotCensorHigh\endcsname{1.998}
\expandafter\def\csname value@RotDeathHR\endcsname{0.973}
\expandafter\def\csname value@RotDeathLow\endcsname{0.950}
\expandafter\def\csname value@RotDeathHigh\endcsname{0.996}
\expandafter\def\csname value@RotFullReference\endcsname{0.459907}

\title{\LARGE\bfseries Model-assisted semi-supervised estimation of the Mann--Whitney probability under right censoring\footnotemark[2]}
\author{\fontsize{13}{15.5}\selectfont Hengyao Xu, Tao Tan\thanks{Corresponding author. Email: \texttt{ttan303@163.com}.}\\College of Mathematics and Systems Science, Xinjiang University,\\Urumqi, Xinjiang, China}
\date{}
\begin{document}
\maketitle
{\renewcommand{\thefootnote}{\fnsymbol{footnote}}\footnotetext[2]{This research was supported by the Natural Science Foundation of Xinjiang Uygur Autonomous Region (No. 2025D14015) and the National Natural Science Foundation of China (No. 72571102).}}
\begin{abstract}
We study semi-supervised estimation of the full-lifetime Mann--Whitney probability under right censoring using finite covariate-only samples from two populations. A model-assisted estimator is constructed by adjusting censored pair comparisons with differences between labeled and auxiliary feature means. Resource weights minimize first-order variance within the working projection class. Under joint independent censoring and regularity conditions, we establish consistency and asymptotic normality through a four-pool influence representation that accounts for censoring distributions estimated from the same labeled records. Consistent variance estimation from all four sample pools yields Wald confidence intervals. The efficiency analysis relates variance reduction to projected comparison signals and auxiliary sample sizes and characterizes the gap to the semiparametric bound. A fixed-horizon extension accommodates half-weighted ties and the terminal mass induced by capping follow-up. Simulations show increasing precision gains with auxiliary sample size under moderate or strong projected signals, with finite-sample losses in some settings with zero or weak projected signals. A five-year Rotterdam breast-cancer comparison illustrates precision gains and variability across label allocations when auxiliary samples are formed by controlled outcome withholding.
\end{abstract}


\section{Introduction}\label{sec:intro}
\subsection{Background and motivation}\label{sec:motivation}
The Mann--Whitney probability compares two event-time distributions through the ordering of independently sampled individuals \citep{mann1947test,thas2012probabilistic}. For event times $T_1$ and $T_2$ from the respective populations, the strict full-lifetime target is
\begin{equation*}
\theta=\Prb(T_1<T_2).
\end{equation*}
For an adverse event such as death, $\theta$ is the probability that an individual from population 1 experiences the event first. Under continuous event-time distributions, values below one half indicate that population 1 tends to experience the event later. The parameter compares marginal event-time distributions without imposing proportional hazards.

Right censoring obscures some pairwise orderings, so identification depends on the follow-up support and censoring mechanism. Under independent censoring and suitable support, inverse probability of censoring weighting (IPCW) and Kaplan--Meier integration recover comparisons from observed event records \citep{datta2010,chen2019,fernandez2020}. Identification of the lifetime comparison requires censoring support over the effective event-time range. A prespecified horizon instead defines a restricted comparison of capped event times, with half weight assigned to ties \citep{dobler2018,zhang2022restricted}; all individuals surviving beyond that horizon occupy the same terminal category.

Semi-supervised inference can use baseline covariates from individuals whose event histories are unavailable. We study two populations, each contributing labeled records and an independent covariate-only sample from the same population. Labeled records estimate how covariates predict pairwise comparison information, and the auxiliary records improve estimation of its population average, as in fully observed two-sample comparisons \citep{tan2025efficient}. Inference must account for variation in the auxiliary feature means and for censoring distributions estimated from the same labeled records used in the pair comparisons.

\subsection{Related work}\label{sec:related}
Auxiliary covariates improve estimation by refining the population average of a covariate-predictable outcome or estimating function. This principle connects model-assisted survey estimation \citep{saerndal1992model} with semi-supervised inference for means, regression parameters and general estimating equations \citep{zhang2019semisupervised,chakrabortty2018efficient,song2024general}. For example, \citet{gronsbell2018semisupervised} combine a calibrated outcome model with the covariate distribution estimated from unlabeled observations, while prediction-powered inference uses a related prediction-plus-correction structure \citep{angelopoulos2023prediction,angelopoulos2026ppi}. General semiparametric theory characterizes the gain through the covariate-predictable component of a supervised influence function and the relative labeled and unlabeled sample sizes \citep{xu2025unified}. The observation model determines which score must be learned and how it enters feasible inference.

For two-sample comparisons, outcome information is carried by pairs. \citet{schisterman2001} developed auxiliary-variable estimation for $K$-sample $U$-statistics with missing outcomes, and \citet{cannings2022correlation} studied correlation-assisted corrections that include $U$-statistic applications. \citet{kim2025} established a general semi-supervised theory for $U$-statistics, including efficiency and degeneracy. The closest two-sample design is that of \citet{tan2025efficient}, who combine semiparametric imputation of pairwise comparison information with weights balancing labeled and unlabeled contributions. Their design has independent labeled and covariate-only samples in each population, with matching covariate distributions within each population while allowing the two populations to differ. \citet{zhang2025semisupervised} extend semi-supervised methods to two-population functionals, while \citet{xu2026grouped} study group-specific functionals through conditional imputation. With right censoring, the supervised pair statistic also depends on the estimated censoring distributions.

Censored pair statistics use IPCW $U$-statistics to recover comparisons obscured by censoring \citep{datta2010,chen2019}; Kaplan--Meier $V$- and $U$-statistic theory characterizes the contribution from estimating event-time distributions \citep{fernandez2020}. General results for $U$-statistics with estimated nuisance parameters describe how nuisance estimation enters their limiting distributions \citep{desgagne2025nuisance}. Censored two-sample alternatives also include empirical likelihood \citep{zhou2005empirical} and semiparametric estimating equations \citep{bai2014semiparametric}. Covariate augmentation for censored Mann--Whitney effects has been developed for causal comparisons \citep{zhang2019censored} and randomized trials \citep{zhang2020trial}. In the latter setting, a Kaplan--Meier plug-in influence function is projected onto covariate adjustment directions under a common covariate distribution. The present sampling design gives each population its own auxiliary pool, so the expansion includes uncertainty in two population-specific auxiliary means alongside the group-specific censoring-estimation contributions.

Recent survival work uses covariates in pseudo-observation regression for conditional Mann--Whitney effects \citep{dobler2026conditional} and in semi-supervised treatment rules \citep{li2026efficient_survival_itr}. Other work addresses restricted benefits and ties in fixed-horizon comparisons \citep{zhang2022restricted,dobler2018}. Related observation models address covariate-dependent labeling under missing at random \citep{su2024mar} and identification under dependent censoring with proxy information \citep{ying2024proximal}. We study marginal comparisons under within-population independence of censoring from event times and covariates.

\subsection{Contributions and organization}\label{sec:contributions}
We construct a model-assisted estimator of the Mann--Whitney probability for two right-censored populations with finite, population-specific auxiliary samples. It combines an IPCW pair statistic with a rectangular cross-fitted covariate adjustment. The asymptotic linear representation includes the contribution of estimating each censoring distribution from the records used in the pair comparisons. The resulting four-pool variance estimator supports feasible Wald inference. The transfer weights balance the sampling variation of the labeled and auxiliary feature means. We characterize the optimal weights within the working projection class and decompose the gap to the canonical efficiency bound into supervised-score and feature-projection components.

For comparisons through a prespecified follow-up horizon, we establish identification and inference for a half-tie probability on a fixed recording grid. The analysis incorporates the terminal atom induced by capping event times and the priority of events at shared event/censoring times. Simulations examine how signal, auxiliary sample size and finite-sample fitting costs affect precision and interval calibration. A five-year Rotterdam comparison illustrates the procedure with covariate-only samples obtained through controlled outcome withholding.

Section~\ref{sec:method} presents the estimator, and Section~\ref{sec:theory} develops its full-lifetime theory. Section~\ref{sec:restricted} gives the fixed-horizon extension. Sections~\ref{sec:numerical} and~\ref{sec:application} report the simulations and Rotterdam analysis, followed by discussion in Section~\ref{sec:discussion}. Appendices~\ref{app:auxiliary} and~\ref{app:proofs} contain auxiliary results and proofs; Appendix~\ref{app:additional} supplies additional numerical results.

\section{Methodology}\label{sec:method}
\subsection{Data structure and assumptions}
For population $k=1,2$, let $O_k=(Y_k,\Delta_k,\bm{X}_k)$, where $Y_k=\min(T_k,C_k)$ and $\Delta_k=\ind(T_k\le C_k)$.

Following the two-population sampling design of \citet{tan2025efficient}, we write the observed data in population $k$ as a labeled sample $\mathcal L_k$ and an unlabeled sample $\mathcal U_k$:
\[
\mathcal L_k=\{(Y_{ki},\Delta_{ki},\bm X_{ki}):i=1,\ldots,n_k\},
\qquad
\mathcal U_k=\{\bm X^U_{kj}:j=1,\ldots,M_k\},
\qquad k=1,2.
\]
The observed data thus comprise $(\mathcal L_1,\mathcal U_1,\mathcal L_2,\mathcal U_2)$. Each labeled record includes an observed follow-up time, an event indicator, and covariates. Each unlabeled record includes only covariates.

Throughout the theory, $n=n_1+n_2$ is the total labeled size. Vectors and matrices are set in bold. The working feature map is denoted by $\bm{u}_k$; expectation, variance and covariance are written as $\E$, $\Var$ and $\Cov$.

The estimator starts from labeled pair comparisons and adjusts for the difference between auxiliary and labeled feature means. Because the auxiliary mean is estimated, the amount of transfer depends on both sample sizes. The following assumptions specify the full-lifetime experiment.
\begin{enumerate}[leftmargin=*,label=\textbf{(A\arabic*)}]
\item\label{ass:design} The four pools are mutually independent and i.i.d. within each pool. Labeled and unlabeled covariates have the same marginal law within a population. As $n\to\infty$, $n_k/n\to\pi_k>0$ and $M_k/n\to\lambda_k\in(0,\infty)$.
\item\label{ass:censor} $C_k\perp(T_k,\bm{X}_k)$; $T_k$ is nonnegative, finite almost surely and marginally nonatomic. On the effective event-time support, the cumulative censoring hazard is absolutely continuous with bounded measurable hazard rate on $[0,b_0]$ for a fixed finite $b_0$, with no effective censoring hazard thereafter. Also $H_k(b_0)=\Prb(Y_k\ge b_0)>0$. In a bounded event-time model, changes to censoring beyond the event support are immaterial; with unbounded event times, surviving censoring mass can be at $C_k=\infty$.
\item\label{ass:features} The feature map $\bm{u}_k(\bm{X}_k)\in\mathbb R^{J_k}$ has fixed dimension, $\E\|\bm{u}_k\|^{2+\eta}<\infty$ for some $\eta>0$, and positive-definite covariance $\bm{\Gamma}_k$.
\item\label{ass:folds} A fixed $K\ge2$ partitions labeled indices into folds $I_{ka}$ independently of their values. All fold and training fractions have positive limiting proportions. The denominator safeguard is $\varepsilon_n=\min(0.01,n^{-1/4})$.
\item\label{ass:variance} The limiting variance of the estimator under consideration is positive when studentized inference is asserted.
\end{enumerate}
Assumptions~(A1)--(A2) describe sampling and identification; Assumptions~(A3)--(A5) control feature fitting, fold stability and studentization. The censoring restriction gives a positive lower bound for $G_k(t)=\Prb(C_k\ge t)$ on the effective event support and for $H_k$ on $[0,b_0]$, supplying the uniform expansion used below. Lifetime identification requires this support beyond the observed sample times.

Within-population equality of the labeled and auxiliary covariate laws centers the feature-mean correction at its target.

\subsection{Semi-supervised estimation under censoring}
For independent draws from the two populations define
\begin{equation*}
q(O_1,O_2)=\frac{\Delta_1\Delta_2\ind(Y_1<Y_2)}{G_1(Y_1)G_2(Y_2)}.
\end{equation*}
Estimate $G_k$ by a group-specific reverse Kaplan--Meier (KM) estimator fitted to all labeled observations in population $k$. The fit pools labeled observations across folds within that population and remains separate between populations. Use $\max\{\widehat G_k(Y_{ki}-),\varepsilon_n\}$ in the fitted kernel $\widehat q_{ij}$. Under Assumption~(A2), finite censoring times do not tie with event times; the left-limit convention is predictable. The clipping safeguard is inactive with probability tending to one.

Let $P_n$ denote the full labeled pair average and $P_{ab}$ the average over $I_{1a}\times I_{2b}$. Write $n_{ka}=|I_{ka}|$ and $w_{ab}=n_{1a}n_{2b}/(n_1n_2)$. For rectangular cross-fitting, $I_{1a}\times I_{2b}$ is the evaluation rectangle and $I_{1a}^c\times I_{2b}^c$ is its training rectangle. On the training rectangle, regress $\widehat q$ on $(1,\bm{u}_1^\top,\bm{u}_2^\top)^\top$ by ordinary least squares (OLS), with slope blocks $\widehat{\bm{\beta}}_{1,ab},\widehat{\bm{\beta}}_{2,ab}$. If an empirical Gram matrix is singular, its Moore--Penrose inverse defines the fitted slopes; under Assumption~(A3), the limiting population Gram matrix is nonsingular. With $\ell_1(a,b)=a$ and $\ell_2(a,b)=b$, define
\begin{align}\label{eq:F1}
\widehat\theta(s_1,s_2)
={}&\sum_{a,b}w_{ab}\left[P_{ab}\widehat q+
\sum_{k=1}^2s_k\widehat{\bm{\beta}}_{k,ab}^{\top}
\{\bar{\bm{u}}_k^U-\bar{\bm{u}}_{k,\ell_k(a,b)}\}\right],\\
c_{k,n}={}&\frac{M_k}{M_k+n_k},\qquad
\widehat\theta_{\mathrm{SS}}=\widehat\theta(c_{1,n},c_{2,n}).\label{eq:actualc}
\end{align}
We denote the proposed semi-supervised estimator by $\widehat\theta_{\mathrm{SS}}$, the supervised estimator $\widehat\theta(0,0)=P_n\widehat q$ by $\widehat\theta_{\mathrm{sup}}$, and the unshrunk semi-supervised estimator $\widehat\theta(1,1)$ by $\widehat\theta_{\mathrm{un}}$. The three procedures use the same labeled observations and censoring estimates. The two semi-supervised procedures also share folds and regressions, differing only in the transfer weights. Weighted uses $s_k=c_{k,n}$ to account for the size of the auxiliary pool, and Unshrunk uses $s_k=1$.

In each population, the adjustment compares the auxiliary feature mean with the mean in the held-out labeled fold. The fitted slope converts this difference into a correction on the scale of the pairwise comparison. When the working features capture variation in the conditional comparison signal, the auxiliary mean can improve estimation of its population average. The correction also reflects uncertainty in that mean: $c_{k,n}$ approaches zero when the auxiliary pool is small relative to the labeled pool and approaches one as its relative size grows. Corollary~\ref{cor:ratio} gives the corresponding first-order variance comparison.

Each auxiliary pool estimates the feature distribution of its own population. Under Assumption~(A1), the labeled and auxiliary feature means within a population estimate the same expectation, while the covariate distributions of the two populations may differ. The two corrections therefore use population-specific information to refine estimation of the marginal ordering probability. Their relative contributions depend on both the precision of the feature means and the extent to which the fitted features explain comparison outcomes involving that population. An asymmetric design can thus provide a precise auxiliary mean for one group and a less precise mean for the other, motivating separate transfer weights and separate contributions to the four-pool variance.

The regression slopes are fitted on pairs outside both held-out folds, whereas each censoring distribution is estimated using all labeled observations in its population. The influence-function calculation below accounts for this shared censoring fit when constructing the standard error.

The intercept cancels from the difference of feature means. An additive fit avoids constructing an unlabeled pair matrix. In a complete training rectangle the two centered feature blocks have exactly zero empirical cross-covariance, so the two slope blocks can be computed from pair row and column means. Unequal folds require the displayed weights: $\sum_{ab}w_{ab}P_{ab}\widehat q=P_n\widehat q$ exactly. Replacing every fitted slope by a fixed population slope similarly aggregates held-out means into the full labeled mean.

Algorithm~\ref{alg:ss} summarizes estimation and inference.

\begin{algorithm}[H]
\caption{Semi-supervised estimation and variance estimation}\label{alg:ss}
\small
\begin{algorithmic}[1]
\Require Four data pools; features $\bm{u}_k$; fold count $K$; full-lifetime or fixed-horizon target.
\Ensure $\widehat\theta_{\mathrm{SS}}$, standard error, Wald interval, and rank, clipping, nonfinite-value and risk-set exhaustion diagnostics.
\State Compute $c_{k,n}=M_k/(M_k+n_k)$ and the prespecified safeguard; partition labeled indices into data-independent folds.
\For{$k=1,2$}
  \State Fit the group-specific censoring curve using all labeled observations in that group, with the full-lifetime rule or the atomic rule in Section~\ref{sec:restricted}.
\EndFor
\State Construct the fitted pairwise kernel $\widehat q$.
\For{each held-out rectangle $(a,b)$}
  \State Fit additive least squares on $I_{1a}^{c}\times I_{2b}^{c}$ and record rank diagnostics.
  \State Compute the held-out feature-mean corrections and weight them by $w_{ab}=n_{1a}n_{2b}/(n_1n_2)$.
\EndFor
\State Evaluate \eqref{eq:F1} at $(c_{1,n},c_{2,n})$; the intercept cancels from the feature-mean contrast.
\For{$k=1,2$}
  \State Compute pair row/column projections and the complete same-sample censoring contribution.
  \State Form labeled scores $\widehat d_k-\widehat h_k$ and unlabeled scores $\widehat h_k$.
\EndFor
\State Compute four separately centered, sample variances with denominator $m-1$, divided by their pool sizes, as in \eqref{eq:SE}.
\State \Return $\widehat\theta_{\mathrm{SS}}$, its standard error, Wald interval and the listed diagnostics.
\end{algorithmic}
\end{algorithm}

\section{Theoretical properties}\label{sec:theory}
Censoring estimation changes the labeled score on which covariate transfer acts. We first identify that complete score, then quantify the variance recovered by finite auxiliary samples and its relation to the canonical benchmark.
\subsection{Identification and asymptotic linearity}
Put $r_1(t)=\Prb(T_2>t)$ and $r_2(t)=\Prb(T_1<t)$. Define
\begin{equation}\label{eq:signals}
\begin{aligned}
g_k(O_k)&=\frac{\Delta_k r_k(Y_k)}{G_k(Y_k)}-\theta,
& b_k(\bm{X}_k)&=\E\{r_k(T_k)\mid \bm{X}_k\}-\theta,\\
\bm{\mu}_k&=\E \bm{u}_k(\bm{X}_k),
& p_k(\bm{X}_k)&=\bm{\beta}_k^\top\{\bm{u}_k(\bm{X}_k)-\bm{\mu}_k\},\\
\bm{\beta}_k&=\bm{\Gamma}_k^{-1}\E[\{\bm{u}_k-\bm{\mu}_k\}b_k(\bm{X}_k)].
\end{aligned}
\end{equation}
The quantity $g_k$ is the first-order projection of the pairwise kernel when the censoring distributions are known. We call $p_k$ the projected conditional comparison signal: it is the $L_2$ projection of $b_k$ onto the centered working-feature space. Independence of the two populations makes the additive population normal equations block diagonal. The projection remains well defined when $b_k$ is not represented by the working features.

Let $R_k(t)=\ind(Y_k\ge t)$, $N_k^C(t)=\ind(Y_k\le t,\Delta_k=0)$ and $M_k^C(t)=N_k^C(t)-\int_0^tR_k(s)d\Lambda_k^C(s)$. Here $M_k^C$ is a martingale; $M_k$ without a superscript remains an auxiliary sample size. For $0\le t\le b_0$, set
\begin{equation*}
\begin{aligned}
a_{0k}(t)&=\E\{r_k(T_k)\mid T_k\ge t\},\\
\psi_k(O_k)&=\int_0^{b_0}\frac{a_{0k}(t)}{G_k(t)}dM_k^C(t),\\
d_k&=g_k+\psi_k.
\end{aligned}
\end{equation*}
The complete supervised influence function is $d_k=g_k+\psi_k$, including the effect of estimating the censoring distribution. The integrand can equivalently be written $A_k(t)/H_k(t)$, where $A_k(t)=\E\{q\ind(Y_k\ge t)\}$. Its covariance with the pair projection determines the net variance effect of estimating censoring.

\begin{lemma}\label{lem:identity}
Suppose that Assumptions~(A1)--(A4) hold. Then $\E q=\theta$ and
\[
\E(q\mid \bm{X}_1,\bm{X}_2)=\Prb(T_1<T_2\mid \bm{X}_1,\bm{X}_2),\qquad
\E(\psi_k\mid T_k,\bm{X}_k)=0.
\]
For each $k=1,2$, $\E(d_k\mid \bm{X}_k)=b_k$, and
\begin{equation*}
\Cov(g_k,p_k)=\Var(p_k),\quad
\Cov(p_k,\psi_k)=0,\quad
\Cov(g_k,\psi_k)=-\Var(\psi_k).
\end{equation*}
Consequently $\Var(d_k)=\Var(g_k)-\Var(\psi_k)$.
\end{lemma}
The proof is given in Appendix~\ref{s:identity}.
\begin{remark}\label{rem:covariance}
The covariance identity in Lemma~\ref{lem:identity} records how estimating censoring weights on the same labeled observations used for the pair comparisons changes the complete labeled score: $\Cov(g_k,\psi_k)=-\Var(\psi_k)$ and hence $\Var(d_k)=\Var(g_k)-\Var(\psi_k)$. The identity uses the unscaled projection score $p_k$; transfer weights enter the four-pool variance separately.
\end{remark}

\begin{theorem}\label{thm:alr}
Suppose that Assumptions~(A1)--(A4) hold. Let $\bm{s}_n=(s_{1,n},s_{2,n})$ be a bounded deterministic sequence with $s_{k,n}\to s_k$ for $k=1,2$, and write $\bm{s}=(s_1,s_2)$. Then
\begin{equation}\label{eq:alr}
\widehat\theta(\bm{s}_n)-\theta
=\sum_{k=1}^2\left[\frac1{n_k}\sum_i\{d_k(O_{ki})-s_kp_k(\bm{X}_{ki})\}
+\frac1{M_k}\sum_js_kp_k(\bm{X}^U_{kj})\right]+\op(n^{-1/2}).
\end{equation}
Its limiting $\sqrt n$ variance is
\begin{equation}\label{eq:Vs}
V(\bm{s})=\sum_k\left\{\pi_k^{-1}\Var(d_k-s_kp_k)+\lambda_k^{-1}s_k^2\Var(p_k)\right\}.
\end{equation}
If Assumption~(A5) also holds, then
\[
\sqrt n\{\widehat\theta(\bm{s}_n)-\theta\}\rightsquigarrow N\{0,V(\bm{s})\}.
\]
For $s_{k,n}=c_{k,n}$, the limiting weights are $s_k=c_k=\lambda_k/(\pi_k+\lambda_k)$ and the asymptotic variance is $V_{\mathrm{SS}}$.
\end{theorem}

The proof is given in Appendix~\ref{s:regression}, using Lemmas~\ref{lem:aux-km} and~\ref{lem:aux-supervised}.
\begin{remark}\label{rem:alr}
The four-pool sampling design follows \citet{tan2025efficient} and shares the residual-plus-transfer structure of semi-supervised $U$-statistics \citep{kim2025}. For right-censored comparisons, the labeled score $d_k=g_k+\psi_k$ includes the effect of estimating each censoring distribution by Kaplan--Meier using the same group-specific labeled records that enter the pair statistic. This contribution and the sampling variation in the two finite auxiliary pools both enter the variance used for inference. Censored Mann--Whitney augmentation in randomized trials \citep{zhang2020trial} uses a common covariate distribution across groups. Here each auxiliary pool represents its own population, allowing distinct covariate distributions and population-specific resource weights. Section~\ref{sec:covariatecomparison} compares the resulting variance reductions in a common submodel.
\end{remark}

The two parts of the expansion distinguish outcome-bearing information from uncertainty in the auxiliary feature means. Their relative sampling rates determine the optimal amount of transfer in the stated class.

\subsection{Optimal transfer weights}
Write $\bm{c}=(c_1,c_2)$, $V_{\mathrm{SS}}=V(c_1,c_2)$, and let $V_{\mathrm{sup}}=\sum_k\pi_k^{-1}\Var(d_k)$. Expanding \eqref{eq:Vs} gives
\begin{equation}\label{eq:quadratic}
V(\bm{s})=V_{\mathrm{sup}}-\sum_k\left\{\frac{2s_k}{\pi_k}-s_k^2\left(\frac1{\pi_k}+\frac1{\lambda_k}\right)\right\}\Var(p_k).
\end{equation}
\begin{corollary}\label{cor:ratio}
Under the assumptions of Theorem~\ref{thm:alr}, the variance $V(\bm{s})$ in \eqref{eq:quadratic} is minimized over $\bm{s}\in\mathbb R^2$ at $s_k=c_k$, $k=1,2$. The minimizer in coordinate $k$ is unique when $\Var(p_k)>0$; when $\Var(p_k)=0$, $V(\bm{s})$ does not depend on that coordinate. In particular,
\begin{align*}
V_{\mathrm{sup}}-V_{\mathrm{SS}}&=\sum_k\frac{c_k}{\pi_k}\Var(p_k)\ge0,\\
V(1,1)-V_{\mathrm{SS}}&=\sum_k(\pi_k^{-1}+\lambda_k^{-1})(1-c_k)^2\Var(p_k)\ge0.
\end{align*}
The unshrunk gain is instead $V_{\mathrm{sup}}-V(1,1)=\sum_k(\pi_k^{-1}-\lambda_k^{-1})\Var(p_k)$.
\end{corollary}
The proof is given in Appendix~\ref{s:corollary}.
\begin{remark}\label{rem:optimalityscope}
At finite limiting ratios, $c_k=\lambda_k/(\pi_k+\lambda_k)$ reflects the relative precision of the labeled and auxiliary means and equals $1/2$ when their asymptotic sizes are equal. For finite positive limiting ratios, weighting reduces the first-order variance relative to unshrunk transfer whenever at least one group has positive projected variance. If $\Var(p_k)=0$, the first-order variance is unchanged by transfer in that coordinate; the finite-sample example in Appendix~\ref{s:zeroprojection} shows how estimating a zero-signal adjustment can still affect precision.
\end{remark}

The relevant broader class fixes the supervised influence function $d_k$. For centered square-integrable functions $h_k(\bm{X}_k)$ define
\begin{equation}\label{eq:Vd}
V_d(h)=\sum_k\{\pi_k^{-1}\Var(d_k-h_k)+\lambda_k^{-1}\Var(h_k)\}.
\end{equation}
Conditional centering and completion of squares yield
\begin{align*}
V_d^*&=\sum_k\left\{\pi_k^{-1}\E\Var(d_k\mid \bm{X}_k)+(\pi_k+\lambda_k)^{-1}\Var(b_k)\right\},\\
V_d(h)-V_d^*&=\sum_k(\pi_k^{-1}+\lambda_k^{-1})\Var(h_k-c_kb_k).
\end{align*}
Thus $c_kb_k$ is optimal within the fixed-$d$ transfer class, and $c_kp_k$ is optimal in the working linear space. The residual-and-transfer principle and its finite-resource variance reduction are instances of general semi-supervised projection theory \citep{xu2025unified}; the present construction supplies the specific censored supervised score and its feasible estimation. Appendix~\ref{s:generalssl} establishes first-order equivalence to an empirically centered projection update in the two-population experiment.

\subsection{Semiparametric efficiency}\label{sec:canonical}
Write $S_k(t\mid \bm{x})=\Prb(T_k\ge t\mid \bm{X}_k=\bm{x})$ and
\begin{equation}\label{eq:eD}
\begin{aligned}
a_{Xk}(t,\bm{x})&=\E\{r_k(T_k)\mid T_k\ge t,\bm{X}_k=\bm{x}\},\\
e_k&=g_k+\int_0^{b_0}\frac{a_{Xk}(t,\bm{X}_k)}{G_k(t)}dM_k^C(t),\\
D_k&=d_k-e_k.
\end{aligned}
\end{equation}
Conditional tails on zero-risk sets may be assigned arbitrary values, since their risk-weighted contributions vanish. The marginal tail in $d_k$ and the conditional tail in $e_k$ are different objects. The efficiency model retains $C_k\perp(T_k,\bm{X}_k)$ from Assumption~(A2); its censoring-score directions depend on time alone. Covariate-dependent censoring would change this tangent space and its canonical gradient.

\begin{theorem}\label{thm:canonical}
Suppose that Assumptions~(A1)--(A4) hold. For each $k=1,2$, the quantities in \eqref{eq:eD} have the following properties.

(i) $e_k$ is the supervised observed-data canonical gradient for the group-$k$ contribution. It satisfies $\E(e_k\mid \bm{X}_k)=b_k$, while $D_k$ is orthogonal to the observed tangent space, to $e_k$, and to square-integrable functions of $\bm{X}_k$.

(ii) In the four-pool experiment, the efficient influence functions in sample-mean form are $(e_k-c_kb_k,c_kb_k)$ for the labeled and auxiliary pools. The efficiency bound under $\sqrt n$ scaling is
\begin{equation*}
V_{\mathrm{eff}}=\sum_k\left\{\pi_k^{-1}\E\Var(e_k\mid \bm{X}_k)+(\pi_k+\lambda_k)^{-1}\Var(b_k)\right\}.
\end{equation*}

(iii) For every centered square-integrable transfer $h_k(\bm{X}_k)$ in the class \eqref{eq:Vd}, the excess variance over the efficiency bound separates into supervised-score and transfer losses:
\begin{equation*}
V_d(h)-V_{\mathrm{eff}}
=\sum_k\left\{\pi_k^{-1}\Var(D_k)+(\pi_k^{-1}+\lambda_k^{-1})\Var(h_k-c_kb_k)\right\}.
\end{equation*}
For the proposed linear transfer, the second term is the loss from approximating the conditional comparison signal in the working feature space:
\begin{equation*}
V_{\mathrm{SS}}-V_{\mathrm{eff}}
=\sum_k\frac{\Var(D_k)}{\pi_k}
+\sum_k(\pi_k^{-1}+\lambda_k^{-1})c_k^2\Var(b_k-p_k).
\end{equation*}
The proposed estimator attains $V_{\mathrm{eff}}$ if and only if $D_k=0$ and $p_k=b_k$ almost surely for both populations.
\end{theorem}

The proof is given in Appendix~\ref{s:canonical}, using the observed-data projection approach to censoring efficiency \citep{tsiatis2006semiparametric}. That section also gives the Riesz normalization corresponding to the sample-mean coordinates in the theorem.
\begin{remark}\label{rem:efficiencyloss}
The two terms in Theorem~\ref{thm:canonical} have distinct interpretations: $\Var(D_k)/\pi_k$ measures the gap between the supervised score and the canonical gradient, while the second term measures approximation of $b_k$ by the working features. Example~\ref{rem:counterexample} sets the latter loss to zero and isolates the supervised-score component.
\end{remark}

\begin{example}\label{rem:counterexample}
Take independent $X_1,X_2\sim\operatorname{Bernoulli}(1/2)$, with $T_1\mid X_1=0\sim\operatorname{Unif}[2,3]$, $T_1\mid X_1=1\sim\operatorname{Unif}[4,5]$, and $T_2\sim\operatorname{Unif}[2,5]$ independent of $(X_1,X_2)$. Let $C_1,C_2$ be independent and identically distributed, independent of all event times and covariates, with hazard $6\log(2)t(1-t)$ on $[0,1]$ and zero thereafter, leaving mass $1/2$ at infinity. The marginal comparison probability is $\theta=1/2$. The conditional comparison signals are $b_1(X_1)=(1-2X_1)/3$ and $b_2(X_2)=0$, so the linear working features $u_k=X_k$ give the exact projections $p_k=b_k$. With $\pi_1=\pi_2=1/2$ and $\lambda_1=\lambda_2=5/2$, the asymptotic variances are
\[
V_{\mathrm{sup}}=19/27,\qquad V_{\mathrm{SS}}=V_d^*=14/27,\qquad V_{\mathrm{eff}}=8/27.
\]
The working-projection loss is zero, while $V_{\mathrm{SS}}-V_{\mathrm{eff}}=2/9$ remains as the censoring-information component.
\end{example}

The full-lifetime results allow event-time atoms when effective censoring remains continuous, with strict functions $r_1(t)=\Prb(T_2>t)$ and $r_2(t)=F_1(t-)$. Event ties receive zero weight. The fixed-grid proof handles shared event--censoring atoms using the finite-product derivative in \eqref{s:atomicderivative}.

\subsection{Variance estimation and inference}\label{sec:inference}
The four independent pools contribute separate sample variances. Within a labeled pool, however, the pair projection and censoring contribution must be combined before taking a variance. The following construction estimates that complete score.

Let $\widehat Q_{ki}$ be a row or column mean of $\widehat q$, and set $\widehat g_{ki}=\widehat Q_{ki}-\widehat\theta_{\mathrm{sup}}$. Estimate $\psi_k$ by
\begin{equation*}
\widehat\psi_{ki}=\int\widehat\eta_k(t)d\widehat M^C_{ki}(t),\qquad
\widehat\eta_k(t)=\frac{P_n\{\widehat q\ind(Y_k\ge t)\}}{n_k^{-1}\sum_i\ind(Y_{ki}\ge t)}.
\end{equation*}
Integrals run over observed censoring jumps; under the full model these lie in $[0,b_0]$. The empirical compensator uses the reverse Nelson--Aalen increments. Define
\begin{equation*}
\begin{aligned}
\bar{\bm{\beta}}_k&=\sum_{ab}w_{ab}\widehat{\bm{\beta}}_{k,ab},\\
\widehat h_k(\bm{x})&=c_{k,n}\bar{\bm{\beta}}_k^\top\{\bm{u}_k(\bm{x})-\bar{\bm{u}}_k^U\},\\
\widehat\phi^L_{ki}&=\widehat g_{ki}+\widehat\psi_{ki}-\widehat h_k(\bm{X}_{ki}),\\
\widehat\phi^U_{kj}&=\widehat h_k(\bm{X}^U_{kj}).
\end{aligned}
\end{equation*}
Each pool variance is invariant to this constant centering by $\bar{\bm{u}}_k^U$. Let $s_m^2(v)=(m-1)^{-1}\sum_{i=1}^m(v_i-\bar v)^2$. The reported squared standard error is
\begin{equation}\label{eq:SE}
\widehat\se_{\mathrm{SS}}^2=\sum_k\left\{\frac{s_{n_k}^2(\widehat\phi^L_k)}{n_k}+\frac{s_{M_k}^2(\widehat\phi^U_k)}{M_k}\right\}.
\end{equation}
\begin{theorem}\label{thm:inference}
Suppose that Assumptions~(A1)--(A5) hold. Let $\widehat\theta_{\mathrm{SS}}$ be defined by \eqref{eq:actualc}, and let $\widehat\se_{\mathrm{SS}}^2$ be the variance estimator in \eqref{eq:SE}. Then, as $n\to\infty$,
\[
n\widehat\se_{\mathrm{SS}}^2\xrightarrow{p}V_{\mathrm{SS}},
\qquad
\frac{\widehat\theta_{\mathrm{SS}}-\theta}{\widehat\se_{\mathrm{SS}}}
\rightsquigarrow N(0,1).
\]
Consequently, for each fixed $\alpha\in(0,1)$,
\[
\Prb\!\left\{
\theta\in\left[
\widehat\theta_{\mathrm{SS}}-z_{1-\alpha/2}\widehat\se_{\mathrm{SS}},\,
\widehat\theta_{\mathrm{SS}}+z_{1-\alpha/2}\widehat\se_{\mathrm{SS}}
\right]\right\}\longrightarrow 1-\alpha,
\]
where $z_{1-\alpha/2}$ is the $(1-\alpha/2)$ quantile of the standard normal distribution.
\end{theorem}
The proof is given in Appendix~\ref{s:variance} and uses empirical $L_2$ convergence of the complete influence-function estimates.
\begin{remark}\label{rem:completeinference}
The Gaussian multiplier in Appendix~\ref{s:variance} perturbs these estimated influence-function scores.
\end{remark}

When clinical follow-up is limited to a specified horizon, the following capped comparison supplies a target supported by that observation window.

\section{Fixed-horizon extension}\label{sec:restricted}
For a fixed integer horizon $\tau$, define recorded-day event times capped at the horizon, $S_k=T_k\wedge\tau$, and
\begin{equation}\label{eq:restricted}
\theta_\tau=\E\{H(S_1,S_2)\},\qquad H(s,t)=\ind(s<t)+\tfrac12\ind(s=t).
\end{equation}
The relation to the full strict target is
\begin{align*}
\theta_\tau-\theta={}&\tfrac12\Prb(T_1=T_2<\tau)
+\tfrac12\Prb(T_1\ge\tau,T_2\ge\tau)\\[-2pt]
&-\Prb(T_1<T_2,T_1\ge\tau,T_2\ge\tau).
\end{align*}
This identity separates ties observed before the horizon from pairs whose ordering is lost beyond it. If event times are almost surely finite, the capped half-tie parameter tends to the full half-tie probability as $\tau$ increases; the limit agrees with the strict target when original ties have probability zero.

Higher values indicate earlier events in population 1. Ties include observed same-day events and the terminal tie when both capped times equal $\tau$. Each capped distribution has a terminal atom formed by event times at or beyond $\tau$. An unobserved ordering caused by early censoring is not assigned a half score.

Let $Z_k=Y_k\wedge\tau$ and
\begin{equation*}
A_k=\ind(\Delta_k=1,Y_k<\tau)+\ind(Y_k\ge\tau).
\end{equation*}
When $A_k=1$, $S_k=Z_k$ is known, including records originally censored exactly at $\tau$. Otherwise $Z_k=C_k<\tau$ and $S_k>Z_k$. For this subsection, impose Assumption \textbf{(H1)} for each population $k$:
\begin{enumerate}[leftmargin=*,label=(\alph*)]
\item The recorded-day grid and horizon $\tau$ are fixed and finite. Both capped event times and pre-horizon censoring times take values on this prescribed grid; the four-pool sampling, feature and fold assumptions are those in Assumptions~(A1), (A3) and (A4).
\item Censoring is jointly independent, $C_k\perp(S_k,\bm{X}_k)$, on the horizon, and $G_{k,-}(\tau)=\Prb(C_k\ge\tau)>0$.
\item The capped event time has positive terminal mass, $\Prb(S_k=\tau)>0$, and each active pre-horizon censoring hazard is less than one.
\item The limiting variance is positive when studentized inference is asserted.
\end{enumerate}
The positive terminal mass and active-risk conditions in (H1) support the finite-grid expansion. Events take priority when event and censoring times share a grid point, as specified by $\Delta=\ind(T\le C)$.

Under event priority, estimate the censoring hazard at $t<\tau$ by
\begin{equation*}
\widehat a_{kt}=\frac{dC_k(t)}{R_k(t)-dE_k(t)},\qquad
\widehat G_{k,-}(s)=\prod_{t<s}(1-\widehat a_{kt}),
\end{equation*}
where $R_k(t)=\sum_i\ind(Z_{ki}\ge t)$, $dE_k(t)=\sum_i\ind(Z_{ki}=t,A_{ki}=1)$ and $dC_k(t)=\sum_i\ind(Z_{ki}=t,A_{ki}=0)$. There is no censoring jump at $\tau$. Events at the same day are removed before estimating the censoring hazard. A zero-risk time with no censoring events contributes no jump.

The fitted pairwise kernel uses $A_1A_2H(Z_1,Z_2)/(\widehat G_{1,-}\widehat G_{2,-})$. Define the corresponding population $q_\tau$, the half-tie functions $r_{1\tau}(s)=\Prb(S_2>s)+\Prb(S_2=s)/2$ and $r_{2\tau}(s)=\Prb(S_1<s)+\Prb(S_1=s)/2$, and $g_{k\tau}=A_kr_{k\tau}(Z_k)/G_{k,-}(Z_k)-\theta_\tau$. For $t<\tau$, put
\begin{equation}\label{eq:atomicpsi}
\begin{aligned}
B_k(t)&=\ind(Z_k>t)+\ind(Z_k=t,A_k=0),\quad D_k^C(t)=\ind(Z_k=t,A_k=0),\\
\mathcal M_k(t)&=D_k^C(t)-a_{kt}B_k(t),\quad h_{kt}=\Prb(S_k>t)G_{k,-}(t),\\
A_{k+}(t)&=\E\{r_{k\tau}(S_k)\ind(S_k>t)\},\\
\psi_{k\tau}&=\sum_{t<\tau}\frac{A_{k+}(t)}{h_{kt}(1-a_{kt})}\mathcal M_k(t).
\end{aligned}
\end{equation}
Differentiating the atomic product yields the strict post-time tail and the factor $(1-a_{kt})^{-1}$ in the discrete nuisance score.

Let $d_{k\tau}=g_{k\tau}+\psi_{k\tau}$, $b_{k\tau}(\bm{X}_k)=\E\{r_{k\tau}(S_k)\mid \bm{X}_k\}-\theta_\tau$, and let $p_{k\tau}$ be the projection of $b_{k\tau}$ onto the centered feature space in \eqref{eq:signals}. Write $\widehat\theta_\tau(\bm{s}_n)$ for \eqref{eq:F1} computed from the fixed-horizon pairwise kernel, and let $\widehat\theta_{\mathrm{SS},\tau}=\widehat\theta_\tau(c_{1,n},c_{2,n})$. For the efficiency comparison, define
\[
\begin{aligned}
a_{Xk,\tau}(t,\bm{x})&=\E\{r_{k\tau}(S_k)\mid S_k>t,\bm{X}_k=\bm{x}\},\\
e_{k\tau}&=g_{k\tau}+\sum_{t<\tau}\frac{a_{Xk,\tau}(t,\bm{X}_k)}{G_{k,-}(t)(1-a_{kt})}\mathcal M_k(t),\qquad
D_{k\tau}=d_{k\tau}-e_{k\tau}.
\end{aligned}
\]
Conditional expectations on zero-risk sets may be assigned arbitrary values.

\begin{theorem}\label{thm:restricted}
Suppose that Assumption~(H1) holds. Then the following statements hold.

(i) The restricted score identifies the capped comparison:
$\E(q_\tau\mid S_1,S_2,\bm{X}_1,\bm{X}_2)=H(S_1,S_2)$ and $\E q_\tau=\theta_\tau$.

(ii) For bounded deterministic $\bm{s}_n\to\bm{s}$, the expansion in \eqref{eq:alr} holds for $\widehat\theta_\tau(\bm{s}_n)-\theta_\tau$ with $d_k,p_k$ replaced by $d_{k\tau},p_{k\tau}$. Its limiting $\sqrt n$ variance is
\[
V_\tau(\bm{s})=\sum_k\left\{\pi_k^{-1}\Var(d_{k\tau}-s_kp_{k\tau})+\lambda_k^{-1}s_k^2\Var(p_{k\tau})\right\}.
\]
It is minimized at $s_k=c_k$ as in Corollary~\ref{cor:ratio}. Let $\widehat\se_{\mathrm{SS},\tau}^2$ denote \eqref{eq:SE} computed using the atomic complete scores. If $V_\tau(\bm{c})>0$, then
\[
n\widehat\se_{\mathrm{SS},\tau}^2\xrightarrow{p}V_\tau(\bm{c}),\qquad
\frac{\widehat\theta_{\mathrm{SS},\tau}-\theta_\tau}{\widehat\se_{\mathrm{SS},\tau}}\rightsquigarrow N(0,1).
\]

(iii) In the capped observed-data experiment, $e_{k\tau}$ is the supervised canonical gradient. The four-pool efficiency bound and the gap for the proposed estimator are
\[
\begin{aligned}
V_{\mathrm{eff},\tau}&=\sum_k\left\{\pi_k^{-1}\E\Var(e_{k\tau}\mid \bm{X}_k)+(\pi_k+\lambda_k)^{-1}\Var(b_{k\tau})\right\},\\
V_\tau(\bm{c})-V_{\mathrm{eff},\tau}&=\sum_k\frac{\Var(D_{k\tau})}{\pi_k}
+\sum_k(\pi_k^{-1}+\lambda_k^{-1})c_k^2\Var(b_{k\tau}-p_{k\tau}).
\end{aligned}
\]
\end{theorem}
Appendix~\ref{s:restricted} gives the proof for the fixed-grid experiment.
\begin{remark}\label{rem:horizon}
Theorem~\ref{thm:restricted} supplies inference for the comparison available at a prespecified follow-up horizon, as used in the Rotterdam application. The recorded-day grid, terminal tie and event-priority rule determine how boundary observations enter the score. Appendix~\ref{s:atoms} gives completion bounds for the remaining unobserved orderings.
\end{remark}

\section{Simulation studies}\label{sec:numerical}
We assess when finite covariate-only samples improve a censored two-sample comparison. Full-lifetime designs vary signal and resource size; fixed-horizon designs add terminal ties, alternative working features and a common submodel for comparison with trial augmentation.

The methods share labeled observations and censoring fits, and the adjusted estimators also share regressions and folds. In balanced designs, $n_0$ and $M_0$ denote per-population sizes; asymmetric designs list all four pool sizes. Resource levels reuse labeled samples and nested prefixes of the covariate-only pools.

Table rows denote Supervised, Unshrunk and Weighted by $\widehat{\theta}_{\mathrm{sup}}$, $\widehat{\theta}_{\mathrm{un}}$ and $\widehat{\theta}_{\mathrm{SS}}$; fixed-horizon rows add the subscript $\tau$. Let $s_A^2$ be the Monte Carlo variance centered at method $A$'s own mean. The relative variance reduction $G_{A:B}=100\{1-s_A^2/s_B^2\}$ compares $A$ with $B$, with positive values favoring $A$; parentheses give paired MCSE in percentage points. The asymptotic variance reduction (AVR) is the first-order working-projection prediction for Weighted relative to Supervised. The RMSE gain is $\mathrm{RG}=100\{1-\mathrm{RMSE}_{\mathrm{SS}}/\mathrm{RMSE}_{\mathrm{sup}}\}$, which also reflects bias.

\subsection{Full-lifetime comparisons}
The full-lifetime laws vary the marginal comparison probability and the working-projection signal. Event times are generated as
\[
T_1=2+\exp\{\mu_1+aX_1+\sqrt{1-a^2}E_1\},\qquad T_2=2+\exp(E_2),
\]
Here $X_k,E_k\stackrel{\mathrm{iid}}{\sim}N(0,1)$, the working feature is $u_k=X_k$, and $\theta=\Phi(-\mu_1/\sqrt2)$. Censoring is independent, with $G(5)=0.60$ except in the heavier-censoring Scenario 8, where $G(5)=0.35$; Appendix~\ref{app:additional} specifies its distribution. Scenarios 1--3 have $\theta=0.50$ and $a=0,0.50,0.85$, corresponding to zero, moderate and strong working-projection signal. Each uses $n_0=200,800$ and $M_0/n_0=1,2,5$. Scenarios 4--8 vary the comparison probability or censoring and are specified in Appendix~\ref{app:additional}.

\begingroup\TableFont
\renewcommand{\arraystretch}{0.88}
\setlength{\LTpre}{10pt}\setlength{\LTpost}{0pt}
\setlength{\tabcolsep}{0pt}\sbox{\UniformMeasure}{\begin{tabular}{@{}lcccccccccr@{}}Scenario & $n_0$ & $M_0$ & Method & Bias & \shortstack{Emp\\SD} & \shortstack{Mean\\SE} & RMSE & CP & \shortstack{$G_{A:\mathrm{sup}}$\\(MCSE)} & AVR (\%) \\
1  &  \tablenum[table-format=3.0]{200}  &  --  &  $\widehat{\theta}_{\mathrm{sup}}$  &  \tablenum[table-format=-1.2]{0.81}  &  \tablenum[table-format=2.2]{32.79}  &  \tablenum[table-format=2.2]{33.04}  &  \tablenum[table-format=2.2]{32.80}  &  \tablenum[table-format=2.2]{95.04}  &  --  &  -- \\
1  &  \tablenum[table-format=3.0]{200}  &  \tablenum[table-format=4.0]{200}  &  $\widehat{\theta}_{\mathrm{un}}$  &  \tablenum[table-format=-1.2]{0.87}  &  \tablenum[table-format=2.2]{33.36}  &  \tablenum[table-format=2.2]{33.23}  &  \tablenum[table-format=2.2]{33.37}  &  \tablenum[table-format=2.2]{94.79}  &  \tablenum[table-format=-2.2]{-3.51}\,(\tablenum[table-format=1.2]{0.41})  &  -- \\
1  &  \tablenum[table-format=3.0]{200}  &  \tablenum[table-format=4.0]{200}  &  $\widehat{\theta}_{\mathrm{SS}}$  &  \tablenum[table-format=-1.2]{0.84}  &  \tablenum[table-format=2.2]{32.95}  &  \tablenum[table-format=2.2]{33.05}  &  \tablenum[table-format=2.2]{32.96}  &  \tablenum[table-format=2.2]{94.92}  &  \tablenum[table-format=-2.2]{-0.96}\,(\tablenum[table-format=1.2]{0.20})  &  \num{0.00} \\
1  &  \tablenum[table-format=3.0]{200}  &  \tablenum[table-format=4.0]{400}  &  $\widehat{\theta}_{\mathrm{un}}$  &  \tablenum[table-format=-1.2]{0.85}  &  \tablenum[table-format=2.2]{33.25}  &  \tablenum[table-format=2.2]{33.14}  &  \tablenum[table-format=2.2]{33.26}  &  \tablenum[table-format=2.2]{94.79}  &  \tablenum[table-format=-2.2]{-2.83}\,(\tablenum[table-format=1.2]{0.37})  &  -- \\
1  &  \tablenum[table-format=3.0]{200}  &  \tablenum[table-format=4.0]{400}  &  $\widehat{\theta}_{\mathrm{SS}}$  &  \tablenum[table-format=-1.2]{0.83}  &  \tablenum[table-format=2.2]{33.00}  &  \tablenum[table-format=2.2]{33.05}  &  \tablenum[table-format=2.2]{33.01}  &  \tablenum[table-format=2.2]{94.83}  &  \tablenum[table-format=-2.2]{-1.28}\,(\tablenum[table-format=1.2]{0.25})  &  \num{0.00} \\
1  &  \tablenum[table-format=3.0]{200}  &  \tablenum[table-format=4.0]{1000}  &  $\widehat{\theta}_{\mathrm{un}}$  &  \tablenum[table-format=-1.2]{0.80}  &  \tablenum[table-format=2.2]{33.20}  &  \tablenum[table-format=2.2]{33.08}  &  \tablenum[table-format=2.2]{33.21}  &  \tablenum[table-format=2.2]{94.77}  &  \tablenum[table-format=-2.2]{-2.50}\,(\tablenum[table-format=1.2]{0.35})  &  -- \\
1  &  \tablenum[table-format=3.0]{200}  &  \tablenum[table-format=4.0]{1000}  &  $\widehat{\theta}_{\mathrm{SS}}$  &  \tablenum[table-format=-1.2]{0.80}  &  \tablenum[table-format=2.2]{33.08}  &  \tablenum[table-format=2.2]{33.05}  &  \tablenum[table-format=2.2]{33.09}  &  \tablenum[table-format=2.2]{94.81}  &  \tablenum[table-format=-2.2]{-1.75}\,(\tablenum[table-format=1.2]{0.29})  &  \num{0.00} \\
\addlinespace[1pt]
1  &  \tablenum[table-format=3.0]{800}  &  --  &  $\widehat{\theta}_{\mathrm{sup}}$  &  \tablenum[table-format=-1.2]{-0.29}  &  \tablenum[table-format=2.2]{16.67}  &  \tablenum[table-format=2.2]{16.47}  &  \tablenum[table-format=2.2]{16.67}  &  \tablenum[table-format=2.2]{94.92}  &  --  &  -- \\
1  &  \tablenum[table-format=3.0]{800}  &  \tablenum[table-format=4.0]{800}  &  $\widehat{\theta}_{\mathrm{un}}$  &  \tablenum[table-format=-1.2]{-0.32}  &  \tablenum[table-format=2.2]{16.74}  &  \tablenum[table-format=2.2]{16.49}  &  \tablenum[table-format=2.2]{16.74}  &  \tablenum[table-format=2.2]{94.73}  &  \tablenum[table-format=-2.2]{-0.84}\,(\tablenum[table-format=1.2]{0.19})  &  -- \\
1  &  \tablenum[table-format=3.0]{800}  &  \tablenum[table-format=4.0]{800}  &  $\widehat{\theta}_{\mathrm{SS}}$  &  \tablenum[table-format=-1.2]{-0.31}  &  \tablenum[table-format=2.2]{16.69}  &  \tablenum[table-format=2.2]{16.47}  &  \tablenum[table-format=2.2]{16.69}  &  \tablenum[table-format=2.2]{94.89}  &  \tablenum[table-format=-2.2]{-0.23}\,(\tablenum[table-format=1.2]{0.09})  &  \num{0.00} \\
1  &  \tablenum[table-format=3.0]{800}  &  \tablenum[table-format=4.0]{1600}  &  $\widehat{\theta}_{\mathrm{un}}$  &  \tablenum[table-format=-1.2]{-0.32}  &  \tablenum[table-format=2.2]{16.73}  &  \tablenum[table-format=2.2]{16.48}  &  \tablenum[table-format=2.2]{16.73}  &  \tablenum[table-format=2.2]{94.76}  &  \tablenum[table-format=-2.2]{-0.70}\,(\tablenum[table-format=1.2]{0.17})  &  -- \\
1  &  \tablenum[table-format=3.0]{800}  &  \tablenum[table-format=4.0]{1600}  &  $\widehat{\theta}_{\mathrm{SS}}$  &  \tablenum[table-format=-1.2]{-0.31}  &  \tablenum[table-format=2.2]{16.70}  &  \tablenum[table-format=2.2]{16.47}  &  \tablenum[table-format=2.2]{16.70}  &  \tablenum[table-format=2.2]{94.84}  &  \tablenum[table-format=-2.2]{-0.33}\,(\tablenum[table-format=1.2]{0.12})  &  \num{0.00} \\
1  &  \tablenum[table-format=3.0]{800}  &  \tablenum[table-format=4.0]{4000}  &  $\widehat{\theta}_{\mathrm{un}}$  &  \tablenum[table-format=-1.2]{-0.32}  &  \tablenum[table-format=2.2]{16.71}  &  \tablenum[table-format=2.2]{16.48}  &  \tablenum[table-format=2.2]{16.71}  &  \tablenum[table-format=2.2]{94.79}  &  \tablenum[table-format=-2.2]{-0.51}\,(\tablenum[table-format=1.2]{0.16})  &  -- \\
1  &  \tablenum[table-format=3.0]{800}  &  \tablenum[table-format=4.0]{4000}  &  $\widehat{\theta}_{\mathrm{SS}}$  &  \tablenum[table-format=-1.2]{-0.31}  &  \tablenum[table-format=2.2]{16.70}  &  \tablenum[table-format=2.2]{16.47}  &  \tablenum[table-format=2.2]{16.70}  &  \tablenum[table-format=2.2]{94.85}  &  \tablenum[table-format=-2.2]{-0.35}\,(\tablenum[table-format=1.2]{0.14})  &  \num{0.00} \\
\addlinespace[1pt]
2  &  \tablenum[table-format=3.0]{200}  &  --  &  $\widehat{\theta}_{\mathrm{sup}}$  &  \tablenum[table-format=-1.2]{0.17}  &  \tablenum[table-format=2.2]{32.93}  &  \tablenum[table-format=2.2]{33.04}  &  \tablenum[table-format=2.2]{32.93}  &  \tablenum[table-format=2.2]{94.76}  &  --  &  -- \\
2  &  \tablenum[table-format=3.0]{200}  &  \tablenum[table-format=4.0]{200}  &  $\widehat{\theta}_{\mathrm{un}}$  &  \tablenum[table-format=-1.2]{0.20}  &  \tablenum[table-format=2.2]{33.68}  &  \tablenum[table-format=2.2]{33.23}  &  \tablenum[table-format=2.2]{33.68}  &  \tablenum[table-format=2.2]{94.58}  &  \tablenum[table-format=-2.2]{-4.57}\,(\tablenum[table-format=1.2]{1.01})  &  -- \\
2  &  \tablenum[table-format=3.0]{200}  &  \tablenum[table-format=4.0]{200}  &  $\widehat{\theta}_{\mathrm{SS}}$  &  \tablenum[table-format=-1.2]{0.19}  &  \tablenum[table-format=2.2]{32.43}  &  \tablenum[table-format=2.2]{32.30}  &  \tablenum[table-format=2.2]{32.43}  &  \tablenum[table-format=2.2]{94.70}  &  \tablenum[table-format=-2.2]{3.04}\,(\tablenum[table-format=1.2]{0.48})  &  \num{4.59} \\
2  &  \tablenum[table-format=3.0]{200}  &  \tablenum[table-format=4.0]{400}  &  $\widehat{\theta}_{\mathrm{un}}$  &  \tablenum[table-format=-1.2]{0.15}  &  \tablenum[table-format=2.2]{32.78}  &  \tablenum[table-format=2.2]{32.39}  &  \tablenum[table-format=2.2]{32.78}  &  \tablenum[table-format=2.2]{94.56}  &  \tablenum[table-format=-2.2]{0.92}\,(\tablenum[table-format=1.2]{0.86})  &  -- \\
2  &  \tablenum[table-format=3.0]{200}  &  \tablenum[table-format=4.0]{400}  &  $\widehat{\theta}_{\mathrm{SS}}$  &  \tablenum[table-format=-1.2]{0.16}  &  \tablenum[table-format=2.2]{32.23}  &  \tablenum[table-format=2.2]{32.04}  &  \tablenum[table-format=2.2]{32.23}  &  \tablenum[table-format=2.2]{94.54}  &  \tablenum[table-format=-2.2]{4.21}\,(\tablenum[table-format=1.2]{0.56})  &  \num{6.12} \\
2  &  \tablenum[table-format=3.0]{200}  &  \tablenum[table-format=4.0]{1000}  &  $\widehat{\theta}_{\mathrm{un}}$  &  \tablenum[table-format=-1.2]{0.17}  &  \tablenum[table-format=2.2]{32.16}  &  \tablenum[table-format=2.2]{31.87}  &  \tablenum[table-format=2.2]{32.16}  &  \tablenum[table-format=2.2]{94.64}  &  \tablenum[table-format=-2.2]{4.65}\,(\tablenum[table-format=1.2]{0.74})  &  -- \\
2  &  \tablenum[table-format=3.0]{200}  &  \tablenum[table-format=4.0]{1000}  &  $\widehat{\theta}_{\mathrm{SS}}$  &  \tablenum[table-format=-1.2]{0.17}  &  \tablenum[table-format=2.2]{31.98}  &  \tablenum[table-format=2.2]{31.79}  &  \tablenum[table-format=2.2]{31.98}  &  \tablenum[table-format=2.2]{94.74}  &  \tablenum[table-format=-2.2]{5.69}\,(\tablenum[table-format=1.2]{0.62})  &  \num{7.65} \\
\addlinespace[1pt]
2  &  \tablenum[table-format=3.0]{800}  &  --  &  $\widehat{\theta}_{\mathrm{sup}}$  &  \tablenum[table-format=-1.2]{0.04}  &  \tablenum[table-format=2.2]{16.52}  &  \tablenum[table-format=2.2]{16.47}  &  \tablenum[table-format=2.2]{16.52}  &  \tablenum[table-format=2.2]{95.05}  &  --  &  -- \\
2  &  \tablenum[table-format=3.0]{800}  &  \tablenum[table-format=4.0]{800}  &  $\widehat{\theta}_{\mathrm{un}}$  &  \tablenum[table-format=-1.2]{0.16}  &  \tablenum[table-format=2.2]{16.60}  &  \tablenum[table-format=2.2]{16.49}  &  \tablenum[table-format=2.2]{16.60}  &  \tablenum[table-format=2.2]{94.94}  &  \tablenum[table-format=-2.2]{-0.97}\,(\tablenum[table-format=1.2]{0.96})  &  -- \\
2  &  \tablenum[table-format=3.0]{800}  &  \tablenum[table-format=4.0]{800}  &  $\widehat{\theta}_{\mathrm{SS}}$  &  \tablenum[table-format=-1.2]{0.10}  &  \tablenum[table-format=2.2]{16.16}  &  \tablenum[table-format=2.2]{16.09}  &  \tablenum[table-format=2.2]{16.16}  &  \tablenum[table-format=2.2]{95.29}  &  \tablenum[table-format=-2.2]{4.34}\,(\tablenum[table-format=1.2]{0.47})  &  \num{4.59} \\
2  &  \tablenum[table-format=3.0]{800}  &  \tablenum[table-format=4.0]{1600}  &  $\widehat{\theta}_{\mathrm{un}}$  &  \tablenum[table-format=-1.2]{0.13}  &  \tablenum[table-format=2.2]{16.13}  &  \tablenum[table-format=2.2]{16.10}  &  \tablenum[table-format=2.2]{16.13}  &  \tablenum[table-format=2.2]{95.00}  &  \tablenum[table-format=-2.2]{4.72}\,(\tablenum[table-format=1.2]{0.79})  &  -- \\
2  &  \tablenum[table-format=3.0]{800}  &  \tablenum[table-format=4.0]{1600}  &  $\widehat{\theta}_{\mathrm{SS}}$  &  \tablenum[table-format=-1.2]{0.10}  &  \tablenum[table-format=2.2]{16.00}  &  \tablenum[table-format=2.2]{15.96}  &  \tablenum[table-format=2.2]{16.00}  &  \tablenum[table-format=2.2]{95.20}  &  \tablenum[table-format=-2.2]{6.28}\,(\tablenum[table-format=1.2]{0.52})  &  \num{6.12} \\
2  &  \tablenum[table-format=3.0]{800}  &  \tablenum[table-format=4.0]{4000}  &  $\widehat{\theta}_{\mathrm{un}}$  &  \tablenum[table-format=-1.2]{0.10}  &  \tablenum[table-format=2.2]{15.92}  &  \tablenum[table-format=2.2]{15.86}  &  \tablenum[table-format=2.2]{15.92}  &  \tablenum[table-format=2.2]{94.92}  &  \tablenum[table-format=-2.2]{7.20}\,(\tablenum[table-format=1.2]{0.70})  &  -- \\
2  &  \tablenum[table-format=3.0]{800}  &  \tablenum[table-format=4.0]{4000}  &  $\widehat{\theta}_{\mathrm{SS}}$  &  \tablenum[table-format=-1.2]{0.09}  &  \tablenum[table-format=2.2]{15.89}  &  \tablenum[table-format=2.2]{15.83}  &  \tablenum[table-format=2.2]{15.88}  &  \tablenum[table-format=2.2]{95.09}  &  \tablenum[table-format=-2.2]{7.59}\,(\tablenum[table-format=1.2]{0.58})  &  \num{7.65} \\
\addlinespace[1pt]
3  &  \tablenum[table-format=3.0]{200}  &  --  &  $\widehat{\theta}_{\mathrm{sup}}$  &  \tablenum[table-format=-1.2]{-0.41}  &  \tablenum[table-format=2.2]{33.22}  &  \tablenum[table-format=2.2]{33.03}  &  \tablenum[table-format=2.2]{33.22}  &  \tablenum[table-format=2.2]{95.04}  &  --  &  -- \\
3  &  \tablenum[table-format=3.0]{200}  &  \tablenum[table-format=4.0]{200}  &  $\widehat{\theta}_{\mathrm{un}}$  &  \tablenum[table-format=-1.2]{-0.43}  &  \tablenum[table-format=2.2]{33.70}  &  \tablenum[table-format=2.2]{33.24}  &  \tablenum[table-format=2.2]{33.70}  &  \tablenum[table-format=2.2]{94.73}  &  \tablenum[table-format=-2.2]{-2.93}\,(\tablenum[table-format=1.2]{1.58})  &  -- \\
3  &  \tablenum[table-format=3.0]{200}  &  \tablenum[table-format=4.0]{200}  &  $\widehat{\theta}_{\mathrm{SS}}$  &  \tablenum[table-format=-1.2]{-0.42}  &  \tablenum[table-format=2.2]{31.11}  &  \tablenum[table-format=2.2]{30.80}  &  \tablenum[table-format=2.2]{31.11}  &  \tablenum[table-format=2.2]{94.81}  &  \tablenum[table-format=-2.2]{12.31}\,(\tablenum[table-format=1.2]{0.73})  &  \num{13.27} \\
3  &  \tablenum[table-format=3.0]{200}  &  \tablenum[table-format=4.0]{400}  &  $\widehat{\theta}_{\mathrm{un}}$  &  \tablenum[table-format=-1.2]{-0.46}  &  \tablenum[table-format=2.2]{31.37}  &  \tablenum[table-format=2.2]{30.92}  &  \tablenum[table-format=2.2]{31.37}  &  \tablenum[table-format=2.2]{94.61}  &  \tablenum[table-format=-2.2]{10.84}\,(\tablenum[table-format=1.2]{1.23})  &  -- \\
3  &  \tablenum[table-format=3.0]{200}  &  \tablenum[table-format=4.0]{400}  &  $\widehat{\theta}_{\mathrm{SS}}$  &  \tablenum[table-format=-1.2]{-0.45}  &  \tablenum[table-format=2.2]{30.35}  &  \tablenum[table-format=2.2]{30.02}  &  \tablenum[table-format=2.2]{30.36}  &  \tablenum[table-format=2.2]{94.75}  &  \tablenum[table-format=-2.2]{16.50}\,(\tablenum[table-format=1.2]{0.79})  &  \num{17.69} \\
3  &  \tablenum[table-format=3.0]{200}  &  \tablenum[table-format=4.0]{1000}  &  $\widehat{\theta}_{\mathrm{un}}$  &  \tablenum[table-format=-1.2]{-0.42}  &  \tablenum[table-format=2.2]{29.93}  &  \tablenum[table-format=2.2]{29.42}  &  \tablenum[table-format=2.2]{29.93}  &  \tablenum[table-format=2.2]{94.24}  &  \tablenum[table-format=-2.2]{18.80}\,(\tablenum[table-format=1.2]{1.05})  &  -- \\
3  &  \tablenum[table-format=3.0]{200}  &  \tablenum[table-format=4.0]{1000}  &  $\widehat{\theta}_{\mathrm{SS}}$  &  \tablenum[table-format=-1.2]{-0.42}  &  \tablenum[table-format=2.2]{29.64}  &  \tablenum[table-format=2.2]{29.21}  &  \tablenum[table-format=2.2]{29.65}  &  \tablenum[table-format=2.2]{94.34}  &  \tablenum[table-format=-2.2]{20.36}\,(\tablenum[table-format=1.2]{0.86})  &  \num{22.11} \\
\addlinespace[1pt]
3  &  \tablenum[table-format=3.0]{800}  &  --  &  $\widehat{\theta}_{\mathrm{sup}}$  &  \tablenum[table-format=-1.2]{0.33}  &  \tablenum[table-format=2.2]{16.56}  &  \tablenum[table-format=2.2]{16.47}  &  \tablenum[table-format=2.2]{16.56}  &  \tablenum[table-format=2.2]{95.05}  &  --  &  -- \\
3  &  \tablenum[table-format=3.0]{800}  &  \tablenum[table-format=4.0]{800}  &  $\widehat{\theta}_{\mathrm{un}}$  &  \tablenum[table-format=-1.2]{0.11}  &  \tablenum[table-format=2.2]{16.44}  &  \tablenum[table-format=2.2]{16.50}  &  \tablenum[table-format=2.2]{16.44}  &  \tablenum[table-format=2.2]{94.91}  &  \tablenum[table-format=-2.2]{1.36}\,(\tablenum[table-format=1.2]{1.50})  &  -- \\
3  &  \tablenum[table-format=3.0]{800}  &  \tablenum[table-format=4.0]{800}  &  $\widehat{\theta}_{\mathrm{SS}}$  &  \tablenum[table-format=-1.2]{0.22}  &  \tablenum[table-format=2.2]{15.37}  &  \tablenum[table-format=2.2]{15.34}  &  \tablenum[table-format=2.2]{15.37}  &  \tablenum[table-format=2.2]{94.81}  &  \tablenum[table-format=-2.2]{13.83}\,(\tablenum[table-format=1.2]{0.70})  &  \num{13.27} \\
3  &  \tablenum[table-format=3.0]{800}  &  \tablenum[table-format=4.0]{1600}  &  $\widehat{\theta}_{\mathrm{un}}$  &  \tablenum[table-format=-1.2]{0.13}  &  \tablenum[table-format=2.2]{15.36}  &  \tablenum[table-format=2.2]{15.36}  &  \tablenum[table-format=2.2]{15.36}  &  \tablenum[table-format=2.2]{94.86}  &  \tablenum[table-format=-2.2]{13.96}\,(\tablenum[table-format=1.2]{1.19})  &  -- \\
3  &  \tablenum[table-format=3.0]{800}  &  \tablenum[table-format=4.0]{1600}  &  $\widehat{\theta}_{\mathrm{SS}}$  &  \tablenum[table-format=-1.2]{0.20}  &  \tablenum[table-format=2.2]{14.98}  &  \tablenum[table-format=2.2]{14.95}  &  \tablenum[table-format=2.2]{14.98}  &  \tablenum[table-format=2.2]{94.65}  &  \tablenum[table-format=-2.2]{18.16}\,(\tablenum[table-format=1.2]{0.77})  &  \num{17.69} \\
3  &  \tablenum[table-format=3.0]{800}  &  \tablenum[table-format=4.0]{4000}  &  $\widehat{\theta}_{\mathrm{un}}$  &  \tablenum[table-format=-1.2]{0.10}  &  \tablenum[table-format=2.2]{14.62}  &  \tablenum[table-format=2.2]{14.63}  &  \tablenum[table-format=2.2]{14.62}  &  \tablenum[table-format=2.2]{94.99}  &  \tablenum[table-format=-2.2]{22.03}\,(\tablenum[table-format=1.2]{0.97})  &  -- \\
3  &  \tablenum[table-format=3.0]{800}  &  \tablenum[table-format=4.0]{4000}  &  $\widehat{\theta}_{\mathrm{SS}}$  &  \tablenum[table-format=-1.2]{0.14}  &  \tablenum[table-format=2.2]{14.55}  &  \tablenum[table-format=2.2]{14.54}  &  \tablenum[table-format=2.2]{14.55}  &  \tablenum[table-format=2.2]{94.89}  &  \tablenum[table-format=-2.2]{22.76}\,(\tablenum[table-format=1.2]{0.81})  &  \num{22.11} \\
\addlinespace[1pt]
\end{tabular}}
\setlength{\UniformWidth}{\linewidth}\setlength{\tabcolsep}{\dimexpr(\linewidth-\wd\UniformMeasure)/20\relax}
\ifdim\tabcolsep<0pt\PackageWarning{UnifiedTables}{Negative column gap}\fi
\setlength{\LTleft}{0pt plus 1fill}\setlength{\LTright}{0pt plus 1fill}
\begin{longtable}{@{}lcccccccccr@{}}
\caption{Full-lifetime estimation with auxiliary samples.}\label{tab:fullMain}\\
\toprule Scenario & $n_0$ & $M_0$ & Method & Bias & \shortstack{Emp\\SD} & \shortstack{Mean\\SE} & RMSE & CP & \shortstack{$G_{A:\mathrm{sup}}$\\(MCSE)} & AVR (\%) \\\midrule\endfirsthead
\multicolumn{11}{l}{\textit{Table \thetable\ continued}}\\\toprule Scenario & $n_0$ & $M_0$ & Method & Bias & \shortstack{Emp\\SD} & \shortstack{Mean\\SE} & RMSE & CP & \shortstack{$G_{A:\mathrm{sup}}$\\(MCSE)} & AVR (\%) \\\midrule\endhead
\midrule\multicolumn{11}{r}{\textit{Continued on next page}}\\\endfoot
\endlastfoot

1  &  \tablenum[table-format=3.0]{200}  &  --  &  $\widehat{\theta}_{\mathrm{sup}}$  &  \tablenum[table-format=-1.2]{0.81}  &  \tablenum[table-format=2.2]{32.79}  &  \tablenum[table-format=2.2]{33.04}  &  \tablenum[table-format=2.2]{32.80}  &  \tablenum[table-format=2.2]{95.04}  &  --  &  -- \\
1  &  \tablenum[table-format=3.0]{200}  &  \tablenum[table-format=4.0]{200}  &  $\widehat{\theta}_{\mathrm{un}}$  &  \tablenum[table-format=-1.2]{0.87}  &  \tablenum[table-format=2.2]{33.36}  &  \tablenum[table-format=2.2]{33.23}  &  \tablenum[table-format=2.2]{33.37}  &  \tablenum[table-format=2.2]{94.79}  &  \tablenum[table-format=-2.2]{-3.51}\,(\tablenum[table-format=1.2]{0.41})  &  -- \\
1  &  \tablenum[table-format=3.0]{200}  &  \tablenum[table-format=4.0]{200}  &  $\widehat{\theta}_{\mathrm{SS}}$  &  \tablenum[table-format=-1.2]{0.84}  &  \tablenum[table-format=2.2]{32.95}  &  \tablenum[table-format=2.2]{33.05}  &  \tablenum[table-format=2.2]{32.96}  &  \tablenum[table-format=2.2]{94.92}  &  \tablenum[table-format=-2.2]{-0.96}\,(\tablenum[table-format=1.2]{0.20})  &  \num{0.00} \\
1  &  \tablenum[table-format=3.0]{200}  &  \tablenum[table-format=4.0]{400}  &  $\widehat{\theta}_{\mathrm{un}}$  &  \tablenum[table-format=-1.2]{0.85}  &  \tablenum[table-format=2.2]{33.25}  &  \tablenum[table-format=2.2]{33.14}  &  \tablenum[table-format=2.2]{33.26}  &  \tablenum[table-format=2.2]{94.79}  &  \tablenum[table-format=-2.2]{-2.83}\,(\tablenum[table-format=1.2]{0.37})  &  -- \\
1  &  \tablenum[table-format=3.0]{200}  &  \tablenum[table-format=4.0]{400}  &  $\widehat{\theta}_{\mathrm{SS}}$  &  \tablenum[table-format=-1.2]{0.83}  &  \tablenum[table-format=2.2]{33.00}  &  \tablenum[table-format=2.2]{33.05}  &  \tablenum[table-format=2.2]{33.01}  &  \tablenum[table-format=2.2]{94.83}  &  \tablenum[table-format=-2.2]{-1.28}\,(\tablenum[table-format=1.2]{0.25})  &  \num{0.00} \\
1  &  \tablenum[table-format=3.0]{200}  &  \tablenum[table-format=4.0]{1000}  &  $\widehat{\theta}_{\mathrm{un}}$  &  \tablenum[table-format=-1.2]{0.80}  &  \tablenum[table-format=2.2]{33.20}  &  \tablenum[table-format=2.2]{33.08}  &  \tablenum[table-format=2.2]{33.21}  &  \tablenum[table-format=2.2]{94.77}  &  \tablenum[table-format=-2.2]{-2.50}\,(\tablenum[table-format=1.2]{0.35})  &  -- \\
1  &  \tablenum[table-format=3.0]{200}  &  \tablenum[table-format=4.0]{1000}  &  $\widehat{\theta}_{\mathrm{SS}}$  &  \tablenum[table-format=-1.2]{0.80}  &  \tablenum[table-format=2.2]{33.08}  &  \tablenum[table-format=2.2]{33.05}  &  \tablenum[table-format=2.2]{33.09}  &  \tablenum[table-format=2.2]{94.81}  &  \tablenum[table-format=-2.2]{-1.75}\,(\tablenum[table-format=1.2]{0.29})  &  \num{0.00} \\
\addlinespace[1pt]
1  &  \tablenum[table-format=3.0]{800}  &  --  &  $\widehat{\theta}_{\mathrm{sup}}$  &  \tablenum[table-format=-1.2]{-0.29}  &  \tablenum[table-format=2.2]{16.67}  &  \tablenum[table-format=2.2]{16.47}  &  \tablenum[table-format=2.2]{16.67}  &  \tablenum[table-format=2.2]{94.92}  &  --  &  -- \\
1  &  \tablenum[table-format=3.0]{800}  &  \tablenum[table-format=4.0]{800}  &  $\widehat{\theta}_{\mathrm{un}}$  &  \tablenum[table-format=-1.2]{-0.32}  &  \tablenum[table-format=2.2]{16.74}  &  \tablenum[table-format=2.2]{16.49}  &  \tablenum[table-format=2.2]{16.74}  &  \tablenum[table-format=2.2]{94.73}  &  \tablenum[table-format=-2.2]{-0.84}\,(\tablenum[table-format=1.2]{0.19})  &  -- \\
1  &  \tablenum[table-format=3.0]{800}  &  \tablenum[table-format=4.0]{800}  &  $\widehat{\theta}_{\mathrm{SS}}$  &  \tablenum[table-format=-1.2]{-0.31}  &  \tablenum[table-format=2.2]{16.69}  &  \tablenum[table-format=2.2]{16.47}  &  \tablenum[table-format=2.2]{16.69}  &  \tablenum[table-format=2.2]{94.89}  &  \tablenum[table-format=-2.2]{-0.23}\,(\tablenum[table-format=1.2]{0.09})  &  \num{0.00} \\
1  &  \tablenum[table-format=3.0]{800}  &  \tablenum[table-format=4.0]{1600}  &  $\widehat{\theta}_{\mathrm{un}}$  &  \tablenum[table-format=-1.2]{-0.32}  &  \tablenum[table-format=2.2]{16.73}  &  \tablenum[table-format=2.2]{16.48}  &  \tablenum[table-format=2.2]{16.73}  &  \tablenum[table-format=2.2]{94.76}  &  \tablenum[table-format=-2.2]{-0.70}\,(\tablenum[table-format=1.2]{0.17})  &  -- \\
1  &  \tablenum[table-format=3.0]{800}  &  \tablenum[table-format=4.0]{1600}  &  $\widehat{\theta}_{\mathrm{SS}}$  &  \tablenum[table-format=-1.2]{-0.31}  &  \tablenum[table-format=2.2]{16.70}  &  \tablenum[table-format=2.2]{16.47}  &  \tablenum[table-format=2.2]{16.70}  &  \tablenum[table-format=2.2]{94.84}  &  \tablenum[table-format=-2.2]{-0.33}\,(\tablenum[table-format=1.2]{0.12})  &  \num{0.00} \\
1  &  \tablenum[table-format=3.0]{800}  &  \tablenum[table-format=4.0]{4000}  &  $\widehat{\theta}_{\mathrm{un}}$  &  \tablenum[table-format=-1.2]{-0.32}  &  \tablenum[table-format=2.2]{16.71}  &  \tablenum[table-format=2.2]{16.48}  &  \tablenum[table-format=2.2]{16.71}  &  \tablenum[table-format=2.2]{94.79}  &  \tablenum[table-format=-2.2]{-0.51}\,(\tablenum[table-format=1.2]{0.16})  &  -- \\
1  &  \tablenum[table-format=3.0]{800}  &  \tablenum[table-format=4.0]{4000}  &  $\widehat{\theta}_{\mathrm{SS}}$  &  \tablenum[table-format=-1.2]{-0.31}  &  \tablenum[table-format=2.2]{16.70}  &  \tablenum[table-format=2.2]{16.47}  &  \tablenum[table-format=2.2]{16.70}  &  \tablenum[table-format=2.2]{94.85}  &  \tablenum[table-format=-2.2]{-0.35}\,(\tablenum[table-format=1.2]{0.14})  &  \num{0.00} \\
\addlinespace[1pt]
2  &  \tablenum[table-format=3.0]{200}  &  --  &  $\widehat{\theta}_{\mathrm{sup}}$  &  \tablenum[table-format=-1.2]{0.17}  &  \tablenum[table-format=2.2]{32.93}  &  \tablenum[table-format=2.2]{33.04}  &  \tablenum[table-format=2.2]{32.93}  &  \tablenum[table-format=2.2]{94.76}  &  --  &  -- \\
2  &  \tablenum[table-format=3.0]{200}  &  \tablenum[table-format=4.0]{200}  &  $\widehat{\theta}_{\mathrm{un}}$  &  \tablenum[table-format=-1.2]{0.20}  &  \tablenum[table-format=2.2]{33.68}  &  \tablenum[table-format=2.2]{33.23}  &  \tablenum[table-format=2.2]{33.68}  &  \tablenum[table-format=2.2]{94.58}  &  \tablenum[table-format=-2.2]{-4.57}\,(\tablenum[table-format=1.2]{1.01})  &  -- \\
2  &  \tablenum[table-format=3.0]{200}  &  \tablenum[table-format=4.0]{200}  &  $\widehat{\theta}_{\mathrm{SS}}$  &  \tablenum[table-format=-1.2]{0.19}  &  \tablenum[table-format=2.2]{32.43}  &  \tablenum[table-format=2.2]{32.30}  &  \tablenum[table-format=2.2]{32.43}  &  \tablenum[table-format=2.2]{94.70}  &  \tablenum[table-format=-2.2]{3.04}\,(\tablenum[table-format=1.2]{0.48})  &  \num{4.59} \\
2  &  \tablenum[table-format=3.0]{200}  &  \tablenum[table-format=4.0]{400}  &  $\widehat{\theta}_{\mathrm{un}}$  &  \tablenum[table-format=-1.2]{0.15}  &  \tablenum[table-format=2.2]{32.78}  &  \tablenum[table-format=2.2]{32.39}  &  \tablenum[table-format=2.2]{32.78}  &  \tablenum[table-format=2.2]{94.56}  &  \tablenum[table-format=-2.2]{0.92}\,(\tablenum[table-format=1.2]{0.86})  &  -- \\
2  &  \tablenum[table-format=3.0]{200}  &  \tablenum[table-format=4.0]{400}  &  $\widehat{\theta}_{\mathrm{SS}}$  &  \tablenum[table-format=-1.2]{0.16}  &  \tablenum[table-format=2.2]{32.23}  &  \tablenum[table-format=2.2]{32.04}  &  \tablenum[table-format=2.2]{32.23}  &  \tablenum[table-format=2.2]{94.54}  &  \tablenum[table-format=-2.2]{4.21}\,(\tablenum[table-format=1.2]{0.56})  &  \num{6.12} \\
2  &  \tablenum[table-format=3.0]{200}  &  \tablenum[table-format=4.0]{1000}  &  $\widehat{\theta}_{\mathrm{un}}$  &  \tablenum[table-format=-1.2]{0.17}  &  \tablenum[table-format=2.2]{32.16}  &  \tablenum[table-format=2.2]{31.87}  &  \tablenum[table-format=2.2]{32.16}  &  \tablenum[table-format=2.2]{94.64}  &  \tablenum[table-format=-2.2]{4.65}\,(\tablenum[table-format=1.2]{0.74})  &  -- \\
2  &  \tablenum[table-format=3.0]{200}  &  \tablenum[table-format=4.0]{1000}  &  $\widehat{\theta}_{\mathrm{SS}}$  &  \tablenum[table-format=-1.2]{0.17}  &  \tablenum[table-format=2.2]{31.98}  &  \tablenum[table-format=2.2]{31.79}  &  \tablenum[table-format=2.2]{31.98}  &  \tablenum[table-format=2.2]{94.74}  &  \tablenum[table-format=-2.2]{5.69}\,(\tablenum[table-format=1.2]{0.62})  &  \num{7.65} \\
\addlinespace[1pt]
2  &  \tablenum[table-format=3.0]{800}  &  --  &  $\widehat{\theta}_{\mathrm{sup}}$  &  \tablenum[table-format=-1.2]{0.04}  &  \tablenum[table-format=2.2]{16.52}  &  \tablenum[table-format=2.2]{16.47}  &  \tablenum[table-format=2.2]{16.52}  &  \tablenum[table-format=2.2]{95.05}  &  --  &  -- \\
2  &  \tablenum[table-format=3.0]{800}  &  \tablenum[table-format=4.0]{800}  &  $\widehat{\theta}_{\mathrm{un}}$  &  \tablenum[table-format=-1.2]{0.16}  &  \tablenum[table-format=2.2]{16.60}  &  \tablenum[table-format=2.2]{16.49}  &  \tablenum[table-format=2.2]{16.60}  &  \tablenum[table-format=2.2]{94.94}  &  \tablenum[table-format=-2.2]{-0.97}\,(\tablenum[table-format=1.2]{0.96})  &  -- \\
2  &  \tablenum[table-format=3.0]{800}  &  \tablenum[table-format=4.0]{800}  &  $\widehat{\theta}_{\mathrm{SS}}$  &  \tablenum[table-format=-1.2]{0.10}  &  \tablenum[table-format=2.2]{16.16}  &  \tablenum[table-format=2.2]{16.09}  &  \tablenum[table-format=2.2]{16.16}  &  \tablenum[table-format=2.2]{95.29}  &  \tablenum[table-format=-2.2]{4.34}\,(\tablenum[table-format=1.2]{0.47})  &  \num{4.59} \\
2  &  \tablenum[table-format=3.0]{800}  &  \tablenum[table-format=4.0]{1600}  &  $\widehat{\theta}_{\mathrm{un}}$  &  \tablenum[table-format=-1.2]{0.13}  &  \tablenum[table-format=2.2]{16.13}  &  \tablenum[table-format=2.2]{16.10}  &  \tablenum[table-format=2.2]{16.13}  &  \tablenum[table-format=2.2]{95.00}  &  \tablenum[table-format=-2.2]{4.72}\,(\tablenum[table-format=1.2]{0.79})  &  -- \\
2  &  \tablenum[table-format=3.0]{800}  &  \tablenum[table-format=4.0]{1600}  &  $\widehat{\theta}_{\mathrm{SS}}$  &  \tablenum[table-format=-1.2]{0.10}  &  \tablenum[table-format=2.2]{16.00}  &  \tablenum[table-format=2.2]{15.96}  &  \tablenum[table-format=2.2]{16.00}  &  \tablenum[table-format=2.2]{95.20}  &  \tablenum[table-format=-2.2]{6.28}\,(\tablenum[table-format=1.2]{0.52})  &  \num{6.12} \\
2  &  \tablenum[table-format=3.0]{800}  &  \tablenum[table-format=4.0]{4000}  &  $\widehat{\theta}_{\mathrm{un}}$  &  \tablenum[table-format=-1.2]{0.10}  &  \tablenum[table-format=2.2]{15.92}  &  \tablenum[table-format=2.2]{15.86}  &  \tablenum[table-format=2.2]{15.92}  &  \tablenum[table-format=2.2]{94.92}  &  \tablenum[table-format=-2.2]{7.20}\,(\tablenum[table-format=1.2]{0.70})  &  -- \\
2  &  \tablenum[table-format=3.0]{800}  &  \tablenum[table-format=4.0]{4000}  &  $\widehat{\theta}_{\mathrm{SS}}$  &  \tablenum[table-format=-1.2]{0.09}  &  \tablenum[table-format=2.2]{15.89}  &  \tablenum[table-format=2.2]{15.83}  &  \tablenum[table-format=2.2]{15.88}  &  \tablenum[table-format=2.2]{95.09}  &  \tablenum[table-format=-2.2]{7.59}\,(\tablenum[table-format=1.2]{0.58})  &  \num{7.65} \\
\addlinespace[1pt]
3  &  \tablenum[table-format=3.0]{200}  &  --  &  $\widehat{\theta}_{\mathrm{sup}}$  &  \tablenum[table-format=-1.2]{-0.41}  &  \tablenum[table-format=2.2]{33.22}  &  \tablenum[table-format=2.2]{33.03}  &  \tablenum[table-format=2.2]{33.22}  &  \tablenum[table-format=2.2]{95.04}  &  --  &  -- \\
3  &  \tablenum[table-format=3.0]{200}  &  \tablenum[table-format=4.0]{200}  &  $\widehat{\theta}_{\mathrm{un}}$  &  \tablenum[table-format=-1.2]{-0.43}  &  \tablenum[table-format=2.2]{33.70}  &  \tablenum[table-format=2.2]{33.24}  &  \tablenum[table-format=2.2]{33.70}  &  \tablenum[table-format=2.2]{94.73}  &  \tablenum[table-format=-2.2]{-2.93}\,(\tablenum[table-format=1.2]{1.58})  &  -- \\
3  &  \tablenum[table-format=3.0]{200}  &  \tablenum[table-format=4.0]{200}  &  $\widehat{\theta}_{\mathrm{SS}}$  &  \tablenum[table-format=-1.2]{-0.42}  &  \tablenum[table-format=2.2]{31.11}  &  \tablenum[table-format=2.2]{30.80}  &  \tablenum[table-format=2.2]{31.11}  &  \tablenum[table-format=2.2]{94.81}  &  \tablenum[table-format=-2.2]{12.31}\,(\tablenum[table-format=1.2]{0.73})  &  \num{13.27} \\
3  &  \tablenum[table-format=3.0]{200}  &  \tablenum[table-format=4.0]{400}  &  $\widehat{\theta}_{\mathrm{un}}$  &  \tablenum[table-format=-1.2]{-0.46}  &  \tablenum[table-format=2.2]{31.37}  &  \tablenum[table-format=2.2]{30.92}  &  \tablenum[table-format=2.2]{31.37}  &  \tablenum[table-format=2.2]{94.61}  &  \tablenum[table-format=-2.2]{10.84}\,(\tablenum[table-format=1.2]{1.23})  &  -- \\
3  &  \tablenum[table-format=3.0]{200}  &  \tablenum[table-format=4.0]{400}  &  $\widehat{\theta}_{\mathrm{SS}}$  &  \tablenum[table-format=-1.2]{-0.45}  &  \tablenum[table-format=2.2]{30.35}  &  \tablenum[table-format=2.2]{30.02}  &  \tablenum[table-format=2.2]{30.36}  &  \tablenum[table-format=2.2]{94.75}  &  \tablenum[table-format=-2.2]{16.50}\,(\tablenum[table-format=1.2]{0.79})  &  \num{17.69} \\
3  &  \tablenum[table-format=3.0]{200}  &  \tablenum[table-format=4.0]{1000}  &  $\widehat{\theta}_{\mathrm{un}}$  &  \tablenum[table-format=-1.2]{-0.42}  &  \tablenum[table-format=2.2]{29.93}  &  \tablenum[table-format=2.2]{29.42}  &  \tablenum[table-format=2.2]{29.93}  &  \tablenum[table-format=2.2]{94.24}  &  \tablenum[table-format=-2.2]{18.80}\,(\tablenum[table-format=1.2]{1.05})  &  -- \\
3  &  \tablenum[table-format=3.0]{200}  &  \tablenum[table-format=4.0]{1000}  &  $\widehat{\theta}_{\mathrm{SS}}$  &  \tablenum[table-format=-1.2]{-0.42}  &  \tablenum[table-format=2.2]{29.64}  &  \tablenum[table-format=2.2]{29.21}  &  \tablenum[table-format=2.2]{29.65}  &  \tablenum[table-format=2.2]{94.34}  &  \tablenum[table-format=-2.2]{20.36}\,(\tablenum[table-format=1.2]{0.86})  &  \num{22.11} \\
\addlinespace[1pt]
3  &  \tablenum[table-format=3.0]{800}  &  --  &  $\widehat{\theta}_{\mathrm{sup}}$  &  \tablenum[table-format=-1.2]{0.33}  &  \tablenum[table-format=2.2]{16.56}  &  \tablenum[table-format=2.2]{16.47}  &  \tablenum[table-format=2.2]{16.56}  &  \tablenum[table-format=2.2]{95.05}  &  --  &  -- \\
3  &  \tablenum[table-format=3.0]{800}  &  \tablenum[table-format=4.0]{800}  &  $\widehat{\theta}_{\mathrm{un}}$  &  \tablenum[table-format=-1.2]{0.11}  &  \tablenum[table-format=2.2]{16.44}  &  \tablenum[table-format=2.2]{16.50}  &  \tablenum[table-format=2.2]{16.44}  &  \tablenum[table-format=2.2]{94.91}  &  \tablenum[table-format=-2.2]{1.36}\,(\tablenum[table-format=1.2]{1.50})  &  -- \\
3  &  \tablenum[table-format=3.0]{800}  &  \tablenum[table-format=4.0]{800}  &  $\widehat{\theta}_{\mathrm{SS}}$  &  \tablenum[table-format=-1.2]{0.22}  &  \tablenum[table-format=2.2]{15.37}  &  \tablenum[table-format=2.2]{15.34}  &  \tablenum[table-format=2.2]{15.37}  &  \tablenum[table-format=2.2]{94.81}  &  \tablenum[table-format=-2.2]{13.83}\,(\tablenum[table-format=1.2]{0.70})  &  \num{13.27} \\
3  &  \tablenum[table-format=3.0]{800}  &  \tablenum[table-format=4.0]{1600}  &  $\widehat{\theta}_{\mathrm{un}}$  &  \tablenum[table-format=-1.2]{0.13}  &  \tablenum[table-format=2.2]{15.36}  &  \tablenum[table-format=2.2]{15.36}  &  \tablenum[table-format=2.2]{15.36}  &  \tablenum[table-format=2.2]{94.86}  &  \tablenum[table-format=-2.2]{13.96}\,(\tablenum[table-format=1.2]{1.19})  &  -- \\
3  &  \tablenum[table-format=3.0]{800}  &  \tablenum[table-format=4.0]{1600}  &  $\widehat{\theta}_{\mathrm{SS}}$  &  \tablenum[table-format=-1.2]{0.20}  &  \tablenum[table-format=2.2]{14.98}  &  \tablenum[table-format=2.2]{14.95}  &  \tablenum[table-format=2.2]{14.98}  &  \tablenum[table-format=2.2]{94.65}  &  \tablenum[table-format=-2.2]{18.16}\,(\tablenum[table-format=1.2]{0.77})  &  \num{17.69} \\
3  &  \tablenum[table-format=3.0]{800}  &  \tablenum[table-format=4.0]{4000}  &  $\widehat{\theta}_{\mathrm{un}}$  &  \tablenum[table-format=-1.2]{0.10}  &  \tablenum[table-format=2.2]{14.62}  &  \tablenum[table-format=2.2]{14.63}  &  \tablenum[table-format=2.2]{14.62}  &  \tablenum[table-format=2.2]{94.99}  &  \tablenum[table-format=-2.2]{22.03}\,(\tablenum[table-format=1.2]{0.97})  &  -- \\
3  &  \tablenum[table-format=3.0]{800}  &  \tablenum[table-format=4.0]{4000}  &  $\widehat{\theta}_{\mathrm{SS}}$  &  \tablenum[table-format=-1.2]{0.14}  &  \tablenum[table-format=2.2]{14.55}  &  \tablenum[table-format=2.2]{14.54}  &  \tablenum[table-format=2.2]{14.55}  &  \tablenum[table-format=2.2]{94.89}  &  \tablenum[table-format=-2.2]{22.76}\,(\tablenum[table-format=1.2]{0.81})  &  \num{22.11} \\*
\TableEndRule \multicolumn{11}{@{}p{\UniformWidth}@{}}{\textit{Note:} Bias, EmpSD, MeanSE and RMSE are scaled by $10^3$; CP is coverage (\%). Gains are relative to Supervised; parentheses give paired MCSE in percentage points. Each cell uses 8,000 repetitions. A dash in $M_0$ denotes no auxiliary pool. AVR is the asymptotic variance reduction of Weighted relative to Supervised.}\\
\end{longtable}\endgroup

\begin{figure}[!htbp]\centering
\makebox[\linewidth][c]{\includegraphics[width=\linewidth]{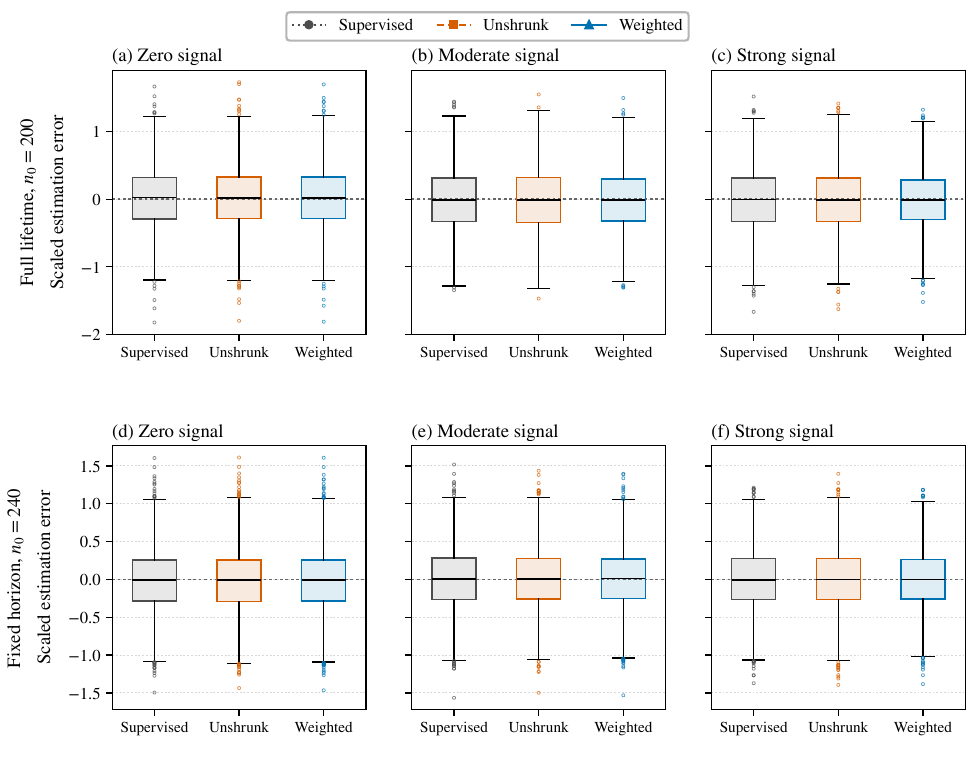}}
\caption{Truth-centered estimation-error distributions under zero, moderate and strong working projection signal. Panels (a)--(c) use full-lifetime Scenarios 1--3, $n_0=M_0=200$, and 8,000 repetitions per cell; panels (d)--(f) use fixed-horizon Scenarios 9, 11 and 12, $n_0=M_0=240$, and 4,000 repetitions. The ordinate is $\sqrt{n_0}(\widehat\theta-\theta)$ with the appropriate target in each row. Supervised uses the labeled observations only. Boxes span the interquartile range; black lines are medians; whiskers extend to observations within 1.5 interquartile ranges; observations beyond the whiskers appear as open circles. Zero is the truth-centered reference, and scales are shared within each row.}\label{fig:errorDistributions}
\end{figure}

Within each repetition, the same event-history sample is used at all auxiliary sample sizes, isolating the contribution of the auxiliary pool. Table~\ref{tab:fullMain} summarizes 8,000 repetitions for each combination of signal, labeled size and auxiliary ratio. At $n_0=200$, increasing $M_0$ from 200 to 1,000 raises the weighted gain from 3.04\% to 5.69\% under moderate signal and from 12.31\% to 20.36\% under strong signal. The zero-signal cells show small losses. Bias is small relative to empirical sampling variation. Figure~\ref{fig:errorDistributions} places these precision differences alongside the zero-signal error distribution.

When $M_0=n_0$, unshrunk transfer has zero first-order gain, while finite-resource weighting recovers half of the transferable projection variance. As auxiliary pools grow, the weighted and unshrunk gains approach one another. At $n_0=800$ under strong signal, the weighted gain rises from 13.83\% (MCSE 0.70) at ratio 1 to 22.76\% (0.81) at ratio 5, close to predictions of 13.27\% and 22.11\%. Zero-projection cells show small negative empirical gains, consistent with slope-estimation costs when no first-order gain is available.

\begin{figure}[!htbp]\centering
\makebox[\linewidth][c]{\includegraphics[width=\linewidth]{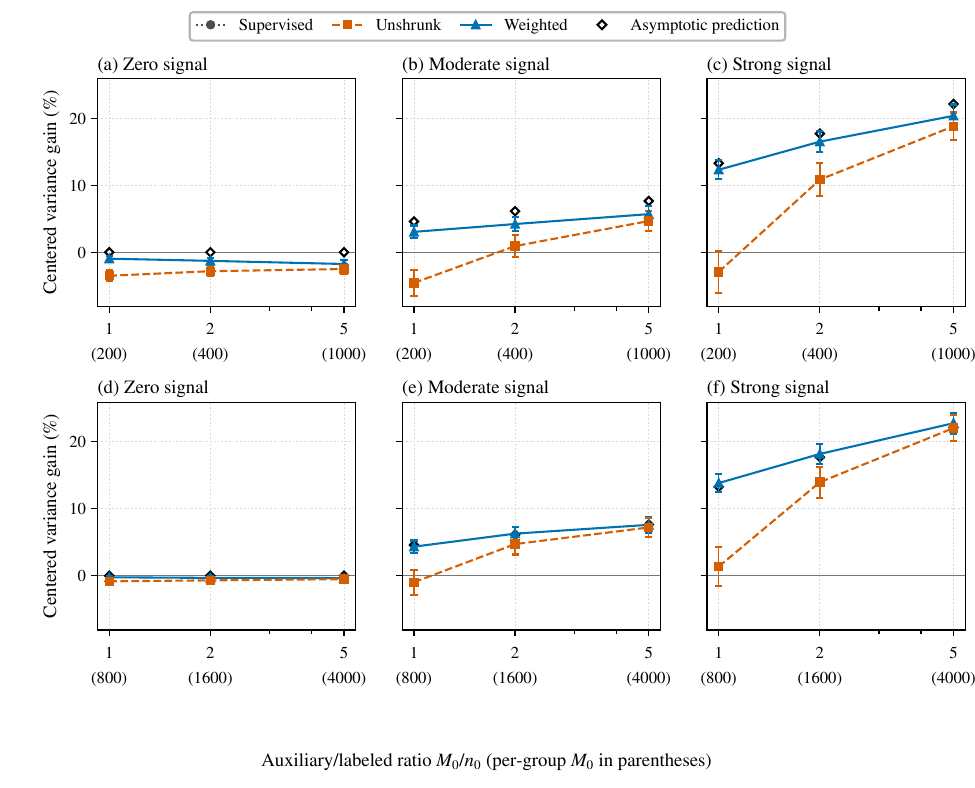}}
\caption{Full-lifetime resource comparisons under Scenarios 1--3. Each population has $n_0$ labeled observations; the ratio $M_0/n_0$ is on the horizontal axis and the per-group auxiliary count $M_0$ appears in parentheses. Points show centered variance gains relative to supervised estimation, with pointwise $\pm1.96$ paired-MCSE intervals; open diamonds show the first-order prediction. Each cell has 8,000 repetitions, with labeled observations and folds shared across resource levels.}\label{fig:resources}
\end{figure}

Four-pool sampling allows the two populations to contribute different amounts of auxiliary information. Scenario 22 places a linear signal in both groups and varies their labeled and auxiliary sizes. Scenarios 23 and 24 share a quadratic event mechanism but use linear versus linear-plus-quadratic features. Scenario 25 uses different covariate distributions across groups, retaining the same covariate distribution within each labeled/auxiliary pair. Their exact laws are given in Appendix~\ref{app:additional}.

Table~\ref{tab:structureMain} shows gains from both auxiliary pools as well as from their allocation. In the balanced dual-signal configuration, the weighted gain rises from 8.86\% to 14.27\% as the ratio increases from one to five. With labeled sizes $(200,400)$, the two asymmetric auxiliary allocations give weighted gains of 7.57\% and 9.87\%; the corresponding unshrunk gains are $-36.53\%$ and $-10.69\%$. Figure~\ref{fig:theoryCalibration}(a) compares these allocations with the first-order prediction. The separate resource weights accommodate the unequal precision of the two auxiliary means.

The same table distinguishes information present in the event mechanism from information represented by the working features. At $n_0=800$, the quadratic law gives a $-0.57\%$ gain with a linear feature and 8.66\% after the quadratic feature is included. Under different population covariate laws, the weighted gains are 11.81\% and 13.94\% at $n_0=200$ and 800. These settings retain matching labeled and auxiliary covariate laws within each population while allowing the two populations to differ. The nonzero-effect rows at $\theta=0.35$ and 0.65 are summarized in Table~\ref{tab:additionalFull}, which also reports method-specific calibration.

\begingroup\TableFont
\setlength{\tabcolsep}{0pt}\sbox{\UniformMeasure}{\begin{tabular}{@{}lrrrrrrrr@{}}Scenario & $(n_1,n_2)$ & $(M_1,M_2)$ & AVR (\%) & $G_{\mathrm{SS:sup}}$ & $G_{\mathrm{un:sup}}$ & $G_{\mathrm{SS:un}}$ & SE/SD & CP \\
8  &  (200,200)  &  (200,200)  &  \tablenum[table-format=2.2]{10.04}  &  \tablenum[table-format=-2.2]{7.65}\,(\tablenum[table-format=1.2]{0.96})  &  \tablenum[table-format=-2.2]{-5.78}\,(\tablenum[table-format=1.2]{2.10})  &  \tablenum[table-format=2.2]{12.70}\,(\tablenum[table-format=1.2]{0.91})  &  \tablenum[table-format=1.3]{0.987}  &  \tablenum[table-format=2.2]{94.73} \\
8  &  (200,200)  &  (1000,1000)  &  \tablenum[table-format=2.2]{16.73}  &  \tablenum[table-format=-2.2]{12.83}\,(\tablenum[table-format=1.2]{1.18})  &  \tablenum[table-format=-2.2]{11.03}\,(\tablenum[table-format=1.2]{1.44})  &  \tablenum[table-format=2.2]{2.02}\,(\tablenum[table-format=1.2]{0.29})  &  \tablenum[table-format=1.3]{0.980}  &  \tablenum[table-format=2.2]{94.58} \\
8  &  (800,800)  &  (800,800)  &  \tablenum[table-format=2.2]{10.04}  &  \tablenum[table-format=-2.2]{9.87}\,(\tablenum[table-format=1.2]{0.90})  &  \tablenum[table-format=-2.2]{-0.29}\,(\tablenum[table-format=1.2]{1.90})  &  \tablenum[table-format=2.2]{10.13}\,(\tablenum[table-format=1.2]{0.89})  &  \tablenum[table-format=1.3]{1.003}  &  \tablenum[table-format=2.2]{95.03} \\
8  &  (800,800)  &  (4000,4000)  &  \tablenum[table-format=2.2]{16.73}  &  \tablenum[table-format=-2.2]{15.64}\,(\tablenum[table-format=1.2]{1.09})  &  \tablenum[table-format=-2.2]{14.77}\,(\tablenum[table-format=1.2]{1.32})  &  \tablenum[table-format=2.2]{1.02}\,(\tablenum[table-format=1.2]{0.28})  &  \tablenum[table-format=1.3]{0.998}  &  \tablenum[table-format=2.2]{94.95} \\
22  &  (200,200)  &  (200,200)  &  \tablenum[table-format=2.2]{9.18}  &  \tablenum[table-format=-2.2]{8.86}\,(\tablenum[table-format=1.2]{0.90})  &  \tablenum[table-format=-2.2]{-1.94}\,(\tablenum[table-format=1.2]{1.93})  &  \tablenum[table-format=2.2]{10.59}\,(\tablenum[table-format=1.2]{0.90})  &  \tablenum[table-format=1.3]{1.000}  &  \tablenum[table-format=2.2]{94.77} \\
22  &  (200,200)  &  (1000,1000)  &  \tablenum[table-format=2.2]{15.30}  &  \tablenum[table-format=-2.2]{14.27}\,(\tablenum[table-format=1.2]{1.09})  &  \tablenum[table-format=-2.2]{13.18}\,(\tablenum[table-format=1.2]{1.32})  &  \tablenum[table-format=2.2]{1.25}\,(\tablenum[table-format=1.2]{0.28})  &  \tablenum[table-format=1.3]{0.996}  &  \tablenum[table-format=2.2]{94.80} \\
22  &  (200,400)  &  (50,2000)  &  \tablenum[table-format=2.2]{7.55}  &  \tablenum[table-format=-2.2]{7.57}\,(\tablenum[table-format=1.2]{0.83})  &  \tablenum[table-format=-2.2]{-36.53}\,(\tablenum[table-format=1.2]{3.14})  &  \tablenum[table-format=2.2]{32.30}\,(\tablenum[table-format=1.2]{1.22})  &  \tablenum[table-format=1.3]{1.011}  &  \tablenum[table-format=2.2]{95.28} \\
22  &  (200,400)  &  (1000,100)  &  \tablenum[table-format=2.2]{11.43}  &  \tablenum[table-format=-2.2]{9.87}\,(\tablenum[table-format=1.2]{1.01})  &  \tablenum[table-format=-2.2]{-10.69}\,(\tablenum[table-format=1.2]{2.29})  &  \tablenum[table-format=2.2]{18.58}\,(\tablenum[table-format=1.2]{1.16})  &  \tablenum[table-format=1.3]{1.002}  &  \tablenum[table-format=2.2]{94.92} \\
23  &  (200,200)  &  (1000,1000)  &  \tablenum[table-format=2.2]{0.00}  &  \tablenum[table-format=-2.2]{-2.02}\,(\tablenum[table-format=1.2]{0.50})  &  \tablenum[table-format=-2.2]{-2.96}\,(\tablenum[table-format=1.2]{0.60})  &  \tablenum[table-format=2.2]{0.92}\,(\tablenum[table-format=1.2]{0.10})  &  \tablenum[table-format=1.3]{0.993}  &  \tablenum[table-format=2.2]{94.77} \\
24  &  (200,200)  &  (1000,1000)  &  \tablenum[table-format=2.2]{11.30}  &  \tablenum[table-format=-2.2]{6.96}\,(\tablenum[table-format=1.2]{1.14})  &  \tablenum[table-format=-2.2]{4.53}\,(\tablenum[table-format=1.2]{1.39})  &  \tablenum[table-format=2.2]{2.55}\,(\tablenum[table-format=1.2]{0.25})  &  \tablenum[table-format=1.3]{0.983}  &  \tablenum[table-format=2.2]{94.80} \\
23  &  (800,800)  &  (4000,4000)  &  \tablenum[table-format=2.2]{0.00}  &  \tablenum[table-format=-2.2]{-0.57}\,(\tablenum[table-format=1.2]{0.25})  &  \tablenum[table-format=-2.2]{-0.82}\,(\tablenum[table-format=1.2]{0.30})  &  \tablenum[table-format=2.2]{0.25}\,(\tablenum[table-format=1.2]{0.05})  &  \tablenum[table-format=1.3]{0.985}  &  \tablenum[table-format=2.2]{94.75} \\
24  &  (800,800)  &  (4000,4000)  &  \tablenum[table-format=2.2]{11.30}  &  \tablenum[table-format=-2.2]{8.66}\,(\tablenum[table-format=1.2]{0.99})  &  \tablenum[table-format=-2.2]{7.47}\,(\tablenum[table-format=1.2]{1.21})  &  \tablenum[table-format=2.2]{1.28}\,(\tablenum[table-format=1.2]{0.23})  &  \tablenum[table-format=1.3]{0.974}  &  \tablenum[table-format=2.2]{94.58} \\
25  &  (200,200)  &  (1000,1000)  &  \tablenum[table-format=2.2]{15.60}  &  \tablenum[table-format=-2.2]{11.81}\,(\tablenum[table-format=1.2]{1.13})  &  \tablenum[table-format=-2.2]{10.16}\,(\tablenum[table-format=1.2]{1.38})  &  \tablenum[table-format=2.2]{1.84}\,(\tablenum[table-format=1.2]{0.27})  &  \tablenum[table-format=1.3]{0.986}  &  \tablenum[table-format=2.2]{94.83} \\
25  &  (800,800)  &  (4000,4000)  &  \tablenum[table-format=2.2]{15.60}  &  \tablenum[table-format=-2.2]{13.94}\,(\tablenum[table-format=1.2]{1.09})  &  \tablenum[table-format=-2.2]{12.77}\,(\tablenum[table-format=1.2]{1.32})  &  \tablenum[table-format=2.2]{1.34}\,(\tablenum[table-format=1.2]{0.27})  &  \tablenum[table-format=1.3]{1.010}  &  \tablenum[table-format=2.2]{94.97} \\
\midrule\multicolumn{9}{l}{Nonzero comparison probabilities: strong signal}\\
5  &  (200,200)  &  (200,200)  &  --  &  \tablenum[table-format=-2.2]{11.05}\,(\tablenum[table-format=1.2]{1.44})  &  \tablenum[table-format=-2.2]{-5.29}\,(\tablenum[table-format=1.2]{3.21})  &  \tablenum[table-format=2.2]{15.52}\,(\tablenum[table-format=1.2]{1.41})  &  \tablenum[table-format=1.3]{0.996}  &  \tablenum[table-format=2.2]{95.00} \\
5  &  (800,800)  &  (800,800)  &  --  &  \tablenum[table-format=-2.2]{13.34}\,(\tablenum[table-format=1.2]{1.38})  &  \tablenum[table-format=-2.2]{1.48}\,(\tablenum[table-format=1.2]{2.88})  &  \tablenum[table-format=2.2]{12.04}\,(\tablenum[table-format=1.2]{1.36})  &  \tablenum[table-format=1.3]{1.009}  &  \tablenum[table-format=2.2]{94.75} \\
7  &  (200,200)  &  (200,200)  &  --  &  \tablenum[table-format=-2.2]{11.44}\,(\tablenum[table-format=1.2]{1.51})  &  \tablenum[table-format=-2.2]{-7.53}\,(\tablenum[table-format=1.2]{3.37})  &  \tablenum[table-format=2.2]{17.64}\,(\tablenum[table-format=1.2]{1.39})  &  \tablenum[table-format=1.3]{1.034}  &  \tablenum[table-format=2.2]{95.80} \\
7  &  (800,800)  &  (800,800)  &  --  &  \tablenum[table-format=-2.2]{14.56}\,(\tablenum[table-format=1.2]{1.39})  &  \tablenum[table-format=-2.2]{3.19}\,(\tablenum[table-format=1.2]{2.97})  &  \tablenum[table-format=2.2]{11.74}\,(\tablenum[table-format=1.2]{1.44})  &  \tablenum[table-format=1.3]{0.989}  &  \tablenum[table-format=2.2]{94.60} \\
\end{tabular}}
\setlength{\UniformWidth}{\linewidth}\setlength{\tabcolsep}{\dimexpr(\linewidth-\wd\UniformMeasure)/16\relax}
\ifdim\tabcolsep<0pt\PackageWarning{UnifiedTables}{Negative column gap}\fi
\setlength{\LTleft}{0pt plus 1fill}\setlength{\LTright}{0pt plus 1fill}
\begin{longtable}{@{}lrrrrrrrr@{}}
\caption{Signal structure and resource allocation.}\label{tab:structureMain}\\
\toprule Scenario & $(n_1,n_2)$ & $(M_1,M_2)$ & AVR (\%) & $G_{\mathrm{SS:sup}}$ & $G_{\mathrm{un:sup}}$ & $G_{\mathrm{SS:un}}$ & SE/SD & CP \\\midrule\endfirsthead
\multicolumn{9}{l}{\textit{Table \thetable\ continued}}\\\toprule Scenario & $(n_1,n_2)$ & $(M_1,M_2)$ & AVR (\%) & $G_{\mathrm{SS:sup}}$ & $G_{\mathrm{un:sup}}$ & $G_{\mathrm{SS:un}}$ & SE/SD & CP \\\midrule\endhead
\midrule\multicolumn{9}{r}{\textit{Continued on next page}}\\\endfoot
\endlastfoot

8  &  (200,200)  &  (200,200)  &  \tablenum[table-format=2.2]{10.04}  &  \tablenum[table-format=-2.2]{7.65}\,(\tablenum[table-format=1.2]{0.96})  &  \tablenum[table-format=-2.2]{-5.78}\,(\tablenum[table-format=1.2]{2.10})  &  \tablenum[table-format=2.2]{12.70}\,(\tablenum[table-format=1.2]{0.91})  &  \tablenum[table-format=1.3]{0.987}  &  \tablenum[table-format=2.2]{94.73} \\
8  &  (200,200)  &  (1000,1000)  &  \tablenum[table-format=2.2]{16.73}  &  \tablenum[table-format=-2.2]{12.83}\,(\tablenum[table-format=1.2]{1.18})  &  \tablenum[table-format=-2.2]{11.03}\,(\tablenum[table-format=1.2]{1.44})  &  \tablenum[table-format=2.2]{2.02}\,(\tablenum[table-format=1.2]{0.29})  &  \tablenum[table-format=1.3]{0.980}  &  \tablenum[table-format=2.2]{94.58} \\
8  &  (800,800)  &  (800,800)  &  \tablenum[table-format=2.2]{10.04}  &  \tablenum[table-format=-2.2]{9.87}\,(\tablenum[table-format=1.2]{0.90})  &  \tablenum[table-format=-2.2]{-0.29}\,(\tablenum[table-format=1.2]{1.90})  &  \tablenum[table-format=2.2]{10.13}\,(\tablenum[table-format=1.2]{0.89})  &  \tablenum[table-format=1.3]{1.003}  &  \tablenum[table-format=2.2]{95.03} \\
8  &  (800,800)  &  (4000,4000)  &  \tablenum[table-format=2.2]{16.73}  &  \tablenum[table-format=-2.2]{15.64}\,(\tablenum[table-format=1.2]{1.09})  &  \tablenum[table-format=-2.2]{14.77}\,(\tablenum[table-format=1.2]{1.32})  &  \tablenum[table-format=2.2]{1.02}\,(\tablenum[table-format=1.2]{0.28})  &  \tablenum[table-format=1.3]{0.998}  &  \tablenum[table-format=2.2]{94.95} \\
22  &  (200,200)  &  (200,200)  &  \tablenum[table-format=2.2]{9.18}  &  \tablenum[table-format=-2.2]{8.86}\,(\tablenum[table-format=1.2]{0.90})  &  \tablenum[table-format=-2.2]{-1.94}\,(\tablenum[table-format=1.2]{1.93})  &  \tablenum[table-format=2.2]{10.59}\,(\tablenum[table-format=1.2]{0.90})  &  \tablenum[table-format=1.3]{1.000}  &  \tablenum[table-format=2.2]{94.77} \\
22  &  (200,200)  &  (1000,1000)  &  \tablenum[table-format=2.2]{15.30}  &  \tablenum[table-format=-2.2]{14.27}\,(\tablenum[table-format=1.2]{1.09})  &  \tablenum[table-format=-2.2]{13.18}\,(\tablenum[table-format=1.2]{1.32})  &  \tablenum[table-format=2.2]{1.25}\,(\tablenum[table-format=1.2]{0.28})  &  \tablenum[table-format=1.3]{0.996}  &  \tablenum[table-format=2.2]{94.80} \\
22  &  (200,400)  &  (50,2000)  &  \tablenum[table-format=2.2]{7.55}  &  \tablenum[table-format=-2.2]{7.57}\,(\tablenum[table-format=1.2]{0.83})  &  \tablenum[table-format=-2.2]{-36.53}\,(\tablenum[table-format=1.2]{3.14})  &  \tablenum[table-format=2.2]{32.30}\,(\tablenum[table-format=1.2]{1.22})  &  \tablenum[table-format=1.3]{1.011}  &  \tablenum[table-format=2.2]{95.28} \\
22  &  (200,400)  &  (1000,100)  &  \tablenum[table-format=2.2]{11.43}  &  \tablenum[table-format=-2.2]{9.87}\,(\tablenum[table-format=1.2]{1.01})  &  \tablenum[table-format=-2.2]{-10.69}\,(\tablenum[table-format=1.2]{2.29})  &  \tablenum[table-format=2.2]{18.58}\,(\tablenum[table-format=1.2]{1.16})  &  \tablenum[table-format=1.3]{1.002}  &  \tablenum[table-format=2.2]{94.92} \\
23  &  (200,200)  &  (1000,1000)  &  \tablenum[table-format=2.2]{0.00}  &  \tablenum[table-format=-2.2]{-2.02}\,(\tablenum[table-format=1.2]{0.50})  &  \tablenum[table-format=-2.2]{-2.96}\,(\tablenum[table-format=1.2]{0.60})  &  \tablenum[table-format=2.2]{0.92}\,(\tablenum[table-format=1.2]{0.10})  &  \tablenum[table-format=1.3]{0.993}  &  \tablenum[table-format=2.2]{94.77} \\
24  &  (200,200)  &  (1000,1000)  &  \tablenum[table-format=2.2]{11.30}  &  \tablenum[table-format=-2.2]{6.96}\,(\tablenum[table-format=1.2]{1.14})  &  \tablenum[table-format=-2.2]{4.53}\,(\tablenum[table-format=1.2]{1.39})  &  \tablenum[table-format=2.2]{2.55}\,(\tablenum[table-format=1.2]{0.25})  &  \tablenum[table-format=1.3]{0.983}  &  \tablenum[table-format=2.2]{94.80} \\
23  &  (800,800)  &  (4000,4000)  &  \tablenum[table-format=2.2]{0.00}  &  \tablenum[table-format=-2.2]{-0.57}\,(\tablenum[table-format=1.2]{0.25})  &  \tablenum[table-format=-2.2]{-0.82}\,(\tablenum[table-format=1.2]{0.30})  &  \tablenum[table-format=2.2]{0.25}\,(\tablenum[table-format=1.2]{0.05})  &  \tablenum[table-format=1.3]{0.985}  &  \tablenum[table-format=2.2]{94.75} \\
24  &  (800,800)  &  (4000,4000)  &  \tablenum[table-format=2.2]{11.30}  &  \tablenum[table-format=-2.2]{8.66}\,(\tablenum[table-format=1.2]{0.99})  &  \tablenum[table-format=-2.2]{7.47}\,(\tablenum[table-format=1.2]{1.21})  &  \tablenum[table-format=2.2]{1.28}\,(\tablenum[table-format=1.2]{0.23})  &  \tablenum[table-format=1.3]{0.974}  &  \tablenum[table-format=2.2]{94.58} \\
25  &  (200,200)  &  (1000,1000)  &  \tablenum[table-format=2.2]{15.60}  &  \tablenum[table-format=-2.2]{11.81}\,(\tablenum[table-format=1.2]{1.13})  &  \tablenum[table-format=-2.2]{10.16}\,(\tablenum[table-format=1.2]{1.38})  &  \tablenum[table-format=2.2]{1.84}\,(\tablenum[table-format=1.2]{0.27})  &  \tablenum[table-format=1.3]{0.986}  &  \tablenum[table-format=2.2]{94.83} \\
25  &  (800,800)  &  (4000,4000)  &  \tablenum[table-format=2.2]{15.60}  &  \tablenum[table-format=-2.2]{13.94}\,(\tablenum[table-format=1.2]{1.09})  &  \tablenum[table-format=-2.2]{12.77}\,(\tablenum[table-format=1.2]{1.32})  &  \tablenum[table-format=2.2]{1.34}\,(\tablenum[table-format=1.2]{0.27})  &  \tablenum[table-format=1.3]{1.010}  &  \tablenum[table-format=2.2]{94.97} \\
\midrule\multicolumn{9}{l}{Nonzero comparison probabilities: strong signal}\\
5  &  (200,200)  &  (200,200)  &  --  &  \tablenum[table-format=-2.2]{11.05}\,(\tablenum[table-format=1.2]{1.44})  &  \tablenum[table-format=-2.2]{-5.29}\,(\tablenum[table-format=1.2]{3.21})  &  \tablenum[table-format=2.2]{15.52}\,(\tablenum[table-format=1.2]{1.41})  &  \tablenum[table-format=1.3]{0.996}  &  \tablenum[table-format=2.2]{95.00} \\
5  &  (800,800)  &  (800,800)  &  --  &  \tablenum[table-format=-2.2]{13.34}\,(\tablenum[table-format=1.2]{1.38})  &  \tablenum[table-format=-2.2]{1.48}\,(\tablenum[table-format=1.2]{2.88})  &  \tablenum[table-format=2.2]{12.04}\,(\tablenum[table-format=1.2]{1.36})  &  \tablenum[table-format=1.3]{1.009}  &  \tablenum[table-format=2.2]{94.75} \\
7  &  (200,200)  &  (200,200)  &  --  &  \tablenum[table-format=-2.2]{11.44}\,(\tablenum[table-format=1.2]{1.51})  &  \tablenum[table-format=-2.2]{-7.53}\,(\tablenum[table-format=1.2]{3.37})  &  \tablenum[table-format=2.2]{17.64}\,(\tablenum[table-format=1.2]{1.39})  &  \tablenum[table-format=1.3]{1.034}  &  \tablenum[table-format=2.2]{95.80} \\
7  &  (800,800)  &  (800,800)  &  --  &  \tablenum[table-format=-2.2]{14.56}\,(\tablenum[table-format=1.2]{1.39})  &  \tablenum[table-format=-2.2]{3.19}\,(\tablenum[table-format=1.2]{2.97})  &  \tablenum[table-format=2.2]{11.74}\,(\tablenum[table-format=1.2]{1.44})  &  \tablenum[table-format=1.3]{0.989}  &  \tablenum[table-format=2.2]{94.60} \\*
\TableEndRule \multicolumn{9}{@{}p{\UniformWidth}@{}}{\textit{Note:} The gain columns are $G_{\mathrm{SS:sup}}$, $G_{\mathrm{un:sup}}$ and $G_{\mathrm{SS:un}}$; parentheses give paired MCSE in percentage points. SE/SD and CP (\%) refer to Weighted. Upper-block cells use 4,000 repetitions; nonzero-effect cells use 2,000. AVR is the asymptotic variance reduction of Weighted relative to Supervised. A dash marks a configuration for which that prediction was not reported. Additional bias and precision results are in Table~\ref{tab:additionalFull}.}\\
\end{longtable}\endgroup

For the usual-censoring, strong-signal Scenario 3 in Table~\ref{tab:fullMain}, Weighted coverage is 94.34\% at $n_0=200$ and $M_0/n_0=5$ (Wilson interval 93.81\%--94.82\%). Under heavier censoring in Scenario 8, Weighted coverage is 94.4\% and 95.1\% at $n_0=200$ and 800, respectively, when $M_0/n_0=1$; at ratio 5 it is 94.58\% and 94.95\% (Table~\ref{tab:additionalFull}). Figure~\ref{fig:theoryCalibration}(c,e) gives the matched coverage and SE-calibration comparisons.

\subsection{Fixed-horizon comparisons}
The fixed-horizon designs examine a different observed-data problem: event and censoring times can coincide, and capping follow-up creates a positive terminal atom. On the grid $\{0,\ldots,4\}$ we use event priority and the half-tie target of Theorem~\ref{thm:restricted}. The 12 law/feature settings vary projected signal, censoring, resource ratio and working features. One setting records a continuous latent event time on a grid before capping; its target is the recorded-grid probability. Scenarios 9--12 have zero, weak, moderate and strong working-projection signal, respectively. Scenarios 13--14 share a quadratic event law and use linear versus linear-plus-quadratic features. Scenarios 15--16 combine weak signal with heavy versus overlap censoring; Scenarios 17--18 use weak signal and auxiliary-to-labeled ratios 0.25 versus 5. Scenario 19 has a larger group contrast ($\theta_\tau=0.599225$); Scenario 20 uses recorded-grid exponential event times ($\theta_\tau=0.593044$). Unless stated otherwise, the ratio is one and censoring is light; Scenario 20 uses overlap censoring. Conditional PMFs are specified in Appendix~\ref{s:restricteddesign}.

The fixed-horizon kernel treats two event times at or beyond the horizon as a terminal tie. Censoring before the horizon can leave the pairwise ordering unresolved. The second row of Figure~\ref{fig:errorDistributions} shows the resulting error distributions, and Table~\ref{tab:restrictedMain} compares all three methods at $n_0=240$. Moderate and strong projections give gains close to the first-order predictions of about 3\% and 8\%. Under weak projection, little variance is available to recover, and estimating the correction can erase that small gain.

\begingroup\TableFont
\setlength{\tabcolsep}{0pt}\sbox{\UniformMeasure}{\begin{tabular}{@{}lcccccccccr@{}}Scenario & $n_0$ & $M_0$ & Method & Bias & \shortstack{Emp\\SD} & \shortstack{Mean\\SE} & RMSE & CP & \shortstack{$G_{A:\mathrm{sup}}$\\(MCSE)} & AVR (\%)\\
9  &  \tablenum[table-format=3.0]{240}  &  \tablenum[table-format=4.0]{240}  &  $\widehat{\theta}_{\mathrm{sup},\tau}$  &  \tablenum[table-format=-1.2]{-0.61}  &  \tablenum[table-format=2.2]{25.77}  &  \tablenum[table-format=2.2]{25.76}  &  \tablenum[table-format=2.2]{25.78}  &  \tablenum[table-format=1.3]{0.951}  &  --  &  --\\*
  &  \tablenum[table-format=3.0]{240}  &  \tablenum[table-format=4.0]{240}  &  $\widehat{\theta}_{\mathrm{un},\tau}$  &  \tablenum[table-format=-1.2]{-0.65}  &  \tablenum[table-format=2.2]{25.98}  &  \tablenum[table-format=2.2]{25.79}  &  \tablenum[table-format=2.2]{25.98}  &  \tablenum[table-format=1.3]{0.951}  &  \tablenum[table-format=-2.2]{-1.60}\,(\tablenum[table-format=1.2]{0.40})  &  --\\*
  &  \tablenum[table-format=3.0]{240}  &  \tablenum[table-format=4.0]{240}  &  $\widehat{\theta}_{\mathrm{SS},\tau}$  &  \tablenum[table-format=-1.2]{-0.63}  &  \tablenum[table-format=2.2]{25.82}  &  \tablenum[table-format=2.2]{25.74}  &  \tablenum[table-format=2.2]{25.83}  &  \tablenum[table-format=1.3]{0.952}  &  \tablenum[table-format=-2.2]{-0.41}\,(\tablenum[table-format=1.2]{0.20})  &  \num{0.00} \\
\addlinespace[2pt]
10  &  \tablenum[table-format=3.0]{240}  &  \tablenum[table-format=4.0]{240}  &  $\widehat{\theta}_{\mathrm{sup},\tau}$  &  \tablenum[table-format=-1.2]{0.21}  &  \tablenum[table-format=2.2]{25.92}  &  \tablenum[table-format=2.2]{25.77}  &  \tablenum[table-format=2.2]{25.92}  &  \tablenum[table-format=1.3]{0.943}  &  --  &  --\\*
  &  \tablenum[table-format=3.0]{240}  &  \tablenum[table-format=4.0]{240}  &  $\widehat{\theta}_{\mathrm{un},\tau}$  &  \tablenum[table-format=-1.2]{0.14}  &  \tablenum[table-format=2.2]{26.13}  &  \tablenum[table-format=2.2]{25.80}  &  \tablenum[table-format=2.2]{26.13}  &  \tablenum[table-format=1.3]{0.942}  &  \tablenum[table-format=-2.2]{-1.64}\,(\tablenum[table-format=1.2]{0.48})  &  --\\*
  &  \tablenum[table-format=3.0]{240}  &  \tablenum[table-format=4.0]{240}  &  $\widehat{\theta}_{\mathrm{SS},\tau}$  &  \tablenum[table-format=-1.2]{0.18}  &  \tablenum[table-format=2.2]{25.95}  &  \tablenum[table-format=2.2]{25.73}  &  \tablenum[table-format=2.2]{25.95}  &  \tablenum[table-format=1.3]{0.941}  &  \tablenum[table-format=-2.2]{-0.24}\,(\tablenum[table-format=1.2]{0.24})  &  \num{0.17} \\
\addlinespace[2pt]
11  &  \tablenum[table-format=3.0]{240}  &  \tablenum[table-format=4.0]{240}  &  $\widehat{\theta}_{\mathrm{sup},\tau}$  &  \tablenum[table-format=-1.2]{0.46}  &  \tablenum[table-format=2.2]{25.55}  &  \tablenum[table-format=2.2]{26.02}  &  \tablenum[table-format=2.2]{25.56}  &  \tablenum[table-format=1.3]{0.955}  &  --  &  --\\*
  &  \tablenum[table-format=3.0]{240}  &  \tablenum[table-format=4.0]{240}  &  $\widehat{\theta}_{\mathrm{un},\tau}$  &  \tablenum[table-format=-1.2]{0.63}  &  \tablenum[table-format=2.2]{25.63}  &  \tablenum[table-format=2.2]{26.05}  &  \tablenum[table-format=2.2]{25.64}  &  \tablenum[table-format=1.3]{0.954}  &  \tablenum[table-format=-2.2]{-0.61}\,(\tablenum[table-format=1.2]{1.17})  &  --\\*
  &  \tablenum[table-format=3.0]{240}  &  \tablenum[table-format=4.0]{240}  &  $\widehat{\theta}_{\mathrm{SS},\tau}$  &  \tablenum[table-format=-1.2]{0.54}  &  \tablenum[table-format=2.2]{25.14}  &  \tablenum[table-format=2.2]{25.61}  &  \tablenum[table-format=2.2]{25.15}  &  \tablenum[table-format=1.3]{0.954}  &  \tablenum[table-format=-2.2]{3.19}\,(\tablenum[table-format=1.2]{0.57})  &  \num{3.00} \\
\addlinespace[2pt]
12  &  \tablenum[table-format=3.0]{240}  &  \tablenum[table-format=4.0]{240}  &  $\widehat{\theta}_{\mathrm{sup},\tau}$  &  \tablenum[table-format=-1.2]{-0.04}  &  \tablenum[table-format=2.2]{25.95}  &  \tablenum[table-format=2.2]{25.97}  &  \tablenum[table-format=2.2]{25.95}  &  \tablenum[table-format=1.3]{0.951}  &  --  &  --\\*
  &  \tablenum[table-format=3.0]{240}  &  \tablenum[table-format=4.0]{240}  &  $\widehat{\theta}_{\mathrm{un},\tau}$  &  \tablenum[table-format=-1.2]{0.14}  &  \tablenum[table-format=2.2]{25.90}  &  \tablenum[table-format=2.2]{26.00}  &  \tablenum[table-format=2.2]{25.89}  &  \tablenum[table-format=1.3]{0.948}  &  \tablenum[table-format=-2.2]{0.44}\,(\tablenum[table-format=1.2]{1.71})  &  --\\*
  &  \tablenum[table-format=3.0]{240}  &  \tablenum[table-format=4.0]{240}  &  $\widehat{\theta}_{\mathrm{SS},\tau}$  &  \tablenum[table-format=-1.2]{0.05}  &  \tablenum[table-format=2.2]{24.83}  &  \tablenum[table-format=2.2]{24.90}  &  \tablenum[table-format=2.2]{24.82}  &  \tablenum[table-format=1.3]{0.951}  &  \tablenum[table-format=-2.2]{8.50}\,(\tablenum[table-format=1.2]{0.82})  &  \num{8.00} \\
\addlinespace[2pt]
13  &  \tablenum[table-format=3.0]{240}  &  \tablenum[table-format=4.0]{240}  &  $\widehat{\theta}_{\mathrm{sup},\tau}$  &  \tablenum[table-format=-1.2]{0.25}  &  \tablenum[table-format=2.2]{26.07}  &  \tablenum[table-format=2.2]{26.02}  &  \tablenum[table-format=2.2]{26.07}  &  \tablenum[table-format=1.3]{0.947}  &  --  &  --\\*
  &  \tablenum[table-format=3.0]{240}  &  \tablenum[table-format=4.0]{240}  &  $\widehat{\theta}_{\mathrm{un},\tau}$  &  \tablenum[table-format=-1.2]{0.20}  &  \tablenum[table-format=2.2]{26.27}  &  \tablenum[table-format=2.2]{26.05}  &  \tablenum[table-format=2.2]{26.27}  &  \tablenum[table-format=1.3]{0.948}  &  \tablenum[table-format=-2.2]{-1.54}\,(\tablenum[table-format=1.2]{0.42})  &  --\\*
  &  \tablenum[table-format=3.0]{240}  &  \tablenum[table-format=4.0]{240}  &  $\widehat{\theta}_{\mathrm{SS},\tau}$  &  \tablenum[table-format=-1.2]{0.23}  &  \tablenum[table-format=2.2]{26.12}  &  \tablenum[table-format=2.2]{26.00}  &  \tablenum[table-format=2.2]{26.12}  &  \tablenum[table-format=1.3]{0.948}  &  \tablenum[table-format=-2.2]{-0.34}\,(\tablenum[table-format=1.2]{0.21})  &  \num{0.00} \\
\addlinespace[2pt]
14  &  \tablenum[table-format=3.0]{240}  &  \tablenum[table-format=4.0]{240}  &  $\widehat{\theta}_{\mathrm{sup},\tau}$  &  \tablenum[table-format=-1.2]{0.25}  &  \tablenum[table-format=2.2]{26.07}  &  \tablenum[table-format=2.2]{26.02}  &  \tablenum[table-format=2.2]{26.07}  &  \tablenum[table-format=1.3]{0.947}  &  --  &  --\\*
  &  \tablenum[table-format=3.0]{240}  &  \tablenum[table-format=4.0]{240}  &  $\widehat{\theta}_{\mathrm{un},\tau}$  &  \tablenum[table-format=-1.2]{0.33}  &  \tablenum[table-format=2.2]{26.83}  &  \tablenum[table-format=2.2]{26.08}  &  \tablenum[table-format=2.2]{26.83}  &  \tablenum[table-format=1.3]{0.943}  &  \tablenum[table-format=-2.2]{-5.89}\,(\tablenum[table-format=1.2]{1.24})  &  --\\*
  &  \tablenum[table-format=3.0]{240}  &  \tablenum[table-format=4.0]{240}  &  $\widehat{\theta}_{\mathrm{SS},\tau}$  &  \tablenum[table-format=-1.2]{0.29}  &  \tablenum[table-format=2.2]{25.96}  &  \tablenum[table-format=2.2]{25.59}  &  \tablenum[table-format=2.2]{25.96}  &  \tablenum[table-format=1.3]{0.946}  &  \tablenum[table-format=-2.2]{0.87}\,(\tablenum[table-format=1.2]{0.60})  &  \num{2.97} \\
\addlinespace[2pt]
15  &  \tablenum[table-format=3.0]{240}  &  \tablenum[table-format=4.0]{240}  &  $\widehat{\theta}_{\mathrm{sup},\tau}$  &  \tablenum[table-format=-1.2]{0.07}  &  \tablenum[table-format=2.2]{28.68}  &  \tablenum[table-format=2.2]{28.38}  &  \tablenum[table-format=2.2]{28.68}  &  \tablenum[table-format=1.3]{0.947}  &  --  &  --\\*
  &  \tablenum[table-format=3.0]{240}  &  \tablenum[table-format=4.0]{240}  &  $\widehat{\theta}_{\mathrm{un},\tau}$  &  \tablenum[table-format=-1.2]{0.03}  &  \tablenum[table-format=2.2]{29.11}  &  \tablenum[table-format=2.2]{28.56}  &  \tablenum[table-format=2.2]{29.10}  &  \tablenum[table-format=1.3]{0.945}  &  \tablenum[table-format=-2.2]{-2.96}\,(\tablenum[table-format=1.2]{0.60})  &  --\\*
  &  \tablenum[table-format=3.0]{240}  &  \tablenum[table-format=4.0]{240}  &  $\widehat{\theta}_{\mathrm{SS},\tau}$  &  \tablenum[table-format=-1.2]{0.05}  &  \tablenum[table-format=2.2]{28.77}  &  \tablenum[table-format=2.2]{28.38}  &  \tablenum[table-format=2.2]{28.77}  &  \tablenum[table-format=1.3]{0.947}  &  \tablenum[table-format=-2.2]{-0.60}\,(\tablenum[table-format=1.2]{0.30})  &  \num{0.14} \\
\addlinespace[2pt]
16  &  \tablenum[table-format=3.0]{240}  &  \tablenum[table-format=4.0]{240}  &  $\widehat{\theta}_{\mathrm{sup},\tau}$  &  \tablenum[table-format=-1.2]{-0.88}  &  \tablenum[table-format=2.2]{27.76}  &  \tablenum[table-format=2.2]{27.52}  &  \tablenum[table-format=2.2]{27.77}  &  \tablenum[table-format=1.3]{0.949}  &  --  &  --\\*
  &  \tablenum[table-format=3.0]{240}  &  \tablenum[table-format=4.0]{240}  &  $\widehat{\theta}_{\mathrm{un},\tau}$  &  \tablenum[table-format=-1.2]{-0.78}  &  \tablenum[table-format=2.2]{28.29}  &  \tablenum[table-format=2.2]{27.74}  &  \tablenum[table-format=2.2]{28.30}  &  \tablenum[table-format=1.3]{0.947}  &  \tablenum[table-format=-2.2]{-3.88}\,(\tablenum[table-format=1.2]{0.64})  &  --\\*
  &  \tablenum[table-format=3.0]{240}  &  \tablenum[table-format=4.0]{240}  &  $\widehat{\theta}_{\mathrm{SS},\tau}$  &  \tablenum[table-format=-1.2]{-0.83}  &  \tablenum[table-format=2.2]{27.88}  &  \tablenum[table-format=2.2]{27.52}  &  \tablenum[table-format=2.2]{27.89}  &  \tablenum[table-format=1.3]{0.949}  &  \tablenum[table-format=-2.2]{-0.90}\,(\tablenum[table-format=1.2]{0.32})  &  \num{0.14} \\
\addlinespace[2pt]
17  &  \tablenum[table-format=3.0]{240}  &  \tablenum[table-format=4.0]{60}  &  $\widehat{\theta}_{\mathrm{sup},\tau}$  &  \tablenum[table-format=-1.2]{-0.00}  &  \tablenum[table-format=2.2]{25.55}  &  \tablenum[table-format=2.2]{25.76}  &  \tablenum[table-format=2.2]{25.55}  &  \tablenum[table-format=1.3]{0.951}  &  --  &  --\\*
  &  \tablenum[table-format=3.0]{240}  &  \tablenum[table-format=4.0]{60}  &  $\widehat{\theta}_{\mathrm{un},\tau}$  &  \tablenum[table-format=-1.2]{-0.06}  &  \tablenum[table-format=2.2]{26.04}  &  \tablenum[table-format=2.2]{26.13}  &  \tablenum[table-format=2.2]{26.03}  &  \tablenum[table-format=1.3]{0.951}  &  \tablenum[table-format=-2.2]{-3.85}\,(\tablenum[table-format=1.2]{0.72})  &  --\\*
  &  \tablenum[table-format=3.0]{240}  &  \tablenum[table-format=4.0]{60}  &  $\widehat{\theta}_{\mathrm{SS},\tau}$  &  \tablenum[table-format=-1.2]{-0.01}  &  \tablenum[table-format=2.2]{25.54}  &  \tablenum[table-format=2.2]{25.75}  &  \tablenum[table-format=2.2]{25.54}  &  \tablenum[table-format=1.3]{0.951}  &  \tablenum[table-format=-2.2]{0.04}\,(\tablenum[table-format=1.2]{0.14})  &  \num{0.07} \\
\addlinespace[2pt]
18  &  \tablenum[table-format=3.0]{240}  &  \tablenum[table-format=4.0]{1200}  &  $\widehat{\theta}_{\mathrm{sup},\tau}$  &  \tablenum[table-format=-1.2]{-0.00}  &  \tablenum[table-format=2.2]{25.55}  &  \tablenum[table-format=2.2]{25.76}  &  \tablenum[table-format=2.2]{25.55}  &  \tablenum[table-format=1.3]{0.951}  &  --  &  --\\*
  &  \tablenum[table-format=3.0]{240}  &  \tablenum[table-format=4.0]{1200}  &  $\widehat{\theta}_{\mathrm{un},\tau}$  &  \tablenum[table-format=-1.2]{-0.03}  &  \tablenum[table-format=2.2]{25.66}  &  \tablenum[table-format=2.2]{25.70}  &  \tablenum[table-format=2.2]{25.66}  &  \tablenum[table-format=1.3]{0.950}  &  \tablenum[table-format=-2.2]{-0.90}\,(\tablenum[table-format=1.2]{0.42})  &  --\\*
  &  \tablenum[table-format=3.0]{240}  &  \tablenum[table-format=4.0]{1200}  &  $\widehat{\theta}_{\mathrm{SS},\tau}$  &  \tablenum[table-format=-1.2]{-0.02}  &  \tablenum[table-format=2.2]{25.62}  &  \tablenum[table-format=2.2]{25.70}  &  \tablenum[table-format=2.2]{25.61}  &  \tablenum[table-format=1.3]{0.950}  &  \tablenum[table-format=-2.2]{-0.52}\,(\tablenum[table-format=1.2]{0.35})  &  \num{0.28} \\
\addlinespace[2pt]
19  &  \tablenum[table-format=3.0]{240}  &  \tablenum[table-format=4.0]{240}  &  $\widehat{\theta}_{\mathrm{sup},\tau}$  &  \tablenum[table-format=-1.2]{-0.22}  &  \tablenum[table-format=2.2]{25.91}  &  \tablenum[table-format=2.2]{25.90}  &  \tablenum[table-format=2.2]{25.91}  &  \tablenum[table-format=1.3]{0.951}  &  --  &  --\\*
  &  \tablenum[table-format=3.0]{240}  &  \tablenum[table-format=4.0]{240}  &  $\widehat{\theta}_{\mathrm{un},\tau}$  &  \tablenum[table-format=-1.2]{-0.13}  &  \tablenum[table-format=2.2]{26.18}  &  \tablenum[table-format=2.2]{25.94}  &  \tablenum[table-format=2.2]{26.17}  &  \tablenum[table-format=1.3]{0.947}  &  \tablenum[table-format=-2.2]{-2.07}\,(\tablenum[table-format=1.2]{0.72})  &  --\\*
  &  \tablenum[table-format=3.0]{240}  &  \tablenum[table-format=4.0]{240}  &  $\widehat{\theta}_{\mathrm{SS},\tau}$  &  \tablenum[table-format=-1.2]{-0.17}  &  \tablenum[table-format=2.2]{25.87}  &  \tablenum[table-format=2.2]{25.76}  &  \tablenum[table-format=2.2]{25.87}  &  \tablenum[table-format=1.3]{0.948}  &  \tablenum[table-format=-2.2]{0.30}\,(\tablenum[table-format=1.2]{0.35})  &  \num{0.93} \\
\addlinespace[2pt]
20  &  \tablenum[table-format=3.0]{240}  &  \tablenum[table-format=4.0]{240}  &  $\widehat{\theta}_{\mathrm{sup},\tau}$  &  \tablenum[table-format=-1.2]{-0.19}  &  \tablenum[table-format=2.2]{25.31}  &  \tablenum[table-format=2.2]{25.56}  &  \tablenum[table-format=2.2]{25.31}  &  \tablenum[table-format=1.3]{0.949}  &  --  &  --\\*
  &  \tablenum[table-format=3.0]{240}  &  \tablenum[table-format=4.0]{240}  &  $\widehat{\theta}_{\mathrm{un},\tau}$  &  \tablenum[table-format=-1.2]{0.09}  &  \tablenum[table-format=2.2]{26.56}  &  \tablenum[table-format=2.2]{26.14}  &  \tablenum[table-format=2.2]{26.56}  &  \tablenum[table-format=1.3]{0.949}  &  \tablenum[table-format=-2.2]{-10.13}\,(\tablenum[table-format=1.2]{1.81})  &  --\\*
  &  \tablenum[table-format=3.0]{240}  &  \tablenum[table-format=4.0]{240}  &  $\widehat{\theta}_{\mathrm{SS},\tau}$  &  \tablenum[table-format=-1.2]{-0.05}  &  \tablenum[table-format=2.2]{24.99}  &  \tablenum[table-format=2.2]{24.99}  &  \tablenum[table-format=2.2]{24.98}  &  \tablenum[table-format=1.3]{0.952}  &  \tablenum[table-format=-2.2]{2.55}\,(\tablenum[table-format=1.2]{0.84})  &  \num{5.23} \\
\addlinespace[2pt]
\end{tabular}}
\setlength{\UniformWidth}{\linewidth}\setlength{\tabcolsep}{\dimexpr(\linewidth-\wd\UniformMeasure)/20\relax}
\ifdim\tabcolsep<0pt\PackageWarning{UnifiedTables}{Negative column gap}\fi
\setlength{\LTleft}{0pt plus 1fill}\setlength{\LTright}{0pt plus 1fill}
\begin{longtable}{@{}lcccccccccr@{}}
\caption{Fixed-horizon estimation.}\label{tab:restrictedMain}\\
\toprule Scenario & $n_0$ & $M_0$ & Method & Bias & \shortstack{Emp\\SD} & \shortstack{Mean\\SE} & RMSE & CP & \shortstack{$G_{A:\mathrm{sup}}$\\(MCSE)} & AVR (\%)\\\midrule\endfirsthead
\multicolumn{11}{l}{\textit{Table \thetable\ continued}}\\\toprule Scenario & $n_0$ & $M_0$ & Method & Bias & \shortstack{Emp\\SD} & \shortstack{Mean\\SE} & RMSE & CP & \shortstack{$G_{A:\mathrm{sup}}$\\(MCSE)} & AVR (\%)\\\midrule\endhead
\midrule\multicolumn{11}{r}{\textit{Continued on next page}}\\\endfoot
\endlastfoot

9  &  \tablenum[table-format=3.0]{240}  &  \tablenum[table-format=4.0]{240}  &  $\widehat{\theta}_{\mathrm{sup},\tau}$  &  \tablenum[table-format=-1.2]{-0.61}  &  \tablenum[table-format=2.2]{25.77}  &  \tablenum[table-format=2.2]{25.76}  &  \tablenum[table-format=2.2]{25.78}  &  \tablenum[table-format=1.3]{0.951}  &  --  &  --\\*
  &  \tablenum[table-format=3.0]{240}  &  \tablenum[table-format=4.0]{240}  &  $\widehat{\theta}_{\mathrm{un},\tau}$  &  \tablenum[table-format=-1.2]{-0.65}  &  \tablenum[table-format=2.2]{25.98}  &  \tablenum[table-format=2.2]{25.79}  &  \tablenum[table-format=2.2]{25.98}  &  \tablenum[table-format=1.3]{0.951}  &  \tablenum[table-format=-2.2]{-1.60}\,(\tablenum[table-format=1.2]{0.40})  &  --\\*
  &  \tablenum[table-format=3.0]{240}  &  \tablenum[table-format=4.0]{240}  &  $\widehat{\theta}_{\mathrm{SS},\tau}$  &  \tablenum[table-format=-1.2]{-0.63}  &  \tablenum[table-format=2.2]{25.82}  &  \tablenum[table-format=2.2]{25.74}  &  \tablenum[table-format=2.2]{25.83}  &  \tablenum[table-format=1.3]{0.952}  &  \tablenum[table-format=-2.2]{-0.41}\,(\tablenum[table-format=1.2]{0.20})  &  \num{0.00} \\
\addlinespace[2pt]
10  &  \tablenum[table-format=3.0]{240}  &  \tablenum[table-format=4.0]{240}  &  $\widehat{\theta}_{\mathrm{sup},\tau}$  &  \tablenum[table-format=-1.2]{0.21}  &  \tablenum[table-format=2.2]{25.92}  &  \tablenum[table-format=2.2]{25.77}  &  \tablenum[table-format=2.2]{25.92}  &  \tablenum[table-format=1.3]{0.943}  &  --  &  --\\*
  &  \tablenum[table-format=3.0]{240}  &  \tablenum[table-format=4.0]{240}  &  $\widehat{\theta}_{\mathrm{un},\tau}$  &  \tablenum[table-format=-1.2]{0.14}  &  \tablenum[table-format=2.2]{26.13}  &  \tablenum[table-format=2.2]{25.80}  &  \tablenum[table-format=2.2]{26.13}  &  \tablenum[table-format=1.3]{0.942}  &  \tablenum[table-format=-2.2]{-1.64}\,(\tablenum[table-format=1.2]{0.48})  &  --\\*
  &  \tablenum[table-format=3.0]{240}  &  \tablenum[table-format=4.0]{240}  &  $\widehat{\theta}_{\mathrm{SS},\tau}$  &  \tablenum[table-format=-1.2]{0.18}  &  \tablenum[table-format=2.2]{25.95}  &  \tablenum[table-format=2.2]{25.73}  &  \tablenum[table-format=2.2]{25.95}  &  \tablenum[table-format=1.3]{0.941}  &  \tablenum[table-format=-2.2]{-0.24}\,(\tablenum[table-format=1.2]{0.24})  &  \num{0.17} \\
\addlinespace[2pt]
11  &  \tablenum[table-format=3.0]{240}  &  \tablenum[table-format=4.0]{240}  &  $\widehat{\theta}_{\mathrm{sup},\tau}$  &  \tablenum[table-format=-1.2]{0.46}  &  \tablenum[table-format=2.2]{25.55}  &  \tablenum[table-format=2.2]{26.02}  &  \tablenum[table-format=2.2]{25.56}  &  \tablenum[table-format=1.3]{0.955}  &  --  &  --\\*
  &  \tablenum[table-format=3.0]{240}  &  \tablenum[table-format=4.0]{240}  &  $\widehat{\theta}_{\mathrm{un},\tau}$  &  \tablenum[table-format=-1.2]{0.63}  &  \tablenum[table-format=2.2]{25.63}  &  \tablenum[table-format=2.2]{26.05}  &  \tablenum[table-format=2.2]{25.64}  &  \tablenum[table-format=1.3]{0.954}  &  \tablenum[table-format=-2.2]{-0.61}\,(\tablenum[table-format=1.2]{1.17})  &  --\\*
  &  \tablenum[table-format=3.0]{240}  &  \tablenum[table-format=4.0]{240}  &  $\widehat{\theta}_{\mathrm{SS},\tau}$  &  \tablenum[table-format=-1.2]{0.54}  &  \tablenum[table-format=2.2]{25.14}  &  \tablenum[table-format=2.2]{25.61}  &  \tablenum[table-format=2.2]{25.15}  &  \tablenum[table-format=1.3]{0.954}  &  \tablenum[table-format=-2.2]{3.19}\,(\tablenum[table-format=1.2]{0.57})  &  \num{3.00} \\
\addlinespace[2pt]
12  &  \tablenum[table-format=3.0]{240}  &  \tablenum[table-format=4.0]{240}  &  $\widehat{\theta}_{\mathrm{sup},\tau}$  &  \tablenum[table-format=-1.2]{-0.04}  &  \tablenum[table-format=2.2]{25.95}  &  \tablenum[table-format=2.2]{25.97}  &  \tablenum[table-format=2.2]{25.95}  &  \tablenum[table-format=1.3]{0.951}  &  --  &  --\\*
  &  \tablenum[table-format=3.0]{240}  &  \tablenum[table-format=4.0]{240}  &  $\widehat{\theta}_{\mathrm{un},\tau}$  &  \tablenum[table-format=-1.2]{0.14}  &  \tablenum[table-format=2.2]{25.90}  &  \tablenum[table-format=2.2]{26.00}  &  \tablenum[table-format=2.2]{25.89}  &  \tablenum[table-format=1.3]{0.948}  &  \tablenum[table-format=-2.2]{0.44}\,(\tablenum[table-format=1.2]{1.71})  &  --\\*
  &  \tablenum[table-format=3.0]{240}  &  \tablenum[table-format=4.0]{240}  &  $\widehat{\theta}_{\mathrm{SS},\tau}$  &  \tablenum[table-format=-1.2]{0.05}  &  \tablenum[table-format=2.2]{24.83}  &  \tablenum[table-format=2.2]{24.90}  &  \tablenum[table-format=2.2]{24.82}  &  \tablenum[table-format=1.3]{0.951}  &  \tablenum[table-format=-2.2]{8.50}\,(\tablenum[table-format=1.2]{0.82})  &  \num{8.00} \\
\addlinespace[2pt]
13  &  \tablenum[table-format=3.0]{240}  &  \tablenum[table-format=4.0]{240}  &  $\widehat{\theta}_{\mathrm{sup},\tau}$  &  \tablenum[table-format=-1.2]{0.25}  &  \tablenum[table-format=2.2]{26.07}  &  \tablenum[table-format=2.2]{26.02}  &  \tablenum[table-format=2.2]{26.07}  &  \tablenum[table-format=1.3]{0.947}  &  --  &  --\\*
  &  \tablenum[table-format=3.0]{240}  &  \tablenum[table-format=4.0]{240}  &  $\widehat{\theta}_{\mathrm{un},\tau}$  &  \tablenum[table-format=-1.2]{0.20}  &  \tablenum[table-format=2.2]{26.27}  &  \tablenum[table-format=2.2]{26.05}  &  \tablenum[table-format=2.2]{26.27}  &  \tablenum[table-format=1.3]{0.948}  &  \tablenum[table-format=-2.2]{-1.54}\,(\tablenum[table-format=1.2]{0.42})  &  --\\*
  &  \tablenum[table-format=3.0]{240}  &  \tablenum[table-format=4.0]{240}  &  $\widehat{\theta}_{\mathrm{SS},\tau}$  &  \tablenum[table-format=-1.2]{0.23}  &  \tablenum[table-format=2.2]{26.12}  &  \tablenum[table-format=2.2]{26.00}  &  \tablenum[table-format=2.2]{26.12}  &  \tablenum[table-format=1.3]{0.948}  &  \tablenum[table-format=-2.2]{-0.34}\,(\tablenum[table-format=1.2]{0.21})  &  \num{0.00} \\
\addlinespace[2pt]
14  &  \tablenum[table-format=3.0]{240}  &  \tablenum[table-format=4.0]{240}  &  $\widehat{\theta}_{\mathrm{sup},\tau}$  &  \tablenum[table-format=-1.2]{0.25}  &  \tablenum[table-format=2.2]{26.07}  &  \tablenum[table-format=2.2]{26.02}  &  \tablenum[table-format=2.2]{26.07}  &  \tablenum[table-format=1.3]{0.947}  &  --  &  --\\*
  &  \tablenum[table-format=3.0]{240}  &  \tablenum[table-format=4.0]{240}  &  $\widehat{\theta}_{\mathrm{un},\tau}$  &  \tablenum[table-format=-1.2]{0.33}  &  \tablenum[table-format=2.2]{26.83}  &  \tablenum[table-format=2.2]{26.08}  &  \tablenum[table-format=2.2]{26.83}  &  \tablenum[table-format=1.3]{0.943}  &  \tablenum[table-format=-2.2]{-5.89}\,(\tablenum[table-format=1.2]{1.24})  &  --\\*
  &  \tablenum[table-format=3.0]{240}  &  \tablenum[table-format=4.0]{240}  &  $\widehat{\theta}_{\mathrm{SS},\tau}$  &  \tablenum[table-format=-1.2]{0.29}  &  \tablenum[table-format=2.2]{25.96}  &  \tablenum[table-format=2.2]{25.59}  &  \tablenum[table-format=2.2]{25.96}  &  \tablenum[table-format=1.3]{0.946}  &  \tablenum[table-format=-2.2]{0.87}\,(\tablenum[table-format=1.2]{0.60})  &  \num{2.97} \\
\addlinespace[2pt]
15  &  \tablenum[table-format=3.0]{240}  &  \tablenum[table-format=4.0]{240}  &  $\widehat{\theta}_{\mathrm{sup},\tau}$  &  \tablenum[table-format=-1.2]{0.07}  &  \tablenum[table-format=2.2]{28.68}  &  \tablenum[table-format=2.2]{28.38}  &  \tablenum[table-format=2.2]{28.68}  &  \tablenum[table-format=1.3]{0.947}  &  --  &  --\\*
  &  \tablenum[table-format=3.0]{240}  &  \tablenum[table-format=4.0]{240}  &  $\widehat{\theta}_{\mathrm{un},\tau}$  &  \tablenum[table-format=-1.2]{0.03}  &  \tablenum[table-format=2.2]{29.11}  &  \tablenum[table-format=2.2]{28.56}  &  \tablenum[table-format=2.2]{29.10}  &  \tablenum[table-format=1.3]{0.945}  &  \tablenum[table-format=-2.2]{-2.96}\,(\tablenum[table-format=1.2]{0.60})  &  --\\*
  &  \tablenum[table-format=3.0]{240}  &  \tablenum[table-format=4.0]{240}  &  $\widehat{\theta}_{\mathrm{SS},\tau}$  &  \tablenum[table-format=-1.2]{0.05}  &  \tablenum[table-format=2.2]{28.77}  &  \tablenum[table-format=2.2]{28.38}  &  \tablenum[table-format=2.2]{28.77}  &  \tablenum[table-format=1.3]{0.947}  &  \tablenum[table-format=-2.2]{-0.60}\,(\tablenum[table-format=1.2]{0.30})  &  \num{0.14} \\
\addlinespace[2pt]
16  &  \tablenum[table-format=3.0]{240}  &  \tablenum[table-format=4.0]{240}  &  $\widehat{\theta}_{\mathrm{sup},\tau}$  &  \tablenum[table-format=-1.2]{-0.88}  &  \tablenum[table-format=2.2]{27.76}  &  \tablenum[table-format=2.2]{27.52}  &  \tablenum[table-format=2.2]{27.77}  &  \tablenum[table-format=1.3]{0.949}  &  --  &  --\\*
  &  \tablenum[table-format=3.0]{240}  &  \tablenum[table-format=4.0]{240}  &  $\widehat{\theta}_{\mathrm{un},\tau}$  &  \tablenum[table-format=-1.2]{-0.78}  &  \tablenum[table-format=2.2]{28.29}  &  \tablenum[table-format=2.2]{27.74}  &  \tablenum[table-format=2.2]{28.30}  &  \tablenum[table-format=1.3]{0.947}  &  \tablenum[table-format=-2.2]{-3.88}\,(\tablenum[table-format=1.2]{0.64})  &  --\\*
  &  \tablenum[table-format=3.0]{240}  &  \tablenum[table-format=4.0]{240}  &  $\widehat{\theta}_{\mathrm{SS},\tau}$  &  \tablenum[table-format=-1.2]{-0.83}  &  \tablenum[table-format=2.2]{27.88}  &  \tablenum[table-format=2.2]{27.52}  &  \tablenum[table-format=2.2]{27.89}  &  \tablenum[table-format=1.3]{0.949}  &  \tablenum[table-format=-2.2]{-0.90}\,(\tablenum[table-format=1.2]{0.32})  &  \num{0.14} \\
\addlinespace[2pt]
17  &  \tablenum[table-format=3.0]{240}  &  \tablenum[table-format=4.0]{60}  &  $\widehat{\theta}_{\mathrm{sup},\tau}$  &  \tablenum[table-format=-1.2]{-0.00}  &  \tablenum[table-format=2.2]{25.55}  &  \tablenum[table-format=2.2]{25.76}  &  \tablenum[table-format=2.2]{25.55}  &  \tablenum[table-format=1.3]{0.951}  &  --  &  --\\*
  &  \tablenum[table-format=3.0]{240}  &  \tablenum[table-format=4.0]{60}  &  $\widehat{\theta}_{\mathrm{un},\tau}$  &  \tablenum[table-format=-1.2]{-0.06}  &  \tablenum[table-format=2.2]{26.04}  &  \tablenum[table-format=2.2]{26.13}  &  \tablenum[table-format=2.2]{26.03}  &  \tablenum[table-format=1.3]{0.951}  &  \tablenum[table-format=-2.2]{-3.85}\,(\tablenum[table-format=1.2]{0.72})  &  --\\*
  &  \tablenum[table-format=3.0]{240}  &  \tablenum[table-format=4.0]{60}  &  $\widehat{\theta}_{\mathrm{SS},\tau}$  &  \tablenum[table-format=-1.2]{-0.01}  &  \tablenum[table-format=2.2]{25.54}  &  \tablenum[table-format=2.2]{25.75}  &  \tablenum[table-format=2.2]{25.54}  &  \tablenum[table-format=1.3]{0.951}  &  \tablenum[table-format=-2.2]{0.04}\,(\tablenum[table-format=1.2]{0.14})  &  \num{0.07} \\
\addlinespace[2pt]
18  &  \tablenum[table-format=3.0]{240}  &  \tablenum[table-format=4.0]{1200}  &  $\widehat{\theta}_{\mathrm{sup},\tau}$  &  \tablenum[table-format=-1.2]{-0.00}  &  \tablenum[table-format=2.2]{25.55}  &  \tablenum[table-format=2.2]{25.76}  &  \tablenum[table-format=2.2]{25.55}  &  \tablenum[table-format=1.3]{0.951}  &  --  &  --\\*
  &  \tablenum[table-format=3.0]{240}  &  \tablenum[table-format=4.0]{1200}  &  $\widehat{\theta}_{\mathrm{un},\tau}$  &  \tablenum[table-format=-1.2]{-0.03}  &  \tablenum[table-format=2.2]{25.66}  &  \tablenum[table-format=2.2]{25.70}  &  \tablenum[table-format=2.2]{25.66}  &  \tablenum[table-format=1.3]{0.950}  &  \tablenum[table-format=-2.2]{-0.90}\,(\tablenum[table-format=1.2]{0.42})  &  --\\*
  &  \tablenum[table-format=3.0]{240}  &  \tablenum[table-format=4.0]{1200}  &  $\widehat{\theta}_{\mathrm{SS},\tau}$  &  \tablenum[table-format=-1.2]{-0.02}  &  \tablenum[table-format=2.2]{25.62}  &  \tablenum[table-format=2.2]{25.70}  &  \tablenum[table-format=2.2]{25.61}  &  \tablenum[table-format=1.3]{0.950}  &  \tablenum[table-format=-2.2]{-0.52}\,(\tablenum[table-format=1.2]{0.35})  &  \num{0.28} \\
\addlinespace[2pt]
19  &  \tablenum[table-format=3.0]{240}  &  \tablenum[table-format=4.0]{240}  &  $\widehat{\theta}_{\mathrm{sup},\tau}$  &  \tablenum[table-format=-1.2]{-0.22}  &  \tablenum[table-format=2.2]{25.91}  &  \tablenum[table-format=2.2]{25.90}  &  \tablenum[table-format=2.2]{25.91}  &  \tablenum[table-format=1.3]{0.951}  &  --  &  --\\*
  &  \tablenum[table-format=3.0]{240}  &  \tablenum[table-format=4.0]{240}  &  $\widehat{\theta}_{\mathrm{un},\tau}$  &  \tablenum[table-format=-1.2]{-0.13}  &  \tablenum[table-format=2.2]{26.18}  &  \tablenum[table-format=2.2]{25.94}  &  \tablenum[table-format=2.2]{26.17}  &  \tablenum[table-format=1.3]{0.947}  &  \tablenum[table-format=-2.2]{-2.07}\,(\tablenum[table-format=1.2]{0.72})  &  --\\*
  &  \tablenum[table-format=3.0]{240}  &  \tablenum[table-format=4.0]{240}  &  $\widehat{\theta}_{\mathrm{SS},\tau}$  &  \tablenum[table-format=-1.2]{-0.17}  &  \tablenum[table-format=2.2]{25.87}  &  \tablenum[table-format=2.2]{25.76}  &  \tablenum[table-format=2.2]{25.87}  &  \tablenum[table-format=1.3]{0.948}  &  \tablenum[table-format=-2.2]{0.30}\,(\tablenum[table-format=1.2]{0.35})  &  \num{0.93} \\
\addlinespace[2pt]
20  &  \tablenum[table-format=3.0]{240}  &  \tablenum[table-format=4.0]{240}  &  $\widehat{\theta}_{\mathrm{sup},\tau}$  &  \tablenum[table-format=-1.2]{-0.19}  &  \tablenum[table-format=2.2]{25.31}  &  \tablenum[table-format=2.2]{25.56}  &  \tablenum[table-format=2.2]{25.31}  &  \tablenum[table-format=1.3]{0.949}  &  --  &  --\\*
  &  \tablenum[table-format=3.0]{240}  &  \tablenum[table-format=4.0]{240}  &  $\widehat{\theta}_{\mathrm{un},\tau}$  &  \tablenum[table-format=-1.2]{0.09}  &  \tablenum[table-format=2.2]{26.56}  &  \tablenum[table-format=2.2]{26.14}  &  \tablenum[table-format=2.2]{26.56}  &  \tablenum[table-format=1.3]{0.949}  &  \tablenum[table-format=-2.2]{-10.13}\,(\tablenum[table-format=1.2]{1.81})  &  --\\*
  &  \tablenum[table-format=3.0]{240}  &  \tablenum[table-format=4.0]{240}  &  $\widehat{\theta}_{\mathrm{SS},\tau}$  &  \tablenum[table-format=-1.2]{-0.05}  &  \tablenum[table-format=2.2]{24.99}  &  \tablenum[table-format=2.2]{24.99}  &  \tablenum[table-format=2.2]{24.98}  &  \tablenum[table-format=1.3]{0.952}  &  \tablenum[table-format=-2.2]{2.55}\,(\tablenum[table-format=1.2]{0.84})  &  \num{5.23} \\*
\TableEndRule \multicolumn{11}{@{}p{\UniformWidth}@{}}{\textit{Note:} $n_0$ and $M_0$ are labeled and covariate-only pool sizes per group; supervised estimation uses the labeled data. EmpSD: empirical SD; MeanSE: mean estimated SE; RMSE: root mean squared error. Bias, EmpSD, MeanSE and RMSE use $10^{-3}$ units; CP: 95\% Wald coverage. The gain column reports $G_{A:\mathrm{sup}}$ for row estimator $A$; parentheses give paired MCSE in percentage points. AVR is the asymptotic variance reduction of Weighted relative to Supervised; $n_0=240$, $R=4,000$.}\\
\end{longtable}\endgroup

The paired working-feature comparison in Figure~\ref{fig:sFeatureBoxes} separates these two sources of information. A linear feature has zero projection under the centered quadratic law, whereas a quadratic feature makes that signal available for transfer. The fits use the same observed data and folds. The larger-sample results in Table~\ref{tab:additionalFixed} show how this feature advantage develops as the labeled sample grows.

\subsection{Auxiliary sample size and finite-sample performance}
\begin{figure}[!htbp]\centering
\makebox[\linewidth][c]{\includegraphics[width=\linewidth]{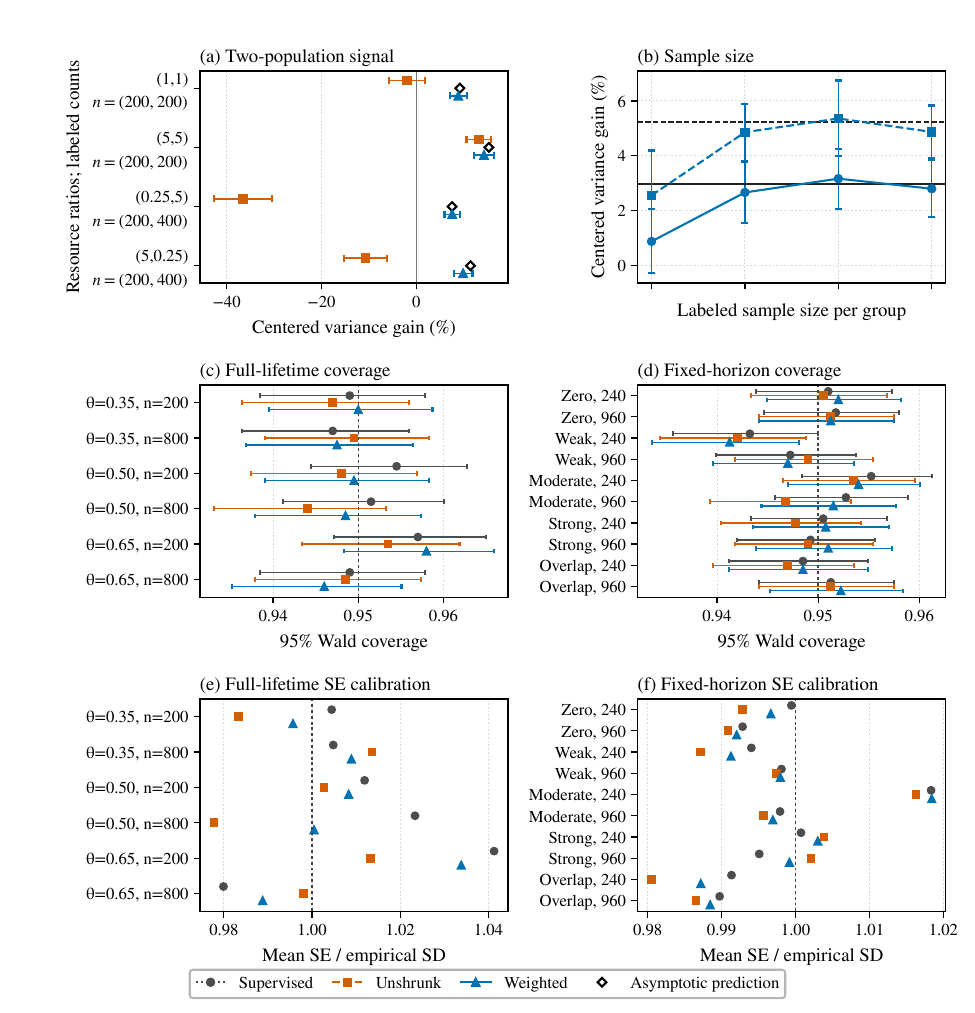}}
\caption{Resource gains and inferential calibration. (a) Scenario 22 compares dual-signal allocations; labels give both auxiliary-to-labeled ratios and labeled counts. Points are empirical gains, vertical bars are pointwise $\pm1.96$ paired MCSEs, and open diamonds are first-order predictions; each cell uses 4,000 repetitions. (b) The quadratic-feature and recorded-grid laws are shown at $n_0=240,480,960,1920$ with $M_0/n_0=1$; circles and squares distinguish the two laws, and horizontal lines mark their first-order predictions. Both laws use 4,000 repetitions at $n_0=240,960$. At $n_0=480,1920$, the quadratic-feature and recorded-grid laws use 4,000 and 8,000 repetitions, respectively. (c,e) Full-lifetime coverage and MeanSE/EmpSD are shown for three values of $\theta$ and $n_0=200,800$ (2,000 repetitions), with heavy-censoring cells at ratio 5 using 4,000. (d,f) Fixed-horizon results cover five signal/censoring settings at $n_0=240,960$ with equal auxiliary and labeled counts and 4,000 repetitions. Coverage intervals are Wilson intervals; reference lines are 0.95 for coverage and 1 for MeanSE/EmpSD.}\label{fig:theoryCalibration}
\end{figure}

The fixed-horizon resource study includes Scenario 21, which combines the strong-signal event law of Scenario 12 with overlap censoring and ratios 0.25, 1 and 5 at $n_0=240,960$. Sample-size paths for Scenarios 14 and 20 use $n_0=480,1920$ under the same respective laws; the zero-, weak- and strong-signal comparisons use $n_0=240$. Table~\ref{tab:confirmationMain} and Figure~\ref{fig:theoryCalibration} report the replication counts and results.

Table~\ref{tab:confirmationMain} quantifies the cost of an imprecise auxiliary mean when the pool is small. At $n_0=240$ and ratio 0.25, unshrunk transfer increases variance by 45.89\%, whereas weighting reduces it by 3.01\%. At ratio five and $n_0=960$, the weighted gain is \result{CresourceGain}\% (MCSE \result{CresourceMCSE}); RMSE changes follow the same direction in these cells. Zero- and weak-signal configurations show small losses, where little projected variance is available to offset regression fitting variability.

\begingroup\TableFont
\setlength{\tabcolsep}{0pt}\sbox{\UniformMeasure}{\begin{tabular}{@{}lrrrrrrrrr@{}}Scenario & $n_0$ & $M_0$ & AVR (\%) & $G_{\mathrm{SS:sup}}$ & $G_{\mathrm{un:sup}}$ & $G_{\mathrm{SS:un}}$ & RG (\%) & SE/SD & CP\\
\multicolumn{10}{l}{\textit{Finite resources under strong signal and overlap censoring}}\\\addlinespace[2pt]
21  &  \tablenum[table-format=4.0]{240}  &  \tablenum[table-format=4.0]{60}  &  \tablenum[table-format=2.2]{2.89}  &  \tablenum[table-format=-2.2]{3.01}\,(\tablenum[table-format=1.2]{0.37})  &  \tablenum[table-format=-2.2]{-45.89}\,(\tablenum[table-format=1.2]{2.34})  &  \tablenum[table-format=2.2]{33.52}\,(\tablenum[table-format=1.2]{0.88})  &  \tablenum[table-format=-1.2]{1.52}\,(\tablenum[table-format=1.2]{0.19})  &  \tablenum[table-format=1.3]{0.996}  &  \tablenum[table-format=1.3]{0.948} \\
21  &  \tablenum[table-format=4.0]{960}  &  \tablenum[table-format=4.0]{240}  &  \tablenum[table-format=2.2]{2.89}  &  \tablenum[table-format=-2.2]{2.22}\,(\tablenum[table-format=1.2]{0.38})  &  \tablenum[table-format=-2.2]{-50.05}\,(\tablenum[table-format=1.2]{2.38})  &  \tablenum[table-format=2.2]{34.83}\,(\tablenum[table-format=1.2]{0.84})  &  \tablenum[table-format=-1.2]{1.12}\,(\tablenum[table-format=1.2]{0.19})  &  \tablenum[table-format=1.3]{1.007}  &  \tablenum[table-format=1.3]{0.954} \\
21  &  \tablenum[table-format=4.0]{240}  &  \tablenum[table-format=4.0]{240}  &  \tablenum[table-format=2.2]{7.21}  &  \tablenum[table-format=-2.2]{6.32}\,(\tablenum[table-format=1.2]{0.58})  &  \tablenum[table-format=-2.2]{-3.61}\,(\tablenum[table-format=1.2]{1.23})  &  \tablenum[table-format=2.2]{9.58}\,(\tablenum[table-format=1.2]{0.56})  &  \tablenum[table-format=-1.2]{3.21}\,(\tablenum[table-format=1.2]{0.30})  &  \tablenum[table-format=1.3]{0.999}  &  \tablenum[table-format=1.3]{0.950} \\
21  &  \tablenum[table-format=4.0]{960}  &  \tablenum[table-format=4.0]{960}  &  \tablenum[table-format=2.2]{7.21}  &  \tablenum[table-format=-2.2]{6.75}\,(\tablenum[table-format=1.2]{0.56})  &  \tablenum[table-format=-2.2]{-0.91}\,(\tablenum[table-format=1.2]{1.17})  &  \tablenum[table-format=2.2]{7.59}\,(\tablenum[table-format=1.2]{0.56})  &  \tablenum[table-format=-1.2]{3.43}\,(\tablenum[table-format=1.2]{0.29})  &  \tablenum[table-format=1.3]{0.997}  &  \tablenum[table-format=1.3]{0.950} \\
21  &  \tablenum[table-format=4.0]{240}  &  \tablenum[table-format=4.0]{1200}  &  \tablenum[table-format=2.2]{12.02}  &  \tablenum[table-format=-2.2]{9.77}\,(\tablenum[table-format=1.2]{0.74})  &  \tablenum[table-format=-2.2]{8.45}\,(\tablenum[table-format=1.2]{0.89})  &  \tablenum[table-format=2.2]{1.45}\,(\tablenum[table-format=1.2]{0.17})  &  \tablenum[table-format=-1.2]{5.01}\,(\tablenum[table-format=1.2]{0.39})  &  \tablenum[table-format=1.3]{0.991}  &  \tablenum[table-format=1.3]{0.948} \\
21  &  \tablenum[table-format=4.0]{960}  &  \tablenum[table-format=4.0]{4800}  &  \tablenum[table-format=2.2]{12.02}  &  \tablenum[table-format=-2.2]{12.53}\,(\tablenum[table-format=1.2]{0.68})  &  \tablenum[table-format=-2.2]{11.98}\,(\tablenum[table-format=1.2]{0.82})  &  \tablenum[table-format=2.2]{0.63}\,(\tablenum[table-format=1.2]{0.17})  &  \tablenum[table-format=-1.2]{6.48}\,(\tablenum[table-format=1.2]{0.36})  &  \tablenum[table-format=1.3]{1.005}  &  \tablenum[table-format=1.3]{0.954} \\
\addlinespace[3pt]
\multicolumn{10}{l}{\textit{Working features and recorded-time comparisons}}\\\addlinespace[2pt]
14  &  \tablenum[table-format=4.0]{480}  &  \tablenum[table-format=4.0]{480}  &  \tablenum[table-format=2.2]{2.97}  &  \tablenum[table-format=-2.2]{2.66}\,(\tablenum[table-format=1.2]{0.56})  &  \tablenum[table-format=-2.2]{-1.63}\,(\tablenum[table-format=1.2]{1.15})  &  \tablenum[table-format=2.2]{4.21}\,(\tablenum[table-format=1.2]{0.55})  &  \tablenum[table-format=-1.2]{1.34}\,(\tablenum[table-format=1.2]{0.28})  &  \tablenum[table-format=1.3]{1.013}  &  \tablenum[table-format=1.3]{0.952} \\
14  &  \tablenum[table-format=4.0]{1920}  &  \tablenum[table-format=4.0]{1920}  &  \tablenum[table-format=2.2]{2.97}  &  \tablenum[table-format=-2.2]{2.79}\,(\tablenum[table-format=1.2]{0.53})  &  \tablenum[table-format=-2.2]{-0.58}\,(\tablenum[table-format=1.2]{1.08})  &  \tablenum[table-format=2.2]{3.35}\,(\tablenum[table-format=1.2]{0.53})  &  \tablenum[table-format=-1.2]{1.40}\,(\tablenum[table-format=1.2]{0.27})  &  \tablenum[table-format=1.3]{0.999}  &  \tablenum[table-format=1.3]{0.950} \\
20  &  \tablenum[table-format=4.0]{480}  &  \tablenum[table-format=4.0]{480}  &  \tablenum[table-format=2.2]{5.23}  &  \tablenum[table-format=-2.2]{4.86}\,(\tablenum[table-format=1.2]{0.53})  &  \tablenum[table-format=-2.2]{-2.49}\,(\tablenum[table-format=1.2]{1.11})  &  \tablenum[table-format=2.2]{7.17}\,(\tablenum[table-format=1.2]{0.52})  &  \tablenum[table-format=-1.2]{2.45}\,(\tablenum[table-format=1.2]{0.27})  &  \tablenum[table-format=1.3]{0.996}  &  \tablenum[table-format=1.3]{0.948} \\
20  &  \tablenum[table-format=4.0]{1920}  &  \tablenum[table-format=4.0]{1920}  &  \tablenum[table-format=2.2]{5.23}  &  \tablenum[table-format=-2.2]{4.87}\,(\tablenum[table-format=1.2]{0.49})  &  \tablenum[table-format=-2.2]{-1.49}\,(\tablenum[table-format=1.2]{1.02})  &  \tablenum[table-format=2.2]{6.26}\,(\tablenum[table-format=1.2]{0.49})  &  \tablenum[table-format=-1.2]{2.46}\,(\tablenum[table-format=1.2]{0.25})  &  \tablenum[table-format=1.3]{1.007}  &  \tablenum[table-format=1.3]{0.952} \\
\addlinespace[3pt]
\multicolumn{10}{l}{\textit{Zero-, weak- and strong-signal settings}}\\\addlinespace[2pt]
9  &  \tablenum[table-format=4.0]{240}  &  \tablenum[table-format=4.0]{240}  &  \tablenum[table-format=2.2]{0.00}  &  \tablenum[table-format=-2.2]{-0.31}\,(\tablenum[table-format=1.2]{0.20})  &  \tablenum[table-format=-2.2]{-1.42}\,(\tablenum[table-format=1.2]{0.40})  &  \tablenum[table-format=2.2]{1.09}\,(\tablenum[table-format=1.2]{0.19})  &  \tablenum[table-format=-1.2]{-0.16}\,(\tablenum[table-format=1.2]{0.10})  &  \tablenum[table-format=1.3]{1.000}  &  \tablenum[table-format=1.3]{0.948} \\
10  &  \tablenum[table-format=4.0]{240}  &  \tablenum[table-format=4.0]{240}  &  \tablenum[table-format=2.2]{0.17}  &  \tablenum[table-format=-2.2]{-0.30}\,(\tablenum[table-format=1.2]{0.23})  &  \tablenum[table-format=-2.2]{-1.68}\,(\tablenum[table-format=1.2]{0.47})  &  \tablenum[table-format=2.2]{1.36}\,(\tablenum[table-format=1.2]{0.23})  &  \tablenum[table-format=-1.2]{-0.15}\,(\tablenum[table-format=1.2]{0.11})  &  \tablenum[table-format=1.3]{0.989}  &  \tablenum[table-format=1.3]{0.945} \\
12  &  \tablenum[table-format=4.0]{240}  &  \tablenum[table-format=4.0]{240}  &  \tablenum[table-format=2.2]{8.00}  &  \tablenum[table-format=-2.2]{7.44}\,(\tablenum[table-format=1.2]{0.59})  &  \tablenum[table-format=-2.2]{-1.97}\,(\tablenum[table-format=1.2]{1.24})  &  \tablenum[table-format=2.2]{9.23}\,(\tablenum[table-format=1.2]{0.58})  &  \tablenum[table-format=-1.2]{3.80}\,(\tablenum[table-format=1.2]{0.31})  &  \tablenum[table-format=1.3]{0.988}  &  \tablenum[table-format=1.3]{0.948} \\
\addlinespace[3pt]
\end{tabular}}
\setlength{\UniformWidth}{\linewidth}\setlength{\tabcolsep}{\dimexpr(\linewidth-\wd\UniformMeasure)/18\relax}
\ifdim\tabcolsep<0pt\PackageWarning{UnifiedTables}{Negative column gap}\fi
\setlength{\LTleft}{0pt plus 1fill}\setlength{\LTright}{0pt plus 1fill}
\begin{longtable}{@{}lrrrrrrrrr@{}}
\caption{Auxiliary sample size and estimation accuracy.}\label{tab:confirmationMain}\\
\toprule Scenario & $n_0$ & $M_0$ & AVR (\%) & $G_{\mathrm{SS:sup}}$ & $G_{\mathrm{un:sup}}$ & $G_{\mathrm{SS:un}}$ & RG (\%) & SE/SD & CP\\\midrule\endfirsthead
\multicolumn{10}{l}{\textit{Table \thetable\ continued}}\\\toprule Scenario & $n_0$ & $M_0$ & AVR (\%) & $G_{\mathrm{SS:sup}}$ & $G_{\mathrm{un:sup}}$ & $G_{\mathrm{SS:un}}$ & RG (\%) & SE/SD & CP\\\midrule\endhead
\midrule\multicolumn{10}{r}{\textit{Continued on next page}}\\\endfoot
\endlastfoot

\multicolumn{10}{l}{\textit{Finite resources under strong signal and overlap censoring}}\\\addlinespace[2pt]
21  &  \tablenum[table-format=4.0]{240}  &  \tablenum[table-format=4.0]{60}  &  \tablenum[table-format=2.2]{2.89}  &  \tablenum[table-format=-2.2]{3.01}\,(\tablenum[table-format=1.2]{0.37})  &  \tablenum[table-format=-2.2]{-45.89}\,(\tablenum[table-format=1.2]{2.34})  &  \tablenum[table-format=2.2]{33.52}\,(\tablenum[table-format=1.2]{0.88})  &  \tablenum[table-format=-1.2]{1.52}\,(\tablenum[table-format=1.2]{0.19})  &  \tablenum[table-format=1.3]{0.996}  &  \tablenum[table-format=1.3]{0.948} \\
21  &  \tablenum[table-format=4.0]{960}  &  \tablenum[table-format=4.0]{240}  &  \tablenum[table-format=2.2]{2.89}  &  \tablenum[table-format=-2.2]{2.22}\,(\tablenum[table-format=1.2]{0.38})  &  \tablenum[table-format=-2.2]{-50.05}\,(\tablenum[table-format=1.2]{2.38})  &  \tablenum[table-format=2.2]{34.83}\,(\tablenum[table-format=1.2]{0.84})  &  \tablenum[table-format=-1.2]{1.12}\,(\tablenum[table-format=1.2]{0.19})  &  \tablenum[table-format=1.3]{1.007}  &  \tablenum[table-format=1.3]{0.954} \\
21  &  \tablenum[table-format=4.0]{240}  &  \tablenum[table-format=4.0]{240}  &  \tablenum[table-format=2.2]{7.21}  &  \tablenum[table-format=-2.2]{6.32}\,(\tablenum[table-format=1.2]{0.58})  &  \tablenum[table-format=-2.2]{-3.61}\,(\tablenum[table-format=1.2]{1.23})  &  \tablenum[table-format=2.2]{9.58}\,(\tablenum[table-format=1.2]{0.56})  &  \tablenum[table-format=-1.2]{3.21}\,(\tablenum[table-format=1.2]{0.30})  &  \tablenum[table-format=1.3]{0.999}  &  \tablenum[table-format=1.3]{0.950} \\
21  &  \tablenum[table-format=4.0]{960}  &  \tablenum[table-format=4.0]{960}  &  \tablenum[table-format=2.2]{7.21}  &  \tablenum[table-format=-2.2]{6.75}\,(\tablenum[table-format=1.2]{0.56})  &  \tablenum[table-format=-2.2]{-0.91}\,(\tablenum[table-format=1.2]{1.17})  &  \tablenum[table-format=2.2]{7.59}\,(\tablenum[table-format=1.2]{0.56})  &  \tablenum[table-format=-1.2]{3.43}\,(\tablenum[table-format=1.2]{0.29})  &  \tablenum[table-format=1.3]{0.997}  &  \tablenum[table-format=1.3]{0.950} \\
21  &  \tablenum[table-format=4.0]{240}  &  \tablenum[table-format=4.0]{1200}  &  \tablenum[table-format=2.2]{12.02}  &  \tablenum[table-format=-2.2]{9.77}\,(\tablenum[table-format=1.2]{0.74})  &  \tablenum[table-format=-2.2]{8.45}\,(\tablenum[table-format=1.2]{0.89})  &  \tablenum[table-format=2.2]{1.45}\,(\tablenum[table-format=1.2]{0.17})  &  \tablenum[table-format=-1.2]{5.01}\,(\tablenum[table-format=1.2]{0.39})  &  \tablenum[table-format=1.3]{0.991}  &  \tablenum[table-format=1.3]{0.948} \\
21  &  \tablenum[table-format=4.0]{960}  &  \tablenum[table-format=4.0]{4800}  &  \tablenum[table-format=2.2]{12.02}  &  \tablenum[table-format=-2.2]{12.53}\,(\tablenum[table-format=1.2]{0.68})  &  \tablenum[table-format=-2.2]{11.98}\,(\tablenum[table-format=1.2]{0.82})  &  \tablenum[table-format=2.2]{0.63}\,(\tablenum[table-format=1.2]{0.17})  &  \tablenum[table-format=-1.2]{6.48}\,(\tablenum[table-format=1.2]{0.36})  &  \tablenum[table-format=1.3]{1.005}  &  \tablenum[table-format=1.3]{0.954} \\
\addlinespace[3pt]
\multicolumn{10}{l}{\textit{Working features and recorded-time comparisons}}\\\addlinespace[2pt]
14  &  \tablenum[table-format=4.0]{480}  &  \tablenum[table-format=4.0]{480}  &  \tablenum[table-format=2.2]{2.97}  &  \tablenum[table-format=-2.2]{2.66}\,(\tablenum[table-format=1.2]{0.56})  &  \tablenum[table-format=-2.2]{-1.63}\,(\tablenum[table-format=1.2]{1.15})  &  \tablenum[table-format=2.2]{4.21}\,(\tablenum[table-format=1.2]{0.55})  &  \tablenum[table-format=-1.2]{1.34}\,(\tablenum[table-format=1.2]{0.28})  &  \tablenum[table-format=1.3]{1.013}  &  \tablenum[table-format=1.3]{0.952} \\
14  &  \tablenum[table-format=4.0]{1920}  &  \tablenum[table-format=4.0]{1920}  &  \tablenum[table-format=2.2]{2.97}  &  \tablenum[table-format=-2.2]{2.79}\,(\tablenum[table-format=1.2]{0.53})  &  \tablenum[table-format=-2.2]{-0.58}\,(\tablenum[table-format=1.2]{1.08})  &  \tablenum[table-format=2.2]{3.35}\,(\tablenum[table-format=1.2]{0.53})  &  \tablenum[table-format=-1.2]{1.40}\,(\tablenum[table-format=1.2]{0.27})  &  \tablenum[table-format=1.3]{0.999}  &  \tablenum[table-format=1.3]{0.950} \\
20  &  \tablenum[table-format=4.0]{480}  &  \tablenum[table-format=4.0]{480}  &  \tablenum[table-format=2.2]{5.23}  &  \tablenum[table-format=-2.2]{4.86}\,(\tablenum[table-format=1.2]{0.53})  &  \tablenum[table-format=-2.2]{-2.49}\,(\tablenum[table-format=1.2]{1.11})  &  \tablenum[table-format=2.2]{7.17}\,(\tablenum[table-format=1.2]{0.52})  &  \tablenum[table-format=-1.2]{2.45}\,(\tablenum[table-format=1.2]{0.27})  &  \tablenum[table-format=1.3]{0.996}  &  \tablenum[table-format=1.3]{0.948} \\
20  &  \tablenum[table-format=4.0]{1920}  &  \tablenum[table-format=4.0]{1920}  &  \tablenum[table-format=2.2]{5.23}  &  \tablenum[table-format=-2.2]{4.87}\,(\tablenum[table-format=1.2]{0.49})  &  \tablenum[table-format=-2.2]{-1.49}\,(\tablenum[table-format=1.2]{1.02})  &  \tablenum[table-format=2.2]{6.26}\,(\tablenum[table-format=1.2]{0.49})  &  \tablenum[table-format=-1.2]{2.46}\,(\tablenum[table-format=1.2]{0.25})  &  \tablenum[table-format=1.3]{1.007}  &  \tablenum[table-format=1.3]{0.952} \\
\addlinespace[3pt]
\multicolumn{10}{l}{\textit{Zero-, weak- and strong-signal settings}}\\\addlinespace[2pt]
9  &  \tablenum[table-format=4.0]{240}  &  \tablenum[table-format=4.0]{240}  &  \tablenum[table-format=2.2]{0.00}  &  \tablenum[table-format=-2.2]{-0.31}\,(\tablenum[table-format=1.2]{0.20})  &  \tablenum[table-format=-2.2]{-1.42}\,(\tablenum[table-format=1.2]{0.40})  &  \tablenum[table-format=2.2]{1.09}\,(\tablenum[table-format=1.2]{0.19})  &  \tablenum[table-format=-1.2]{-0.16}\,(\tablenum[table-format=1.2]{0.10})  &  \tablenum[table-format=1.3]{1.000}  &  \tablenum[table-format=1.3]{0.948} \\
10  &  \tablenum[table-format=4.0]{240}  &  \tablenum[table-format=4.0]{240}  &  \tablenum[table-format=2.2]{0.17}  &  \tablenum[table-format=-2.2]{-0.30}\,(\tablenum[table-format=1.2]{0.23})  &  \tablenum[table-format=-2.2]{-1.68}\,(\tablenum[table-format=1.2]{0.47})  &  \tablenum[table-format=2.2]{1.36}\,(\tablenum[table-format=1.2]{0.23})  &  \tablenum[table-format=-1.2]{-0.15}\,(\tablenum[table-format=1.2]{0.11})  &  \tablenum[table-format=1.3]{0.989}  &  \tablenum[table-format=1.3]{0.945} \\
12  &  \tablenum[table-format=4.0]{240}  &  \tablenum[table-format=4.0]{240}  &  \tablenum[table-format=2.2]{8.00}  &  \tablenum[table-format=-2.2]{7.44}\,(\tablenum[table-format=1.2]{0.59})  &  \tablenum[table-format=-2.2]{-1.97}\,(\tablenum[table-format=1.2]{1.24})  &  \tablenum[table-format=2.2]{9.23}\,(\tablenum[table-format=1.2]{0.58})  &  \tablenum[table-format=-1.2]{3.80}\,(\tablenum[table-format=1.2]{0.31})  &  \tablenum[table-format=1.3]{0.988}  &  \tablenum[table-format=1.3]{0.948} \\*
\TableEndRule \multicolumn{10}{@{}p{\UniformWidth}@{}}{\textit{Note:} $n_0$ and $M_0$ are the labeled and covariate-only sizes per group. The gain columns compare Weighted and Unshrunk with the stated reference; $G_{A:B}=100(1-s_A^2/s_B^2)$. RG is $100(1-\mathrm{RMSE}_{\mathrm{SS}}/\mathrm{RMSE}_{\mathrm{sup}})$; parentheses give paired MCSE in percentage points. SE/SD is MeanSE/EmpSD; CP is 95\% Wald coverage. AVR is the asymptotic variance reduction of Weighted relative to Supervised; SE/SD and CP describe Weighted. Scenario 14 uses 4,000 repetitions; the other scenarios use 8,000 per cell.}\\
\end{longtable}\endgroup

The sample-size paths in Figure~\ref{fig:theoryCalibration}(b) follow the quadratic-feature and recorded-grid laws from $n_0=240$ to 1,920. At the larger sizes, empirical gains approach the first-order values of 2.97\% and 5.23\%, respectively, with pointwise intervals showing the paired Monte Carlo uncertainty. In the matched calibration panels (c)--(f), MeanSE/EmpSD remains near one and coverage varies around 95\%; the cell-wise Wilson intervals show where small departures are most apparent.

Coverage can also fall below the nominal level in full-lifetime designs that satisfy the model assumptions. Table~\ref{tab:full_lowcoverage} reports the settings with the lowest coverage and shows how the estimated standard errors compare with the observed dispersion.

\subsection{Comparison with covariate augmentation}\label{sec:covariatecomparison}
To compare the two adjustments, we use the same supervised score and fixed working space, a common covariate law and balanced labeled counts. At $M_k=n_k=n_0$, trial augmentation reduces first-order variance by $\Var(p_1-p_2)/(2n_0)$, while four-pool transfer reduces it by $\{\Var(p_1)+\Var(p_2)\}/(2n_0)$. Subtracting the trial reduction from the four-pool reduction gives $\Cov(p_1,p_2)/n_0$. Appendix~\ref{s:commoncomparison} derives the comparison for general auxiliary sizes and relates it to trial augmentation on a fixed grid.

In the moderate- and strong-signal examples, the trial-augmentation direction cancels. In the recorded-grid example, its theoretical reduction is about 10.37\%, compared with 5.23\% for Weighted. The paired results show that the relative value of labeled imbalance adjustment and auxiliary-pool transfer depends on the direction as well as the amount of covariate information.

\begin{figure}[!htbp]\centering
\makebox[\linewidth][c]{\includegraphics[width=\linewidth]{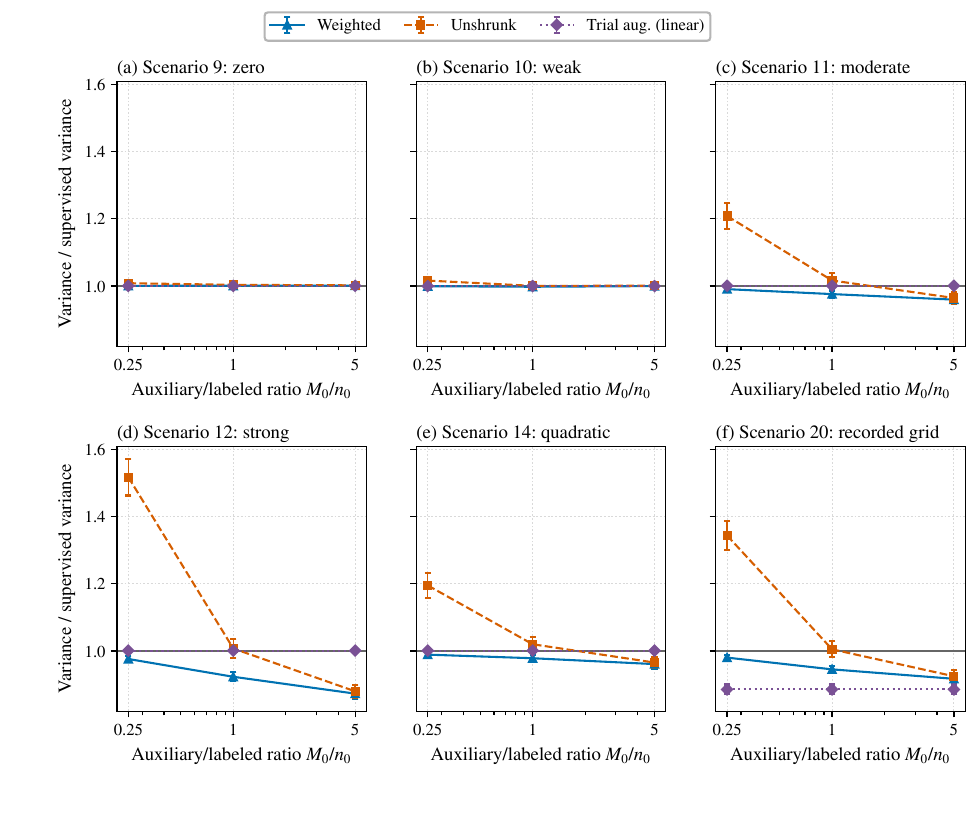}}
\caption{Fixed-horizon comparisons across auxiliary sample sizes. Each population has 960 labeled observations; $M_0/n_0\in\{0.25,1,5\}$. The series show centered variance divided by supervised variance for Weighted, Unshrunk and fixed-linear-basis trial augmentation; the horizontal reference is one. Pointwise intervals use paired Monte Carlo uncertainty. Scenarios 9, 10, 11 and 14 use 4,000 repetitions per cell; Scenarios 12 and 20 use 6,000. The trial comparison requires the common-covariate-distribution submodel.}\label{fig:competitiveResources}
\end{figure}

The numerical comparison retains the common covariate law and balanced labeled counts. It varies the auxiliary ratio across three levels for each of six fixed-horizon mechanisms, with 960 labeled observations per population. Trial augmentation follows the labeled-covariate adjustment structure studied for censored Mann--Whitney effects in randomized trials \citep{zhang2020trial}; the weighted estimator also uses the specified covariate-only pool. Figure~\ref{fig:competitiveResources} expresses each variance relative to the same supervised variance: values below one indicate a precision gain. This common scale shows both the resource response of the weighted procedure and the precision available from labeled imbalance adjustment.

Table~\ref{tab:competitiveResources} reports the direct paired contrasts. In Scenario 12, weighted variance is lower than that of fixed-linear-basis trial augmentation by 2.49\%, 7.77\% and 12.78\% as the auxiliary ratio increases. In Scenario 20 the corresponding gains are $-10.68\%$, $-6.73\%$ and $-3.56\%$, with trial augmentation more precise. Its nonzero imbalance direction already captures useful information from the labeled samples; expanding the auxiliary pool narrows the difference. Together with the common-score formula, these comparisons identify signal direction and resource availability as distinct determinants of the preferred adjustment. The numerical comparison uses fixed-basis learners; additional precision and calibration results for both trial-augmentation bases appear in Table~\ref{tab:competitiveMethods}, where the fixed-basis procedures are labeled Trial aug. (linear) and Trial aug. (categorical).

\FloatBarrier
\begingroup\TableFont
\setlength{\tabcolsep}{0pt}\sbox{\UniformMeasure}{\begin{tabular}{@{}llrrrrrr@{}}Scenario & $n_0,M_0$ & $G_{\mathrm{SS:sup}}$ & $G_{\mathrm{SS:un}}$ & $G_{\mathrm{SS:linear}}$ & SE/SD & CP (\%) & $R$ \\
9  &  960, 240  &  \tablenum[table-format=-2.2]{-0.03}\,(\tablenum[table-format=1.2]{0.06})  &  \tablenum[table-format=2.2]{0.77}\,(\tablenum[table-format=1.2]{0.22})  &  \tablenum[table-format=-2.2]{-0.01}\,(\tablenum[table-format=1.2]{0.09})  &  \tablenum[table-format=1.3]{1.012}  &  \tablenum[table-format=2.2]{95.15}  &  \tablenum[table-format=4.0]{4000} \\
9  &  960, 960  &  \tablenum[table-format=-2.2]{-0.06}\,(\tablenum[table-format=1.2]{0.10})  &  \tablenum[table-format=2.2]{0.27}\,(\tablenum[table-format=1.2]{0.10})  &  \tablenum[table-format=-2.2]{-0.04}\,(\tablenum[table-format=1.2]{0.10})  &  \tablenum[table-format=1.3]{1.012}  &  \tablenum[table-format=2.2]{95.03}  &  \tablenum[table-format=4.0]{4000} \\
9  &  960, 4800  &  \tablenum[table-format=-2.2]{-0.10}\,(\tablenum[table-format=1.2]{0.14})  &  \tablenum[table-format=2.2]{0.07}\,(\tablenum[table-format=1.2]{0.03})  &  \tablenum[table-format=-2.2]{-0.08}\,(\tablenum[table-format=1.2]{0.14})  &  \tablenum[table-format=1.3]{1.011}  &  \tablenum[table-format=2.2]{95.12}  &  \tablenum[table-format=4.0]{4000} \\
10  &  960, 240  &  \tablenum[table-format=-2.2]{0.08}\,(\tablenum[table-format=1.2]{0.10})  &  \tablenum[table-format=2.2]{1.64}\,(\tablenum[table-format=1.2]{0.39})  &  \tablenum[table-format=-2.2]{0.04}\,(\tablenum[table-format=1.2]{0.12})  &  \tablenum[table-format=1.3]{0.990}  &  \tablenum[table-format=2.2]{94.62}  &  \tablenum[table-format=4.0]{4000} \\
10  &  960, 960  &  \tablenum[table-format=-2.2]{0.25}\,(\tablenum[table-format=1.2]{0.16})  &  \tablenum[table-format=2.2]{0.27}\,(\tablenum[table-format=1.2]{0.16})  &  \tablenum[table-format=-2.2]{0.22}\,(\tablenum[table-format=1.2]{0.16})  &  \tablenum[table-format=1.3]{0.991}  &  \tablenum[table-format=2.2]{94.45}  &  \tablenum[table-format=4.0]{4000} \\
10  &  960, 4800  &  \tablenum[table-format=-2.2]{0.04}\,(\tablenum[table-format=1.2]{0.21})  &  \tablenum[table-format=2.2]{0.10}\,(\tablenum[table-format=1.2]{0.04})  &  \tablenum[table-format=-2.2]{0.01}\,(\tablenum[table-format=1.2]{0.21})  &  \tablenum[table-format=1.3]{0.989}  &  \tablenum[table-format=2.2]{94.53}  &  \tablenum[table-format=4.0]{4000} \\
11  &  960, 240  &  \tablenum[table-format=-2.2]{0.96}\,(\tablenum[table-format=1.2]{0.35})  &  \tablenum[table-format=2.2]{18.01}\,(\tablenum[table-format=1.2]{1.07})  &  \tablenum[table-format=-2.2]{1.00}\,(\tablenum[table-format=1.2]{0.36})  &  \tablenum[table-format=1.3]{1.001}  &  \tablenum[table-format=2.2]{94.90}  &  \tablenum[table-format=4.0]{4000} \\
11  &  960, 960  &  \tablenum[table-format=-2.2]{2.43}\,(\tablenum[table-format=1.2]{0.55})  &  \tablenum[table-format=2.2]{3.99}\,(\tablenum[table-format=1.2]{0.54})  &  \tablenum[table-format=-2.2]{2.47}\,(\tablenum[table-format=1.2]{0.55})  &  \tablenum[table-format=1.3]{0.999}  &  \tablenum[table-format=2.2]{94.97}  &  \tablenum[table-format=4.0]{4000} \\
11  &  960, 4800  &  \tablenum[table-format=-2.2]{4.08}\,(\tablenum[table-format=1.2]{0.71})  &  \tablenum[table-format=2.2]{0.50}\,(\tablenum[table-format=1.2]{0.15})  &  \tablenum[table-format=-2.2]{4.12}\,(\tablenum[table-format=1.2]{0.71})  &  \tablenum[table-format=1.3]{0.997}  &  \tablenum[table-format=2.2]{94.95}  &  \tablenum[table-format=4.0]{4000} \\
12  &  960, 240  &  \tablenum[table-format=-2.2]{2.41}\,(\tablenum[table-format=1.2]{0.45})  &  \tablenum[table-format=2.2]{35.66}\,(\tablenum[table-format=1.2]{0.96})  &  \tablenum[table-format=-2.2]{2.49}\,(\tablenum[table-format=1.2]{0.45})  &  \tablenum[table-format=1.3]{0.997}  &  \tablenum[table-format=2.2]{95.17}  &  \tablenum[table-format=4.0]{6000} \\
12  &  960, 960  &  \tablenum[table-format=-2.2]{7.69}\,(\tablenum[table-format=1.2]{0.68})  &  \tablenum[table-format=2.2]{8.34}\,(\tablenum[table-format=1.2]{0.68})  &  \tablenum[table-format=-2.2]{7.77}\,(\tablenum[table-format=1.2]{0.68})  &  \tablenum[table-format=1.3]{0.999}  &  \tablenum[table-format=2.2]{94.97}  &  \tablenum[table-format=4.0]{6000} \\
12  &  960, 4800  &  \tablenum[table-format=-2.2]{12.71}\,(\tablenum[table-format=1.2]{0.84})  &  \tablenum[table-format=2.2]{0.78}\,(\tablenum[table-format=1.2]{0.20})  &  \tablenum[table-format=-2.2]{12.78}\,(\tablenum[table-format=1.2]{0.84})  &  \tablenum[table-format=1.3]{0.997}  &  \tablenum[table-format=2.2]{94.78}  &  \tablenum[table-format=4.0]{6000} \\
14  &  960, 240  &  \tablenum[table-format=-2.2]{1.11}\,(\tablenum[table-format=1.2]{0.34})  &  \tablenum[table-format=2.2]{17.23}\,(\tablenum[table-format=1.2]{1.07})  &  \tablenum[table-format=-2.2]{1.21}\,(\tablenum[table-format=1.2]{0.36})  &  \tablenum[table-format=1.3]{0.994}  &  \tablenum[table-format=2.2]{95.00}  &  \tablenum[table-format=4.0]{4000} \\
14  &  960, 960  &  \tablenum[table-format=-2.2]{2.18}\,(\tablenum[table-format=1.2]{0.55})  &  \tablenum[table-format=2.2]{4.08}\,(\tablenum[table-format=1.2]{0.54})  &  \tablenum[table-format=-2.2]{2.28}\,(\tablenum[table-format=1.2]{0.56})  &  \tablenum[table-format=1.3]{0.990}  &  \tablenum[table-format=2.2]{94.97}  &  \tablenum[table-format=4.0]{4000} \\
14  &  960, 4800  &  \tablenum[table-format=-2.2]{3.92}\,(\tablenum[table-format=1.2]{0.70})  &  \tablenum[table-format=2.2]{0.47}\,(\tablenum[table-format=1.2]{0.15})  &  \tablenum[table-format=-2.2]{4.02}\,(\tablenum[table-format=1.2]{0.69})  &  \tablenum[table-format=1.3]{0.989}  &  \tablenum[table-format=2.2]{95.00}  &  \tablenum[table-format=4.0]{4000} \\
20  &  960, 240  &  \tablenum[table-format=-2.2]{1.99}\,(\tablenum[table-format=1.2]{0.38})  &  \tablenum[table-format=2.2]{27.03}\,(\tablenum[table-format=1.2]{0.98})  &  \tablenum[table-format=-2.2]{-10.68}\,(\tablenum[table-format=1.2]{0.85})  &  \tablenum[table-format=1.3]{0.994}  &  \tablenum[table-format=2.2]{94.53}  &  \tablenum[table-format=4.0]{6000} \\
20  &  960, 960  &  \tablenum[table-format=-2.2]{5.49}\,(\tablenum[table-format=1.2]{0.58})  &  \tablenum[table-format=2.2]{5.94}\,(\tablenum[table-format=1.2]{0.58})  &  \tablenum[table-format=-2.2]{-6.73}\,(\tablenum[table-format=1.2]{0.68})  &  \tablenum[table-format=1.3]{0.997}  &  \tablenum[table-format=2.2]{94.68}  &  \tablenum[table-format=4.0]{6000} \\
20  &  960, 4800  &  \tablenum[table-format=-2.2]{8.29}\,(\tablenum[table-format=1.2]{0.75})  &  \tablenum[table-format=2.2]{0.81}\,(\tablenum[table-format=1.2]{0.17})  &  \tablenum[table-format=-2.2]{-3.56}\,(\tablenum[table-format=1.2]{0.47})  &  \tablenum[table-format=1.3]{0.994}  &  \tablenum[table-format=2.2]{94.63}  &  \tablenum[table-format=4.0]{6000} \\
\end{tabular}}
\setlength{\UniformWidth}{\linewidth}\setlength{\tabcolsep}{\dimexpr(\linewidth-\wd\UniformMeasure)/14\relax}
\ifdim\tabcolsep<0pt\PackageWarning{UnifiedTables}{Negative column gap}\fi
\setlength{\LTleft}{0pt plus 1fill}\setlength{\LTright}{0pt plus 1fill}
\begin{longtable}{@{}llrrrrrr@{}}
\caption{Fixed-horizon resource and augmentation comparisons.}\label{tab:competitiveResources}\\
\toprule Scenario & $n_0,M_0$ & $G_{\mathrm{SS:sup}}$ & $G_{\mathrm{SS:un}}$ & $G_{\mathrm{SS:linear}}$ & SE/SD & CP (\%) & $R$ \\\midrule\endfirsthead
\multicolumn{8}{l}{\textit{Table \thetable\ continued}}\\\toprule Scenario & $n_0,M_0$ & $G_{\mathrm{SS:sup}}$ & $G_{\mathrm{SS:un}}$ & $G_{\mathrm{SS:linear}}$ & SE/SD & CP (\%) & $R$ \\\midrule\endhead
\midrule\multicolumn{8}{r}{\textit{Continued on next page}}\\\endfoot
\endlastfoot

9  &  960, 240  &  \tablenum[table-format=-2.2]{-0.03}\,(\tablenum[table-format=1.2]{0.06})  &  \tablenum[table-format=2.2]{0.77}\,(\tablenum[table-format=1.2]{0.22})  &  \tablenum[table-format=-2.2]{-0.01}\,(\tablenum[table-format=1.2]{0.09})  &  \tablenum[table-format=1.3]{1.012}  &  \tablenum[table-format=2.2]{95.15}  &  \tablenum[table-format=4.0]{4000} \\
9  &  960, 960  &  \tablenum[table-format=-2.2]{-0.06}\,(\tablenum[table-format=1.2]{0.10})  &  \tablenum[table-format=2.2]{0.27}\,(\tablenum[table-format=1.2]{0.10})  &  \tablenum[table-format=-2.2]{-0.04}\,(\tablenum[table-format=1.2]{0.10})  &  \tablenum[table-format=1.3]{1.012}  &  \tablenum[table-format=2.2]{95.03}  &  \tablenum[table-format=4.0]{4000} \\
9  &  960, 4800  &  \tablenum[table-format=-2.2]{-0.10}\,(\tablenum[table-format=1.2]{0.14})  &  \tablenum[table-format=2.2]{0.07}\,(\tablenum[table-format=1.2]{0.03})  &  \tablenum[table-format=-2.2]{-0.08}\,(\tablenum[table-format=1.2]{0.14})  &  \tablenum[table-format=1.3]{1.011}  &  \tablenum[table-format=2.2]{95.12}  &  \tablenum[table-format=4.0]{4000} \\
10  &  960, 240  &  \tablenum[table-format=-2.2]{0.08}\,(\tablenum[table-format=1.2]{0.10})  &  \tablenum[table-format=2.2]{1.64}\,(\tablenum[table-format=1.2]{0.39})  &  \tablenum[table-format=-2.2]{0.04}\,(\tablenum[table-format=1.2]{0.12})  &  \tablenum[table-format=1.3]{0.990}  &  \tablenum[table-format=2.2]{94.62}  &  \tablenum[table-format=4.0]{4000} \\
10  &  960, 960  &  \tablenum[table-format=-2.2]{0.25}\,(\tablenum[table-format=1.2]{0.16})  &  \tablenum[table-format=2.2]{0.27}\,(\tablenum[table-format=1.2]{0.16})  &  \tablenum[table-format=-2.2]{0.22}\,(\tablenum[table-format=1.2]{0.16})  &  \tablenum[table-format=1.3]{0.991}  &  \tablenum[table-format=2.2]{94.45}  &  \tablenum[table-format=4.0]{4000} \\
10  &  960, 4800  &  \tablenum[table-format=-2.2]{0.04}\,(\tablenum[table-format=1.2]{0.21})  &  \tablenum[table-format=2.2]{0.10}\,(\tablenum[table-format=1.2]{0.04})  &  \tablenum[table-format=-2.2]{0.01}\,(\tablenum[table-format=1.2]{0.21})  &  \tablenum[table-format=1.3]{0.989}  &  \tablenum[table-format=2.2]{94.53}  &  \tablenum[table-format=4.0]{4000} \\
11  &  960, 240  &  \tablenum[table-format=-2.2]{0.96}\,(\tablenum[table-format=1.2]{0.35})  &  \tablenum[table-format=2.2]{18.01}\,(\tablenum[table-format=1.2]{1.07})  &  \tablenum[table-format=-2.2]{1.00}\,(\tablenum[table-format=1.2]{0.36})  &  \tablenum[table-format=1.3]{1.001}  &  \tablenum[table-format=2.2]{94.90}  &  \tablenum[table-format=4.0]{4000} \\
11  &  960, 960  &  \tablenum[table-format=-2.2]{2.43}\,(\tablenum[table-format=1.2]{0.55})  &  \tablenum[table-format=2.2]{3.99}\,(\tablenum[table-format=1.2]{0.54})  &  \tablenum[table-format=-2.2]{2.47}\,(\tablenum[table-format=1.2]{0.55})  &  \tablenum[table-format=1.3]{0.999}  &  \tablenum[table-format=2.2]{94.97}  &  \tablenum[table-format=4.0]{4000} \\
11  &  960, 4800  &  \tablenum[table-format=-2.2]{4.08}\,(\tablenum[table-format=1.2]{0.71})  &  \tablenum[table-format=2.2]{0.50}\,(\tablenum[table-format=1.2]{0.15})  &  \tablenum[table-format=-2.2]{4.12}\,(\tablenum[table-format=1.2]{0.71})  &  \tablenum[table-format=1.3]{0.997}  &  \tablenum[table-format=2.2]{94.95}  &  \tablenum[table-format=4.0]{4000} \\
12  &  960, 240  &  \tablenum[table-format=-2.2]{2.41}\,(\tablenum[table-format=1.2]{0.45})  &  \tablenum[table-format=2.2]{35.66}\,(\tablenum[table-format=1.2]{0.96})  &  \tablenum[table-format=-2.2]{2.49}\,(\tablenum[table-format=1.2]{0.45})  &  \tablenum[table-format=1.3]{0.997}  &  \tablenum[table-format=2.2]{95.17}  &  \tablenum[table-format=4.0]{6000} \\
12  &  960, 960  &  \tablenum[table-format=-2.2]{7.69}\,(\tablenum[table-format=1.2]{0.68})  &  \tablenum[table-format=2.2]{8.34}\,(\tablenum[table-format=1.2]{0.68})  &  \tablenum[table-format=-2.2]{7.77}\,(\tablenum[table-format=1.2]{0.68})  &  \tablenum[table-format=1.3]{0.999}  &  \tablenum[table-format=2.2]{94.97}  &  \tablenum[table-format=4.0]{6000} \\
12  &  960, 4800  &  \tablenum[table-format=-2.2]{12.71}\,(\tablenum[table-format=1.2]{0.84})  &  \tablenum[table-format=2.2]{0.78}\,(\tablenum[table-format=1.2]{0.20})  &  \tablenum[table-format=-2.2]{12.78}\,(\tablenum[table-format=1.2]{0.84})  &  \tablenum[table-format=1.3]{0.997}  &  \tablenum[table-format=2.2]{94.78}  &  \tablenum[table-format=4.0]{6000} \\
14  &  960, 240  &  \tablenum[table-format=-2.2]{1.11}\,(\tablenum[table-format=1.2]{0.34})  &  \tablenum[table-format=2.2]{17.23}\,(\tablenum[table-format=1.2]{1.07})  &  \tablenum[table-format=-2.2]{1.21}\,(\tablenum[table-format=1.2]{0.36})  &  \tablenum[table-format=1.3]{0.994}  &  \tablenum[table-format=2.2]{95.00}  &  \tablenum[table-format=4.0]{4000} \\
14  &  960, 960  &  \tablenum[table-format=-2.2]{2.18}\,(\tablenum[table-format=1.2]{0.55})  &  \tablenum[table-format=2.2]{4.08}\,(\tablenum[table-format=1.2]{0.54})  &  \tablenum[table-format=-2.2]{2.28}\,(\tablenum[table-format=1.2]{0.56})  &  \tablenum[table-format=1.3]{0.990}  &  \tablenum[table-format=2.2]{94.97}  &  \tablenum[table-format=4.0]{4000} \\
14  &  960, 4800  &  \tablenum[table-format=-2.2]{3.92}\,(\tablenum[table-format=1.2]{0.70})  &  \tablenum[table-format=2.2]{0.47}\,(\tablenum[table-format=1.2]{0.15})  &  \tablenum[table-format=-2.2]{4.02}\,(\tablenum[table-format=1.2]{0.69})  &  \tablenum[table-format=1.3]{0.989}  &  \tablenum[table-format=2.2]{95.00}  &  \tablenum[table-format=4.0]{4000} \\
20  &  960, 240  &  \tablenum[table-format=-2.2]{1.99}\,(\tablenum[table-format=1.2]{0.38})  &  \tablenum[table-format=2.2]{27.03}\,(\tablenum[table-format=1.2]{0.98})  &  \tablenum[table-format=-2.2]{-10.68}\,(\tablenum[table-format=1.2]{0.85})  &  \tablenum[table-format=1.3]{0.994}  &  \tablenum[table-format=2.2]{94.53}  &  \tablenum[table-format=4.0]{6000} \\
20  &  960, 960  &  \tablenum[table-format=-2.2]{5.49}\,(\tablenum[table-format=1.2]{0.58})  &  \tablenum[table-format=2.2]{5.94}\,(\tablenum[table-format=1.2]{0.58})  &  \tablenum[table-format=-2.2]{-6.73}\,(\tablenum[table-format=1.2]{0.68})  &  \tablenum[table-format=1.3]{0.997}  &  \tablenum[table-format=2.2]{94.68}  &  \tablenum[table-format=4.0]{6000} \\
20  &  960, 4800  &  \tablenum[table-format=-2.2]{8.29}\,(\tablenum[table-format=1.2]{0.75})  &  \tablenum[table-format=2.2]{0.81}\,(\tablenum[table-format=1.2]{0.17})  &  \tablenum[table-format=-2.2]{-3.56}\,(\tablenum[table-format=1.2]{0.47})  &  \tablenum[table-format=1.3]{0.994}  &  \tablenum[table-format=2.2]{94.63}  &  \tablenum[table-format=4.0]{6000} \\*
\TableEndRule \multicolumn{8}{@{}p{\UniformWidth}@{}}{\textit{Note:} Columns report $G_{\mathrm{SS:sup}}$, $G_{\mathrm{SS:un}}$ and $G_{\mathrm{SS:linear}}$; parentheses give paired MCSE in percentage points. Trial aug. (linear) is a fixed-basis comparison in the common-covariate-distribution submodel. CP denotes coverage. Each group has $n_0$ labeled and $M_0$ X-only observations; $R$ is the number of repetitions.}\\
\end{longtable}\endgroup

\FloatBarrier

\section{Application to the Rotterdam cohort}\label{sec:application}
\subsection{Data and analysis design}
The Rotterdam tumor-bank cohort pairs baseline prognostic information with right-censored mortality records. We use the \texttt{survival} 3.8-3 dataset, with death measured from surgery (\texttt{dtime/death}) \citep{royston2013,rotterdamdata}. The eligible cohort contains 2,982 unique patients: 1,312 premenopausal patients in population 1 and 1,670 postmenopausal patients in population 2. Their age and disease-characteristic distributions are summarized in Table~\ref{tab:cohortMain}; the target is a descriptive comparison of these populations.

The target is the five-year recorded-day half-tie probability $\theta_{1825}$ in \eqref{eq:restricted}, covered by Theorem~\ref{thm:restricted}. Values below 0.5 favor later capped death times in population 1. Figure~\ref{fig:rotterdam}(a) shows the full-cohort mortality curves, and Table~\ref{tab:cohortMain} reports each patient's observation status at five years.

At a prespecified allocation, half of each population contributes $(Y,\Delta,\bm X)$ and the other half contributes baseline covariates only. The labeled sizes are 656 and 835, respectively. Primary features are age/10, log positive nodes, two tumor-size indicators and ordinal grade; the basic feature set keeps age and log nodes. We form the covariate-only samples by withholding outcome records for the remaining cohort members. The retained features are measured at or around surgery, although individual pathology-report availability dates are unavailable.

\begin{table}[!htbp]\centering\TableFont
\caption{Rotterdam cohort and five-year follow-up.}\label{tab:cohortMain}
\setlength{\tabcolsep}{0pt}\sbox{\UniformMeasure}{\begin{tabular}{@{}lrr@{}}
Characteristic & Premenopausal & Postmenopausal \\
\multicolumn{3}{l}{\textit{Baseline characteristics}}\\
Patients  &  1,312  &  1,670 \\
Age, years  &  43.3\,(6.4)  &  64.3\,(8.6) \\
Positive nodes  &  0 [0,3]  &  1 [0,5]\\
Tumor size $\leq20$ mm  &  654\,(49.8\%)  &  733\,(43.9\%) \\
Tumor size 20--50 mm  &  553\,(42.1\%)  &  738\,(44.2\%) \\
Tumor size $>50$ mm  &  105\,(8.0\%)  &  199\,(11.9\%) \\
Differentiation grade 2  &  392\,(29.9\%)  &  402\,(24.1\%) \\
Differentiation grade 3  &  920\,(70.1\%)  &  1268\,(75.9\%) \\
\addlinespace\multicolumn{3}{l}{\textit{Observation before or through day 1825}}\\
Observed death before five years  &  277\,(21.1\%)  &  476\,(28.5\%) \\
Censored before five years  &  62\,(4.7\%)  &  83\,(5.0\%) \\
Observed through five years  &  973\,(74.2\%)  &  1111\,(66.5\%) \\
\addlinespace\multicolumn{3}{l}{\textit{Primary 50\% outcome-withholding allocation}}\\
Labeled patients  &  656  &  835 \\
Covariate-only patients  &  656  &  835 \\
Labeled risk set at five years  &  472  &  582 \\
Labeled $\widehat G(1825-)$  &  0.9402  &  0.9483 \\
\end{tabular}}
\setlength{\tabcolsep}{5pt}\setlength{\UniformWidth}{\dimexpr\wd\UniformMeasure+20pt\relax}
\begin{tabular}{@{}lrr@{}}\toprule 
Characteristic & Premenopausal & Postmenopausal \\\midrule 
\multicolumn{3}{l}{\textit{Baseline characteristics}}\\
Patients  &  1,312  &  1,670 \\
Age, years  &  43.3\,(6.4)  &  64.3\,(8.6) \\
Positive nodes  &  0 [0,3]  &  1 [0,5]\\
Tumor size $\leq20$ mm  &  654\,(49.8\%)  &  733\,(43.9\%) \\
Tumor size 20--50 mm  &  553\,(42.1\%)  &  738\,(44.2\%) \\
Tumor size $>50$ mm  &  105\,(8.0\%)  &  199\,(11.9\%) \\
Differentiation grade 2  &  392\,(29.9\%)  &  402\,(24.1\%) \\
Differentiation grade 3  &  920\,(70.1\%)  &  1268\,(75.9\%) \\
\addlinespace\multicolumn{3}{l}{\textit{Observation before or through day 1825}}\\
Observed death before five years  &  277\,(21.1\%)  &  476\,(28.5\%) \\
Censored before five years  &  62\,(4.7\%)  &  83\,(5.0\%) \\
Observed through five years  &  973\,(74.2\%)  &  1111\,(66.5\%) \\
\addlinespace\multicolumn{3}{l}{\textit{Primary 50\% outcome-withholding allocation}}\\
Labeled patients  &  656  &  835 \\
Covariate-only patients  &  656  &  835 \\
Labeled risk set at five years  &  472  &  582 \\
Labeled $\widehat G(1825-)$  &  0.9402  &  0.9483 \\
\bottomrule\end{tabular}
\par\smallskip\noindent\begin{minipage}{\UniformWidth}\TableFont\textit{Note:} Age is mean (SD), nodes are median [quartiles], and other entries are counts (percent). Five-year statuses are mutually exclusive observed categories. $\widehat G$ is event-priority censoring survival in the primary labeled samples.\end{minipage}\end{table}

The primary labeled partition has 472 and 582 patients at risk at five years, with event-priority censoring-survival estimates of 0.9402 and 0.9483. Figure~\ref{fig:sSupport} summarizes the corresponding risk sets and weights across all 1,000 allocations at each budget.

\subsection{Estimates and precision}
At the prespecified primary allocation, the weighted estimate is 0.4795 with SE 0.0110 and a 95\% interval of [0.4578, 0.5011] (Table~\ref{tab:rotterdamPrimary}(A)). The estimate is below 0.5, while its interval includes 0.5. Weighting reduces the fitted SE relative to supervised estimation without materially separating the three point estimates at this allocation.

\begin{table}[!htbp]\centering\TableFont
\caption{Five-year estimates and precision.}\label{tab:rotterdamPrimary}
\setlength{\tabcolsep}{0pt}\sbox{\UniformMeasure}{\begin{tabular}{@{}lrrl@{}}
\multicolumn{4}{l}{\textit{A. Prespecified 50\% labeled allocation}}\\
Method  &  Estimate  &  IF-SE  &  95\% Wald interval \\\midrule
$\widehat{\theta}_{\mathrm{sup},\tau}$  &  \tablenum[table-format=1.4]{0.4815}  &  \tablenum[table-format=1.4]{0.0115}  &  [0.4590, 0.5039] \\
$\widehat{\theta}_{\mathrm{un},\tau}$  &  \tablenum[table-format=1.4]{0.4775}  &  \tablenum[table-format=1.4]{0.0115}  &  [0.4549, 0.5000] \\
$\widehat{\theta}_{\mathrm{SS},\tau}$  &  \tablenum[table-format=1.4]{0.4795}  &  \tablenum[table-format=1.4]{0.0110}  &  [0.4578, 0.5011] \\
\end{tabular}}
\setlength{\UniformWidth}{\linewidth}\setlength{\tabcolsep}{\dimexpr(\linewidth-\wd\UniformMeasure)/6\relax}
\ifdim\tabcolsep<0pt\PackageWarning{UnifiedTables}{Negative column gap}\fi
\begin{tabular}{@{}lrrl@{}}\toprule 
\multicolumn{4}{l}{\textit{A. Prespecified 50\% labeled allocation}}\\
Method  &  Estimate  &  IF-SE  &  95\% Wald interval \\\midrule
$\widehat{\theta}_{\mathrm{sup},\tau}$  &  \tablenum[table-format=1.4]{0.4815}  &  \tablenum[table-format=1.4]{0.0115}  &  [0.4590, 0.5039] \\
$\widehat{\theta}_{\mathrm{un},\tau}$  &  \tablenum[table-format=1.4]{0.4775}  &  \tablenum[table-format=1.4]{0.0115}  &  [0.4549, 0.5000] \\
$\widehat{\theta}_{\mathrm{SS},\tau}$  &  \tablenum[table-format=1.4]{0.4795}  &  \tablenum[table-format=1.4]{0.0110}  &  [0.4578, 0.5011] \\
\bottomrule\end{tabular}

\setlength{\tabcolsep}{0pt}\sbox{\UniformMeasure}{\begin{tabular}{@{}lrrl@{}}\multicolumn{4}{l}{\textit{B. Variance relative to supervised estimation}}\\
Assessment  &  Gain (\%)  &  MCSE (pp)  &  Source of variation \\\midrule
$\widehat{\theta}_{\mathrm{SS},\tau}$ (fitted IF)  &  \tablenum[table-format=-2.2]{7.15}  &  --  &  Primary allocation \\
$\widehat{\theta}_{\mathrm{SS},\tau}$ (withholding)  &  \tablenum[table-format=-2.2]{13.83}  &  \tablenum[table-format=1.2]{2.19}  &  Label allocation \\
$\widehat{\theta}_{\mathrm{SS},\tau}$ (patient bootstrap)  &  \tablenum[table-format=-2.2]{4.55}  &  \tablenum[table-format=1.2]{1.67}  &  Patients and allocation \\
$\widehat{\theta}_{\mathrm{un},\tau}$ (fitted IF)  &  \tablenum[table-format=-2.2]{-0.90}  &  --  &  Primary allocation \\
$\widehat{\theta}_{\mathrm{un},\tau}$ (withholding)  &  \tablenum[table-format=-2.2]{-4.98}  &  \tablenum[table-format=1.2]{4.77}  &  Label allocation \\
$\widehat{\theta}_{\mathrm{un},\tau}$ (patient bootstrap)  &  \tablenum[table-format=-2.2]{-6.01}  &  \tablenum[table-format=1.2]{3.56}  &  Patients and allocation \\
\end{tabular}}
\setlength{\UniformWidth}{\linewidth}\setlength{\tabcolsep}{\dimexpr(\linewidth-\wd\UniformMeasure)/6\relax}
\ifdim\tabcolsep<0pt\PackageWarning{UnifiedTables}{Negative column gap}\fi
\begin{tabular}{@{}lrrl@{}}\toprule \multicolumn{4}{l}{\textit{B. Variance relative to supervised estimation}}\\
Assessment  &  Gain (\%)  &  MCSE (pp)  &  Source of variation \\\midrule
$\widehat{\theta}_{\mathrm{SS},\tau}$ (fitted IF)  &  \tablenum[table-format=-2.2]{7.15}  &  --  &  Primary allocation \\
$\widehat{\theta}_{\mathrm{SS},\tau}$ (withholding)  &  \tablenum[table-format=-2.2]{13.83}  &  \tablenum[table-format=1.2]{2.19}  &  Label allocation \\
$\widehat{\theta}_{\mathrm{SS},\tau}$ (patient bootstrap)  &  \tablenum[table-format=-2.2]{4.55}  &  \tablenum[table-format=1.2]{1.67}  &  Patients and allocation \\
$\widehat{\theta}_{\mathrm{un},\tau}$ (fitted IF)  &  \tablenum[table-format=-2.2]{-0.90}  &  --  &  Primary allocation \\
$\widehat{\theta}_{\mathrm{un},\tau}$ (withholding)  &  \tablenum[table-format=-2.2]{-4.98}  &  \tablenum[table-format=1.2]{4.77}  &  Label allocation \\
$\widehat{\theta}_{\mathrm{un},\tau}$ (patient bootstrap)  &  \tablenum[table-format=-2.2]{-6.01}  &  \tablenum[table-format=1.2]{3.56}  &  Patients and allocation \\
\bottomrule\end{tabular}

\setlength{\tabcolsep}{0pt}\sbox{\UniformMeasure}{\begin{tabular}{@{}lrrl@{}}\multicolumn{4}{l}{\textit{C. Label budgets and allocation precision}}\\
Labels  &  $(n_1,n_2)$; $(M_1,M_2)$  &  $G_{\mathrm{SS:sup}}$ (MCSE)  &  $G_{\mathrm{un:sup}}$ (MCSE)\\\midrule
25\%  &  (328,417); (984,1253)  &  \tablenum[table-format=2.2]{10.41}\,(\tablenum[table-format=1.2]{2.27})  &  \tablenum[table-format=-3.2]{5.47}\,(\tablenum[table-format=2.2]{3.13}) \\
50\%  &  (656,835); (656,835)  &  \tablenum[table-format=2.2]{13.83}\,(\tablenum[table-format=1.2]{2.19})  &  \tablenum[table-format=-3.2]{-4.98}\,(\tablenum[table-format=2.2]{4.77}) \\
75\%  &  (984,1252); (328,418)  &  \tablenum[table-format=2.2]{13.50}\,(\tablenum[table-format=1.2]{2.22})  &  \tablenum[table-format=-3.2]{-139.14}\,(\tablenum[table-format=2.2]{14.90}) \\
\end{tabular}}
\setlength{\UniformWidth}{\linewidth}\setlength{\tabcolsep}{\dimexpr(\linewidth-\wd\UniformMeasure)/6\relax}
\ifdim\tabcolsep<0pt\PackageWarning{UnifiedTables}{Negative column gap}\fi
\begin{tabular}{@{}lrrl@{}}\toprule \multicolumn{4}{l}{\textit{C. Label budgets and allocation precision}}\\
Labels  &  $(n_1,n_2)$; $(M_1,M_2)$  &  $G_{\mathrm{SS:sup}}$ (MCSE)  &  $G_{\mathrm{un:sup}}$ (MCSE)\\\midrule
25\%  &  (328,417); (984,1253)  &  \tablenum[table-format=2.2]{10.41}\,(\tablenum[table-format=1.2]{2.27})  &  \tablenum[table-format=-3.2]{5.47}\,(\tablenum[table-format=2.2]{3.13}) \\
50\%  &  (656,835); (656,835)  &  \tablenum[table-format=2.2]{13.83}\,(\tablenum[table-format=1.2]{2.19})  &  \tablenum[table-format=-3.2]{-4.98}\,(\tablenum[table-format=2.2]{4.77}) \\
75\%  &  (984,1252); (328,418)  &  \tablenum[table-format=2.2]{13.50}\,(\tablenum[table-format=1.2]{2.22})  &  \tablenum[table-format=-3.2]{-139.14}\,(\tablenum[table-format=2.2]{14.90}) \\
\bottomrule\end{tabular}

\par\smallskip\noindent\begin{minipage}{\linewidth}\TableFont\textit{Note:} The 50\% allocation has labeled sizes $(656,835)$ and covariate-only sizes $(656,835)$. IF denotes influence function and SE denotes standard error. Panel A uses the fixed primary allocation. In panel B, gains are relative to supervised variance: fitted IF uses squared SEs; withholding and patient bootstrap use 1,000 paired replicates each. Bootstrap samples receive new allocations. In panel C, the columns report $G_{\mathrm{SS:sup}}$ and $G_{\mathrm{un:sup}}$; MCSE is in percentage points.\end{minipage}\end{table}

\begin{figure}[!htb]\centering
\makebox[\linewidth][c]{\includegraphics[width=\linewidth]{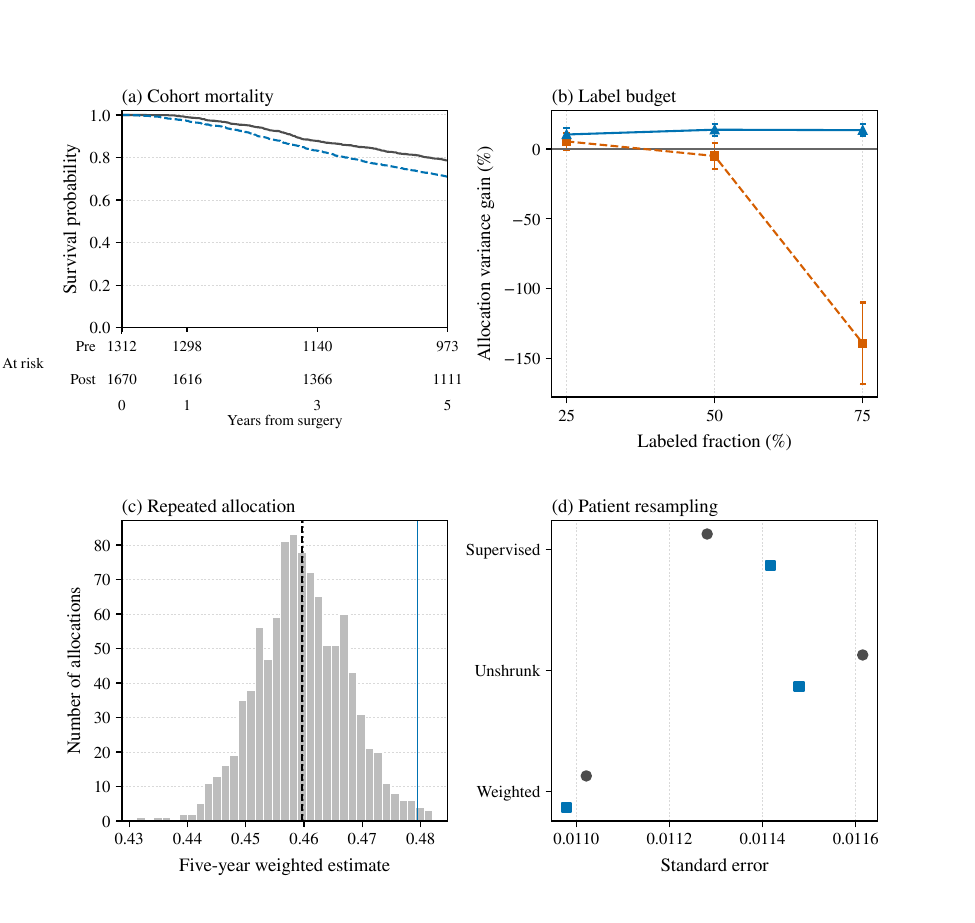}}
\caption{Five-year Rotterdam comparison. (a) Full-cohort death KM curves; solid gray and dashed blue denote premenopausal and postmenopausal groups. Risk counts below the curves are aligned at 0, 1, 3 and 5 years. (b) Allocation variance gains relative to supervised estimation under the primary features at three label budgets; orange dashed squares and blue solid triangles denote Unshrunk and Weighted, with pointwise $\pm1.96$ paired MCSE from 1,000 allocations. (c) All 1,000 weighted estimates at 50\% labeled; solid blue, dashed black and dotted gray lines mark the primary allocation, allocation mean and descriptive full-record reference. (d) Patient-level bootstrap SD (gray circles) and mean IF-SE (blue squares) for the three methods from 1,000 resamples with new allocations. Allocation variability conditions on the cohort, whereas patient resampling also varies the cohort.}\label{fig:rotterdam}
\end{figure}

The allocation distribution places the primary estimate in context. The weighted estimate lies 2.57 allocation SDs above the repeated-allocation mean of 0.4597, with 994 of the 1,000 repeated estimates below it (Figure~\ref{fig:rotterdam}(c)). The descriptive full-record supervised estimate is 0.4599.

The three precision summaries in Table~\ref{tab:rotterdamPrimary}(B) correspond to different sampling procedures. At the primary allocation, the fitted influence-function variance is 7.15\% lower for Weighted than for Supervised. Repeated outcome withholding varies the label allocation conditional on the observed cohort and gives a 13.83\% variance gain (MCSE 2.19 percentage points). Patient-level bootstrap resampling stratified by population and followed by a new allocation varies the sampled patients as well as the allocation and gives a 4.55\% gain (MCSE 1.67 percentage points).

Figure~\ref{fig:rotterdam}(b) and Table~\ref{tab:rotterdamPrimary}(C) show how this allocation precision changes with the label budget. At 25\% and 75\% labeled, the group-specific labeled sizes are $(328,417)$ and $(984,1252)$, with covariate-only sizes $(984,1253)$ and $(328,418)$. A larger labeled pool leaves a smaller auxiliary pool within this fixed cohort. Weighted gains remain positive at the three budgets; unshrunk transfer increases allocation variance by 139.14\% at 75\% labeled. This contrast illustrates why the precision of the remaining auxiliary mean matters alongside the number of outcome-bearing records.

In Figure~\ref{fig:rotterdam}(d), mean influence-function SE and bootstrap SD are close for all three methods; their ratio is 0.996 for Weighted. Table~\ref{tab:rotterdamBootstrap} gives the method-specific summaries.

Feature ablation shows an expanded-minus-basic allocation-gain difference of 0.24 percentage points (MCSE 0.93) at the 50\% budget. Paired intervals include zero at all three budgets (Table~\ref{tab:rotMasks}), so the expanded feature set has no resolved incremental gain in this cohort.

\FloatBarrier
\subsection{Sensitivity to censoring assumptions}
A five-year group-stratified model relates surgery year to observed follow-up and gives a censoring hazard ratio of \result{RotCensorHR} per surgery year (95\% interval [\result{RotCensorLow}, \result{RotCensorHigh}]); the corresponding observed-death hazard ratio is \result{RotDeathHR} ([\result{RotDeathLow}, \result{RotDeathHigh}]). The data documentation notes that death information may be updated after recurrence follow-up has ended \citep{rotterdamdata}. Individual administrative cutoffs and censoring reasons are unavailable, and surgery year is absent from the primary working features. This calendar-time structure poses a substantive risk to the working assumption $C_k\perp(S_k,X_k)$.

At the primary allocation, the prespecified alternatives differ from supervised marginal IPCW (0.481482) by approximately $-0.0023$ to $0.0017$ (Table~\ref{tab:rotterdamSensitivity}). Period-stratified IPCW estimates the censoring distribution separately for surgeries through 1987 and from 1988 onward. Within each group, we combine the period-specific event distributions using the period proportions of either the labeled patients or the complete cohort. Conditional IPCW uses Cox censoring models with linear or spline terms for surgery year. Outcome-regression standardization averages the fitted outcome distributions over the baseline covariates of either the labeled patients or the complete cohort. All outcome and censoring models use labeled records; cohort standardization additionally uses baseline information from the covariate-only patients.

\begin{table}[!htbp]\centering\TableFont
\caption{Five-year point sensitivity.}\label{tab:rotterdamSensitivity}
\setlength{\tabcolsep}{0pt}\sbox{\UniformMeasure}{\begin{tabular}{@{}>{\raggedright\arraybackslash}p{0.54\linewidth}rr@{}}
Procedure & Point estimate & Shift from reference \\
Marginal IPCW  &  \tablenum[table-format=1.6]{0.481482}  &  \tablenum[table-format=+1.6]{+0.000000} \\
Period-stratified IPCW, labeled standardization  &  \tablenum[table-format=1.6]{0.481699}  &  \tablenum[table-format=+1.6]{+0.000216} \\
Period-stratified IPCW, cohort standardization  &  \tablenum[table-format=1.6]{0.483228}  &  \tablenum[table-format=+1.6]{+0.001745} \\
Conditional IPCW, linear calendar-year effect  &  \tablenum[table-format=1.6]{0.480852}  &  \tablenum[table-format=+1.6]{-0.000630} \\
Conditional IPCW, spline calendar-year effect  &  \tablenum[table-format=1.6]{0.480882}  &  \tablenum[table-format=+1.6]{-0.000601} \\
Outcome-regression standardization, labeled covariates  &  \tablenum[table-format=1.6]{0.481608}  &  \tablenum[table-format=+1.6]{+0.000125} \\
Outcome-regression standardization, cohort covariates  &  \tablenum[table-format=1.6]{0.479157}  &  \tablenum[table-format=+1.6]{-0.002325} \\
\end{tabular}}
\setlength{\UniformWidth}{\linewidth}\setlength{\tabcolsep}{\dimexpr(\linewidth-\wd\UniformMeasure)/4\relax}
\ifdim\tabcolsep<0pt\PackageWarning{UnifiedTables}{Negative column gap}\fi
\begin{tabular}{@{}>{\raggedright\arraybackslash}p{0.54\linewidth}rr@{}}\toprule 
Procedure & Point estimate & Shift from reference \\\midrule 
Marginal IPCW  &  \tablenum[table-format=1.6]{0.481482}  &  \tablenum[table-format=+1.6]{+0.000000} \\
Period-stratified IPCW, labeled standardization  &  \tablenum[table-format=1.6]{0.481699}  &  \tablenum[table-format=+1.6]{+0.000216} \\
Period-stratified IPCW, cohort standardization  &  \tablenum[table-format=1.6]{0.483228}  &  \tablenum[table-format=+1.6]{+0.001745} \\
Conditional IPCW, linear calendar-year effect  &  \tablenum[table-format=1.6]{0.480852}  &  \tablenum[table-format=+1.6]{-0.000630} \\
Conditional IPCW, spline calendar-year effect  &  \tablenum[table-format=1.6]{0.480882}  &  \tablenum[table-format=+1.6]{-0.000601} \\
Outcome-regression standardization, labeled covariates  &  \tablenum[table-format=1.6]{0.481608}  &  \tablenum[table-format=+1.6]{+0.000125} \\
Outcome-regression standardization, cohort covariates  &  \tablenum[table-format=1.6]{0.479157}  &  \tablenum[table-format=+1.6]{-0.002325} \\
\bottomrule\end{tabular}
\par\smallskip\noindent\begin{minipage}{\UniformWidth}\TableFont\textit{Note:} The 50\% allocation has labeled sizes $(656,835)$ and covariate-only sizes $(656,835)$. Shifts are point-estimate differences from supervised marginal IPCW (0.481482); conditional censoring weights are estimated from Cox models.\end{minipage}\end{table}

Across 100 prespecified allocations, outcome standardization produced larger shifts than at the primary allocation and more variable shifts than the censoring-model alternatives. The spline censoring model failed to fit in 3 allocations (Table~\ref{tab:rotSensitivityRepeated}).
\FloatBarrier

\section{Discussion}\label{sec:discussion}
Auxiliary covariates improve first-order precision when they explain variation in the score for ordering event times across populations. The estimator adjusts the comparisons available from censored records by using auxiliary samples to estimate the feature means in each population. The variance calculation retains the covariance induced by using the same labeled records to form pairs and estimate the censoring distributions.

The resource weights matter most when the feature means in the auxiliary pools are imprecise. In the strong-signal resource experiments, the variance gains increase with the auxiliary sample sizes and track the first-order predictions. Weighted and unshrunk adjustments become closer as auxiliary information grows, while at zero or weak projection even a modest cost of fitting the slopes can offset the available gain. A contrast between the two labeled groups can itself capture a useful signal, so the relative value of auxiliary sampling depends on both the size of the pools and the direction of the two group-specific projections.

The five-year Rotterdam comparison illustrates how these gains should be interpreted when the follow-up window is part of the scientific question. The fixed allocation describes the precision of one analysis, whereas reallocating the labels shows how that precision depends on which patients contribute outcomes. The patient bootstrap puts this allocation variation in the context of patient sampling. The spread across allocations also puts the relatively high point estimate from the primary allocation in perspective.

The association between surgery year and follow-up in Rotterdam motivates extending the method to conditional censoring, which would change both identification and the influence function used for inference. Studies that obtain auxiliary covariates from external cohorts face a second design question: how to use those measurements when their distribution differs from that of the patients with observed follow-up. In either setting, the benefit of richer working features must be weighed against the accuracy with which their effects can be estimated from the labeled records.

\FloatBarrier

\section*{Acknowledgement}
This research was supported by the Natural Science Foundation of Xinjiang Uygur Autonomous Region (No. 2025D14015) and the National Natural Science Foundation of China (No. 72571102).

\section*{Data and code availability}
The Rotterdam data are distributed with the R package \texttt{survival} \citep{rotterdamdata}. The computational archive contains the simulation and application scripts, prespecified configurations, and numerical results.

\FloatBarrier
\bibliographystyle{publicationnat}
\bibliography{references}
\clearpage
\begin{appendices}
\numberwithin{equation}{section}
\numberwithin{table}{section}
\numberwithin{figure}{section}
\section{Auxiliary results}\label{app:auxiliary}
The following two lemmas collect the inverse-censoring-weight expansion and supervised pair-statistic expansion used in the proofs of the main results. Both retain the dependence induced by estimating the censoring distribution on the labeled observations.

\begin{auxlemma}\label{lem:aux-km}\label{s:KM}
Suppose that Assumptions~(A1)--(A2) hold. For population $k$, let
\[
L_{ki}(t)=\int_0^{t\wedge b_0}\frac{dM^C_{ki}(s)}{H_k(s)}.
\]
Then, uniformly over the effective event-time support,
\begin{equation*}
\frac{G_k(t)}{\widehat G_k(t-)}-1=\frac1{n_k}\sum_{i=1}^{n_k}L_{ki}(t)+\Op(n_k^{-1}).
\end{equation*}
\end{auxlemma}
\begin{proof}
Suppress the population index and put $m=n_k$. On $[0,b_0]$ define $\bar R=m^{-1}\sum_iR_i$, $\bar M^C=m^{-1}\sum_iM_i^C$ and $H=\E R$. The empirical distribution bound gives $\sup_t|\bar R(t)-H(t)|=\Op(m^{-1/2})$. Since $H(b_0)>0$, the event $\inf_{t\le b_0}\bar R(t)>H(b_0)/2$ has probability tending to one.

The Nelson--Aalen estimator obeys the exact identity on this event
\[
\widehat\Lambda^C(t)-\Lambda^C(t)=\int_0^t\bar R(s)^{-1}d\bar M^C(s).
\]
Replacing $\bar R^{-1}$ by $H^{-1}$ has a uniform $\Op(m^{-1})$ remainder. Localize the integral when the risk falls below $H(b_0)/2$ or the inverse-risk error first exceeds $A m^{-1/2}$. Its predictable bracket is bounded in probability by a constant times
\[
m^{-1}\int_0^{b_0}(\bar R^{-1}-H^{-1})^2\bar R\,d\Lambda^C=\Op(m^{-2}).
\]
The martingale maximal inequality followed by removal of localization, first $m\to\infty$ and then $A\to\infty$, gives the asserted remainder. With
\[
L_i(t)=\int_0^{t\wedge b_0}\frac{dM_i^C(s)}{H(s)},
\]
we obtain $\widehat\Lambda^C-\Lambda^C=m^{-1}\sum_iL_i+\Op(m^{-1})$ uniformly. Also $\sup_i\|L_i\|_\infty$ is bounded by $H(b_0)^{-1}\{1+\Lambda^C(b_0)\}$.

Continuity of effective censoring times makes each censoring jump of the empirical hazard $O_p(m^{-1})$. Its sum of squared jumps is $\Op(m^{-1})$. Consequently the logarithm of the product-limit differs from the negative Nelson--Aalen estimator by $\Op(m^{-1})$. Taylor expansion yields
\begin{equation}\label{s:inverseKM}
\frac{G(t)}{\widehat G(t-)}-1=\frac1m\sum_iL_i(t)+\Op(m^{-1})
\end{equation}
uniformly on the effective event support. Beyond $b_0$, both the true and observable censoring contributions are constant; the martingale integral is stopped at $b_0$. If event support is bounded, censoring changes beyond it are unobservable and irrelevant. With unbounded $T$, a surviving censoring mass at infinity allows positive $G$ throughout. Since the denominator lower bound is positive and $\varepsilon_n\to0$, clipping is inactive with probability tending to one.

The expansion uses the positive effective risk in (A2). When censoring hazards have fixed positive jumps, the finite-product derivative in Section~\ref{s:restricted} gives the corresponding expansion.
\end{proof}

\begin{auxlemma}\label{lem:aux-supervised}\label{s:triples}
Suppose that Assumptions~(A1)--(A4) hold. The supervised estimator satisfies
\begin{equation*}
\widehat\theta_{\mathrm{sup}}-\theta=\sum_{k=1}^2P_{L,k}\{g_k+\psi_k\}+\Op(n^{-1}).
\end{equation*}
\end{auxlemma}
\begin{proof}
We derive the supervised expansion, accounting for observations shared between the comparison pairs and the censoring estimator.
Let $q^\circ=q-\theta-g_1-g_2$. Its conditional mean is zero given either pair observation. In the variance of $P_nq^\circ$, pairs sharing only one index have zero covariance after conditioning on that observation, and disjoint pairs are independent. Only pairs sharing both indices contribute. Thus
\[
\Var(P_nq^\circ)=\frac{\E\{(q^\circ)^2\}}{n_1n_2}=O(n^{-2}),\qquad
P_nq-\theta=\sum_kP_{L,k}g_k+\Op(n^{-1}).
\]

Using \eqref{s:inverseKM}, the first-order contribution from estimating $G_1$ is
\begin{equation}\label{s:triple}
B_{1n}=\frac1{n_1^2n_2}\sum_{i,l,j}q(O_{1i},O_{2j})L_{1l}(Y_{1i}).
\end{equation}
Uniform inverse-weight products and their quadratic errors have total remainder $\Op(n^{-1})$, because $q$ is bounded. In \eqref{s:triple}, $l=i$ contributes only $n_1n_2$ bounded terms and is $\Op(n_1^{-1})$. Its expectation need not be zero.

For $i\ne l$ there are three independent units, with two roles from the first population. Let $f(i,l;j)=q(i,j)L_l(Y_i)$. Fixing the pair role $i$ or $j$ gives zero one-unit projection, because $\E L_l(t)=0$ at deterministic $t$. Fixing the nuisance role $l$ and exchanging bounded integrals gives
\[
\E\{f(i,l;j)\mid O_{1l}\}=\int_0^{b_0}\frac{A_1(s)}{H_1(s)}dM^C_{1l}(s)=\psi_1(O_{1l}).
\]
After subtracting $\psi_l$, the residual is degenerate in every one-unit role. Two residual triples with at most one shared unit have zero covariance, including cross-role matches of the first-population units. Nonzero covariances require at least two shared units. Their number is $O(n^4)$ against a squared normalizer of order $n^6$; bounded summands give residual variance $O(n^{-2})$. This count expressly includes index interchange between the pair and nuisance roles. Therefore
\[
B_{1n}=\frac{n_1-1}{n_1}P_{L,1}\psi_1+\Op(n^{-1})=P_{L,1}\psi_1+\Op(n^{-1}).
\]
The second population is identical with its own roles. Combining both contributions with the ordinary projection proves
\begin{equation}\label{s:supervised}
\widehat\theta_{\mathrm{sup}}-\theta=\sum_kP_{L,k}(g_k+\psi_k)+\Op(n^{-1}).
\end{equation}
The nuisance triples account for dependence between the group-specific censoring fits using all labeled folds and the pair evaluations. The $O_p(n^{-1})$ remainder follows from the compact effective-censoring assumptions in (A2), in line with same-sample expansions for censored $U$-statistics \citep{datta2010,chen2019,fernandez2020}.
\end{proof}

\section{Proofs of the main results}\label{app:proofs}
The proofs follow the order of the full-lifetime results, then establish the fixed-horizon extension. Technical expansions used repeatedly are stated separately in Appendix~\ref{app:auxiliary}.

\subsection{Proof of Lemma 1}\label{s:identity}\label{s:cov}
\begin{proof}[Proof of Lemma~\ref{lem:identity}]
We first establish identification and the projection relations used in Lemma~\ref{lem:identity}.
We use Assumptions~(A1)--(A4) from the main text, including joint independent censoring, the effective censor-support restriction and positive risk. Fix $n=n_1+n_2$, $n_k/n\to\pi_k>0$, $M_k/n\to\lambda_k>0$. Feature dimensions and fold count are fixed. Every probabilistic order below is under a fixed population law satisfying these assumptions. Pool sizes are treated as fixed design quantities.

The full target is $\theta=\Prb(T_1<T_2)$ and the kernel is $q=\Delta_1\Delta_2\ind(Y_1<Y_2)/\{G_1(Y_1)G_2(Y_2)\}$. Given $(T_k,\bm{X}_k)$,
\[
\E\left\{\frac{\Delta_k}{G_k(T_k)}\,\middle|\,T_k,\bm{X}_k\right\}=1.
\]
Population independence implies $\E(q\mid T_1,T_2,\bm{X}_1,\bm{X}_2)=\ind(T_1<T_2)$. Integrating proves unconditional and conditional identification. Under the effective-support assumptions $q$ is bounded by a constant. Integrating the other population gives $\E(q\mid O_k)=\Delta_kr_k(Y_k)/G_k(Y_k)$, with $r_1(t)=\Prb(T_2>t)$ and $r_2(t)=\Prb(T_1<t)$.

Let $\bm{v}_k=\bm{u}_k-\bm{\mu}_k$ and $b_k(\bm{X}_k)=\E\{r_k(T_k)\mid \bm{X}_k\}-\theta$. In the centered basis $(1,\bm{v}_1^\top,\bm{v}_2^\top)$ the additive population Gram matrix is $\operatorname{diag}(1,\bm{\Gamma}_1,\bm{\Gamma}_2)$. Hence the intercept is $\theta$ and $\bm{\beta}_k=\bm{\Gamma}_k^{-1}\E(\bm{v}_kb_k)$. Therefore $p_k=\bm{\beta}_k^\top \bm{v}_k$ is the orthogonal projection of $b_k$ and
\[
\E p_k=0,\qquad \E\{(b_k-p_k)\bm{v}_k\}=0,\qquad
\Cov(g_k,p_k)=\E(b_kp_k)=\Var(p_k).
\]
Finite second moments suffice for these projection identities. In the regression expansion below, slope consistency suffices because the slope error multiplies a centered feature-mean difference of order $O_p(n^{-1/2})$.

Conditional on $(T,\bm{X})$, the independent censoring process is stopped at the fixed event time. Its compensator is still $\int R\,d\Lambda^C$. Thus for a bounded predictable integrand $a(t,\bm{X})$,
\begin{equation}\label{s:condmart}
\E\left\{\int_0^{b_0}a(t,\bm{X})dM^C(t)\,\middle|\,T,\bm{X}\right\}=0.
\end{equation}
This gives $\E(\psi\mid T,\bm{X})=0$, $\E(d\mid \bm{X})=b$, and $\Cov(\psi,p)=0$. Joint independent censoring, $C\perp(T,\bm{X})$, gives the martingale centering conditional on $(T,\bm{X})$.
For one group suppress subscripts, let $S_X(t)=\Prb(T\ge t\mid \bm{X})$ and $a_X(t,\bm{X})=\E\{r(T)\mid T\ge t,\bm{X}\}$. For bounded predictable functions define $L_a=\int_0^{b_0}a(t,\bm{X})G(t)^{-1}dM^C(t)$. The martingale isometry and conditioning on $(T,\bm{X})$ give
\begin{align}\label{s:isometry}
\Cov(L_a,L_{a'})&=\int_0^{b_0}\frac{\E\{S_X(t)a(t,\bm{X})a'(t,\bm{X})\}}{G(t)}d\Lambda^C(t),\\
\Cov(g,L_a)&=-\int_0^{b_0}\frac{\E\{S_X(t)a_X(t,\bm{X})a(t,\bm{X})\}}{G(t)}d\Lambda^C(t).\nonumber
\end{align}
For the second identity, on the complete-event path there is no censor jump before $T$, so only the negative compensator enters the product with $\Delta r(T)/G(T)$. Independent censoring cancels its inclusion probability; Fubini gives the displayed conditional tail. The centering constant has zero covariance with a martingale integral.

Taking $a=a_0(t)=\E\{r(T)\mid T\ge t\}$ and using $\E(S_Xa_X)=S a_0$ gives $\Cov(g,\psi)=-\Var(\psi)$. Combining this identity with the conditional martingale centering and the projection equations established above gives $\Cov(p,\psi)=0$, $\Cov(g,p)=\Var(p)$ and $\Var(d)=\Var(g)-\Var(\psi)$.

\end{proof}

\subsection{Proof of Theorem 1}\label{s:regression}
\begin{proof}[Proof of Theorem~\ref{thm:alr}]
We now combine the supervised expansion with the fitted covariate adjustment.
Fix a fold pair $(a,b)$; its complementary indices are selected independently of values and retain positive asymptotic fractions. Two-sample LLNs apply to each integrable Gram or response kernel. They follow for bounded kernels from the index-sharing variance argument, and for general integrable kernels by truncation in $L_1$. The training empirical Gram converges to the positive-definite population Gram. For the fitted response,
\[
\|P_{-a,-b}\{\bm{z}(\widehat q-q)\}\|
\le\sup_{i,j}|\widehat q_{ij}-q_{ij}|P_{-a,-b}\|\bm{z}\|=\op(1),
\]
where $\bm{z}=(1,\bm{u}_1^\top,\bm{u}_2^\top)^\top$. Hence $\widehat{\bm{\beta}}_{ab}\to_p\beta$. A fixed finite number of rectangles implies simultaneous slope consistency. Positive limiting Gram eigenvalues also make the finite-sample generalized inverse agree with the ordinary inverse with probability tending to one.

For each fold, finite feature second moments and the i.i.d. pool CLTs give $\bar{\bm{u}}^U_k-\bar{\bm{u}}_{k,\ell_k(a,b)}=\Op(n^{-1/2})$. Thus
\[
\sum_{ab}w_{ab}(\widehat{\bm{\beta}}_{k,ab}-\bm{\beta}_k)^\top(\bar{\bm{u}}^U_k-\bar{\bm{u}}_{k,\ell_k(a,b)})=\op(n^{-1/2}).
\]
The product is $o_p(n^{-1/2})$ from the $o_p(1)$ slope error and the $O_p(n^{-1/2})$ feature-mean difference.

The exact identities $\sum_bw_{ab}=n_{1a}/n_1$ and $\sum_aw_{ab}=n_{2b}/n_2$ show that population-slope corrections aggregate to $\sum_ks_{k,n}\bm{\beta}_k^\top(\bar{\bm{u}}^U_k-\bar{\bm{u}}^L_k)$. Also $\sum_{ab}w_{ab}P_{ab}\widehat q=P_n\widehat q$ exactly for unequal folds. If bounded $s_{k,n}\to s_k$, their differences from their limits multiply centered $\Op(n^{-1/2})$ means, giving another $\op(n^{-1/2})$ remainder. Combine this with \eqref{s:supervised} to obtain the main asymptotic linear representation. The four pools are independent and their limiting influence functions have finite second moments, so the ordinary multivariate CLT gives the variance $V(\bm{s})$ in the main text. Actual $c_{k,n}=M_k/(M_k+n_k)$ meets the required convergence.

Thus consistency of the finite-dimensional regression suffices; the conditional comparison signal need not be correctly specified by the working linear model.
\end{proof}

\subsection{Proof of Corollary 1}\label{s:corollary}
\begin{proof}[Proof of Corollary~\ref{cor:ratio}]
Expand the variance with $h=sp$ to obtain
\[
V(\bm{s})=V_{\mathrm{sup}}-\sum_k2s_k\pi_k^{-1}\Var(p_k)+\sum_ks_k^2(\pi_k^{-1}+\lambda_k^{-1})\Var(p_k).
\]
Each nonzero-signal coordinate is strictly convex and is minimized at $c_k=\lambda_k/(\pi_k+\lambda_k)$. Completing squares proves the gain and comparison with the unshrunk choice. Zero-signal coordinates are flat. The labeled-score covariance is $\Cov(g,cp)=c\Var(p)$, whereas the transferred-score variance is $\Var(cp)=c^2\Var(p)$.

For unrestricted centered $L_2$ transfers, put $\epsilon=d-b$, so $\E(\epsilon\mid \bm{X})=0$. Then $\Var(d-h)=\E\Var(d\mid \bm{X})+\Var(b-h)$. The identity
\begin{equation}\label{s:complete}
\pi^{-1}\Var(b-h)+\lambda^{-1}\Var(h)
=(\pi+\lambda)^{-1}\Var(b)+(\pi^{-1}+\lambda^{-1})\Var(h-cb)
\end{equation}
proves the fixed-$d$ optimum $cb$. Orthogonal decomposition $b=p+(b-p)$ then gives the optimal working-space transfer $cp$ and excess variance $(\pi^{-1}+\lambda^{-1})c^2\Var(b-p)$. Full-model efficiency follows from the separate tangent-space argument in the proof of Theorem~\ref{thm:canonical}; the projection calculation gives the transfer optimum. Related auxiliary and semi-supervised projection results include \citet{schisterman2001,song2024general,xu2025unified}.
\end{proof}

\subsection{Proof of Theorem 2}\label{s:canonical}
\begin{proof}[Proof of Theorem~\ref{thm:canonical}]
We establish canonical membership before deriving the product-experiment bound and evaluating Example~\ref{rem:counterexample}.
The latent model factors as $P_XP_{T\mid \bm{X}}P_C$. Bounded scores can be written $u(\bm{X})$, $v(T,\bm{X})$ and $w(C)$ with $\E u=0$, $\E(v\mid \bm{X})=0$ and $\E w=0$. Their observed scores are $u$, $Bv=\E(v\mid O)$ and $Qw=\E(w\mid O)$. Effective censoring-score directions have the closure generated by deterministic-time integrals $\int\ell(t)dM^C(t)$. Censoring changes beyond the effective event support are invisible in the observed law.

The groupwise derivative of the target is $\E[\{r(T)-\theta\}\{u(\bm{X})+v(T,\bm{X})\}]$, and its derivative in a censoring direction is zero. Define $e=g+L_{a_X}$. Equation~\eqref{s:condmart} implies $\E(e\mid T,\bm{X})=r(T)-\theta$. Hence $e$ represents the $\bm{X}$ and conditional-event derivatives, because $e$ is observed and $\E(eBv)=\E(ev)$. Equation~\eqref{s:isometry} proves $e$ orthogonal to every censoring score. To conclude that $e$ is canonical, membership in the observed tangent closure must also be established.

Put $m(\bm{x})=\E\{r(T)\mid \bm{X}=\bm{x}\}$ and
\[
J(t,\bm{x})=\int_0^{t\wedge b_0}\frac{a_X(s,\bm{x})}{G(s)}d\Lambda^C(s),\qquad
v_*(t,\bm{x})=\frac{r(t)}{G(t)}-J(t,\bm{x})-m(\bm{x}).
\]
These functions are bounded under the support and hazard assumptions even when $T$ is unbounded. Using $d(1/G)=G^{-1}d\Lambda^C$ and Fubini gives $\E(v_*\mid \bm{X})=0$. On an event record, $Bv_*=r(Y)/G(Y)-J(Y,\bm{X})-m(\bm{X})$. On a censor record at $y\le b_0$, condition on $T>y,\bm{X}=\bm{x}$. The identities
\begin{align*}
\E\{r(T)/G(T)\mid T>y,\bm{x}\}
&=\frac{a_X(y,\bm{x})}{G(y)}+
\frac1{S_X(y)}\int_y^{b_0}\frac{S_X(s)a_X(s,\bm{x})}{G(s)}d\Lambda^C(s),\\
\E\{J(T,\bm{x})\mid T>y,\bm{x}\}
&=J(y,\bm{x})+\frac1{S_X(y)}\int_y^{b_0}\frac{S_X(s)a_X(s,\bm{x})}{G(s)}d\Lambda^C(s)
\end{align*}
hold for censoring-almost-every $y$. The future integrals cancel. It follows that
\begin{equation}\label{s:membership}
e=b(\bm{X})+Bv_*(O).
\end{equation}
The two terms are bounded legitimate $\bm{X}$ and conditional-event score directions. Bounded exponential tilts construct dominated submodels, preserve censoring independence and finite moments, and locally preserve positivity and the fixed effective support. Thus $e$ is a tangent member as well as a derivative representer and is the canonical gradient. This membership argument identifies the canonical gradient for the full observed-data model, beyond the minimum within the transfer class. The conditional risk-tail structure is in line with standard censoring efficiency theory \citep{tsiatis2006semiparametric,zhang2019censored}.

To identify the component outside this tangent space, let $D=d-e=L_{a_0-a_X}$. Its conditional expectation given $(T,\bm{X})$ is zero, so it is orthogonal to $\bm{X}$ and conditional-event observed scores. Orthogonality to censor scores follows from $\E\{S_X(t)(a_0-a_X)\}=0$. Therefore $D$ is in the orthogonal complement of the observed tangent space; in particular $D\perp e$ and $D\perp f(\bm{X})$ for any square-integrable $f$. Its variance is
\begin{equation*}
\Var(D)=\int_0^{b_0}\frac{\E[S_X(t)\{a_0(t)-a_X(t,\bm{X})\}^2]}{G(t)}d\Lambda^C(t).
\end{equation*}
It can be positive even when the comparison signal is exactly linear.

We now incorporate the covariate-only pools. At base sample size $n$, the product score inner product is $\sum_k\{\pi_k\E(S^L_kT^L_k)+\lambda_k\E(S^U_kT^U_k)\}$. An ALR coordinate pair $(\phi^L,\phi^U)$ corresponds to Riesz coordinates $(\phi^L/\pi,\phi^U/\lambda)$. The $\bm{X}$ score is shared by the labeled and unlabeled pools; the conditional-outcome and censor scores occur only in the labeled pool. From \eqref{s:membership}, the efficient Riesz components are $b/(\pi+\lambda)$ in both pools for the covariate direction and $Bv_*/\pi$ in the labeled pool for the conditional-outcome direction. The labeled conditional component has conditional mean zero given $\bm{X}$, so it is orthogonal to the shared covariate direction. For any centered covariate score $u$, the two weighted inner products sum to $\E(bu)$; the labeled conditional component represents the remaining derivative through $Bv_*$. Thus these coordinates are tangent members and satisfy the Riesz equations in the product experiment. In ALR form they are
\[
(\phi^L_{\mathrm{eff}},\phi^U_{\mathrm{eff}})=(e-cb,cb).
\]
Their variance is $\pi^{-1}\E\Var(e\mid \bm{X})+(\pi+\lambda)^{-1}\Var(b)$, summed across populations. Because $d=e+D$, orthogonality gives
\[
V_d(h)=V_e(h)+\sum_k\pi_k^{-1}\Var(D_k).
\]
Apply \eqref{s:complete} to $e$ to obtain the complete gap formula. For $h=cp$, the second term is the working-projection loss; the first remains $\sum_k\pi_k^{-1}\Var(D_k)$.
\end{proof}

For Example~\ref{rem:counterexample}, write $R=\ind(C>1)$ and $p=1/2$. The event times exceed 1, and the censoring hazard integrates to $\log 2$ on $[0,1]$, so $R$ is an independent Bernoulli$(1/2)$ observation indicator. The target is $\theta=1/2$. The conditional comparison signals are $b_1(X_1)=p_1(X_1)=(1-2X_1)/3$ and $b_2(X_2)=p_2(X_2)=0$. The complete and canonical scores and their difference reduce to
\[
d_k=\frac{R}{p}(r_k-\theta),\quad e_k=b_k+\frac{R}{p}(r_k-m_k),\quad D_k=(R/p-1)b_k.
\]
For group 1, $r_1=(5-T_1)/3$, $\Var(r_1)=13/108$, $\Var(b_1)=1/9$, and $\E\Var(r_1\mid X_1)=1/108$. For group 2, $\Var(r_2)=1/18$. These moments give
\[
\begin{array}{c|ccc}
&\Var(d_k)&\Var(e_k)&\Var(D_k)\\\hline
1&13/54&7/54&1/9\\
2&1/9&1/9&0
\end{array}
\]
With $\pi_k=1/2$, $\lambda_k=5/2$ and $c_k=5/6$, the resulting variances are
\[
V_{\mathrm{sup}}=19/27,\qquad V_{\mathrm{SS}}=V_d^*=14/27,\qquad V_{\mathrm{eff}}=8/27.
\]
Thus $V_{\mathrm{SS}}-V_{\mathrm{eff}}=2/9$ is entirely due to $\Var(D_1)/\pi_1$; the working-projection term vanishes because $p_k=b_k$ in both groups.

\subsection{Proof of Theorem 3}\label{s:variance}

\begin{proof}[Proof of Theorem~\ref{thm:inference}]
We establish variance consistency by empirical convergence of the complete scores.
For known $G$, a pair row mean $Q_{1i}=n_2^{-1}\sum_jq(O_{1i},O_{2j})$ has conditional expectation $g_1(O_{1i})+\theta$ and conditional mean-squared error at most $C/n_2$, because the other-population observations are independent after conditioning on $O_{1i}$. Averaging over $i$ gives
\[
\E\left[n_1^{-1}\sum_i\{Q_{1i}-g_1(O_{1i})-\theta\}^2\right]\le C/n_2.
\]
Conditioning row by row and averaging the conditional error bound gives empirical $L_2$ consistency of $\widehat g$; uniform $\widehat q-q=\op(1)$ and consistency of $\widehat\theta_{\mathrm{sup}}$ transfer this result to the fitted kernel.

For the nuisance score, $\widehat A(t)=P_n\{\widehat q\ind(Y\ge t)\}\to A(t)$ uniformly. For the bounded known kernel, finite brackets for the weighted threshold measure give a uniform LLN; fitted-weight replacement is uniformly negligible. Also $\widehat H\to H$ uniformly and $\inf H>0$, so $\widehat\eta=\widehat A/\widehat H\to\eta=A/H$ uniformly. Both $A$ and $H$ have bounded variation, hence $\eta$ has bounded variation. Similarly $\widehat\Lambda^C\to\Lambda^C$ uniformly and its total mass is $\Op(1)$.

Write $\widehat\psi_i=(1-\Delta_i)\widehat\eta(Y_i)\ind(Y_i\le b_0)-\int_0^{Y_i\wedge b_0}\widehat\eta\,d\widehat\Lambda^C$, with compatible predictable conventions. The jump error is bounded by $\sup|\widehat\eta-\eta|$. The compensator error is controlled by this supremum times the fitted hazard mass and by the bounded-variation Stieltjes bound
\[
\sup_t\left|\int_0^t\eta\,d(\widehat\Lambda^C-\Lambda^C)\right|
\le C\{\|\eta\|_\infty+\operatorname{TV}(\eta)\}\|\widehat\Lambda^C-\Lambda^C\|_\infty=\op(1).
\]
Thus $\sup_i|\widehat\psi_i-\psi_i|=\op(1)$. Uniform convergence of the fitted integrand and the bounded-variation compensator control the same-sample construction.

Weighted slope consistency and feature moments imply, for either labeled or unlabeled empirical measure,
\[
P_{\mathrm{pool}}|\widehat h-cp|^2
\le C\|c_n\bar{\bm{\beta}}-c\beta\|^2P_{\mathrm{pool}}\|\bm{u}\|^2
+C\|c_n\bar{\bm{\beta}}\|^2\|\bar{\bm{u}}^U-\bm{\mu}\|^2=\op(1).
\]
Combining the three estimates proves empirical $L_2$ convergence of complete scores to $d-cp$ and $cp$. The LLN for square-integrable true scores and Cauchy--Schwarz control of score differences give convergence of the four separately centered sample variances. Bessel denominators have the same limit, so $n\widehat\se^2\to V_{\mathrm{SS}}$. The four-pool CLT and Slutsky prove Wald validity.
\end{proof}

The continuous-censoring nuisance score is a Nelson--Aalen plug-in. At an empirical censoring jump with hazard $a_j$, the exact product-limit derivative includes the factor $(1-a_j)^{-1}$. Under the risk lower bound, $\max_j a_j=O_p(n^{-1})$ and the total fitted hazard is $O_p(1)$, so the two score vectors differ uniformly by $O_p(n^{-1})$. The empirical $L_2$ convergence and variance consistency therefore hold for the reported score estimator.

For a pool of size $m$, let $\xi_i$ be independent $N(0,1)$ multipliers. Its perturbation is $\sum_i\xi_i(\widehat\phi_i-\bar{\widehat\phi})/\sqrt{m(m-1)}$, whose conditional variance is $s_m^2(\widehat\phi)/m$. Four independent perturbations sum to exactly $N(0,\widehat\se^2)$. This influence-function multiplier represents the Gaussian limit of the four estimated pool scores.

\subsection{Proof of Theorem 4}\label{s:restricted}
\begin{proof}[Proof of Theorem~\ref{thm:restricted}]
We prove Theorem~\ref{thm:restricted} in the capped finite-grid experiment. The population index is suppressed when considering one group.
Fix a finite integer $\tau$ and let $S=T\wedge\tau$ lie in $\{0,\ldots,\tau\}$. The target is $\theta_\tau=\E H(S_1,S_2)$, with $H(s,t)=\ind(s<t)+\ind(s=t)/2$. The four-pool, fixed-feature and fold assumptions are as in the main text. Within each population, assume $C\perp(S,\bm{X})$ on the relevant horizon, $G_-(\tau)=\Prb(C\ge\tau)\ge g_0>0$, and $\Prb(S=\tau)\ge s_0>0$. Each active hazard $a_t=\Prb(C=t\mid C\ge t)$, $t<\tau$, is less than one. Zero-hazard cells remain zero in dominated submodels. Positive limiting variance is required for studentization.

The observed map is $Z=Y\wedge\tau$, $A=\ind(\Delta=1,Y<\tau)+\ind(Y\ge\tau)$. If $A=1$ then $S=Z$, including $Z=\tau$; if $A=0$ then $C=Z<\tau$ and $S>Z$. A censoring time at $\tau$ is complete for the capped variable and generates no terminal censor jump. A death exactly at $\tau$ belongs to the same capped terminal atom as a later death. Together, these rules identify when the capped event time $S$ is observed under the event-priority convention.

Given $S=s,\bm{X}$, complete capped observation occurs exactly with probability $G_-(s)$. Hence $\E\{A/G_-(S)\mid S,\bm{X}\}=1$, and independence across populations gives
\[
\E(q_\tau\mid S_1,S_2,\bm{X}_1,\bm{X}_2)=H(S_1,S_2),\qquad
q_\tau=\frac{A_1A_2H(Z_1,Z_2)}{G_{1,-}(Z_1)G_{2,-}(Z_2)}.
\]
This proves unconditional and conditional identification. Let $r_1(s)=\Prb(S_2>s)+\Prb(S_2=s)/2$, $r_2(s)=\Prb(S_1<s)+\Prb(S_1=s)/2$, $g=Ar(Z)/G_-(Z)-\theta_\tau$, and $b(\bm{X})=\E\{r(S)\mid \bm{X}\}-\theta_\tau$. The centering $-\theta_\tau$ applies to incomplete as well as complete records and is part of the observed-data score.

Split pairs into those with both times below $\tau$, exactly one reaching $\tau$, and both reaching $\tau$. The first class differs from the strict full parameter only by half of its original ties; the last contributes half instead of its uncapped ordering. This proves the main exact target-relation formula. As $\tau\to\infty$ and $T$ is finite almost surely, the limit is $\Prb(T_1<T_2)+\Prb(T_1=T_2)/2$. The capped half-tie parameter therefore differs in general from the strict full-lifetime parameter.

For $t<\tau$ define $B_i(t)=\ind(Z_i>t)+\ind(Z_i=t,A_i=0)$, $D_i^C(t)=\ind(Z_i=t,A_i=0)$ and $\mathcal M_i(t)=D_i^C(t)-a_tB_i(t)$. Events at $t$ precede censoring, so $\sum_iB_i=R_t-dE_t$. Its expectation is $h_t=\Prb(S>t)G_-(t)\ge s_0g_0$ and $\E D^C=h_ta_t$. Conditional on $(S,\bm{X})$, $\E\mathcal M(t)=0$: if $S\le t$ both indicators vanish, and if $S>t$ independent censoring gives the usual hazard identity. Distinct time increments are martingale differences under the event-then-censor filtration and have covariance zero; $\E\mathcal M(t)^2=h_ta_t(1-a_t)$.

The empirical hazard is $\widehat a_t=P_mD^C(t)/P_mB(t)$ with zero-count cells interpreted as in the main text. Fixed finite-dimensional Taylor expansion gives $\widehat a_t-a_t=P_m\mathcal M(t)/h_t+\Op(m^{-1})$ simultaneously in $t$. Differentiating the finite product exactly yields
\begin{equation}\label{s:atomicderivative}
\frac{G_-(s)}{\widehat G_-(s)}-1=P_mL_i(s)+\Op(m^{-1}),\qquad
L_i(s)=\sum_{t<s}\frac{\mathcal M_i(t)}{h_t(1-a_t)}.
\end{equation}
The factor $1/(1-a_t)$ is the derivative contribution from the reciprocal product-limit factor. Positivity and fixed finite $\tau$ bound all derivatives locally. Clipping becomes inactive with probability tending to one. A terminal $C=\tau$ atom has no factor in $G_-(\tau)$; moving censor mass between $\tau$ and later times leaves the capped observed functional unchanged.

For the bounded two-sample kernel $q_\tau$, the ordinary degenerate remainder is $\Op(n^{-1})$. Insert \eqref{s:atomicderivative} into the pair statistic. The nuisance triples have the same unit roles as Section~\ref{s:triples}. Equal-index terms contribute $O_p(n^{-1})$. On distinct-unit triples, the only nonzero single-role projection is the nuisance role, now equal to
\begin{equation*}
\psi(O)=\sum_{t<\tau}\frac{A_+(t)}{h_t(1-a_t)}\mathcal M(t),\qquad
A_+(t)=\E\{r(S)\ind(S>t)\}.
\end{equation*}
The strict inequality follows from $t<Z$ in \eqref{s:atomicderivative}. The same shared-unit covariance count gives a residual of order $\Op(n^{-1})$. Therefore $\widehat\theta_{\mathrm{sup},\tau}-\theta_\tau=\sum_kP_{L,k}d_k+\Op(n^{-1})$, with $d=g+\psi$. Also $\E(\psi\mid S,\bm{X})=0$ and $\E(d\mid \bm{X})=b$.

The additive response projection has slopes $\bm{\Gamma}_k^{-1}\E(\bm{v}_kb_k)$ as before, now with restricted $b_k$. Finite-grid uniform censoring consistency and the integrable two-sample Gram LLN prove training slope consistency. The exact rectangle identities and coefficient--mean product argument in Section~\ref{s:regression} give the restricted ALR with labeled $d_k-s_kp_k$ and auxiliary $s_kp_k$. Conditional martingale centering gives $\Cov(d,p)=\Var(p)$; expanding variance proves $c_k=\lambda_k/(\pi_k+\lambda_k)$, the gain formulas and the fixed-$d$ optimum. The same variance decomposition therefore applies to the restricted scores.

To obtain the canonical bound for the capped experiment, write $S_X^+(t)=\Prb(S>t\mid \bm{X})$, $G_{\mathrm{after}}(t)=G_-(t)(1-a_t)$, $a_X(t,\bm{x})=\E\{r(S)\mid S>t,\bm{X}=\bm{x}\}$ and $a_0(t)=\E\{r(S)\mid S>t\}$. Define $L_f=\sum_{t<\tau}f(t,\bm{X})\mathcal M(t)/G_{\mathrm{after}}(t)$. Direct finite-sum isometry gives
\begin{align}\label{s:atomiso}
\Cov(L_f,L_{f'})&=\sum_{t<\tau}\E\{S_X^+(t)f(t,\bm{X})f'(t,\bm{X})\}\frac{a_t}{G_{\mathrm{after}}(t)},\\
\Cov(g,L_f)&=-\sum_{t<\tau}\E\{S_X^+(t)a_X(t,\bm{X})f(t,\bm{X})\}\frac{a_t}{G_{\mathrm{after}}(t)}.\nonumber
\end{align}
For the second expression, on a complete record and at $t<S$ there is no censoring jump and $\mathcal M(t)=-a_t$. Integrating its product with the complete-record IPCW term cancels the inclusion probability. The first uses $h_ta_t(1-a_t)$ and conditional orthogonality of the increments. Set $e=g+L_{a_X}$ and $D=d-e=L_{a_0-a_X}$.

Canonical membership requires a discrete analogue of Section~\ref{s:canonical}. Put
\[
m(\bm{x})=\E\{r(S)\mid \bm{X}=\bm{x}\},\qquad
J(s,\bm{x})=\sum_{t<s}\frac{a_X(t,\bm{x})a_t}{G_{\mathrm{after}}(t)},\qquad
v_*(s,\bm{x})=\frac{r(s)}{G_-(s)}-J(s,\bm{x})-m(\bm{x}).
\]
The exact reciprocal-product telescope $G_{\mathrm{after}}(t)^{-1}-G_-(t)^{-1}=a_t/G_{\mathrm{after}}(t)$ gives $\E(v_*\mid \bm{X})=0$. At a complete record, the conditional score is $v_*(S,\bm{X})$. At a censor record $Z=t,S>t$, expand the reciprocal product after $t$ to obtain
\begin{align*}
\E\{r(S)/G_-(S)\mid S>t,\bm{x}\}
&=\frac{a_X(t,\bm{x})}{G_{\mathrm{after}}(t)}+
\sum_{v>t}\frac{S_X^+(v)}{S_X^+(t)}\frac{a_X(v,\bm{x})a_v}{G_{\mathrm{after}}(v)},\\
\E\{J(S,\bm{x})\mid S>t,\bm{x}\}
&=J(t,\bm{x})+\frac{a_X(t,\bm{x})a_t}{G_{\mathrm{after}}(t)}+
\sum_{v>t}\frac{S_X^+(v)}{S_X^+(t)}\frac{a_X(v,\bm{x})a_v}{G_{\mathrm{after}}(v)}.
\end{align*}
Subtracting gives $\E(v_*\mid O)=a_X(t,\bm{X})/G_-(t)-J(t,\bm{X})-m(\bm{X})$, hence $e=b+\E(v_*\mid O)$. Bounded conditional score tilts establish tangent membership in $P_XP_{S\mid \bm{X}}P_C$, preserving support and local positivity. Conditional centering gives derivative representation, and \eqref{s:atomiso} cancels the deterministic censor-hazard scores. Thus $e$ is canonical for this finite-grid model.

For $D$, conditional centering handles covariate and outcome scores, and $\E\{S_X^+(a_0-a_X)\}=0$ handles censor scores. Therefore $D\perp e$, $D\perp b$, and
\[
\Var(D)=\sum_{t<\tau}\E[S_X^+(t)\{a_0(t)-a_X(t,\bm{X})\}^2]\frac{a_t}{G_{\mathrm{after}}(t)}.
\]
The product-experiment Riesz argument is identical after these restricted identities are established. Its efficient ALR is $(e-cb,cb)$ and its complete gap is $\sum_k\{\pi_k^{-1}\Var(D_k)+(\pi_k^{-1}+\lambda_k^{-1})\Var(h_k-c_kb_k)\}$. The discrete tangent-space and covariance calculations establish canonical membership for the capped experiment.

It remains to establish consistency of the feasible variance estimator. Let $\widehat Q_{ki}$ be the restricted pair row means, $\widehat g_{ki}=\widehat Q_{ki}-\widehat\theta_{\mathrm{sup},\tau}$. Estimate $A_+(t)$ by $n_k^{-1}\sum_i\widehat Q_{ki}\ind(Z_{ki}>t)$, $h_t$ by $(R_t-dE_t)/n_k$, and use
\[
\widehat\psi_{ki}=\sum_{t<\tau}\widehat\eta_t\{D^C_{ki}(t)-\widehat a_tB_{ki}(t)\},\qquad
\widehat\eta_t=\frac{\widehat A_+(t)}{\widehat h_t(1-\widehat a_t)}.
\]
Equivalently $\widehat\eta_t$ is the sum of pair rows strictly after $t$ divided by the number of records strictly after $t$. At a censored individual's own day the increment is $1-\widehat a_t$; earlier increments are $-\widehat a_t$. 

Known-$G$ row empirical $L_2$ error is $O_p(n_{\mathrm{other}}^{-1})$ in mean squared norm. Finite-grid uniform weight consistency transfers that bound to fitted rows. Every tail, hazard and risk estimate is a smooth function of finitely many sample proportions with positive population denominators, so all nuisance estimates converge simultaneously. These bounds prove empirical $L_2$ convergence of $\widehat\psi$ for the fitted same-sample nuisance. Weighted slopes and feature moments give the same convergence for $\widehat h=c_n\bar{\bm{\beta}}^\top(\bm{u}-\bar{\bm{u}}^U)$. Four separate sample variances with denominator $m-1$ of $\widehat g+\widehat\psi-\widehat h$ and $\widehat h$, divided by their actual pool sizes, consistently estimate the ALR variance. Wald and IF-only Gaussian statements follow as in Section~\ref{s:variance}. 

\end{proof}

\subsection{Further mathematical details}\label{s:atoms}
This appendix distinguishes event-time atoms from atomic censoring and treats auxiliary-sample ratios outside the finite positive-ratio regime.

To allow event-time atoms, drop marginal nonatomicity of $T$ but retain continuous effective censoring and all other full assumptions. The kernel remains strict, with $r_2(t)=F_1(t-)$; changing to midranks would change the parameter. Censor times almost surely do not coincide with any event times, even when event times have atoms. Risk DKW bounds require no continuity. Each censor jump still has size $O_p(n_k^{-1})$; multiple event jumps do not enter the censor product. The argument for Lemma~\ref{lem:aux-km} applies because censoring times are distinct almost surely and have no simultaneous event jumps, even when event times have atoms.

Index coincidence in Section~\ref{s:triples} concerns observations, not equality of their numerical event times. The counting argument, regression LLN and mean-product bound are unchanged. Isometry identities hold $d\Lambda^C$-almost everywhere: conditional or marginal event atoms cause no common jump with continuous censoring. Canonical membership uses the same $v_*$ and bounded tilts; no conditional density of $T\mid \bm{X}$ is required.

For discontinuous weighted tails, construct a uniform LLN using the finite measure $\mu(B)=\E\{q\ind(Y\in B)\}$. Isolate finitely many atoms exceeding any prescribed bracket mass, include both open and closed threshold kernels at those atoms, and partition the remainder into intervals of small $\mu$ mass. Endpoint two-sample LLNs and bracketing give uniform convergence. The bounded-variation variance proof then applies with compatible side conventions. These arguments cover event atoms, not atomic censoring; the latter needs Section~\ref{s:restricted}.

The following triangular regimes describe auxiliary-to-labeled sample-size ratios that diverge or vanish. If $M_k/n_k\to\infty$ while $n_k/n\to\pi_k>0$, fixed feature moments give $\sqrt n(\bar{\bm{u}}^U-\bm{\mu})\to_p0$, $c_{k,n}\to1$, and the regression product remains negligible. The variance is $\sum_k\pi_k^{-1}\Var(d_k-p_k)$. The $D$ component remains.

If $M_k/n_k\to0$ with $M_k\to\infty$, $c_{k,n}\sim M_k/n_k$. Its scaled auxiliary fluctuation has order $\sqrt n\,c_{k,n}/\sqrt{M_k}=O(\sqrt{M_k/n_k})\to0$; the labeled shift also vanishes. The primary estimator's first-order variance tends to $V_{\mathrm{sup}}$. The main theorems concern finite positive limiting ratios.

For completion bounds, consider a recorded capped observation and let $(l,u)=(Z,Z)$ when $A=1$, and $(Z+1,\tau)$ when $A=0$. Then $l\le S\le u$. Monotonicity of the comparison score in its two arguments gives
\[
L(O_1,O_2)=H(u_1,l_2)\le H(S_1,S_2)\le U(O_1,O_2)=H(l_1,u_2).
\]
These bounds are sharp for the fixed-cohort pair average: placing all unknown first-population times at their upper endpoints and second-population times at their lower endpoints jointly achieves the lower bound; reverse placements achieve the upper. The resulting interval bounds the pair average in the observed cohort over all compatible completions.

For two independent i.i.d. patient samples, taking expectations gives $\theta_L\le\theta_\tau\le\theta_U$. The observable bounded pair kernels $L,U$ have standard two-sample Hoeffding expansions. Their SEs are the square roots of row-variance/$N_1$ plus column-variance/$N_2$, with sample centering and variance denominator $m-1$. When each endpoint has a positive first-order projection variance, two marginal one-sided 97.5\% confidence limits and a union bound give the asymptotic conservative envelope $[\widehat\theta_L-1.959964\widehat\se_L,\widehat\theta_U+1.959964\widehat\se_U]$, including when the endpoint estimators are correlated.

A constant endpoint kernel can instead be treated exactly. A nonconstant endpoint with a degenerate first-order projection is not covered by this normal approximation.

\subsection{Common-score comparisons and Kaplan--Meier mass identities}
\label{s:commoncomparison}
The following common-score comparison fixes the supervised score and adjustment space, separating the contribution of independent auxiliary covariates from labeled covariate-imbalance adjustment.

\begin{proposition}[Common-score linear comparison]\label{prop:commonscore}
Suppose the two populations have the same covariate law, $n_1=n_2=m$,
$M_1=M_2=M$, and a common fixed centered finite-dimensional feature space
$\mathcal L$. Suppose the supervised estimator has group scores $d_1,d_2$
with mean zero and finite second moments. Put
$p_k=\Pi_{\mathcal L}\E(d_k\mid \bm{X})$ and $c=M/(m+M)$.
Assume the fitted adjustments have negligible first-order remainders.
Write $\operatorname{AV}$ for the actual, unscaled leading variance and
$\operatorname{AV}_{\rm sup}=\{\Var(d_1)+\Var(d_2)\}/m$.
The four-pool correction has scores $(d_1-cp_1,d_2-cp_2)$ in the labeled
pools and $(cp_1,cp_2)$ in the auxiliary pools.
The optimal labeled-only linear imbalance correction has scores
$d_1-a,d_2+a$, where $a=(p_1-p_2)/2$. Their variance decreases are
\begin{align*}
\Delta_B&=\frac{c}{m}\{\Var(p_1)+\Var(p_2)\},\\
\Delta_L&=\frac{1}{2m}\Var(p_1-p_2),\\
\Delta_B-\Delta_L&=\frac{1}{m}\left[
(c-\tfrac12)\{\Var(p_1)+\Var(p_2)\}+\Cov(p_1,p_2)\right].
\end{align*}
Here the same supervised scores, including censoring estimation, are used
in both adjustments.
\end{proposition}
\begin{proof}
Put $A=\Var(p_1)+\Var(p_2)$.
The normal equations imply $\Cov(d_k,v)=\Cov(p_k,v)$ for every
$v\in\mathcal L$. Independence of the four pools gives
\[
\operatorname{AV}_B=\frac1m\sum_k\Var(d_k-cp_k)
 +\frac{c^2}{M}\sum_k\Var(p_k).
\]
Subtracting from the supervised variance gives
$\{2c-c^2(1+m/M)\}A/m=cA/m$.
For an arbitrary centered imbalance function $v\in\mathcal L$, the decrease is
$\{2\Cov(p_1-p_2,v)-2\Var(v)\}/m$.
Completing the square gives its unique function-space maximizer
$v=a=(p_1-p_2)/2$ and maximum $\Var(p_1-p_2)/(2m)$.
Expansion of this last variance proves the difference formula.
\end{proof}

For $A>0$, put $r=\Cov(p_1,p_2)/A\in[-1/2,1/2]$ and $t=M/m$.
For $-1/2<r<1/2$, the four-pool decrease is larger exactly when
$t>(1-2r)/(1+2r)$; equality holds at the threshold.
This follows by solving $t/(1+t)>1/2-r$.
At $r=1/2$, every positive $t$ gives the larger four-pool decrease;
at $r=-1/2$, every finite $t$ gives the smaller decrease.
If $A=0$, both decreases vanish. For unequal fixed group sizes, with the
same common covariate law and score assumptions, the optimal coefficient
of the labeled feature-mean difference is
$(n_2\bm{\beta}_1-n_1\bm{\beta}_2)/(n_1+n_2)$, and its decrease is
$\Var(n_2p_1-n_1p_2)/\{n_1n_2(n_1+n_2)\}$.
Indeed the feature-mean difference has covariance
$\bm{\Gamma}(1/n_1+1/n_2)$, and its covariance with the supervised estimator is
$\bm{\Gamma}(\bm{\beta}_1/n_1-\bm{\beta}_2/n_2)$; the projection quadratic form gives
this expression. These mean differences cease to be centered when the
population covariate distributions differ.

The relation between capped Kaplan--Meier masses and IPCW weights follows from the event and censoring factors. At a group-specific observed time $t$, let $R_t,E_t,C_t$ be the risk,
event and censoring counts. With event priority, the event survival factor
is $1-E_t/R_t$ and the censoring factor is $1-C_t/(R_t-E_t)$.
Inactive zero-risk cells contribute no factor. Their product is
$(R_t-E_t-C_t)/R_t$; telescoping over observed times gives
$\widehat S_T(t-)\widehat G_C(t-)=R_t/m$.
Thus each event mass is
\[
\Delta\widehat F_T(t)=\widehat S_T(t-)E_t/R_t
=E_t/\{m\widehat G_C(t-)\}.
\]
At a fixed cap, the residual event survival is
$\widehat S_T(\tau-)=R_\tau/\{m\widehat G_C(\tau-)\}$,
which equals the total restricted IPCW mass contributed by records
reaching $\tau$. It follows that the capped KM double integral for
\eqref{eq:restricted} equals the unclipped restricted IPCW pair statistic
when the risk sets, priority and kernel agree. Clipping can destroy this
finite-sample identity; Assumption (H1) makes it inactive asymptotically.
For the full target, the identity applies to KM event masses. A survival
integral retaining an unspecified terminal residual mass is a different
finite-sample statistic. For example, one event at time 1 in group 1 and
an event at 2 followed by a censor at 3 in group 2 give event--event pair
value $1/2$, while $\int\widehat S_2(t)\,d\widehat F_1(t)=1$.

For comparison with trial augmentation, the capped plug-in and augmentation in \citet{zhang2020trial},
Section 2.2, use a common covariate law induced by randomization.
Its Supplement B implements capped KM masses and a half-tie kernel;
using the complementary half-tie kernel aligns the comparison direction.
For direct comparison with its stated support condition we additionally require $\Prb(C_k>\tau)>0$; (H1) alone permits only a non-strict terminal survival condition. On the common fixed finite recorded grid of Assumption (H1), with positive
active risks, the capped plug-in is a smooth function $F(P_1,P_2)$ of
the two empirical observed-category probability vectors.
The KM--IPCW identity and the expansion in Section~\ref{s:restricted}
give its group derivatives $d_{1\tau},d_{2\tau}$.

Write $N=2m$. A pooled delete-one perturbation in group
$k$ changes $F$ by $-d_k/(m-1)+o_p(m^{-1})$; the standard pseudo-value
normalization $N-1$ therefore tends to $2d_k$. Add-one pseudo-values have
the same limit. Smoothness on the fixed support, positive denominators
and a fixed full-rank feature basis give consistency of the associated
linear regression and empirical second moments. Under balanced fixed
counts, the pooled supervised score is
$2A d_1+2(1-A)d_2$. The limiting linear trial learner is
$2(p_1-p_2)$; its correction
$N^{-1}\sum_i(A_i-1/2)\,2(p_1-p_2)(\bm{X}_i)$ is exactly the imbalance
correction in Proposition~\ref{prop:commonscore} at first order.
Its conditional group means are zero, so fixing the group counts removes
no first-order allocation-mean variance. Data-independent folds retain
positive group proportions. Consistent fitted coefficients multiply
held-out centered means of order $O_p(m^{-1/2})$, and hence contribute
$o_p(m^{-1/2})$ remainders. Separately, the empirical cross-validated scores
converge in squared mean, establishing the corresponding variance limit.
The calculation establishes the first-order comparison for fixed-basis learners on the common finite recorded grid.

\subsection{A finite-sample cost at zero projection}
\label{s:zeroprojection}
The first-order gain can vanish while fitting still has a finite-sample
cost. Consider the uncensored strict target. Let $B\sim
\operatorname{Bernoulli}(q)$, $0<q<1$, and take $T_1$ uniform on $(0,1)$
if $B=1$ and on $(2,3)$ otherwise. Independently, let $T_2$ be uniform
on $(1.2,1.8)$. Choose mutually independent standard-normal covariates in
all four pools, independent of event times, and use $\bm{u}_k(X)=X$.
Then every pair response equals its group-1 label $B_i$, the target is
$q$, the supervised statistic is $\bar B$, and $p_1=p_2=0$.
For equal two-fold group-1 splits of size at least four, the group-2
training slopes vanish and the correction is
\[
C=\frac c2\{\widehat\beta_{-1}(\bar X^U-\bar X_1)
             +\widehat\beta_{-2}(\bar X^U-\bar X_2)\}.
\]
Here $c=M_1/(n_1+M_1)$ and the fitted slopes regress the training $B$ on
training $X$ with an intercept. Conditional on all $B$ labels and the
fold allocation, each held-out and auxiliary mean is independent of its
training slope and has expectation zero. Consequently
$\E(C\mid B_1,\ldots,B_{n_1},\text{folds})=0$ and
$\Cov(\bar B,C)=0$. The coefficient of the independent auxiliary mean,
$c(\widehat\beta_{-1}+\widehat\beta_{-2})/2$, is nonzero with positive
probability, so $\Var(C)>0$. Gaussian centered training sums of squares
have $m_f-1$ chi-squared degrees of freedom; $m_f\ge4$ gives finite
inverse first moments and hence finite slope second moments. Independence
of training and held-out feature means then gives $\E C^2<\infty$.
Therefore $\Var(\bar B+C)>\Var(\bar B)$, despite equality of first-order
variances. More generally, for $E=\widehat\theta_{\rm fit}-
\widehat\theta_{\rm oracle}$, the exact identity is
\[
\Var(\widehat\theta_{\rm fit})=\Var(\widehat\theta_{\rm oracle})
+\Var(E)+2\Cov(\widehat\theta_{\rm oracle},E).
\]
The MSE identity replaces the last two terms by
$\E E^2+2\E\{(\widehat\theta_{\rm oracle}-\theta)E\}$.
The covariance term determines the sign of the variance contrast, and the cross-moment enters the corresponding MSE contrast.

\subsection{Relation to a general semi-supervised projection update}
\label{s:generalssl}
General semi-supervised theory provides the abstract residual-and-transfer
principle used here \citep{xu2025unified}. The notation follows arXiv:2502.17741v2, Section 4.2.1, equations (26)--(31).
For the present two-population construction, define

For group $k$, define $\bar{\bm{u}}_k^A=(n_k\bar{\bm{u}}_k^L+M_k\bar{\bm{u}}_k^U)/(n_k+M_k)$.
Let $\widetilde{\bm{\Gamma}}_k$ be the covariance matrix of all $n_k+M_k$
covariate records, centered at $\bar{\bm{u}}_k^A$, with divisor $n_k+M_k$, and set
\[
\widetilde{\bm{\beta}}_k=\widetilde{\bm{\Gamma}}_k^{-1}
\frac1{n_k}\sum_i\{\bm{u}_k(\bm{X}_{ki})-\bar{\bm{u}}_k^A\}\widehat d_{ki}.
\]
An inverse is needed only on the event where the empirical Gram is
nonsingular, whose probability tends to one; a generalized inverse can
be used elsewhere. By the empirical $L_2$ consistency of $\widehat d_k$
proved in Section~\ref{s:variance}, finite feature second moments and
Cauchy--Schwarz,
\[
\frac1{n_k}\sum_i(\bm{u}_{ki}-\bar{\bm{u}}_k^A)(\widehat d_{ki}-d_{ki})=o_p(1).
\]
The true-score law of large numbers, Gram consistency and
$\E(d_k\mid \bm{X}_k)=b_k$ therefore imply
$\widetilde{\bm{\beta}}_k\to_p\bm{\Gamma}_k^{-1}\E\{(\bm{u}_k-\bm{\mu}_k)b_k\}=\bm{\beta}_k$.
The same argument uses Section~\ref{s:restricted} for the restricted score.

The empirically centered update has the exact algebraic form
\[
\widetilde\theta=\widehat\theta_{\rm sup}
-\sum_k\widetilde{\bm{\beta}}_k^\top(\bar{\bm{u}}_k^L-\bar{\bm{u}}_k^A)
=\widehat\theta_{\rm sup}
+\sum_k c_{k,n}\widetilde{\bm{\beta}}_k^\top(\bar{\bm{u}}_k^U-\bar{\bm{u}}_k^L).
\]
The second equality follows from
$\bar{\bm{u}}_k^L-\bar{\bm{u}}_k^A=c_{k,n}(\bar{\bm{u}}_k^L-\bar{\bm{u}}_k^U)$.
After population-slope replacement, the rectangular correction in
\eqref{eq:F1} aggregates to the same full feature-mean difference.
Both $\widetilde{\bm{\beta}}_k$ and the finite set of rectangular slopes converge
to $\bm{\beta}_k$. Each corresponding feature-mean difference is
$O_p(n^{-1/2})$ under the finite positive sampling ratios. Consequently
\[
\widehat\theta_{\rm SS}-\widetilde\theta=o_p(n^{-1/2}).
\]
The present rectangular estimator and the empirical-centering projection update are first-order equivalent when supplied with the model-specific supervised scores. The censored pair and atomic expansions above provide the same-sample IPCW score $d_k=g_k+\psi_k$ used in this comparison.

The efficient version of the product experiment uses $e_k$ instead of
$d_k$. In the v2 convention, the auxiliary IF under $\sqrt{n_k}$ scaling
is $\sqrt{c_k(1-c_k)}b_k$ when its sum is divided by $\sqrt{M_k}$.
In sample-mean coordinates it is $c_kb_k$, since
$\sqrt{n_k/M_k}\,c_k=\sqrt{c_k(1-c_k)}$.
Summing the two independent population contributions gives the bound in
Theorem~\ref{thm:canonical}. The product-experiment tangent argument in
Section~\ref{s:canonical} supplies the two-population normalization.
An arbitrary supervised score $d_k=e_k+D_k$ retains
$\sum_k\pi_k^{-1}\Var(D_k)$ after projection. Thus general projection
theory explains the transfer principle, while canonical efficiency still
requires an efficient supervised score as well as the full conditional
projection.

\subsection{A fixed bounded comparison kernel}\label{s:fixed-kernel}

Let $h$ be a scalar-valued, bounded measurable function of two event times,
fixed independently of the observed samples. It need not be symmetric or
nonnegative. For the full-lifetime experiment, retain Assumptions~(A1)--(A4), including
the marginal non-atomicity requirement in Assumption~(A2). For the fixed-horizon
experiment, retain Assumption~(H1), with its grid and horizon fixed. Thus the full and
capped experiments use their respective independent-censoring, support,
feature, fold, and sample-ratio assumptions. The kernel and, in the capped experiment, the horizon are fixed.

Write $W_k=T_k$ in the full experiment and $W_k=S_k$ in the capped
experiment, and let $\theta_h=\E h(W_1,W_2)$. The population pair score is
\[
q_h=\begin{cases}
\displaystyle \frac{\Delta_1\Delta_2h(Y_1,Y_2)}{G_1(Y_1)G_2(Y_2)},
&\text{full-lifetime},\\[6pt]
\displaystyle \frac{A_1A_2h(Z_1,Z_2)}{G_{1,-}(Z_1)G_{2,-}(Z_2)},
&\text{fixed-horizon},
\end{cases}
\]
where $A_k$ and $Z_k$ are defined in Section~\ref{sec:restricted}.
Let $\widehat q_h$ replace the population censoring survivals by the
corresponding fitted Kaplan--Meier quantities, using the left-limit and
denominator-safeguard conventions of Sections~\ref{sec:method} and
\ref{sec:restricted}. Define $\widehat\theta_h(\bm{s}_n)$ by the rectangular
estimator in \eqref{eq:F1}, with $\widehat q_h$ and its corresponding
training-rectangle slopes. For functions of labeled and auxiliary records,
write $P_{L,k}$ and $P_{U,k}$ for their respective empirical means.

Put $r_{1h}(w)=\E h(w,W_2)$ and $r_{2h}(w)=\E h(W_1,w)$. For each group let
\[
\begin{aligned}
 b_{kh}(\bm{X}_k)&=\E\{r_{kh}(W_k)\mid \bm{X}_k\}-\theta_h,\\
 \bm{\mu}_k&=\E \bm{u}_k(\bm{X}_k),\qquad \bm{\Gamma}_k=\Var\{\bm{u}_k(\bm{X}_k)\},\\
 \bm{\beta}_{kh}&=\bm{\Gamma}_k^{-1}\E[\{\bm{u}_k(\bm{X}_k)-\bm{\mu}_k\}b_{kh}(\bm{X}_k)],\\
 p_{kh}(\bm{X}_k)&=\bm{\beta}_{kh}^{\mathsf T}\{\bm{u}_k(\bm{X}_k)-\bm{\mu}_k\}.
\end{aligned}
\]
Thus $p_{kh}$ is the $L_2$ projection of $b_{kh}$ onto the centered feature
space. Define the labeled pair projection and its censoring contribution by
\[
 g_{kh}=\begin{cases}
 \displaystyle\frac{\Delta_k r_{kh}(Y_k)}{G_k(Y_k)}-\theta_h,&\text{full-lifetime},\\[6pt]
 \displaystyle\frac{A_k r_{kh}(Z_k)}{G_{k,-}(Z_k)}-\theta_h,&\text{fixed-horizon},
 \end{cases}
 \qquad d_{kh}=g_{kh}+\psi_{kh},
\]
where
\[
 \psi_{kh}=\begin{cases}
 \displaystyle\int_0^{b_0}\frac{a_{0,kh}(t)}{G_k(t)}\,dM_k^C(t),
 &\text{full-lifetime},\\[8pt]
 \displaystyle\sum_{t<\tau}
 \frac{\E\{r_{kh}(S_k)\ind(S_k>t)\}}
 {h_{kt}(1-a_{kt})}\mathcal M_k(t),&\text{fixed-horizon},
 \end{cases}
\]
with $a_{0,kh}(t)=\E\{r_{kh}(T_k)\mid T_k\ge t\}$ in the full experiment;
the quantities in the capped sum are those in \eqref{eq:atomicpsi}.
For the conditional-tail score, put
\[
 a_{\bm{X},kh}(t,\bm{x})=\begin{cases}
 \E\{r_{kh}(T_k)\mid T_k\ge t,\bm{X}_k=\bm{x}\},&\text{full-lifetime},\\
 \E\{r_{kh}(S_k)\mid S_k>t,\bm{X}_k=\bm{x}\},&\text{fixed-horizon},
 \end{cases}
\]
and define
\[
 e_{kh}=\begin{cases}
 \displaystyle g_{kh}+\int_0^{b_0}\frac{a_{\bm{X},kh}(t,\bm{X}_k)}{G_k(t)}\,dM_k^C(t),
 &\text{full-lifetime},\\[8pt]
 \displaystyle g_{kh}+\sum_{t<\tau}
 \frac{a_{\bm{X},kh}(t,\bm{X}_k)}{G_{k,-}(t)(1-a_{kt})}\mathcal M_k(t),
 &\text{fixed-horizon},
 \end{cases}
 \qquad D_{kh}=d_{kh}-e_{kh}.
\]
Conditional expectations on zero-risk sets may be assigned arbitrary values.
\renewcommand{\theproposition}{\thesection.\arabic{proposition}}
\renewcommand{\theHproposition}{\theHsection.\arabic{proposition}}
\setcounter{proposition}{0}
\begin{proposition}\label{prop:fixed-kernel}
For each fixed $h$ and either specified observation model, let
$\bm{s}_n=(s_{1,n},s_{2,n})$ be a bounded deterministic sequence with
$s_{k,n}\to s_k$. Then
\[
 \widehat\theta_h(\bm{s}_n)-\theta_h
 =\sum_{k=1}^2\left[P_{L,k}(d_{kh}-s_kp_{kh})
                 +P_{U,k}s_kp_{kh}\right]+o_p(n^{-1/2}).
\]
Its limiting $n$-scaled variance is
\[
 V_h(\bm{s})=\sum_{k=1}^2\left\{\pi_k^{-1}\Var(d_{kh}-s_kp_{kh})
 +\lambda_k^{-1}s_k^2\Var(p_{kh})\right\}.
\]
This variance is minimized at $c_k=\lambda_k/(\pi_k+\lambda_k)$.
The minimum in coordinate $k$ is unique when $\Var(p_{kh})>0$; when
$\Var(p_{kh})=0$, that coordinate does not affect $V_h(\bm{s})$.

In the corresponding nonparametric latent product model
$P_{\bm{X}_k}P_{W_k\mid \bm{X}_k}P_{C_k}$ for each group, with censoring independent
of $(W_k,\bm{X}_k)$ and subject to the same observation-model assumptions, the
canonical-gradient coordinates at $c=(c_1,c_2)$ are
$(e_{kh}-c_kb_{kh},c_kb_{kh})$. The efficiency bound is
\[
 V_{\mathrm{eff},h}=\sum_{k=1}^2\left\{
 \pi_k^{-1}\E\Var(e_{kh}\mid \bm{X}_k)
 + (\pi_k+\lambda_k)^{-1}\Var(b_{kh})\right\},
\]
and
\[
 V_h(\bm{c})-V_{\mathrm{eff},h}
 =\sum_{k=1}^2\left\{\frac{\Var(D_{kh})}{\pi_k}
 + (\pi_k^{-1}+\lambda_k^{-1})c_k^2
          \Var(b_{kh}-p_{kh})\right\}.
\]
At the proposed weight $c_{k,n}=M_k/(M_k+n_k)$, the four-pool variance
estimator in \eqref{eq:SE}, formed from the corresponding complete labeled
scores $d_{kh}-c_kp_{kh}$ and auxiliary scores $c_kp_{kh}$, is consistent
under $n$ scaling. If $V_h(\bm{c})>0$, studentization is asymptotically normal.
\end{proposition}

\begin{proof}
Conditional on $(W_1,W_2,\bm{X}_1,\bm{X}_2)$, each complete-record multiplier has
expectation one. Their product identifies $h(W_1,W_2)$, including signed
values. Positivity gives a fixed bound on the weighted kernel. Let
$q_h^\circ=q_h-\E(q_h\mid O_1)-\E(q_h\mid O_2)+\theta_h$ be its
fully degenerate two-sample projection. Its U-statistic remainder has
variance $\E(q_h^\circ)^2/(n_1n_2)$. For the full model, Lemma~\ref{lem:aux-km}
does not involve the comparison kernel. Substitution produces the same
three roles as in Lemma~\ref{lem:aux-supervised}. Coincident pair and
nuisance indices have $O(n^2)$ bounded terms over an $O(n^3)$
normalizer. On distinct indices, averaging the nuisance observation
gives zero at a deterministic evaluation time, whereas conditioning on
that observation and averaging the pair gives $\psi_{kh}$. Subtracting
this projection leaves zero single-unit conditional means. Residual
triples sharing at most one unit have zero covariance; those sharing
two or more have $O(n^4)$ terms over an $O(n^6)$ squared normalizer.
Thus the residual is $O_p(n^{-1})$, also for signed $q_h$.

For the capped model, each pre-horizon risk has a positive population
lower bound. Let $P_m$ denote empirical averaging over the labeled pool
of size $m$. Taylor differentiation of the fixed finite hazard vector
gives $\widehat a_t-a_t=P_m\mathcal M(t)/h_t+O_p(m^{-1})$.
Differentiating $\prod_{t<s}(1-a_t)^{-1}$ gives
$\sum_{t<s}\mathcal M(t)/\{h_t(1-a_t)\}$, with a simultaneous
$O_p(m^{-1})$ remainder. Averaging the pair now selects the strict tail
$W>t$; the preceding index count gives the capped supervised expansion.

The training Gram and response kernels are integrable because features
have finite second moments and $q_h$ is bounded. Truncation followed by
the two-sample LLN gives Gram convergence and slope consistency for each
training rectangle. A finite number of rectangles gives simultaneous
consistency. Each coefficient error is multiplied by an
$O_p(n^{-1/2})$ feature-mean difference. The exact identities
$\sum_b w_{ab}=n_{1a}/n_1$ and $\sum_a w_{ab}=n_{2b}/n_2$ aggregate the
population-slope terms. Combining the resulting expansion with the
supervised expansion proves the asserted ALR. This argument allows the
held-out records to contribute to the group-specific censoring fit across labeled folds.

Conditional censoring-increment centering gives
$\E(\psi_{kh}\mid W_k,\bm{X}_k)=0$ and hence
$\E(d_{kh}\mid \bm{X}_k)=b_{kh}$. Projection gives
$\Cov(d_{kh},p_{kh})=\Var(p_{kh})$. The independent four-pool CLT
provides the variance. Completing its quadratic in each $s_k$ gives
the stated optimum, including the flat zero-signal case.

For completeness, canonical membership is established within the stated
product model. Suppress $k$ and write $m(\bm{x})=\E\{r_h(W)\mid \bm{X}=\bm{x}\}$,
$S_X(t)=\Prb(W\ge t\mid \bm{X})$, and $S_X^+(t)=\Prb(W>t\mid \bm{X})$.
For the chosen experiment define $G^*(w)=G(w)$ in the full model and
$G^*(w)=G_-(w)$ in the capped model. Set
\[
 J(w,\bm{x})=\begin{cases}
 \displaystyle\int_0^{w\wedge b_0}\frac{a_X(t,\bm{x})}{G(t)}d\Lambda^C(t),
 &\text{full-lifetime},\\[8pt]
 \displaystyle\sum_{t<w}\frac{a_X(t,\bm{x})a_t}{G_-(t)(1-a_t)},
 &\text{fixed-horizon},
 \end{cases}
 \qquad v_*(w,\bm{x})=\frac{r_h(w)}{G^*(w)}-J(w,\bm{x})-m(\bm{x}).
\]
The reciprocal-survival integral identity, or its exact finite-product
telescope, gives $\E(v_*\mid \bm{X})=0$. On a complete record,
$b(\bm{X})+\E(v_*\mid O)=r_h(W)/G^*(W)-\theta_h-J(W,\bm{X})$.
On a censored record at $t$, the conditional expectation of the future
part of $r_h(W)/G^*(W)$ cancels the future part of $J(W,\bm{X})$, leaving
$a_X(t,\bm{X})/G^*(t)-J(t,\bm{X})-\theta_h$. These are exactly the event and
censor values of $e_h$. Therefore $e_h=b+\E(v_*\mid O)$.
Both $b$ and $v_*$ are bounded. Bounded tilts of $P_X$ and $P_{W\mid \bm{X}}$
preserve the model and its local positivity, proving tangent membership.
The conditional identity $\E(e_h\mid W,\bm{X})=r_h(W)-\theta_h$ represents
all outcome and covariate derivatives. Censoring isometry cancels
every independent censoring score: in the full model its integrand is
$\E\{S_X(t)a_X(t,\bm{X})\}$ and in the capped model it is
$\E\{S_X^+(t)a_X(t,\bm{X})\}$, which equal the respective marginal tails.
This establishes the derivative and nuisance conditions separately
from membership.

The same identities show $D_h\perp e_h$ and $D_h\perp L_2(\bm{X})$:
conditional increment means handle outcome and covariate scores, and
the marginal identity
$\E\{S_X^+(a_0-a_X)\}=0$ (or its full analogue) handles censoring scores.
Under the product-experiment inner product, the shared covariate Riesz
coordinate is $b/(\pi+\lambda)$ in both pools and the conditional-outcome
coordinate is $\E(v_*\mid O)/\pi$ only in the labeled pool. Converting
to sample-mean coordinates gives $(e_h-cb,cb)$. Orthogonality and
completion of squares then give the two nonnegative losses displayed
above.

Finally, for known censoring weights the average row mean-squared
error is bounded by a constant times $1/n_{\mathrm{other}}$.
Uniform weight consistency transfers this to fitted rows. In the full
model the signed weighted tail measure has total variation at most
$\E|q_h|$. Its threshold averages converge uniformly by finite
partitions of that variation measure. The limiting nuisance coefficient
has bounded variation, so Stieltjes integration by parts controls the
fitted same-sample hazard error as in Section~\ref{s:variance}.
In the capped model only a fixed finite vector of risk, tail and hazard
averages is involved; positive denominators permit simultaneous
convergence. Consistent slopes and empirical feature second moments
give adjustment-score empirical $L_2$ convergence in either model.
The four separately centered variances therefore converge to their ALR
variances. Slutsky's theorem gives studentization when the limit is
positive.
\end{proof}

The proposition is pointwise for each prespecified fixed kernel $h$.

\FloatBarrier
\section{Additional numerical results}\label{app:additional}
\subsection{Additional simulation results}
\subsubsection{Event and censoring laws}\label{s:restricteddesign}

The recorded grid is $\{0,1,2,3,4\}$ and $\tau=4$. Covariates have two or three equally likely levels, identically distributed within each labeled/auxiliary pair. Each group has $n_k=240$ or 960, with auxiliary sizes rounded from the specified ratio. 

For binary $x$, the event PMF on the first three event times has baseline $(1-m)(.325,.35,.325)$ and terminal mass $m$, with opposite shifts $\pm\delta(2x-1)(1,0,-1)$. The nonzero-effect baseline uses $(1-m)(.325+.075a,.35,.325-.075a)$ for group 1 and the reversed first/last components for group 2. For three levels, the linear signal is $\pm\delta(x-1)(1,0,-1)$. Scenarios 13--14 use the centered quadratic $q(x)=(x-1)^2-2/3$ and shift $\pm .31183q(x)(1,0,-1)$, with $m=.3$ and zero baseline contrast. Scenario 13 uses $x$, Scenario 14 uses $(x,x^2)$; the data and random splits are shared.

Light censoring is supported on $(0,1,2,4,8)$, with masses $(.02,.03,.05,.20,.70)$. Heavy censoring uses the same support and masses $(.10,.15,.15,.20,.40)$. The overlap-censoring law has support $(0,1,2,3,4,8)$ and masses $(.05,.15,.15,.15,.10,.40)$. Censoring is independent of $(S,X)$. A terminal latent time is 4 or 6 with equal probability, without changing the capped parameter. In Scenario 20, the conditional recorded PMF is that of $\min\{\lfloor4T\rfloor,4\}$ for exponential $T$ with rate $\exp\{.6(x-1)\}$ in group 1 and $.6\exp\{.6(x-1)\}$ in group 2. Its terminal atom is the residual survival mass, so the comparison concerns capped recorded times.

The remaining amplitudes, baseline contrasts and terminal masses are: Scenarios 9, 10 and 15--18 have binary $X$, $m=.4$, baseline contrast $a=1$, with $\delta=0$ in Scenario 9 and .04 otherwise; Scenario 11 has three levels, $m=.3$, $a=0$, $\delta=.18081$; Scenario 12 has binary $X$, $m=.2$, $a=0$, $\delta=.21058$; Scenario 19 has three levels, $m=.3$, $a=2$, $\delta=.10$. Scenarios 17 and 18 use ratios .25 and 5. The initial fixed-horizon grid uses 4,000 repetitions per cell.

The same event laws define the resource configurations. Scenarios 14 and 20 are evaluated at per-group labeled sizes 480 and 1,920, with equal auxiliary sizes. Scenario 21 combines the strong-signal event law of Scenario 12 with overlap censoring and resource ratios 0.25, 1 and 5 at labeled sizes 240 and 960. Scenarios 9, 10 and 12 are also evaluated with 240 labeled observations per group.

\subsubsection{Precision, sample size and working features}
The nonzero-effect configurations show small estimation bias, while the heavier-censoring results retain near-nominal coverage across the two labeled sample sizes. For example, the weighted estimator's coverage in Scenario 8 is 94.4\% and 95.1\% at $n_0=200$ and 800 when $M_0/n_0=1$; Table~\ref{tab:additionalFull} also gives the corresponding calibration under larger auxiliary pools.
\begingroup\TableFont
\setlength{\tabcolsep}{0pt}\sbox{\UniformMeasure}{\begin{tabular}{@{}lrrlrrrrr@{}}Scenario & $(n_1,n_2)$ & $(M_1,M_2)$ & Method & Bias & EmpSD & MeanSE & RMSE & CP (\%) \\
\multicolumn{9}{l}{\textit{A. Nonzero effects and heavier censoring}}\\\addlinespace[2pt]
4 & 200,200 & 200,200 & $\widehat{\theta}_{\mathrm{sup}}$ & \tablenum[table-format=-1.5]{-0.00035} & \tablenum[table-format=1.5]{0.03076} & \tablenum[table-format=1.5]{0.03140} & \tablenum[table-format=1.5]{0.03076} & \tablenum[table-format=2.2]{95.4} \\
4 & 200,200 & 200,200 & $\widehat{\theta}_{\mathrm{un}}$ & \tablenum[table-format=-1.5]{-0.00033} & \tablenum[table-format=1.5]{0.03099} & \tablenum[table-format=1.5]{0.03151} & \tablenum[table-format=1.5]{0.03098} & \tablenum[table-format=2.2]{95.3} \\
4 & 200,200 & 200,200 & $\widehat{\theta}_{\mathrm{SS}}$ & \tablenum[table-format=-1.5]{-0.00034} & \tablenum[table-format=1.5]{0.03078} & \tablenum[table-format=1.5]{0.03139} & \tablenum[table-format=1.5]{0.03077} & \tablenum[table-format=2.2]{95.3} \\
5 & 200,200 & 200,200 & $\widehat{\theta}_{\mathrm{sup}}$ & \tablenum[table-format=-1.5]{0.00149} & \tablenum[table-format=1.5]{0.03130} & \tablenum[table-format=1.5]{0.03144} & \tablenum[table-format=1.5]{0.03133} & \tablenum[table-format=2.2]{94.9} \\
5 & 200,200 & 200,200 & $\widehat{\theta}_{\mathrm{un}}$ & \tablenum[table-format=-1.5]{0.00134} & \tablenum[table-format=1.5]{0.03212} & \tablenum[table-format=1.5]{0.03158} & \tablenum[table-format=1.5]{0.03214} & \tablenum[table-format=2.2]{94.7} \\
5 & 200,200 & 200,200 & $\widehat{\theta}_{\mathrm{SS}}$ & \tablenum[table-format=-1.5]{0.00142} & \tablenum[table-format=1.5]{0.02952} & \tablenum[table-format=1.5]{0.02939} & \tablenum[table-format=1.5]{0.02955} & \tablenum[table-format=2.2]{95} \\
6 & 200,200 & 200,200 & $\widehat{\theta}_{\mathrm{sup}}$ & \tablenum[table-format=-1.5]{0.00086} & \tablenum[table-format=1.5]{0.03105} & \tablenum[table-format=1.5]{0.03088} & \tablenum[table-format=1.5]{0.03105} & \tablenum[table-format=2.2]{95.4} \\
6 & 200,200 & 200,200 & $\widehat{\theta}_{\mathrm{un}}$ & \tablenum[table-format=-1.5]{0.00088} & \tablenum[table-format=1.5]{0.03185} & \tablenum[table-format=1.5]{0.03116} & \tablenum[table-format=1.5]{0.03185} & \tablenum[table-format=2.2]{94.6} \\
6 & 200,200 & 200,200 & $\widehat{\theta}_{\mathrm{SS}}$ & \tablenum[table-format=-1.5]{0.00087} & \tablenum[table-format=1.5]{0.03128} & \tablenum[table-format=1.5]{0.03091} & \tablenum[table-format=1.5]{0.03129} & \tablenum[table-format=2.2]{95.2} \\
7 & 200,200 & 200,200 & $\widehat{\theta}_{\mathrm{sup}}$ & \tablenum[table-format=-1.5]{-0.00080} & \tablenum[table-format=1.5]{0.02968} & \tablenum[table-format=1.5]{0.03090} & \tablenum[table-format=1.5]{0.02968} & \tablenum[table-format=2.2]{95.7} \\
7 & 200,200 & 200,200 & $\widehat{\theta}_{\mathrm{un}}$ & \tablenum[table-format=-1.5]{-0.00049} & \tablenum[table-format=1.5]{0.03077} & \tablenum[table-format=1.5]{0.03118} & \tablenum[table-format=1.5]{0.03077} & \tablenum[table-format=2.2]{95.4} \\
7 & 200,200 & 200,200 & $\widehat{\theta}_{\mathrm{SS}}$ & \tablenum[table-format=-1.5]{-0.00065} & \tablenum[table-format=1.5]{0.02793} & \tablenum[table-format=1.5]{0.02887} & \tablenum[table-format=1.5]{0.02793} & \tablenum[table-format=2.2]{95.8} \\
8 & 200,200 & 200,200 & $\widehat{\theta}_{\mathrm{sup}}$ & \tablenum[table-format=-1.5]{0.00010} & \tablenum[table-format=1.5]{0.03754} & \tablenum[table-format=1.5]{0.03809} & \tablenum[table-format=1.5]{0.03753} & \tablenum[table-format=2.2]{95.2} \\
8 & 200,200 & 200,200 & $\widehat{\theta}_{\mathrm{un}}$ & \tablenum[table-format=-1.5]{-0.00005} & \tablenum[table-format=1.5]{0.03956} & \tablenum[table-format=1.5]{0.03850} & \tablenum[table-format=1.5]{0.03955} & \tablenum[table-format=2.2]{93.4} \\
8 & 200,200 & 200,200 & $\widehat{\theta}_{\mathrm{SS}}$ & \tablenum[table-format=-1.5]{0.00003} & \tablenum[table-format=1.5]{0.03633} & \tablenum[table-format=1.5]{0.03622} & \tablenum[table-format=1.5]{0.03632} & \tablenum[table-format=2.2]{94.4} \\
4 & 800,800 & 800,800 & $\widehat{\theta}_{\mathrm{sup}}$ & \tablenum[table-format=-1.5]{0.00060} & \tablenum[table-format=1.5]{0.01530} & \tablenum[table-format=1.5]{0.01566} & \tablenum[table-format=1.5]{0.01531} & \tablenum[table-format=2.2]{95.4} \\
4 & 800,800 & 800,800 & $\widehat{\theta}_{\mathrm{un}}$ & \tablenum[table-format=-1.5]{0.00060} & \tablenum[table-format=1.5]{0.01537} & \tablenum[table-format=1.5]{0.01568} & \tablenum[table-format=1.5]{0.01538} & \tablenum[table-format=2.2]{95.5} \\
4 & 800,800 & 800,800 & $\widehat{\theta}_{\mathrm{SS}}$ & \tablenum[table-format=-1.5]{0.00060} & \tablenum[table-format=1.5]{0.01532} & \tablenum[table-format=1.5]{0.01566} & \tablenum[table-format=1.5]{0.01533} & \tablenum[table-format=2.2]{95.6} \\
5 & 800,800 & 800,800 & $\widehat{\theta}_{\mathrm{sup}}$ & \tablenum[table-format=-1.5]{0.00014} & \tablenum[table-format=1.5]{0.01558} & \tablenum[table-format=1.5]{0.01566} & \tablenum[table-format=1.5]{0.01558} & \tablenum[table-format=2.2]{94.7} \\
5 & 800,800 & 800,800 & $\widehat{\theta}_{\mathrm{un}}$ & \tablenum[table-format=-1.5]{-0.00039} & \tablenum[table-format=1.5]{0.01547} & \tablenum[table-format=1.5]{0.01568} & \tablenum[table-format=1.5]{0.01547} & \tablenum[table-format=2.2]{95} \\
5 & 800,800 & 800,800 & $\widehat{\theta}_{\mathrm{SS}}$ & \tablenum[table-format=-1.5]{-0.00012} & \tablenum[table-format=1.5]{0.01451} & \tablenum[table-format=1.5]{0.01464} & \tablenum[table-format=1.5]{0.01450} & \tablenum[table-format=2.2]{94.8} \\
6 & 800,800 & 800,800 & $\widehat{\theta}_{\mathrm{sup}}$ & \tablenum[table-format=-1.5]{-0.00021} & \tablenum[table-format=1.5]{0.01535} & \tablenum[table-format=1.5]{0.01541} & \tablenum[table-format=1.5]{0.01535} & \tablenum[table-format=2.2]{94.7} \\
6 & 800,800 & 800,800 & $\widehat{\theta}_{\mathrm{un}}$ & \tablenum[table-format=-1.5]{-0.00020} & \tablenum[table-format=1.5]{0.01544} & \tablenum[table-format=1.5]{0.01544} & \tablenum[table-format=1.5]{0.01544} & \tablenum[table-format=2.2]{94.7} \\
6 & 800,800 & 800,800 & $\widehat{\theta}_{\mathrm{SS}}$ & \tablenum[table-format=-1.5]{-0.00020} & \tablenum[table-format=1.5]{0.01538} & \tablenum[table-format=1.5]{0.01541} & \tablenum[table-format=1.5]{0.01537} & \tablenum[table-format=2.2]{94.8} \\
7 & 800,800 & 800,800 & $\widehat{\theta}_{\mathrm{sup}}$ & \tablenum[table-format=-1.5]{-0.00006} & \tablenum[table-format=1.5]{0.01572} & \tablenum[table-format=1.5]{0.01541} & \tablenum[table-format=1.5]{0.01572} & \tablenum[table-format=2.2]{94.9} \\
7 & 800,800 & 800,800 & $\widehat{\theta}_{\mathrm{un}}$ & \tablenum[table-format=-1.5]{-0.00010} & \tablenum[table-format=1.5]{0.01547} & \tablenum[table-format=1.5]{0.01544} & \tablenum[table-format=1.5]{0.01547} & \tablenum[table-format=2.2]{94.9} \\
7 & 800,800 & 800,800 & $\widehat{\theta}_{\mathrm{SS}}$ & \tablenum[table-format=-1.5]{-0.00008} & \tablenum[table-format=1.5]{0.01453} & \tablenum[table-format=1.5]{0.01437} & \tablenum[table-format=1.5]{0.01453} & \tablenum[table-format=2.2]{94.6} \\
8 & 800,800 & 800,800 & $\widehat{\theta}_{\mathrm{sup}}$ & \tablenum[table-format=-1.5]{0.00012} & \tablenum[table-format=1.5]{0.01929} & \tablenum[table-format=1.5]{0.01895} & \tablenum[table-format=1.5]{0.01928} & \tablenum[table-format=2.2]{94.9} \\
8 & 800,800 & 800,800 & $\widehat{\theta}_{\mathrm{un}}$ & \tablenum[table-format=-1.5]{-0.00006} & \tablenum[table-format=1.5]{0.01900} & \tablenum[table-format=1.5]{0.01901} & \tablenum[table-format=1.5]{0.01899} & \tablenum[table-format=2.2]{95.2} \\
8 & 800,800 & 800,800 & $\widehat{\theta}_{\mathrm{SS}}$ & \tablenum[table-format=-1.5]{0.00003} & \tablenum[table-format=1.5]{0.01817} & \tablenum[table-format=1.5]{0.01799} & \tablenum[table-format=1.5]{0.01817} & \tablenum[table-format=2.2]{95.1} \\
\addlinespace
\multicolumn{9}{l}{\textit{B. Additional resource and signal structures}}\\\addlinespace[2pt]
8 & 200,200 & 1000,1000 & $\widehat{\theta}_{\mathrm{sup}}$ & \tablenum[table-format=-1.5]{-0.0002} & \tablenum[table-format=1.5]{0.0382} & \tablenum[table-format=1.5]{0.0381} & \tablenum[table-format=1.5]{0.0382} & \tablenum[table-format=2.2]{94.8} \\
8 & 200,200 & 1000,1000 & $\widehat{\theta}_{\mathrm{un}}$ & \tablenum[table-format=-1.5]{-0.0006} & \tablenum[table-format=1.5]{0.0360} & \tablenum[table-format=1.5]{0.0351} & \tablenum[table-format=1.5]{0.0360} & \tablenum[table-format=2.2]{94.47} \\
8 & 200,200 & 1000,1000 & $\widehat{\theta}_{\mathrm{SS}}$ & \tablenum[table-format=-1.5]{-0.0005} & \tablenum[table-format=1.5]{0.0356} & \tablenum[table-format=1.5]{0.0349} & \tablenum[table-format=1.5]{0.0356} & \tablenum[table-format=2.2]{94.58} \\
8 & 800,800 & 4000,4000 & $\widehat{\theta}_{\mathrm{sup}}$ & \tablenum[table-format=-1.5]{0.0004} & \tablenum[table-format=1.5]{0.0189} & \tablenum[table-format=1.5]{0.0189} & \tablenum[table-format=1.5]{0.0189} & \tablenum[table-format=2.2]{94.92} \\
8 & 800,800 & 4000,4000 & $\widehat{\theta}_{\mathrm{un}}$ & \tablenum[table-format=-1.5]{0.0004} & \tablenum[table-format=1.5]{0.0174} & \tablenum[table-format=1.5]{0.0174} & \tablenum[table-format=1.5]{0.0174} & \tablenum[table-format=2.2]{95.08} \\
8 & 800,800 & 4000,4000 & $\widehat{\theta}_{\mathrm{SS}}$ & \tablenum[table-format=-1.5]{0.0004} & \tablenum[table-format=1.5]{0.0174} & \tablenum[table-format=1.5]{0.0173} & \tablenum[table-format=1.5]{0.0174} & \tablenum[table-format=2.2]{94.95} \\
22 & 200,200 & 200,200 & $\widehat{\theta}_{\mathrm{SS}}$ & \tablenum[table-format=-1.5]{-0.0006} & \tablenum[table-format=1.5]{0.0315} & \tablenum[table-format=1.5]{0.0315} & \tablenum[table-format=1.5]{0.0315} & \tablenum[table-format=2.2]{94.77} \\
22 & 200,200 & 1000,1000 & $\widehat{\theta}_{\mathrm{SS}}$ & \tablenum[table-format=-1.5]{-0.0007} & \tablenum[table-format=1.5]{0.0306} & \tablenum[table-format=1.5]{0.0304} & \tablenum[table-format=1.5]{0.0306} & \tablenum[table-format=2.2]{94.8} \\
22 & 200,400 & 50,2000 & $\widehat{\theta}_{\mathrm{SS}}$ & \tablenum[table-format=-1.5]{-0.0005} & \tablenum[table-format=1.5]{0.0272} & \tablenum[table-format=1.5]{0.0275} & \tablenum[table-format=1.5]{0.0272} & \tablenum[table-format=2.2]{95.28} \\
22 & 200,400 & 1000,100 & $\widehat{\theta}_{\mathrm{SS}}$ & \tablenum[table-format=-1.5]{-0.0003} & \tablenum[table-format=1.5]{0.0268} & \tablenum[table-format=1.5]{0.0269} & \tablenum[table-format=1.5]{0.0268} & \tablenum[table-format=2.2]{94.92} \\
23 & 200,200 & 1000,1000 & $\widehat{\theta}_{\mathrm{SS}}$ & \tablenum[table-format=-1.5]{-0.0003} & \tablenum[table-format=1.5]{0.0333} & \tablenum[table-format=1.5]{0.0330} & \tablenum[table-format=1.5]{0.0333} & \tablenum[table-format=2.2]{94.77} \\
24 & 200,200 & 1000,1000 & $\widehat{\theta}_{\mathrm{SS}}$ & \tablenum[table-format=-1.5]{-0.0003} & \tablenum[table-format=1.5]{0.0318} & \tablenum[table-format=1.5]{0.0312} & \tablenum[table-format=1.5]{0.0318} & \tablenum[table-format=2.2]{94.8} \\
23 & 800,800 & 4000,4000 & $\widehat{\theta}_{\mathrm{SS}}$ & \tablenum[table-format=-1.5]{0.0000} & \tablenum[table-format=1.5]{0.0167} & \tablenum[table-format=1.5]{0.0165} & \tablenum[table-format=1.5]{0.0167} & \tablenum[table-format=2.2]{94.75} \\
24 & 800,800 & 4000,4000 & $\widehat{\theta}_{\mathrm{SS}}$ & \tablenum[table-format=-1.5]{0.0000} & \tablenum[table-format=1.5]{0.0159} & \tablenum[table-format=1.5]{0.0155} & \tablenum[table-format=1.5]{0.0159} & \tablenum[table-format=2.2]{94.58} \\
25 & 200,200 & 1000,1000 & $\widehat{\theta}_{\mathrm{SS}}$ & \tablenum[table-format=-1.5]{-0.0005} & \tablenum[table-format=1.5]{0.0308} & \tablenum[table-format=1.5]{0.0304} & \tablenum[table-format=1.5]{0.0308} & \tablenum[table-format=2.2]{94.83} \\
25 & 800,800 & 4000,4000 & $\widehat{\theta}_{\mathrm{SS}}$ & \tablenum[table-format=-1.5]{0.0008} & \tablenum[table-format=1.5]{0.0150} & \tablenum[table-format=1.5]{0.0151} & \tablenum[table-format=1.5]{0.0150} & \tablenum[table-format=2.2]{94.97} \\
\addlinespace
\end{tabular}}
\setlength{\UniformWidth}{\linewidth}\setlength{\tabcolsep}{\dimexpr(\linewidth-\wd\UniformMeasure)/16\relax}
\ifdim\tabcolsep<0pt\PackageWarning{CompactTables}{Negative column gap}\fi
\setlength{\LTleft}{0pt plus 1fill}\setlength{\LTright}{0pt plus 1fill}
\begin{longtable}{@{}lrrlrrrrr@{}}
\caption{Additional full-lifetime comparisons.}\label{tab:additionalFull}\\\toprule
Scenario & $(n_1,n_2)$ & $(M_1,M_2)$ & Method & Bias & EmpSD & MeanSE & RMSE & CP (\%) \\
\midrule\endfirsthead
\multicolumn{9}{l}{\textit{Table \thetable\ continued}}\\\toprule
Scenario & $(n_1,n_2)$ & $(M_1,M_2)$ & Method & Bias & EmpSD & MeanSE & RMSE & CP (\%) \\
\midrule\endhead
\midrule\multicolumn{9}{r}{\textit{Continued on next page}}\\\endfoot
\endlastfoot
\multicolumn{9}{l}{\textit{A. Nonzero effects and heavier censoring}}\\\addlinespace[2pt]
4 & 200,200 & 200,200 & $\widehat{\theta}_{\mathrm{sup}}$ & \tablenum[table-format=-1.5]{-0.00035} & \tablenum[table-format=1.5]{0.03076} & \tablenum[table-format=1.5]{0.03140} & \tablenum[table-format=1.5]{0.03076} & \tablenum[table-format=2.2]{95.4} \\
4 & 200,200 & 200,200 & $\widehat{\theta}_{\mathrm{un}}$ & \tablenum[table-format=-1.5]{-0.00033} & \tablenum[table-format=1.5]{0.03099} & \tablenum[table-format=1.5]{0.03151} & \tablenum[table-format=1.5]{0.03098} & \tablenum[table-format=2.2]{95.3} \\
4 & 200,200 & 200,200 & $\widehat{\theta}_{\mathrm{SS}}$ & \tablenum[table-format=-1.5]{-0.00034} & \tablenum[table-format=1.5]{0.03078} & \tablenum[table-format=1.5]{0.03139} & \tablenum[table-format=1.5]{0.03077} & \tablenum[table-format=2.2]{95.3} \\
5 & 200,200 & 200,200 & $\widehat{\theta}_{\mathrm{sup}}$ & \tablenum[table-format=-1.5]{0.00149} & \tablenum[table-format=1.5]{0.03130} & \tablenum[table-format=1.5]{0.03144} & \tablenum[table-format=1.5]{0.03133} & \tablenum[table-format=2.2]{94.9} \\
5 & 200,200 & 200,200 & $\widehat{\theta}_{\mathrm{un}}$ & \tablenum[table-format=-1.5]{0.00134} & \tablenum[table-format=1.5]{0.03212} & \tablenum[table-format=1.5]{0.03158} & \tablenum[table-format=1.5]{0.03214} & \tablenum[table-format=2.2]{94.7} \\
5 & 200,200 & 200,200 & $\widehat{\theta}_{\mathrm{SS}}$ & \tablenum[table-format=-1.5]{0.00142} & \tablenum[table-format=1.5]{0.02952} & \tablenum[table-format=1.5]{0.02939} & \tablenum[table-format=1.5]{0.02955} & \tablenum[table-format=2.2]{95} \\
6 & 200,200 & 200,200 & $\widehat{\theta}_{\mathrm{sup}}$ & \tablenum[table-format=-1.5]{0.00086} & \tablenum[table-format=1.5]{0.03105} & \tablenum[table-format=1.5]{0.03088} & \tablenum[table-format=1.5]{0.03105} & \tablenum[table-format=2.2]{95.4} \\
6 & 200,200 & 200,200 & $\widehat{\theta}_{\mathrm{un}}$ & \tablenum[table-format=-1.5]{0.00088} & \tablenum[table-format=1.5]{0.03185} & \tablenum[table-format=1.5]{0.03116} & \tablenum[table-format=1.5]{0.03185} & \tablenum[table-format=2.2]{94.6} \\
6 & 200,200 & 200,200 & $\widehat{\theta}_{\mathrm{SS}}$ & \tablenum[table-format=-1.5]{0.00087} & \tablenum[table-format=1.5]{0.03128} & \tablenum[table-format=1.5]{0.03091} & \tablenum[table-format=1.5]{0.03129} & \tablenum[table-format=2.2]{95.2} \\
7 & 200,200 & 200,200 & $\widehat{\theta}_{\mathrm{sup}}$ & \tablenum[table-format=-1.5]{-0.00080} & \tablenum[table-format=1.5]{0.02968} & \tablenum[table-format=1.5]{0.03090} & \tablenum[table-format=1.5]{0.02968} & \tablenum[table-format=2.2]{95.7} \\
7 & 200,200 & 200,200 & $\widehat{\theta}_{\mathrm{un}}$ & \tablenum[table-format=-1.5]{-0.00049} & \tablenum[table-format=1.5]{0.03077} & \tablenum[table-format=1.5]{0.03118} & \tablenum[table-format=1.5]{0.03077} & \tablenum[table-format=2.2]{95.4} \\
7 & 200,200 & 200,200 & $\widehat{\theta}_{\mathrm{SS}}$ & \tablenum[table-format=-1.5]{-0.00065} & \tablenum[table-format=1.5]{0.02793} & \tablenum[table-format=1.5]{0.02887} & \tablenum[table-format=1.5]{0.02793} & \tablenum[table-format=2.2]{95.8} \\
8 & 200,200 & 200,200 & $\widehat{\theta}_{\mathrm{sup}}$ & \tablenum[table-format=-1.5]{0.00010} & \tablenum[table-format=1.5]{0.03754} & \tablenum[table-format=1.5]{0.03809} & \tablenum[table-format=1.5]{0.03753} & \tablenum[table-format=2.2]{95.2} \\
8 & 200,200 & 200,200 & $\widehat{\theta}_{\mathrm{un}}$ & \tablenum[table-format=-1.5]{-0.00005} & \tablenum[table-format=1.5]{0.03956} & \tablenum[table-format=1.5]{0.03850} & \tablenum[table-format=1.5]{0.03955} & \tablenum[table-format=2.2]{93.4} \\
8 & 200,200 & 200,200 & $\widehat{\theta}_{\mathrm{SS}}$ & \tablenum[table-format=-1.5]{0.00003} & \tablenum[table-format=1.5]{0.03633} & \tablenum[table-format=1.5]{0.03622} & \tablenum[table-format=1.5]{0.03632} & \tablenum[table-format=2.2]{94.4} \\
4 & 800,800 & 800,800 & $\widehat{\theta}_{\mathrm{sup}}$ & \tablenum[table-format=-1.5]{0.00060} & \tablenum[table-format=1.5]{0.01530} & \tablenum[table-format=1.5]{0.01566} & \tablenum[table-format=1.5]{0.01531} & \tablenum[table-format=2.2]{95.4} \\
4 & 800,800 & 800,800 & $\widehat{\theta}_{\mathrm{un}}$ & \tablenum[table-format=-1.5]{0.00060} & \tablenum[table-format=1.5]{0.01537} & \tablenum[table-format=1.5]{0.01568} & \tablenum[table-format=1.5]{0.01538} & \tablenum[table-format=2.2]{95.5} \\
4 & 800,800 & 800,800 & $\widehat{\theta}_{\mathrm{SS}}$ & \tablenum[table-format=-1.5]{0.00060} & \tablenum[table-format=1.5]{0.01532} & \tablenum[table-format=1.5]{0.01566} & \tablenum[table-format=1.5]{0.01533} & \tablenum[table-format=2.2]{95.6} \\
5 & 800,800 & 800,800 & $\widehat{\theta}_{\mathrm{sup}}$ & \tablenum[table-format=-1.5]{0.00014} & \tablenum[table-format=1.5]{0.01558} & \tablenum[table-format=1.5]{0.01566} & \tablenum[table-format=1.5]{0.01558} & \tablenum[table-format=2.2]{94.7} \\
5 & 800,800 & 800,800 & $\widehat{\theta}_{\mathrm{un}}$ & \tablenum[table-format=-1.5]{-0.00039} & \tablenum[table-format=1.5]{0.01547} & \tablenum[table-format=1.5]{0.01568} & \tablenum[table-format=1.5]{0.01547} & \tablenum[table-format=2.2]{95} \\
5 & 800,800 & 800,800 & $\widehat{\theta}_{\mathrm{SS}}$ & \tablenum[table-format=-1.5]{-0.00012} & \tablenum[table-format=1.5]{0.01451} & \tablenum[table-format=1.5]{0.01464} & \tablenum[table-format=1.5]{0.01450} & \tablenum[table-format=2.2]{94.8} \\
6 & 800,800 & 800,800 & $\widehat{\theta}_{\mathrm{sup}}$ & \tablenum[table-format=-1.5]{-0.00021} & \tablenum[table-format=1.5]{0.01535} & \tablenum[table-format=1.5]{0.01541} & \tablenum[table-format=1.5]{0.01535} & \tablenum[table-format=2.2]{94.7} \\
6 & 800,800 & 800,800 & $\widehat{\theta}_{\mathrm{un}}$ & \tablenum[table-format=-1.5]{-0.00020} & \tablenum[table-format=1.5]{0.01544} & \tablenum[table-format=1.5]{0.01544} & \tablenum[table-format=1.5]{0.01544} & \tablenum[table-format=2.2]{94.7} \\
6 & 800,800 & 800,800 & $\widehat{\theta}_{\mathrm{SS}}$ & \tablenum[table-format=-1.5]{-0.00020} & \tablenum[table-format=1.5]{0.01538} & \tablenum[table-format=1.5]{0.01541} & \tablenum[table-format=1.5]{0.01537} & \tablenum[table-format=2.2]{94.8} \\
7 & 800,800 & 800,800 & $\widehat{\theta}_{\mathrm{sup}}$ & \tablenum[table-format=-1.5]{-0.00006} & \tablenum[table-format=1.5]{0.01572} & \tablenum[table-format=1.5]{0.01541} & \tablenum[table-format=1.5]{0.01572} & \tablenum[table-format=2.2]{94.9} \\
7 & 800,800 & 800,800 & $\widehat{\theta}_{\mathrm{un}}$ & \tablenum[table-format=-1.5]{-0.00010} & \tablenum[table-format=1.5]{0.01547} & \tablenum[table-format=1.5]{0.01544} & \tablenum[table-format=1.5]{0.01547} & \tablenum[table-format=2.2]{94.9} \\
7 & 800,800 & 800,800 & $\widehat{\theta}_{\mathrm{SS}}$ & \tablenum[table-format=-1.5]{-0.00008} & \tablenum[table-format=1.5]{0.01453} & \tablenum[table-format=1.5]{0.01437} & \tablenum[table-format=1.5]{0.01453} & \tablenum[table-format=2.2]{94.6} \\
8 & 800,800 & 800,800 & $\widehat{\theta}_{\mathrm{sup}}$ & \tablenum[table-format=-1.5]{0.00012} & \tablenum[table-format=1.5]{0.01929} & \tablenum[table-format=1.5]{0.01895} & \tablenum[table-format=1.5]{0.01928} & \tablenum[table-format=2.2]{94.9} \\
8 & 800,800 & 800,800 & $\widehat{\theta}_{\mathrm{un}}$ & \tablenum[table-format=-1.5]{-0.00006} & \tablenum[table-format=1.5]{0.01900} & \tablenum[table-format=1.5]{0.01901} & \tablenum[table-format=1.5]{0.01899} & \tablenum[table-format=2.2]{95.2} \\
8 & 800,800 & 800,800 & $\widehat{\theta}_{\mathrm{SS}}$ & \tablenum[table-format=-1.5]{0.00003} & \tablenum[table-format=1.5]{0.01817} & \tablenum[table-format=1.5]{0.01799} & \tablenum[table-format=1.5]{0.01817} & \tablenum[table-format=2.2]{95.1} \\
\addlinespace
\multicolumn{9}{l}{\textit{B. Additional resource and signal structures}}\\\addlinespace[2pt]
8 & 200,200 & 1000,1000 & $\widehat{\theta}_{\mathrm{sup}}$ & \tablenum[table-format=-1.5]{-0.0002} & \tablenum[table-format=1.5]{0.0382} & \tablenum[table-format=1.5]{0.0381} & \tablenum[table-format=1.5]{0.0382} & \tablenum[table-format=2.2]{94.8} \\
8 & 200,200 & 1000,1000 & $\widehat{\theta}_{\mathrm{un}}$ & \tablenum[table-format=-1.5]{-0.0006} & \tablenum[table-format=1.5]{0.0360} & \tablenum[table-format=1.5]{0.0351} & \tablenum[table-format=1.5]{0.0360} & \tablenum[table-format=2.2]{94.47} \\
8 & 200,200 & 1000,1000 & $\widehat{\theta}_{\mathrm{SS}}$ & \tablenum[table-format=-1.5]{-0.0005} & \tablenum[table-format=1.5]{0.0356} & \tablenum[table-format=1.5]{0.0349} & \tablenum[table-format=1.5]{0.0356} & \tablenum[table-format=2.2]{94.58} \\
8 & 800,800 & 4000,4000 & $\widehat{\theta}_{\mathrm{sup}}$ & \tablenum[table-format=-1.5]{0.0004} & \tablenum[table-format=1.5]{0.0189} & \tablenum[table-format=1.5]{0.0189} & \tablenum[table-format=1.5]{0.0189} & \tablenum[table-format=2.2]{94.92} \\
8 & 800,800 & 4000,4000 & $\widehat{\theta}_{\mathrm{un}}$ & \tablenum[table-format=-1.5]{0.0004} & \tablenum[table-format=1.5]{0.0174} & \tablenum[table-format=1.5]{0.0174} & \tablenum[table-format=1.5]{0.0174} & \tablenum[table-format=2.2]{95.08} \\
8 & 800,800 & 4000,4000 & $\widehat{\theta}_{\mathrm{SS}}$ & \tablenum[table-format=-1.5]{0.0004} & \tablenum[table-format=1.5]{0.0174} & \tablenum[table-format=1.5]{0.0173} & \tablenum[table-format=1.5]{0.0174} & \tablenum[table-format=2.2]{94.95} \\
22 & 200,200 & 200,200 & $\widehat{\theta}_{\mathrm{SS}}$ & \tablenum[table-format=-1.5]{-0.0006} & \tablenum[table-format=1.5]{0.0315} & \tablenum[table-format=1.5]{0.0315} & \tablenum[table-format=1.5]{0.0315} & \tablenum[table-format=2.2]{94.77} \\
22 & 200,200 & 1000,1000 & $\widehat{\theta}_{\mathrm{SS}}$ & \tablenum[table-format=-1.5]{-0.0007} & \tablenum[table-format=1.5]{0.0306} & \tablenum[table-format=1.5]{0.0304} & \tablenum[table-format=1.5]{0.0306} & \tablenum[table-format=2.2]{94.8} \\
22 & 200,400 & 50,2000 & $\widehat{\theta}_{\mathrm{SS}}$ & \tablenum[table-format=-1.5]{-0.0005} & \tablenum[table-format=1.5]{0.0272} & \tablenum[table-format=1.5]{0.0275} & \tablenum[table-format=1.5]{0.0272} & \tablenum[table-format=2.2]{95.28} \\
22 & 200,400 & 1000,100 & $\widehat{\theta}_{\mathrm{SS}}$ & \tablenum[table-format=-1.5]{-0.0003} & \tablenum[table-format=1.5]{0.0268} & \tablenum[table-format=1.5]{0.0269} & \tablenum[table-format=1.5]{0.0268} & \tablenum[table-format=2.2]{94.92} \\
23 & 200,200 & 1000,1000 & $\widehat{\theta}_{\mathrm{SS}}$ & \tablenum[table-format=-1.5]{-0.0003} & \tablenum[table-format=1.5]{0.0333} & \tablenum[table-format=1.5]{0.0330} & \tablenum[table-format=1.5]{0.0333} & \tablenum[table-format=2.2]{94.77} \\
24 & 200,200 & 1000,1000 & $\widehat{\theta}_{\mathrm{SS}}$ & \tablenum[table-format=-1.5]{-0.0003} & \tablenum[table-format=1.5]{0.0318} & \tablenum[table-format=1.5]{0.0312} & \tablenum[table-format=1.5]{0.0318} & \tablenum[table-format=2.2]{94.8} \\
23 & 800,800 & 4000,4000 & $\widehat{\theta}_{\mathrm{SS}}$ & \tablenum[table-format=-1.5]{0.0000} & \tablenum[table-format=1.5]{0.0167} & \tablenum[table-format=1.5]{0.0165} & \tablenum[table-format=1.5]{0.0167} & \tablenum[table-format=2.2]{94.75} \\
24 & 800,800 & 4000,4000 & $\widehat{\theta}_{\mathrm{SS}}$ & \tablenum[table-format=-1.5]{0.0000} & \tablenum[table-format=1.5]{0.0159} & \tablenum[table-format=1.5]{0.0155} & \tablenum[table-format=1.5]{0.0159} & \tablenum[table-format=2.2]{94.58} \\
25 & 200,200 & 1000,1000 & $\widehat{\theta}_{\mathrm{SS}}$ & \tablenum[table-format=-1.5]{-0.0005} & \tablenum[table-format=1.5]{0.0308} & \tablenum[table-format=1.5]{0.0304} & \tablenum[table-format=1.5]{0.0308} & \tablenum[table-format=2.2]{94.83} \\
25 & 800,800 & 4000,4000 & $\widehat{\theta}_{\mathrm{SS}}$ & \tablenum[table-format=-1.5]{0.0008} & \tablenum[table-format=1.5]{0.0150} & \tablenum[table-format=1.5]{0.0151} & \tablenum[table-format=1.5]{0.0150} & \tablenum[table-format=2.2]{94.97} \\*
\TableEndRule \multicolumn{9}{@{}p{\UniformWidth}@{}}{\textit{Note:} Bias, EmpSD, MeanSE and RMSE are on the probability scale; CP is 95\% Wald coverage. Method rows use $\widehat{\theta}_{\mathrm{sup}}$, $\widehat{\theta}_{\mathrm{un}}$ and $\widehat{\theta}_{\mathrm{SS}}$ for Supervised, Unshrunk and Weighted. A uses 2,000 repetitions per cell; B uses 4,000 repetitions. Three-method gain comparisons for the structural extensions are in Table~\ref{tab:structureMain}.}\\
\end{longtable}\endgroup

Table~\ref{tab:additionalFixed} reports the weighted estimator's bias, spread, estimated uncertainty and coverage at larger sample sizes and in additional fixed-horizon settings. Zero and weak signal appear alongside the quadratic and recorded-grid mechanisms.
\begingroup\TableFont
\setlength{\tabcolsep}{0pt}\sbox{\UniformMeasure}{\begin{tabular}{@{}lrrrrrrr@{}}Scenario & $n_0$ & $M_0$ & Bias & EmpSD & MeanSE & RMSE & CP (\%) \\
\multicolumn{8}{l}{\textit{A. Larger labeled samples}}\\\addlinespace[2pt]
9 & \tablenum[table-format=4.0]{960} & \tablenum[table-format=4.0]{960} & \tablenum[table-format=-1.4]{0.0000} & \tablenum[table-format=1.4]{0.0130} & \tablenum[table-format=1.4]{0.0129} & \tablenum[table-format=1.4]{0.0130} & \tablenum[table-format=2.2]{95.13} \\
10 & \tablenum[table-format=4.0]{960} & \tablenum[table-format=4.0]{960} & \tablenum[table-format=-1.4]{0.0001} & \tablenum[table-format=1.4]{0.0129} & \tablenum[table-format=1.4]{0.0129} & \tablenum[table-format=1.4]{0.0129} & \tablenum[table-format=2.2]{94.7} \\
11 & \tablenum[table-format=4.0]{960} & \tablenum[table-format=4.0]{960} & \tablenum[table-format=-1.4]{0.0004} & \tablenum[table-format=1.4]{0.0128} & \tablenum[table-format=1.4]{0.0128} & \tablenum[table-format=1.4]{0.0128} & \tablenum[table-format=2.2]{95.15} \\
12 & \tablenum[table-format=4.0]{960} & \tablenum[table-format=4.0]{960} & \tablenum[table-format=-1.4]{-0.0003} & \tablenum[table-format=1.4]{0.0125} & \tablenum[table-format=1.4]{0.0124} & \tablenum[table-format=1.4]{0.0125} & \tablenum[table-format=2.2]{95.1} \\
13 & \tablenum[table-format=4.0]{960} & \tablenum[table-format=4.0]{960} & \tablenum[table-format=-1.4]{-0.0001} & \tablenum[table-format=1.4]{0.0130} & \tablenum[table-format=1.4]{0.0130} & \tablenum[table-format=1.4]{0.0130} & \tablenum[table-format=2.2]{95.23} \\
14 & \tablenum[table-format=4.0]{960} & \tablenum[table-format=4.0]{960} & \tablenum[table-format=-1.4]{-0.0001} & \tablenum[table-format=1.4]{0.0128} & \tablenum[table-format=1.4]{0.0128} & \tablenum[table-format=1.4]{0.0128} & \tablenum[table-format=2.2]{94.93} \\
15 & \tablenum[table-format=4.0]{960} & \tablenum[table-format=4.0]{960} & \tablenum[table-format=-1.4]{0.0003} & \tablenum[table-format=1.4]{0.0139} & \tablenum[table-format=1.4]{0.0142} & \tablenum[table-format=1.4]{0.0139} & \tablenum[table-format=2.2]{95.85} \\
16 & \tablenum[table-format=4.0]{960} & \tablenum[table-format=4.0]{960} & \tablenum[table-format=-1.4]{0.0001} & \tablenum[table-format=1.4]{0.0139} & \tablenum[table-format=1.4]{0.0137} & \tablenum[table-format=1.4]{0.0139} & \tablenum[table-format=2.2]{95.23} \\
17 & \tablenum[table-format=4.0]{960} & \tablenum[table-format=4.0]{240} & \tablenum[table-format=-1.4]{0.0002} & \tablenum[table-format=1.4]{0.0130} & \tablenum[table-format=1.4]{0.0129} & \tablenum[table-format=1.4]{0.0130} & \tablenum[table-format=2.2]{94.17} \\
18 & \tablenum[table-format=4.0]{960} & \tablenum[table-format=4.0]{4800} & \tablenum[table-format=-1.4]{0.0002} & \tablenum[table-format=1.4]{0.0130} & \tablenum[table-format=1.4]{0.0129} & \tablenum[table-format=1.4]{0.0130} & \tablenum[table-format=2.2]{94.27} \\
19 & \tablenum[table-format=4.0]{960} & \tablenum[table-format=4.0]{960} & \tablenum[table-format=-1.4]{-0.0000} & \tablenum[table-format=1.4]{0.0128} & \tablenum[table-format=1.4]{0.0129} & \tablenum[table-format=1.4]{0.0128} & \tablenum[table-format=2.2]{95.2} \\
20 & \tablenum[table-format=4.0]{960} & \tablenum[table-format=4.0]{960} & \tablenum[table-format=-1.4]{0.0003} & \tablenum[table-format=1.4]{0.0125} & \tablenum[table-format=1.4]{0.0124} & \tablenum[table-format=1.4]{0.0125} & \tablenum[table-format=2.2]{94.83} \\
\addlinespace
\multicolumn{8}{l}{\textit{B. Independent comparisons}}\\\addlinespace[2pt]
9 & \tablenum[table-format=4.0]{240} & \tablenum[table-format=4.0]{240} & \tablenum[table-format=-1.4]{-0.0009} & \tablenum[table-format=1.4]{0.0257} & \tablenum[table-format=1.4]{0.0257} & \tablenum[table-format=1.4]{0.0257} & \tablenum[table-format=2.2]{94.75} \\
10 & \tablenum[table-format=4.0]{240} & \tablenum[table-format=4.0]{240} & \tablenum[table-format=-1.4]{-0.0001} & \tablenum[table-format=1.4]{0.0260} & \tablenum[table-format=1.4]{0.0257} & \tablenum[table-format=1.4]{0.0260} & \tablenum[table-format=2.2]{94.47} \\
12 & \tablenum[table-format=4.0]{240} & \tablenum[table-format=4.0]{240} & \tablenum[table-format=-1.4]{0.0001} & \tablenum[table-format=1.4]{0.0252} & \tablenum[table-format=1.4]{0.0249} & \tablenum[table-format=1.4]{0.0252} & \tablenum[table-format=2.2]{94.8} \\
14 & \tablenum[table-format=4.0]{480} & \tablenum[table-format=4.0]{480} & \tablenum[table-format=-1.4]{0.0001} & \tablenum[table-format=1.4]{0.0179} & \tablenum[table-format=1.4]{0.0181} & \tablenum[table-format=1.4]{0.0179} & \tablenum[table-format=2.2]{95.23} \\
14 & \tablenum[table-format=4.0]{1920} & \tablenum[table-format=4.0]{1920} & \tablenum[table-format=-1.4]{-0.0001} & \tablenum[table-format=1.4]{0.0091} & \tablenum[table-format=1.4]{0.0091} & \tablenum[table-format=1.4]{0.0091} & \tablenum[table-format=2.2]{94.97} \\
20 & \tablenum[table-format=4.0]{480} & \tablenum[table-format=4.0]{480} & \tablenum[table-format=-1.4]{0.0005} & \tablenum[table-format=1.4]{0.0177} & \tablenum[table-format=1.4]{0.0176} & \tablenum[table-format=1.4]{0.0177} & \tablenum[table-format=2.2]{94.81} \\
20 & \tablenum[table-format=4.0]{1920} & \tablenum[table-format=4.0]{1920} & \tablenum[table-format=-1.4]{0.0001} & \tablenum[table-format=1.4]{0.0087} & \tablenum[table-format=1.4]{0.0088} & \tablenum[table-format=1.4]{0.0087} & \tablenum[table-format=2.2]{95.24} \\
21 & \tablenum[table-format=4.0]{240} & \tablenum[table-format=4.0]{60} & \tablenum[table-format=-1.4]{-0.0001} & \tablenum[table-format=1.4]{0.0271} & \tablenum[table-format=1.4]{0.0270} & \tablenum[table-format=1.4]{0.0271} & \tablenum[table-format=2.2]{94.77} \\
21 & \tablenum[table-format=4.0]{240} & \tablenum[table-format=4.0]{240} & \tablenum[table-format=-1.4]{0.0005} & \tablenum[table-format=1.4]{0.0264} & \tablenum[table-format=1.4]{0.0264} & \tablenum[table-format=1.4]{0.0264} & \tablenum[table-format=2.2]{94.96} \\
21 & \tablenum[table-format=4.0]{240} & \tablenum[table-format=4.0]{1200} & \tablenum[table-format=-1.4]{0.0001} & \tablenum[table-format=1.4]{0.0259} & \tablenum[table-format=1.4]{0.0257} & \tablenum[table-format=1.4]{0.0259} & \tablenum[table-format=2.2]{94.76} \\
21 & \tablenum[table-format=4.0]{960} & \tablenum[table-format=4.0]{240} & \tablenum[table-format=-1.4]{-0.0002} & \tablenum[table-format=1.4]{0.0134} & \tablenum[table-format=1.4]{0.0135} & \tablenum[table-format=1.4]{0.0134} & \tablenum[table-format=2.2]{95.36} \\
21 & \tablenum[table-format=4.0]{960} & \tablenum[table-format=4.0]{960} & \tablenum[table-format=-1.4]{-0.0002} & \tablenum[table-format=1.4]{0.0132} & \tablenum[table-format=1.4]{0.0132} & \tablenum[table-format=1.4]{0.0132} & \tablenum[table-format=2.2]{94.96} \\
21 & \tablenum[table-format=4.0]{960} & \tablenum[table-format=4.0]{4800} & \tablenum[table-format=-1.4]{0.0001} & \tablenum[table-format=1.4]{0.0128} & \tablenum[table-format=1.4]{0.0128} & \tablenum[table-format=1.4]{0.0128} & \tablenum[table-format=2.2]{95.36} \\
\addlinespace
\end{tabular}}
\setlength{\UniformWidth}{\linewidth}\setlength{\tabcolsep}{\dimexpr(\linewidth-\wd\UniformMeasure)/14\relax}
\ifdim\tabcolsep<0pt\PackageWarning{CompactTables}{Negative column gap}\fi
\setlength{\LTleft}{0pt plus 1fill}\setlength{\LTright}{0pt plus 1fill}
\begin{longtable}{@{}lrrrrrrr@{}}
\caption{Additional fixed-horizon calibration.}\label{tab:additionalFixed}\\\toprule
Scenario & $n_0$ & $M_0$ & Bias & EmpSD & MeanSE & RMSE & CP (\%) \\
\midrule\endfirsthead
\multicolumn{8}{l}{\textit{Table \thetable\ continued}}\\\toprule
Scenario & $n_0$ & $M_0$ & Bias & EmpSD & MeanSE & RMSE & CP (\%) \\
\midrule\endhead
\midrule\multicolumn{8}{r}{\textit{Continued on next page}}\\\endfoot
\endlastfoot
\multicolumn{8}{l}{\textit{A. Larger labeled samples}}\\\addlinespace[2pt]
9 & \tablenum[table-format=4.0]{960} & \tablenum[table-format=4.0]{960} & \tablenum[table-format=-1.4]{0.0000} & \tablenum[table-format=1.4]{0.0130} & \tablenum[table-format=1.4]{0.0129} & \tablenum[table-format=1.4]{0.0130} & \tablenum[table-format=2.2]{95.13} \\
10 & \tablenum[table-format=4.0]{960} & \tablenum[table-format=4.0]{960} & \tablenum[table-format=-1.4]{0.0001} & \tablenum[table-format=1.4]{0.0129} & \tablenum[table-format=1.4]{0.0129} & \tablenum[table-format=1.4]{0.0129} & \tablenum[table-format=2.2]{94.7} \\
11 & \tablenum[table-format=4.0]{960} & \tablenum[table-format=4.0]{960} & \tablenum[table-format=-1.4]{0.0004} & \tablenum[table-format=1.4]{0.0128} & \tablenum[table-format=1.4]{0.0128} & \tablenum[table-format=1.4]{0.0128} & \tablenum[table-format=2.2]{95.15} \\
12 & \tablenum[table-format=4.0]{960} & \tablenum[table-format=4.0]{960} & \tablenum[table-format=-1.4]{-0.0003} & \tablenum[table-format=1.4]{0.0125} & \tablenum[table-format=1.4]{0.0124} & \tablenum[table-format=1.4]{0.0125} & \tablenum[table-format=2.2]{95.1} \\
13 & \tablenum[table-format=4.0]{960} & \tablenum[table-format=4.0]{960} & \tablenum[table-format=-1.4]{-0.0001} & \tablenum[table-format=1.4]{0.0130} & \tablenum[table-format=1.4]{0.0130} & \tablenum[table-format=1.4]{0.0130} & \tablenum[table-format=2.2]{95.23} \\
14 & \tablenum[table-format=4.0]{960} & \tablenum[table-format=4.0]{960} & \tablenum[table-format=-1.4]{-0.0001} & \tablenum[table-format=1.4]{0.0128} & \tablenum[table-format=1.4]{0.0128} & \tablenum[table-format=1.4]{0.0128} & \tablenum[table-format=2.2]{94.93} \\
15 & \tablenum[table-format=4.0]{960} & \tablenum[table-format=4.0]{960} & \tablenum[table-format=-1.4]{0.0003} & \tablenum[table-format=1.4]{0.0139} & \tablenum[table-format=1.4]{0.0142} & \tablenum[table-format=1.4]{0.0139} & \tablenum[table-format=2.2]{95.85} \\
16 & \tablenum[table-format=4.0]{960} & \tablenum[table-format=4.0]{960} & \tablenum[table-format=-1.4]{0.0001} & \tablenum[table-format=1.4]{0.0139} & \tablenum[table-format=1.4]{0.0137} & \tablenum[table-format=1.4]{0.0139} & \tablenum[table-format=2.2]{95.23} \\
17 & \tablenum[table-format=4.0]{960} & \tablenum[table-format=4.0]{240} & \tablenum[table-format=-1.4]{0.0002} & \tablenum[table-format=1.4]{0.0130} & \tablenum[table-format=1.4]{0.0129} & \tablenum[table-format=1.4]{0.0130} & \tablenum[table-format=2.2]{94.17} \\
18 & \tablenum[table-format=4.0]{960} & \tablenum[table-format=4.0]{4800} & \tablenum[table-format=-1.4]{0.0002} & \tablenum[table-format=1.4]{0.0130} & \tablenum[table-format=1.4]{0.0129} & \tablenum[table-format=1.4]{0.0130} & \tablenum[table-format=2.2]{94.27} \\
19 & \tablenum[table-format=4.0]{960} & \tablenum[table-format=4.0]{960} & \tablenum[table-format=-1.4]{-0.0000} & \tablenum[table-format=1.4]{0.0128} & \tablenum[table-format=1.4]{0.0129} & \tablenum[table-format=1.4]{0.0128} & \tablenum[table-format=2.2]{95.2} \\
20 & \tablenum[table-format=4.0]{960} & \tablenum[table-format=4.0]{960} & \tablenum[table-format=-1.4]{0.0003} & \tablenum[table-format=1.4]{0.0125} & \tablenum[table-format=1.4]{0.0124} & \tablenum[table-format=1.4]{0.0125} & \tablenum[table-format=2.2]{94.83} \\
\addlinespace
\multicolumn{8}{l}{\textit{B. Independent comparisons}}\\\addlinespace[2pt]
9 & \tablenum[table-format=4.0]{240} & \tablenum[table-format=4.0]{240} & \tablenum[table-format=-1.4]{-0.0009} & \tablenum[table-format=1.4]{0.0257} & \tablenum[table-format=1.4]{0.0257} & \tablenum[table-format=1.4]{0.0257} & \tablenum[table-format=2.2]{94.75} \\
10 & \tablenum[table-format=4.0]{240} & \tablenum[table-format=4.0]{240} & \tablenum[table-format=-1.4]{-0.0001} & \tablenum[table-format=1.4]{0.0260} & \tablenum[table-format=1.4]{0.0257} & \tablenum[table-format=1.4]{0.0260} & \tablenum[table-format=2.2]{94.47} \\
12 & \tablenum[table-format=4.0]{240} & \tablenum[table-format=4.0]{240} & \tablenum[table-format=-1.4]{0.0001} & \tablenum[table-format=1.4]{0.0252} & \tablenum[table-format=1.4]{0.0249} & \tablenum[table-format=1.4]{0.0252} & \tablenum[table-format=2.2]{94.8} \\
14 & \tablenum[table-format=4.0]{480} & \tablenum[table-format=4.0]{480} & \tablenum[table-format=-1.4]{0.0001} & \tablenum[table-format=1.4]{0.0179} & \tablenum[table-format=1.4]{0.0181} & \tablenum[table-format=1.4]{0.0179} & \tablenum[table-format=2.2]{95.23} \\
14 & \tablenum[table-format=4.0]{1920} & \tablenum[table-format=4.0]{1920} & \tablenum[table-format=-1.4]{-0.0001} & \tablenum[table-format=1.4]{0.0091} & \tablenum[table-format=1.4]{0.0091} & \tablenum[table-format=1.4]{0.0091} & \tablenum[table-format=2.2]{94.97} \\
20 & \tablenum[table-format=4.0]{480} & \tablenum[table-format=4.0]{480} & \tablenum[table-format=-1.4]{0.0005} & \tablenum[table-format=1.4]{0.0177} & \tablenum[table-format=1.4]{0.0176} & \tablenum[table-format=1.4]{0.0177} & \tablenum[table-format=2.2]{94.81} \\
20 & \tablenum[table-format=4.0]{1920} & \tablenum[table-format=4.0]{1920} & \tablenum[table-format=-1.4]{0.0001} & \tablenum[table-format=1.4]{0.0087} & \tablenum[table-format=1.4]{0.0088} & \tablenum[table-format=1.4]{0.0087} & \tablenum[table-format=2.2]{95.24} \\
21 & \tablenum[table-format=4.0]{240} & \tablenum[table-format=4.0]{60} & \tablenum[table-format=-1.4]{-0.0001} & \tablenum[table-format=1.4]{0.0271} & \tablenum[table-format=1.4]{0.0270} & \tablenum[table-format=1.4]{0.0271} & \tablenum[table-format=2.2]{94.77} \\
21 & \tablenum[table-format=4.0]{240} & \tablenum[table-format=4.0]{240} & \tablenum[table-format=-1.4]{0.0005} & \tablenum[table-format=1.4]{0.0264} & \tablenum[table-format=1.4]{0.0264} & \tablenum[table-format=1.4]{0.0264} & \tablenum[table-format=2.2]{94.96} \\
21 & \tablenum[table-format=4.0]{240} & \tablenum[table-format=4.0]{1200} & \tablenum[table-format=-1.4]{0.0001} & \tablenum[table-format=1.4]{0.0259} & \tablenum[table-format=1.4]{0.0257} & \tablenum[table-format=1.4]{0.0259} & \tablenum[table-format=2.2]{94.76} \\
21 & \tablenum[table-format=4.0]{960} & \tablenum[table-format=4.0]{240} & \tablenum[table-format=-1.4]{-0.0002} & \tablenum[table-format=1.4]{0.0134} & \tablenum[table-format=1.4]{0.0135} & \tablenum[table-format=1.4]{0.0134} & \tablenum[table-format=2.2]{95.36} \\
21 & \tablenum[table-format=4.0]{960} & \tablenum[table-format=4.0]{960} & \tablenum[table-format=-1.4]{-0.0002} & \tablenum[table-format=1.4]{0.0132} & \tablenum[table-format=1.4]{0.0132} & \tablenum[table-format=1.4]{0.0132} & \tablenum[table-format=2.2]{94.96} \\
21 & \tablenum[table-format=4.0]{960} & \tablenum[table-format=4.0]{4800} & \tablenum[table-format=-1.4]{0.0001} & \tablenum[table-format=1.4]{0.0128} & \tablenum[table-format=1.4]{0.0128} & \tablenum[table-format=1.4]{0.0128} & \tablenum[table-format=2.2]{95.36} \\*
\TableEndRule \multicolumn{8}{@{}p{\UniformWidth}@{}}{\textit{Note:} Rows report $\widehat{\theta}_{\mathrm{SS},\tau}$ (Weighted); $n_0$ and $M_0$ are pool sizes per population. Bias, EmpSD, MeanSE and RMSE are on the probability scale. A uses 4,000 repetitions per cell; B uses 4,000 for Scenarios 9, 10 and 14, and 8,000 for Scenarios 12, 20 and 21.  Paired gains and MCSE for B are given in Table~\ref{tab:confirmationMain}.}\\
\end{longtable}\endgroup

Across nine fixed-horizon settings with first-order variance gains of 0 to 0.29\%, all nine empirical gains were negative.

In the 36 full-lifetime settings satisfying the model assumptions, Weighted coverage ranged from 92.6\% to 96.1\%. The seven settings with coverage below 94\% had mean estimated standard errors below the empirical SD (Table~\ref{tab:full_lowcoverage}). The remaining 29 settings had coverage between 94.0\% and 96.1\%. In two additional settings, coverage was 0.5\% when the auxiliary covariate law differed from the labeled law and 1.4\% when censoring curtailed lifetime support.

\begin{table}[htbp]\centering
\TableFont
\caption{Full-lifetime settings with Weighted coverage below 94\%.}\label{tab:full_lowcoverage}
\begin{tabular}{@{}lrrrrrrr@{}}
\toprule
Setting & CP & Wilson 95\% interval & SE/SD & Mean $Z$ & SD $Z$ & $\Pr(Z<-1.96)$ & $\Pr(Z>1.96)$ \\
\midrule
3  & 0.934 & [0.9169, 0.9478] & 0.943 & -0.024 & 1.067 & 0.029 & 0.037 \\
4  & 0.926 & [0.9081, 0.9406] & 0.942 & -0.091 & 1.069 & 0.044 & 0.030 \\
15 & 0.929 & [0.9114, 0.9433] & 0.962 & 0.010 & 1.044 & 0.031 & 0.040 \\
24 & 0.929 & [0.9114, 0.9433] & 0.964 & 0.007 & 1.039 & 0.043 & 0.028 \\
26 & 0.938 & [0.9213, 0.9513] & 0.972 & 0.024 & 1.037 & 0.028 & 0.034 \\
31 & 0.931 & [0.9136, 0.9451] & 0.952 & -0.029 & 1.060 & 0.035 & 0.034 \\
33 & 0.937 & [0.9202, 0.9505] & 0.959 & -0.020 & 1.044 & 0.033 & 0.030 \\
\bottomrule
\end{tabular}
\tblnote{Each setting has 1,000 repetitions. CP is the empirical coverage proportion; Wilson limits are pointwise 95\% intervals; SE/SD is the mean estimated SE divided by empirical SD. $Z=(\widehat\theta-\theta)/\widehat{\mathrm{SE}}$. Settings 3, 4, 15, 24, 31 and 33 have balanced per-group labeled and auxiliary sizes $(200,1000)$, $(200,4000)$, $(800,4000)$, $(800,4000)$, $(200,1000)$ and $(800,16000)$, respectively. Setting 26 has labeled sizes $(170,230)$ and auxiliary sizes $(119,529)$. Setting 24 uses features with zero projection. Censoring overlaps event times in setting 31. The censoring survival at the end of the censoring interval is 0.35 in settings 15 and 31 and 0.60 otherwise.}
\end{table}

\begin{figure}[!htbp]\centering
\makebox[\linewidth][c]{\includegraphics[width=\linewidth]{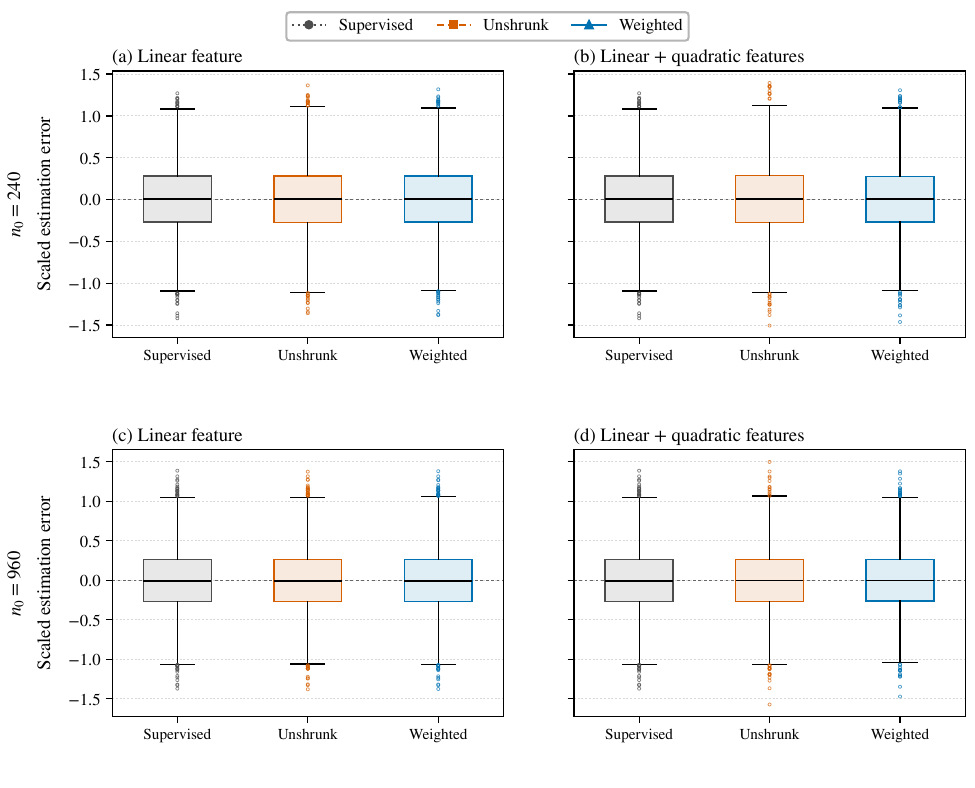}}
\caption{Working-feature comparison under the centered quadratic law. Columns use a linear feature (Scenario 13) or linear and quadratic features (Scenario 14); rows use $n_0=M_0=240$ or 960. The two feature specifications share the generated observations and splits within each repetition. All three feasible methods are shown, with 4,000 repetitions per cell. The ordinate is $\sqrt{n_0}(\widehat\theta-\theta)$. Boxes and outliers follow the conventions of Figure~\ref{fig:errorDistributions}.}\label{fig:sFeatureBoxes}
\end{figure}

Figure~\ref{fig:sFeatureBoxes} isolates the role of the working features. Linear and quadratic-feature fits use the same observations and folds. The linear projection has zero signal, whereas the quadratic term makes the comparison signal available to the adjustment; the smaller labeled sample also exposes its fitting cost.
\subsubsection{Auxiliary resources and independent populations}
The full-lifetime resource configurations use the continuous event and independent censoring laws in Section~\ref{sec:numerical}. Scenarios 1--3 set $\theta=0.50$ and $a=0,0.50,0.85$, representing zero, moderate, and strong projected signal. Scenarios 4--5 set $\theta=0.35$ and $a=0,0.85$; Scenarios 6--7 use the same amplitudes at $\theta=0.65$. Scenario 8 has $\theta=0.50$, $a=0.85$ and heavier censoring. An exponential censoring draw with rate $-\log(p)/5$ is retained when it is at most 5 and otherwise recorded as $C=\infty$; $p=0.35$ in Scenario 8 and $p=0.60$ otherwise. The baseline configurations use $n_0=M_0=200$ or 800, and the resource study varies $M_0/n_0=1,2,5$. Scenarios 1--3 use the one-group signal mechanism.

\FloatBarrier
Scenarios 22--25 examine two-group signal and resource structures. In Scenario 22 both groups have $T_k=2+\exp\{0.5X_k+\sqrt{0.75}E_k\}$ with independent standard-normal $X_k,E_k$. The balanced configurations use 200 labeled observations per group and ratios 1 and 5. The remaining configurations use $(n_1,n_2)=(200,400)$ and resource ratios $(0.25,5)$ or $(5,0.25)$. In Scenarios 23--24, both groups instead use $0.5q(X_k)$ in the exponent, where $q(x)=(x^2-1)/\sqrt2$; their working features are $X$ and $(X,q(X))$, respectively, and paired feature fits use the same data. Scenario 25 uses the linear law with $X_1$ standard normal and $X_2$ uniform on $[-\sqrt3,\sqrt3]$. These four scenarios have $\theta=0.5$ and maintain within-group agreement between the labeled and auxiliary covariate distributions. Scenarios 23--25 use $M_k/n_k=5$ and 200 or 800 labeled records per group. Scenario 8 uses $G(5)=0.35$ and Scenarios 1--7 and 22--25 use $G(5)=0.60$.

Table~\ref{tab:additionalFull} reports the bias, empirical spread, estimated standard error and coverage for these configurations.

\subsubsection{Covariate augmentation}
Table~\ref{tab:competitiveMethods} compares the calibration of Weighted with fixed linear- and categorical-basis trial augmentation over every resource level in the common-distribution experiment. The paired efficiency contrasts with supervised and unshrunk estimation are in Table~\ref{tab:competitiveResources}. The two fixed bases permit comparison of precision and calibration as the outcome law changes.
\begingroup\TableFont
\setlength{\tabcolsep}{0pt}\sbox{\UniformMeasure}{\begin{tabular}{@{}llrrrrrrr@{}}Scenario & Method & $(n_1,n_2)$ & $(M_1,M_2)$ & Bias & EmpSD & MeanSE & RMSE & CP (\%) \\
9 & $\widehat{\theta}_{\mathrm{SS},\tau}$ & 960,960 & 240,240 & \tablenum[table-format=-1.4]{0.0002} & \tablenum[table-format=1.4]{0.0127} & \tablenum[table-format=1.4]{0.0129} & \tablenum[table-format=1.4]{0.0127} & \tablenum[table-format=2.2]{95.15} \\
9 & \shortstack[l]{Trial aug.\\(linear)} & 960,960 & 240,240 & \tablenum[table-format=-1.4]{0.0002} & \tablenum[table-format=1.4]{0.0127} & \tablenum[table-format=1.4]{0.0129} & \tablenum[table-format=1.4]{0.0127} & \tablenum[table-format=2.2]{95.12} \\
9 & \shortstack[l]{Trial aug.\\(categorical)} & 960,960 & 240,240 & \tablenum[table-format=-1.4]{0.0002} & \tablenum[table-format=1.4]{0.0127} & \tablenum[table-format=1.4]{0.0129} & \tablenum[table-format=1.4]{0.0127} & \tablenum[table-format=2.2]{95.12} \\
9 & $\widehat{\theta}_{\mathrm{SS},\tau}$ & 960,960 & 960,960 & \tablenum[table-format=-1.4]{0.0002} & \tablenum[table-format=1.4]{0.0127} & \tablenum[table-format=1.4]{0.0129} & \tablenum[table-format=1.4]{0.0127} & \tablenum[table-format=2.2]{95.03} \\
9 & \shortstack[l]{Trial aug.\\(linear)} & 960,960 & 960,960 & \tablenum[table-format=-1.4]{0.0002} & \tablenum[table-format=1.4]{0.0127} & \tablenum[table-format=1.4]{0.0129} & \tablenum[table-format=1.4]{0.0127} & \tablenum[table-format=2.2]{95.12} \\
9 & \shortstack[l]{Trial aug.\\(categorical)} & 960,960 & 960,960 & \tablenum[table-format=-1.4]{0.0002} & \tablenum[table-format=1.4]{0.0127} & \tablenum[table-format=1.4]{0.0129} & \tablenum[table-format=1.4]{0.0127} & \tablenum[table-format=2.2]{95.12} \\
9 & $\widehat{\theta}_{\mathrm{SS},\tau}$ & 960,960 & 4800,4800 & \tablenum[table-format=-1.4]{0.0002} & \tablenum[table-format=1.4]{0.0127} & \tablenum[table-format=1.4]{0.0129} & \tablenum[table-format=1.4]{0.0127} & \tablenum[table-format=2.2]{95.12} \\
9 & \shortstack[l]{Trial aug.\\(linear)} & 960,960 & 4800,4800 & \tablenum[table-format=-1.4]{0.0002} & \tablenum[table-format=1.4]{0.0127} & \tablenum[table-format=1.4]{0.0129} & \tablenum[table-format=1.4]{0.0127} & \tablenum[table-format=2.2]{95.12} \\
9 & \shortstack[l]{Trial aug.\\(categorical)} & 960,960 & 4800,4800 & \tablenum[table-format=-1.4]{0.0002} & \tablenum[table-format=1.4]{0.0127} & \tablenum[table-format=1.4]{0.0129} & \tablenum[table-format=1.4]{0.0127} & \tablenum[table-format=2.2]{95.12} \\
10 & $\widehat{\theta}_{\mathrm{SS},\tau}$ & 960,960 & 240,240 & \tablenum[table-format=-1.4]{0.0003} & \tablenum[table-format=1.4]{0.0130} & \tablenum[table-format=1.4]{0.0129} & \tablenum[table-format=1.4]{0.0130} & \tablenum[table-format=2.2]{94.62} \\
10 & \shortstack[l]{Trial aug.\\(linear)} & 960,960 & 240,240 & \tablenum[table-format=-1.4]{0.0003} & \tablenum[table-format=1.4]{0.0130} & \tablenum[table-format=1.4]{0.0129} & \tablenum[table-format=1.4]{0.0130} & \tablenum[table-format=2.2]{94.60} \\
10 & \shortstack[l]{Trial aug.\\(categorical)} & 960,960 & 240,240 & \tablenum[table-format=-1.4]{0.0003} & \tablenum[table-format=1.4]{0.0130} & \tablenum[table-format=1.4]{0.0129} & \tablenum[table-format=1.4]{0.0130} & \tablenum[table-format=2.2]{94.60} \\
10 & $\widehat{\theta}_{\mathrm{SS},\tau}$ & 960,960 & 960,960 & \tablenum[table-format=-1.4]{0.0003} & \tablenum[table-format=1.4]{0.0130} & \tablenum[table-format=1.4]{0.0129} & \tablenum[table-format=1.4]{0.0130} & \tablenum[table-format=2.2]{94.45} \\
10 & \shortstack[l]{Trial aug.\\(linear)} & 960,960 & 960,960 & \tablenum[table-format=-1.4]{0.0003} & \tablenum[table-format=1.4]{0.0130} & \tablenum[table-format=1.4]{0.0129} & \tablenum[table-format=1.4]{0.0130} & \tablenum[table-format=2.2]{94.60} \\
10 & \shortstack[l]{Trial aug.\\(categorical)} & 960,960 & 960,960 & \tablenum[table-format=-1.4]{0.0003} & \tablenum[table-format=1.4]{0.0130} & \tablenum[table-format=1.4]{0.0129} & \tablenum[table-format=1.4]{0.0130} & \tablenum[table-format=2.2]{94.60} \\
10 & $\widehat{\theta}_{\mathrm{SS},\tau}$ & 960,960 & 4800,4800 & \tablenum[table-format=-1.4]{0.0003} & \tablenum[table-format=1.4]{0.0130} & \tablenum[table-format=1.4]{0.0129} & \tablenum[table-format=1.4]{0.0130} & \tablenum[table-format=2.2]{94.53} \\
10 & \shortstack[l]{Trial aug.\\(linear)} & 960,960 & 4800,4800 & \tablenum[table-format=-1.4]{0.0003} & \tablenum[table-format=1.4]{0.0130} & \tablenum[table-format=1.4]{0.0129} & \tablenum[table-format=1.4]{0.0130} & \tablenum[table-format=2.2]{94.60} \\
10 & \shortstack[l]{Trial aug.\\(categorical)} & 960,960 & 4800,4800 & \tablenum[table-format=-1.4]{0.0003} & \tablenum[table-format=1.4]{0.0130} & \tablenum[table-format=1.4]{0.0129} & \tablenum[table-format=1.4]{0.0130} & \tablenum[table-format=2.2]{94.60} \\
11 & $\widehat{\theta}_{\mathrm{SS},\tau}$ & 960,960 & 240,240 & \tablenum[table-format=-1.4]{-0.0002} & \tablenum[table-format=1.4]{0.0129} & \tablenum[table-format=1.4]{0.0129} & \tablenum[table-format=1.4]{0.0129} & \tablenum[table-format=2.2]{94.90} \\
11 & \shortstack[l]{Trial aug.\\(linear)} & 960,960 & 240,240 & \tablenum[table-format=-1.4]{-0.0002} & \tablenum[table-format=1.4]{0.0130} & \tablenum[table-format=1.4]{0.0130} & \tablenum[table-format=1.4]{0.0130} & \tablenum[table-format=2.2]{94.80} \\
11 & \shortstack[l]{Trial aug.\\(categorical)} & 960,960 & 240,240 & \tablenum[table-format=-1.4]{-0.0002} & \tablenum[table-format=1.4]{0.0130} & \tablenum[table-format=1.4]{0.0130} & \tablenum[table-format=1.4]{0.0130} & \tablenum[table-format=2.2]{94.75} \\
11 & $\widehat{\theta}_{\mathrm{SS},\tau}$ & 960,960 & 960,960 & \tablenum[table-format=-1.4]{-0.0003} & \tablenum[table-format=1.4]{0.0128} & \tablenum[table-format=1.4]{0.0128} & \tablenum[table-format=1.4]{0.0128} & \tablenum[table-format=2.2]{94.97} \\
11 & \shortstack[l]{Trial aug.\\(linear)} & 960,960 & 960,960 & \tablenum[table-format=-1.4]{-0.0002} & \tablenum[table-format=1.4]{0.0130} & \tablenum[table-format=1.4]{0.0130} & \tablenum[table-format=1.4]{0.0130} & \tablenum[table-format=2.2]{94.80} \\
11 & \shortstack[l]{Trial aug.\\(categorical)} & 960,960 & 960,960 & \tablenum[table-format=-1.4]{-0.0002} & \tablenum[table-format=1.4]{0.0130} & \tablenum[table-format=1.4]{0.0130} & \tablenum[table-format=1.4]{0.0130} & \tablenum[table-format=2.2]{94.75} \\
11 & $\widehat{\theta}_{\mathrm{SS},\tau}$ & 960,960 & 4800,4800 & \tablenum[table-format=-1.4]{-0.0002} & \tablenum[table-format=1.4]{0.0127} & \tablenum[table-format=1.4]{0.0127} & \tablenum[table-format=1.4]{0.0127} & \tablenum[table-format=2.2]{94.95} \\
11 & \shortstack[l]{Trial aug.\\(linear)} & 960,960 & 4800,4800 & \tablenum[table-format=-1.4]{-0.0002} & \tablenum[table-format=1.4]{0.0130} & \tablenum[table-format=1.4]{0.0130} & \tablenum[table-format=1.4]{0.0130} & \tablenum[table-format=2.2]{94.80} \\
11 & \shortstack[l]{Trial aug.\\(categorical)} & 960,960 & 4800,4800 & \tablenum[table-format=-1.4]{-0.0002} & \tablenum[table-format=1.4]{0.0130} & \tablenum[table-format=1.4]{0.0130} & \tablenum[table-format=1.4]{0.0130} & \tablenum[table-format=2.2]{94.75} \\
12 & $\widehat{\theta}_{\mathrm{SS},\tau}$ & 960,960 & 240,240 & \tablenum[table-format=-1.4]{0.0003} & \tablenum[table-format=1.4]{0.0128} & \tablenum[table-format=1.4]{0.0128} & \tablenum[table-format=1.4]{0.0128} & \tablenum[table-format=2.2]{95.17} \\
12 & \shortstack[l]{Trial aug.\\(linear)} & 960,960 & 240,240 & \tablenum[table-format=-1.4]{0.0002} & \tablenum[table-format=1.4]{0.0130} & \tablenum[table-format=1.4]{0.0130} & \tablenum[table-format=1.4]{0.0130} & \tablenum[table-format=2.2]{95.33} \\
12 & \shortstack[l]{Trial aug.\\(categorical)} & 960,960 & 240,240 & \tablenum[table-format=-1.4]{0.0002} & \tablenum[table-format=1.4]{0.0130} & \tablenum[table-format=1.4]{0.0130} & \tablenum[table-format=1.4]{0.0130} & \tablenum[table-format=2.2]{95.33} \\
12 & $\widehat{\theta}_{\mathrm{SS},\tau}$ & 960,960 & 960,960 & \tablenum[table-format=-1.4]{0.0003} & \tablenum[table-format=1.4]{0.0125} & \tablenum[table-format=1.4]{0.0124} & \tablenum[table-format=1.4]{0.0125} & \tablenum[table-format=2.2]{94.97} \\
12 & \shortstack[l]{Trial aug.\\(linear)} & 960,960 & 960,960 & \tablenum[table-format=-1.4]{0.0002} & \tablenum[table-format=1.4]{0.0130} & \tablenum[table-format=1.4]{0.0130} & \tablenum[table-format=1.4]{0.0130} & \tablenum[table-format=2.2]{95.33} \\
12 & \shortstack[l]{Trial aug.\\(categorical)} & 960,960 & 960,960 & \tablenum[table-format=-1.4]{0.0002} & \tablenum[table-format=1.4]{0.0130} & \tablenum[table-format=1.4]{0.0130} & \tablenum[table-format=1.4]{0.0130} & \tablenum[table-format=2.2]{95.33} \\
12 & $\widehat{\theta}_{\mathrm{SS},\tau}$ & 960,960 & 4800,4800 & \tablenum[table-format=-1.4]{0.0002} & \tablenum[table-format=1.4]{0.0121} & \tablenum[table-format=1.4]{0.0121} & \tablenum[table-format=1.4]{0.0121} & \tablenum[table-format=2.2]{94.78} \\
12 & \shortstack[l]{Trial aug.\\(linear)} & 960,960 & 4800,4800 & \tablenum[table-format=-1.4]{0.0002} & \tablenum[table-format=1.4]{0.0130} & \tablenum[table-format=1.4]{0.0130} & \tablenum[table-format=1.4]{0.0130} & \tablenum[table-format=2.2]{95.33} \\
12 & \shortstack[l]{Trial aug.\\(categorical)} & 960,960 & 4800,4800 & \tablenum[table-format=-1.4]{0.0002} & \tablenum[table-format=1.4]{0.0130} & \tablenum[table-format=1.4]{0.0130} & \tablenum[table-format=1.4]{0.0130} & \tablenum[table-format=2.2]{95.33} \\
14 & $\widehat{\theta}_{\mathrm{SS},\tau}$ & 960,960 & 240,240 & \tablenum[table-format=-1.4]{0.0000} & \tablenum[table-format=1.4]{0.0130} & \tablenum[table-format=1.4]{0.0129} & \tablenum[table-format=1.4]{0.0130} & \tablenum[table-format=2.2]{95.00} \\
14 & \shortstack[l]{Trial aug.\\(linear)} & 960,960 & 240,240 & \tablenum[table-format=-1.4]{-0.0001} & \tablenum[table-format=1.4]{0.0131} & \tablenum[table-format=1.4]{0.0130} & \tablenum[table-format=1.4]{0.0131} & \tablenum[table-format=2.2]{95.08} \\
14 & \shortstack[l]{Trial aug.\\(categorical)} & 960,960 & 240,240 & \tablenum[table-format=-1.4]{-0.0001} & \tablenum[table-format=1.4]{0.0131} & \tablenum[table-format=1.4]{0.0130} & \tablenum[table-format=1.4]{0.0131} & \tablenum[table-format=2.2]{95.08} \\
14 & $\widehat{\theta}_{\mathrm{SS},\tau}$ & 960,960 & 960,960 & \tablenum[table-format=-1.4]{0.0000} & \tablenum[table-format=1.4]{0.0129} & \tablenum[table-format=1.4]{0.0128} & \tablenum[table-format=1.4]{0.0129} & \tablenum[table-format=2.2]{94.97} \\
14 & \shortstack[l]{Trial aug.\\(linear)} & 960,960 & 960,960 & \tablenum[table-format=-1.4]{-0.0001} & \tablenum[table-format=1.4]{0.0131} & \tablenum[table-format=1.4]{0.0130} & \tablenum[table-format=1.4]{0.0131} & \tablenum[table-format=2.2]{95.08} \\
14 & \shortstack[l]{Trial aug.\\(categorical)} & 960,960 & 960,960 & \tablenum[table-format=-1.4]{-0.0001} & \tablenum[table-format=1.4]{0.0131} & \tablenum[table-format=1.4]{0.0130} & \tablenum[table-format=1.4]{0.0131} & \tablenum[table-format=2.2]{95.08} \\
14 & $\widehat{\theta}_{\mathrm{SS},\tau}$ & 960,960 & 4800,4800 & \tablenum[table-format=-1.4]{0.0000} & \tablenum[table-format=1.4]{0.0128} & \tablenum[table-format=1.4]{0.0127} & \tablenum[table-format=1.4]{0.0128} & \tablenum[table-format=2.2]{95.00} \\
14 & \shortstack[l]{Trial aug.\\(linear)} & 960,960 & 4800,4800 & \tablenum[table-format=-1.4]{-0.0001} & \tablenum[table-format=1.4]{0.0131} & \tablenum[table-format=1.4]{0.0130} & \tablenum[table-format=1.4]{0.0131} & \tablenum[table-format=2.2]{95.08} \\
14 & \shortstack[l]{Trial aug.\\(categorical)} & 960,960 & 4800,4800 & \tablenum[table-format=-1.4]{-0.0001} & \tablenum[table-format=1.4]{0.0131} & \tablenum[table-format=1.4]{0.0130} & \tablenum[table-format=1.4]{0.0131} & \tablenum[table-format=2.2]{95.08} \\
20 & $\widehat{\theta}_{\mathrm{SS},\tau}$ & 960,960 & 240,240 & \tablenum[table-format=-1.4]{-0.0002} & \tablenum[table-format=1.4]{0.0127} & \tablenum[table-format=1.4]{0.0126} & \tablenum[table-format=1.4]{0.0127} & \tablenum[table-format=2.2]{94.53} \\
20 & \shortstack[l]{Trial aug.\\(linear)} & 960,960 & 240,240 & \tablenum[table-format=-1.4]{-0.0002} & \tablenum[table-format=1.4]{0.0121} & \tablenum[table-format=1.4]{0.0121} & \tablenum[table-format=1.4]{0.0121} & \tablenum[table-format=2.2]{94.73} \\
20 & \shortstack[l]{Trial aug.\\(categorical)} & 960,960 & 240,240 & \tablenum[table-format=-1.4]{-0.0002} & \tablenum[table-format=1.4]{0.0121} & \tablenum[table-format=1.4]{0.0121} & \tablenum[table-format=1.4]{0.0121} & \tablenum[table-format=2.2]{94.73} \\
20 & $\widehat{\theta}_{\mathrm{SS},\tau}$ & 960,960 & 960,960 & \tablenum[table-format=-1.4]{-0.0002} & \tablenum[table-format=1.4]{0.0125} & \tablenum[table-format=1.4]{0.0124} & \tablenum[table-format=1.4]{0.0125} & \tablenum[table-format=2.2]{94.68} \\
20 & \shortstack[l]{Trial aug.\\(linear)} & 960,960 & 960,960 & \tablenum[table-format=-1.4]{-0.0002} & \tablenum[table-format=1.4]{0.0121} & \tablenum[table-format=1.4]{0.0121} & \tablenum[table-format=1.4]{0.0121} & \tablenum[table-format=2.2]{94.73} \\
20 & \shortstack[l]{Trial aug.\\(categorical)} & 960,960 & 960,960 & \tablenum[table-format=-1.4]{-0.0002} & \tablenum[table-format=1.4]{0.0121} & \tablenum[table-format=1.4]{0.0121} & \tablenum[table-format=1.4]{0.0121} & \tablenum[table-format=2.2]{94.73} \\
20 & $\widehat{\theta}_{\mathrm{SS},\tau}$ & 960,960 & 4800,4800 & \tablenum[table-format=-1.4]{-0.0002} & \tablenum[table-format=1.4]{0.0123} & \tablenum[table-format=1.4]{0.0122} & \tablenum[table-format=1.4]{0.0123} & \tablenum[table-format=2.2]{94.63} \\
20 & \shortstack[l]{Trial aug.\\(linear)} & 960,960 & 4800,4800 & \tablenum[table-format=-1.4]{-0.0002} & \tablenum[table-format=1.4]{0.0121} & \tablenum[table-format=1.4]{0.0121} & \tablenum[table-format=1.4]{0.0121} & \tablenum[table-format=2.2]{94.73} \\
20 & \shortstack[l]{Trial aug.\\(categorical)} & 960,960 & 4800,4800 & \tablenum[table-format=-1.4]{-0.0002} & \tablenum[table-format=1.4]{0.0121} & \tablenum[table-format=1.4]{0.0121} & \tablenum[table-format=1.4]{0.0121} & \tablenum[table-format=2.2]{94.73} \\
\addlinespace
\end{tabular}}
\setlength{\UniformWidth}{\linewidth}\setlength{\tabcolsep}{\dimexpr(\linewidth-\wd\UniformMeasure)/16\relax}
\ifdim\tabcolsep<0pt\PackageWarning{CompactTables}{Negative column gap}\fi
\setlength{\LTleft}{0pt plus 1fill}\setlength{\LTright}{0pt plus 1fill}
\begin{longtable}{@{}llrrrrrrr@{}}
\caption{Calibration of covariate-augmented comparisons.}\label{tab:competitiveMethods}\\\toprule
Scenario & Method & $(n_1,n_2)$ & $(M_1,M_2)$ & Bias & EmpSD & MeanSE & RMSE & CP (\%) \\
\midrule\endfirsthead
\multicolumn{9}{l}{\textit{Table \thetable\ continued}}\\\toprule
Scenario & Method & $(n_1,n_2)$ & $(M_1,M_2)$ & Bias & EmpSD & MeanSE & RMSE & CP (\%) \\
\midrule\endhead
\midrule\multicolumn{9}{r}{\textit{Continued on next page}}\\\endfoot
\endlastfoot
9 & $\widehat{\theta}_{\mathrm{SS},\tau}$ & 960,960 & 240,240 & \tablenum[table-format=-1.4]{0.0002} & \tablenum[table-format=1.4]{0.0127} & \tablenum[table-format=1.4]{0.0129} & \tablenum[table-format=1.4]{0.0127} & \tablenum[table-format=2.2]{95.15} \\
9 & \shortstack[l]{Trial aug.\\(linear)} & 960,960 & 240,240 & \tablenum[table-format=-1.4]{0.0002} & \tablenum[table-format=1.4]{0.0127} & \tablenum[table-format=1.4]{0.0129} & \tablenum[table-format=1.4]{0.0127} & \tablenum[table-format=2.2]{95.12} \\
9 & \shortstack[l]{Trial aug.\\(categorical)} & 960,960 & 240,240 & \tablenum[table-format=-1.4]{0.0002} & \tablenum[table-format=1.4]{0.0127} & \tablenum[table-format=1.4]{0.0129} & \tablenum[table-format=1.4]{0.0127} & \tablenum[table-format=2.2]{95.12} \\
9 & $\widehat{\theta}_{\mathrm{SS},\tau}$ & 960,960 & 960,960 & \tablenum[table-format=-1.4]{0.0002} & \tablenum[table-format=1.4]{0.0127} & \tablenum[table-format=1.4]{0.0129} & \tablenum[table-format=1.4]{0.0127} & \tablenum[table-format=2.2]{95.03} \\
9 & \shortstack[l]{Trial aug.\\(linear)} & 960,960 & 960,960 & \tablenum[table-format=-1.4]{0.0002} & \tablenum[table-format=1.4]{0.0127} & \tablenum[table-format=1.4]{0.0129} & \tablenum[table-format=1.4]{0.0127} & \tablenum[table-format=2.2]{95.12} \\
9 & \shortstack[l]{Trial aug.\\(categorical)} & 960,960 & 960,960 & \tablenum[table-format=-1.4]{0.0002} & \tablenum[table-format=1.4]{0.0127} & \tablenum[table-format=1.4]{0.0129} & \tablenum[table-format=1.4]{0.0127} & \tablenum[table-format=2.2]{95.12} \\
9 & $\widehat{\theta}_{\mathrm{SS},\tau}$ & 960,960 & 4800,4800 & \tablenum[table-format=-1.4]{0.0002} & \tablenum[table-format=1.4]{0.0127} & \tablenum[table-format=1.4]{0.0129} & \tablenum[table-format=1.4]{0.0127} & \tablenum[table-format=2.2]{95.12} \\
9 & \shortstack[l]{Trial aug.\\(linear)} & 960,960 & 4800,4800 & \tablenum[table-format=-1.4]{0.0002} & \tablenum[table-format=1.4]{0.0127} & \tablenum[table-format=1.4]{0.0129} & \tablenum[table-format=1.4]{0.0127} & \tablenum[table-format=2.2]{95.12} \\
9 & \shortstack[l]{Trial aug.\\(categorical)} & 960,960 & 4800,4800 & \tablenum[table-format=-1.4]{0.0002} & \tablenum[table-format=1.4]{0.0127} & \tablenum[table-format=1.4]{0.0129} & \tablenum[table-format=1.4]{0.0127} & \tablenum[table-format=2.2]{95.12} \\
10 & $\widehat{\theta}_{\mathrm{SS},\tau}$ & 960,960 & 240,240 & \tablenum[table-format=-1.4]{0.0003} & \tablenum[table-format=1.4]{0.0130} & \tablenum[table-format=1.4]{0.0129} & \tablenum[table-format=1.4]{0.0130} & \tablenum[table-format=2.2]{94.62} \\
10 & \shortstack[l]{Trial aug.\\(linear)} & 960,960 & 240,240 & \tablenum[table-format=-1.4]{0.0003} & \tablenum[table-format=1.4]{0.0130} & \tablenum[table-format=1.4]{0.0129} & \tablenum[table-format=1.4]{0.0130} & \tablenum[table-format=2.2]{94.60} \\
10 & \shortstack[l]{Trial aug.\\(categorical)} & 960,960 & 240,240 & \tablenum[table-format=-1.4]{0.0003} & \tablenum[table-format=1.4]{0.0130} & \tablenum[table-format=1.4]{0.0129} & \tablenum[table-format=1.4]{0.0130} & \tablenum[table-format=2.2]{94.60} \\
10 & $\widehat{\theta}_{\mathrm{SS},\tau}$ & 960,960 & 960,960 & \tablenum[table-format=-1.4]{0.0003} & \tablenum[table-format=1.4]{0.0130} & \tablenum[table-format=1.4]{0.0129} & \tablenum[table-format=1.4]{0.0130} & \tablenum[table-format=2.2]{94.45} \\
10 & \shortstack[l]{Trial aug.\\(linear)} & 960,960 & 960,960 & \tablenum[table-format=-1.4]{0.0003} & \tablenum[table-format=1.4]{0.0130} & \tablenum[table-format=1.4]{0.0129} & \tablenum[table-format=1.4]{0.0130} & \tablenum[table-format=2.2]{94.60} \\
10 & \shortstack[l]{Trial aug.\\(categorical)} & 960,960 & 960,960 & \tablenum[table-format=-1.4]{0.0003} & \tablenum[table-format=1.4]{0.0130} & \tablenum[table-format=1.4]{0.0129} & \tablenum[table-format=1.4]{0.0130} & \tablenum[table-format=2.2]{94.60} \\
10 & $\widehat{\theta}_{\mathrm{SS},\tau}$ & 960,960 & 4800,4800 & \tablenum[table-format=-1.4]{0.0003} & \tablenum[table-format=1.4]{0.0130} & \tablenum[table-format=1.4]{0.0129} & \tablenum[table-format=1.4]{0.0130} & \tablenum[table-format=2.2]{94.53} \\
10 & \shortstack[l]{Trial aug.\\(linear)} & 960,960 & 4800,4800 & \tablenum[table-format=-1.4]{0.0003} & \tablenum[table-format=1.4]{0.0130} & \tablenum[table-format=1.4]{0.0129} & \tablenum[table-format=1.4]{0.0130} & \tablenum[table-format=2.2]{94.60} \\
10 & \shortstack[l]{Trial aug.\\(categorical)} & 960,960 & 4800,4800 & \tablenum[table-format=-1.4]{0.0003} & \tablenum[table-format=1.4]{0.0130} & \tablenum[table-format=1.4]{0.0129} & \tablenum[table-format=1.4]{0.0130} & \tablenum[table-format=2.2]{94.60} \\
11 & $\widehat{\theta}_{\mathrm{SS},\tau}$ & 960,960 & 240,240 & \tablenum[table-format=-1.4]{-0.0002} & \tablenum[table-format=1.4]{0.0129} & \tablenum[table-format=1.4]{0.0129} & \tablenum[table-format=1.4]{0.0129} & \tablenum[table-format=2.2]{94.90} \\
11 & \shortstack[l]{Trial aug.\\(linear)} & 960,960 & 240,240 & \tablenum[table-format=-1.4]{-0.0002} & \tablenum[table-format=1.4]{0.0130} & \tablenum[table-format=1.4]{0.0130} & \tablenum[table-format=1.4]{0.0130} & \tablenum[table-format=2.2]{94.80} \\
11 & \shortstack[l]{Trial aug.\\(categorical)} & 960,960 & 240,240 & \tablenum[table-format=-1.4]{-0.0002} & \tablenum[table-format=1.4]{0.0130} & \tablenum[table-format=1.4]{0.0130} & \tablenum[table-format=1.4]{0.0130} & \tablenum[table-format=2.2]{94.75} \\
11 & $\widehat{\theta}_{\mathrm{SS},\tau}$ & 960,960 & 960,960 & \tablenum[table-format=-1.4]{-0.0003} & \tablenum[table-format=1.4]{0.0128} & \tablenum[table-format=1.4]{0.0128} & \tablenum[table-format=1.4]{0.0128} & \tablenum[table-format=2.2]{94.97} \\
11 & \shortstack[l]{Trial aug.\\(linear)} & 960,960 & 960,960 & \tablenum[table-format=-1.4]{-0.0002} & \tablenum[table-format=1.4]{0.0130} & \tablenum[table-format=1.4]{0.0130} & \tablenum[table-format=1.4]{0.0130} & \tablenum[table-format=2.2]{94.80} \\
11 & \shortstack[l]{Trial aug.\\(categorical)} & 960,960 & 960,960 & \tablenum[table-format=-1.4]{-0.0002} & \tablenum[table-format=1.4]{0.0130} & \tablenum[table-format=1.4]{0.0130} & \tablenum[table-format=1.4]{0.0130} & \tablenum[table-format=2.2]{94.75} \\
11 & $\widehat{\theta}_{\mathrm{SS},\tau}$ & 960,960 & 4800,4800 & \tablenum[table-format=-1.4]{-0.0002} & \tablenum[table-format=1.4]{0.0127} & \tablenum[table-format=1.4]{0.0127} & \tablenum[table-format=1.4]{0.0127} & \tablenum[table-format=2.2]{94.95} \\
11 & \shortstack[l]{Trial aug.\\(linear)} & 960,960 & 4800,4800 & \tablenum[table-format=-1.4]{-0.0002} & \tablenum[table-format=1.4]{0.0130} & \tablenum[table-format=1.4]{0.0130} & \tablenum[table-format=1.4]{0.0130} & \tablenum[table-format=2.2]{94.80} \\
11 & \shortstack[l]{Trial aug.\\(categorical)} & 960,960 & 4800,4800 & \tablenum[table-format=-1.4]{-0.0002} & \tablenum[table-format=1.4]{0.0130} & \tablenum[table-format=1.4]{0.0130} & \tablenum[table-format=1.4]{0.0130} & \tablenum[table-format=2.2]{94.75} \\
12 & $\widehat{\theta}_{\mathrm{SS},\tau}$ & 960,960 & 240,240 & \tablenum[table-format=-1.4]{0.0003} & \tablenum[table-format=1.4]{0.0128} & \tablenum[table-format=1.4]{0.0128} & \tablenum[table-format=1.4]{0.0128} & \tablenum[table-format=2.2]{95.17} \\
12 & \shortstack[l]{Trial aug.\\(linear)} & 960,960 & 240,240 & \tablenum[table-format=-1.4]{0.0002} & \tablenum[table-format=1.4]{0.0130} & \tablenum[table-format=1.4]{0.0130} & \tablenum[table-format=1.4]{0.0130} & \tablenum[table-format=2.2]{95.33} \\
12 & \shortstack[l]{Trial aug.\\(categorical)} & 960,960 & 240,240 & \tablenum[table-format=-1.4]{0.0002} & \tablenum[table-format=1.4]{0.0130} & \tablenum[table-format=1.4]{0.0130} & \tablenum[table-format=1.4]{0.0130} & \tablenum[table-format=2.2]{95.33} \\
12 & $\widehat{\theta}_{\mathrm{SS},\tau}$ & 960,960 & 960,960 & \tablenum[table-format=-1.4]{0.0003} & \tablenum[table-format=1.4]{0.0125} & \tablenum[table-format=1.4]{0.0124} & \tablenum[table-format=1.4]{0.0125} & \tablenum[table-format=2.2]{94.97} \\
12 & \shortstack[l]{Trial aug.\\(linear)} & 960,960 & 960,960 & \tablenum[table-format=-1.4]{0.0002} & \tablenum[table-format=1.4]{0.0130} & \tablenum[table-format=1.4]{0.0130} & \tablenum[table-format=1.4]{0.0130} & \tablenum[table-format=2.2]{95.33} \\
12 & \shortstack[l]{Trial aug.\\(categorical)} & 960,960 & 960,960 & \tablenum[table-format=-1.4]{0.0002} & \tablenum[table-format=1.4]{0.0130} & \tablenum[table-format=1.4]{0.0130} & \tablenum[table-format=1.4]{0.0130} & \tablenum[table-format=2.2]{95.33} \\
12 & $\widehat{\theta}_{\mathrm{SS},\tau}$ & 960,960 & 4800,4800 & \tablenum[table-format=-1.4]{0.0002} & \tablenum[table-format=1.4]{0.0121} & \tablenum[table-format=1.4]{0.0121} & \tablenum[table-format=1.4]{0.0121} & \tablenum[table-format=2.2]{94.78} \\
12 & \shortstack[l]{Trial aug.\\(linear)} & 960,960 & 4800,4800 & \tablenum[table-format=-1.4]{0.0002} & \tablenum[table-format=1.4]{0.0130} & \tablenum[table-format=1.4]{0.0130} & \tablenum[table-format=1.4]{0.0130} & \tablenum[table-format=2.2]{95.33} \\
12 & \shortstack[l]{Trial aug.\\(categorical)} & 960,960 & 4800,4800 & \tablenum[table-format=-1.4]{0.0002} & \tablenum[table-format=1.4]{0.0130} & \tablenum[table-format=1.4]{0.0130} & \tablenum[table-format=1.4]{0.0130} & \tablenum[table-format=2.2]{95.33} \\
14 & $\widehat{\theta}_{\mathrm{SS},\tau}$ & 960,960 & 240,240 & \tablenum[table-format=-1.4]{0.0000} & \tablenum[table-format=1.4]{0.0130} & \tablenum[table-format=1.4]{0.0129} & \tablenum[table-format=1.4]{0.0130} & \tablenum[table-format=2.2]{95.00} \\
14 & \shortstack[l]{Trial aug.\\(linear)} & 960,960 & 240,240 & \tablenum[table-format=-1.4]{-0.0001} & \tablenum[table-format=1.4]{0.0131} & \tablenum[table-format=1.4]{0.0130} & \tablenum[table-format=1.4]{0.0131} & \tablenum[table-format=2.2]{95.08} \\
14 & \shortstack[l]{Trial aug.\\(categorical)} & 960,960 & 240,240 & \tablenum[table-format=-1.4]{-0.0001} & \tablenum[table-format=1.4]{0.0131} & \tablenum[table-format=1.4]{0.0130} & \tablenum[table-format=1.4]{0.0131} & \tablenum[table-format=2.2]{95.08} \\
14 & $\widehat{\theta}_{\mathrm{SS},\tau}$ & 960,960 & 960,960 & \tablenum[table-format=-1.4]{0.0000} & \tablenum[table-format=1.4]{0.0129} & \tablenum[table-format=1.4]{0.0128} & \tablenum[table-format=1.4]{0.0129} & \tablenum[table-format=2.2]{94.97} \\
14 & \shortstack[l]{Trial aug.\\(linear)} & 960,960 & 960,960 & \tablenum[table-format=-1.4]{-0.0001} & \tablenum[table-format=1.4]{0.0131} & \tablenum[table-format=1.4]{0.0130} & \tablenum[table-format=1.4]{0.0131} & \tablenum[table-format=2.2]{95.08} \\
14 & \shortstack[l]{Trial aug.\\(categorical)} & 960,960 & 960,960 & \tablenum[table-format=-1.4]{-0.0001} & \tablenum[table-format=1.4]{0.0131} & \tablenum[table-format=1.4]{0.0130} & \tablenum[table-format=1.4]{0.0131} & \tablenum[table-format=2.2]{95.08} \\
14 & $\widehat{\theta}_{\mathrm{SS},\tau}$ & 960,960 & 4800,4800 & \tablenum[table-format=-1.4]{0.0000} & \tablenum[table-format=1.4]{0.0128} & \tablenum[table-format=1.4]{0.0127} & \tablenum[table-format=1.4]{0.0128} & \tablenum[table-format=2.2]{95.00} \\
14 & \shortstack[l]{Trial aug.\\(linear)} & 960,960 & 4800,4800 & \tablenum[table-format=-1.4]{-0.0001} & \tablenum[table-format=1.4]{0.0131} & \tablenum[table-format=1.4]{0.0130} & \tablenum[table-format=1.4]{0.0131} & \tablenum[table-format=2.2]{95.08} \\
14 & \shortstack[l]{Trial aug.\\(categorical)} & 960,960 & 4800,4800 & \tablenum[table-format=-1.4]{-0.0001} & \tablenum[table-format=1.4]{0.0131} & \tablenum[table-format=1.4]{0.0130} & \tablenum[table-format=1.4]{0.0131} & \tablenum[table-format=2.2]{95.08} \\
20 & $\widehat{\theta}_{\mathrm{SS},\tau}$ & 960,960 & 240,240 & \tablenum[table-format=-1.4]{-0.0002} & \tablenum[table-format=1.4]{0.0127} & \tablenum[table-format=1.4]{0.0126} & \tablenum[table-format=1.4]{0.0127} & \tablenum[table-format=2.2]{94.53} \\
20 & \shortstack[l]{Trial aug.\\(linear)} & 960,960 & 240,240 & \tablenum[table-format=-1.4]{-0.0002} & \tablenum[table-format=1.4]{0.0121} & \tablenum[table-format=1.4]{0.0121} & \tablenum[table-format=1.4]{0.0121} & \tablenum[table-format=2.2]{94.73} \\
20 & \shortstack[l]{Trial aug.\\(categorical)} & 960,960 & 240,240 & \tablenum[table-format=-1.4]{-0.0002} & \tablenum[table-format=1.4]{0.0121} & \tablenum[table-format=1.4]{0.0121} & \tablenum[table-format=1.4]{0.0121} & \tablenum[table-format=2.2]{94.73} \\
20 & $\widehat{\theta}_{\mathrm{SS},\tau}$ & 960,960 & 960,960 & \tablenum[table-format=-1.4]{-0.0002} & \tablenum[table-format=1.4]{0.0125} & \tablenum[table-format=1.4]{0.0124} & \tablenum[table-format=1.4]{0.0125} & \tablenum[table-format=2.2]{94.68} \\
20 & \shortstack[l]{Trial aug.\\(linear)} & 960,960 & 960,960 & \tablenum[table-format=-1.4]{-0.0002} & \tablenum[table-format=1.4]{0.0121} & \tablenum[table-format=1.4]{0.0121} & \tablenum[table-format=1.4]{0.0121} & \tablenum[table-format=2.2]{94.73} \\
20 & \shortstack[l]{Trial aug.\\(categorical)} & 960,960 & 960,960 & \tablenum[table-format=-1.4]{-0.0002} & \tablenum[table-format=1.4]{0.0121} & \tablenum[table-format=1.4]{0.0121} & \tablenum[table-format=1.4]{0.0121} & \tablenum[table-format=2.2]{94.73} \\
20 & $\widehat{\theta}_{\mathrm{SS},\tau}$ & 960,960 & 4800,4800 & \tablenum[table-format=-1.4]{-0.0002} & \tablenum[table-format=1.4]{0.0123} & \tablenum[table-format=1.4]{0.0122} & \tablenum[table-format=1.4]{0.0123} & \tablenum[table-format=2.2]{94.63} \\
20 & \shortstack[l]{Trial aug.\\(linear)} & 960,960 & 4800,4800 & \tablenum[table-format=-1.4]{-0.0002} & \tablenum[table-format=1.4]{0.0121} & \tablenum[table-format=1.4]{0.0121} & \tablenum[table-format=1.4]{0.0121} & \tablenum[table-format=2.2]{94.73} \\
20 & \shortstack[l]{Trial aug.\\(categorical)} & 960,960 & 4800,4800 & \tablenum[table-format=-1.4]{-0.0002} & \tablenum[table-format=1.4]{0.0121} & \tablenum[table-format=1.4]{0.0121} & \tablenum[table-format=1.4]{0.0121} & \tablenum[table-format=2.2]{94.73} \\*
\TableEndRule \multicolumn{9}{@{}p{\UniformWidth}@{}}{\textit{Note:} Bias, EmpSD, MeanSE and RMSE are on the probability scale; CP is 95\% Wald coverage. Method rows use $\widehat{\theta}_{\mathrm{SS},\tau}$ (Weighted), Trial aug. (linear) and Trial aug. (categorical), fixed-basis comparisons in the common-covariate-distribution submodel described in Section~\ref{sec:covariatecomparison}. Each population has 960 labeled observations; $M_1=M_2$ varies. Scenarios 9--11 and 14 use 4,000 repetitions per cell; Scenarios 12 and 20 use 6,000.}\\
\end{longtable}\endgroup

\FloatBarrier
\subsection{Additional Rotterdam analyses}
\subsubsection{Label resources and patient sampling}
Table~\ref{tab:rotMasks} compares the prespecified label budgets, feature sets and follow-up windows. The unshrunk losses at 75\% labels show why uncertainty in the remaining auxiliary pool matters even when many outcomes are available. Its bootstrap panel assesses patient sampling with a new allocation, complementing the fixed-cohort allocation comparisons in the first panel.
\begingroup\TableFont
\setlength{\tabcolsep}{0pt}\sbox{\UniformMeasure}{\begin{tabular}{@{}lrllllll@{}}Features & $\tau$ & Labels & $(n_1,n_2)$ & $(M_1,M_2)$ & $G_{\mathrm{SS:sup}}$ & $G_{\mathrm{un:sup}}$ & $G_{\mathrm{SS:un}}$\\
Basic  &  \tablenum[table-format=4.0]{1825}  &  \tablenum[table-format=1.2]{0.25}  &  328/417  &  984/1253  &  \tablenum[table-format=2.2]{9.87}\,(\tablenum[table-format=1.2]{2.10})  &  \tablenum[table-format=-3.2]{6.40}\,(\tablenum[table-format=2.2]{2.86})  &  \tablenum[table-format=2.2]{3.71}\,(\tablenum[table-format=1.2]{0.78}) \\
Basic  &  \tablenum[table-format=4.0]{1825}  &  \tablenum[table-format=1.2]{0.50}  &  656/835  &  656/835  &  \tablenum[table-format=2.2]{13.60}\,(\tablenum[table-format=1.2]{1.99})  &  \tablenum[table-format=-3.2]{0.43}\,(\tablenum[table-format=2.2]{4.26})  &  \tablenum[table-format=2.2]{13.22}\,(\tablenum[table-format=1.2]{1.98}) \\
Basic  &  \tablenum[table-format=4.0]{1825}  &  \tablenum[table-format=1.2]{0.75}  &  984/1252  &  328/418  &  \tablenum[table-format=2.2]{13.07}\,(\tablenum[table-format=1.2]{2.05})  &  \tablenum[table-format=-3.2]{-116.26}\,(\tablenum[table-format=2.2]{13.40})  &  \tablenum[table-format=2.2]{59.81}\,(\tablenum[table-format=1.2]{2.03}) \\
Primary  &  \tablenum[table-format=4.0]{1825}  &  \tablenum[table-format=1.2]{0.25}  &  328/417  &  984/1253  &  \tablenum[table-format=2.2]{10.41}\,(\tablenum[table-format=1.2]{2.27})  &  \tablenum[table-format=-3.2]{5.47}\,(\tablenum[table-format=2.2]{3.13})  &  \tablenum[table-format=2.2]{5.23}\,(\tablenum[table-format=1.2]{0.86}) \\
Primary  &  \tablenum[table-format=4.0]{1825}  &  \tablenum[table-format=1.2]{0.50}  &  656/835  &  656/835  &  \tablenum[table-format=2.2]{13.83}\,(\tablenum[table-format=1.2]{2.19})  &  \tablenum[table-format=-3.2]{-4.98}\,(\tablenum[table-format=2.2]{4.77})  &  \tablenum[table-format=2.2]{17.92}\,(\tablenum[table-format=1.2]{1.99}) \\
Primary  &  \tablenum[table-format=4.0]{1825}  &  \tablenum[table-format=1.2]{0.75}  &  984/1252  &  328/418  &  \tablenum[table-format=2.2]{13.50}\,(\tablenum[table-format=1.2]{2.22})  &  \tablenum[table-format=-3.2]{-139.14}\,(\tablenum[table-format=2.2]{14.90})  &  \tablenum[table-format=2.2]{63.83}\,(\tablenum[table-format=1.2]{1.81}) \\
Primary  &  \tablenum[table-format=4.0]{2555}  &  \tablenum[table-format=1.2]{0.25}  &  328/417  &  984/1253  &  \tablenum[table-format=2.2]{15.40}\,(\tablenum[table-format=1.2]{2.26})  &  \tablenum[table-format=-3.2]{11.00}\,(\tablenum[table-format=2.2]{3.10})  &  \tablenum[table-format=2.2]{4.93}\,(\tablenum[table-format=1.2]{0.95}) \\
Primary  &  \tablenum[table-format=4.0]{2555}  &  \tablenum[table-format=1.2]{0.50}  &  656/835  &  656/835  &  \tablenum[table-format=2.2]{17.69}\,(\tablenum[table-format=1.2]{2.30})  &  \tablenum[table-format=-3.2]{-3.80}\,(\tablenum[table-format=2.2]{5.25})  &  \tablenum[table-format=2.2]{20.71}\,(\tablenum[table-format=1.2]{2.24}) \\
Primary  &  \tablenum[table-format=4.0]{2555}  &  \tablenum[table-format=1.2]{0.75}  &  984/1252  &  328/418  &  \tablenum[table-format=2.2]{16.20}\,(\tablenum[table-format=1.2]{2.21})  &  \tablenum[table-format=-3.2]{-160.12}\,(\tablenum[table-format=2.2]{16.31})  &  \tablenum[table-format=2.2]{67.78}\,(\tablenum[table-format=1.2]{1.72}) \\
\midrule
\multicolumn{8}{l}{Patient bootstrap (50\% labels, five years)}\\
Method & \multicolumn{2}{c}{Mean} & SD & Mean IF-SE & SE/SD & \multicolumn{2}{c}{95\% quantiles}\\\midrule
$\widehat{\theta}_{\mathrm{sup},\tau}$ & \multicolumn{2}{c}{0.4596} & 0.0113 & 0.0114 & 1.012 & \multicolumn{2}{c}{[0.4380, 0.4823]} \\
$\widehat{\theta}_{\mathrm{un},\tau}$ & \multicolumn{2}{c}{0.4599} & 0.0116 & 0.0115 & 0.988 & \multicolumn{2}{c}{[0.4389, 0.4827]} \\
$\widehat{\theta}_{\mathrm{SS},\tau}$ & \multicolumn{2}{c}{0.4597} & 0.0110 & 0.0110 & 0.996 & \multicolumn{2}{c}{[0.4395, 0.4821]} \\
\end{tabular}}
\setlength{\UniformWidth}{\linewidth}\setlength{\tabcolsep}{\dimexpr(\linewidth-\wd\UniformMeasure)/14\relax}
\ifdim\tabcolsep<0pt\PackageWarning{UnifiedTables}{Negative column gap}\fi
\setlength{\LTleft}{0pt plus 1fill}\setlength{\LTright}{0pt plus 1fill}
\setlength{\LTpre}{5pt}\setlength{\LTpost}{4pt}
\begin{longtable}{@{}lrllllll@{}}
\caption{Additional Rotterdam precision results.}\label{tab:rotMasks}\label{tab:rotterdamBootstrap}\\
\toprule Features & $\tau$ & Labels & $(n_1,n_2)$ & $(M_1,M_2)$ & $G_{\mathrm{SS:sup}}$ & $G_{\mathrm{un:sup}}$ & $G_{\mathrm{SS:un}}$\\\midrule\endfirsthead
\multicolumn{8}{l}{\textit{Table \thetable\ continued}}\\\toprule Features & $\tau$ & Labels & $(n_1,n_2)$ & $(M_1,M_2)$ & $G_{\mathrm{SS:sup}}$ & $G_{\mathrm{un:sup}}$ & $G_{\mathrm{SS:un}}$\\\midrule\endhead
\midrule\multicolumn{8}{r}{\textit{Continued on next page}}\\\endfoot
\endlastfoot

Basic  &  \tablenum[table-format=4.0]{1825}  &  \tablenum[table-format=1.2]{0.25}  &  328/417  &  984/1253  &  \tablenum[table-format=2.2]{9.87}\,(\tablenum[table-format=1.2]{2.10})  &  \tablenum[table-format=-3.2]{6.40}\,(\tablenum[table-format=2.2]{2.86})  &  \tablenum[table-format=2.2]{3.71}\,(\tablenum[table-format=1.2]{0.78}) \\
Basic  &  \tablenum[table-format=4.0]{1825}  &  \tablenum[table-format=1.2]{0.50}  &  656/835  &  656/835  &  \tablenum[table-format=2.2]{13.60}\,(\tablenum[table-format=1.2]{1.99})  &  \tablenum[table-format=-3.2]{0.43}\,(\tablenum[table-format=2.2]{4.26})  &  \tablenum[table-format=2.2]{13.22}\,(\tablenum[table-format=1.2]{1.98}) \\
Basic  &  \tablenum[table-format=4.0]{1825}  &  \tablenum[table-format=1.2]{0.75}  &  984/1252  &  328/418  &  \tablenum[table-format=2.2]{13.07}\,(\tablenum[table-format=1.2]{2.05})  &  \tablenum[table-format=-3.2]{-116.26}\,(\tablenum[table-format=2.2]{13.40})  &  \tablenum[table-format=2.2]{59.81}\,(\tablenum[table-format=1.2]{2.03}) \\
Primary  &  \tablenum[table-format=4.0]{1825}  &  \tablenum[table-format=1.2]{0.25}  &  328/417  &  984/1253  &  \tablenum[table-format=2.2]{10.41}\,(\tablenum[table-format=1.2]{2.27})  &  \tablenum[table-format=-3.2]{5.47}\,(\tablenum[table-format=2.2]{3.13})  &  \tablenum[table-format=2.2]{5.23}\,(\tablenum[table-format=1.2]{0.86}) \\
Primary  &  \tablenum[table-format=4.0]{1825}  &  \tablenum[table-format=1.2]{0.50}  &  656/835  &  656/835  &  \tablenum[table-format=2.2]{13.83}\,(\tablenum[table-format=1.2]{2.19})  &  \tablenum[table-format=-3.2]{-4.98}\,(\tablenum[table-format=2.2]{4.77})  &  \tablenum[table-format=2.2]{17.92}\,(\tablenum[table-format=1.2]{1.99}) \\
Primary  &  \tablenum[table-format=4.0]{1825}  &  \tablenum[table-format=1.2]{0.75}  &  984/1252  &  328/418  &  \tablenum[table-format=2.2]{13.50}\,(\tablenum[table-format=1.2]{2.22})  &  \tablenum[table-format=-3.2]{-139.14}\,(\tablenum[table-format=2.2]{14.90})  &  \tablenum[table-format=2.2]{63.83}\,(\tablenum[table-format=1.2]{1.81}) \\
Primary  &  \tablenum[table-format=4.0]{2555}  &  \tablenum[table-format=1.2]{0.25}  &  328/417  &  984/1253  &  \tablenum[table-format=2.2]{15.40}\,(\tablenum[table-format=1.2]{2.26})  &  \tablenum[table-format=-3.2]{11.00}\,(\tablenum[table-format=2.2]{3.10})  &  \tablenum[table-format=2.2]{4.93}\,(\tablenum[table-format=1.2]{0.95}) \\
Primary  &  \tablenum[table-format=4.0]{2555}  &  \tablenum[table-format=1.2]{0.50}  &  656/835  &  656/835  &  \tablenum[table-format=2.2]{17.69}\,(\tablenum[table-format=1.2]{2.30})  &  \tablenum[table-format=-3.2]{-3.80}\,(\tablenum[table-format=2.2]{5.25})  &  \tablenum[table-format=2.2]{20.71}\,(\tablenum[table-format=1.2]{2.24}) \\
Primary  &  \tablenum[table-format=4.0]{2555}  &  \tablenum[table-format=1.2]{0.75}  &  984/1252  &  328/418  &  \tablenum[table-format=2.2]{16.20}\,(\tablenum[table-format=1.2]{2.21})  &  \tablenum[table-format=-3.2]{-160.12}\,(\tablenum[table-format=2.2]{16.31})  &  \tablenum[table-format=2.2]{67.78}\,(\tablenum[table-format=1.2]{1.72}) \\
\midrule
\multicolumn{8}{l}{Patient bootstrap (50\% labels, five years)}\\
Method & \multicolumn{2}{c}{Mean} & SD & Mean IF-SE & SE/SD & \multicolumn{2}{c}{95\% quantiles}\\\midrule
$\widehat{\theta}_{\mathrm{sup},\tau}$ & \multicolumn{2}{c}{0.4596} & 0.0113 & 0.0114 & 1.012 & \multicolumn{2}{c}{[0.4380, 0.4823]} \\
$\widehat{\theta}_{\mathrm{un},\tau}$ & \multicolumn{2}{c}{0.4599} & 0.0116 & 0.0115 & 0.988 & \multicolumn{2}{c}{[0.4389, 0.4827]} \\
$\widehat{\theta}_{\mathrm{SS},\tau}$ & \multicolumn{2}{c}{0.4597} & 0.0110 & 0.0110 & 0.996 & \multicolumn{2}{c}{[0.4395, 0.4821]} \\*
\TableEndRule \multicolumn{8}{@{}p{\UniformWidth}@{}}{\textit{Note:} $n_k$ and $M_k$ are labeled and covariate-only counts for group $k$; pairs are premenopausal/postmenopausal. The gain columns are $G_{\mathrm{SS:sup}}$, $G_{\mathrm{un:sup}}$ and $G_{\mathrm{SS:un}}$ (percent; paired MCSE in percentage points); each analysis uses 1,000 allocations. Bootstrap results use 1,000 patient-level samples stratified by population, each with a new allocation.}\\
\end{longtable}\endgroup

\subsubsection{Sensitivity and labeled-sample support}
Across 100 allocations, outcome-model standardization produces larger shifts than at the primary partition, and the spline-year censoring model fails in 3 fits (Table~\ref{tab:rotSensitivityRepeated}).

\enlargethispage{12pt}
\begingroup\TableFont
\setlength{\tabcolsep}{0pt}\sbox{\UniformMeasure}{\begin{tabular}{@{}>{\raggedright\arraybackslash}p{0.32\linewidth}lrrrrr@{}}Procedure & Usable & Fail & \shortstack{Mean\\shift} & MCSE & \shortstack{Central 95\%\\range} & \shortstack{Max. abs.\\shift} \\
Marginal IPCW & 100/100 & \tablenum[table-format=1.0]{0} & \tablenum[table-format=-1.3]{0.000} & \tablenum[table-format=1.4]{0.0000} & [0.000, 0.000] & \tablenum[table-format=2.3]{0.000} \\
Period-stratified IPCW, labeled standardization & 100/100 & \tablenum[table-format=1.0]{0} & \tablenum[table-format=-1.3]{0.241} & \tablenum[table-format=1.4]{0.0114} & [0.019, 0.456] & \tablenum[table-format=2.3]{0.502} \\
Period-stratified IPCW, cohort standardization & 100/100 & \tablenum[table-format=1.0]{0} & \tablenum[table-format=-1.3]{0.256} & \tablenum[table-format=1.4]{0.0491} & [-0.824, 1.114] & \tablenum[table-format=2.3]{1.852} \\
Conditional IPCW, linear calendar-year effect & 100/100 & \tablenum[table-format=1.0]{0} & \tablenum[table-format=-1.3]{-0.556} & \tablenum[table-format=1.4]{0.0427} & [-1.450, 0.271] & \tablenum[table-format=2.3]{1.550} \\
Conditional IPCW, spline calendar-year effect & 97/100 & \tablenum[table-format=1.0]{3} & \tablenum[table-format=-1.3]{-0.550} & \tablenum[table-format=1.4]{0.0455} & [-1.407, 0.322] & \tablenum[table-format=2.3]{1.685} \\
Outcome-regression standardization, labeled covariates & 100/100 & \tablenum[table-format=1.0]{0} & \tablenum[table-format=-1.3]{-0.647} & \tablenum[table-format=1.4]{0.0753} & [-2.018, 0.632] & \tablenum[table-format=2.3]{2.168} \\
Outcome-regression standardization, cohort covariates & 100/100 & \tablenum[table-format=1.0]{0} & \tablenum[table-format=-1.3]{-0.205} & \tablenum[table-format=1.4]{0.3564} & [-6.953, 7.136] & \tablenum[table-format=2.3]{10.205} \\
Reverse-KM tie convention & 100/100 & \tablenum[table-format=1.0]{0} & \tablenum[table-format=-1.3]{-0.001} & \tablenum[table-format=1.4]{0.0001} & [-0.002, 0.000] & \tablenum[table-format=2.3]{0.003} \\
\addlinespace
\end{tabular}}
\setlength{\UniformWidth}{\linewidth}\setlength{\tabcolsep}{\dimexpr(\linewidth-\wd\UniformMeasure)/12\relax}
\ifdim\tabcolsep<0pt\PackageWarning{CompactTables}{Negative column gap}\fi
\setlength{\LTleft}{0pt plus 1fill}\setlength{\LTright}{0pt plus 1fill}

\par\noindent
\begin{minipage}{\linewidth}
\centering
\captionsetup{skip=6pt}
\captionof{table}{Sensitivity across label allocations.}\label{tab:rotSensitivityRepeated}
\begin{tabular}{@{}>{\raggedright\arraybackslash}p{0.32\linewidth}lrrrrr@{}}
\toprule
Procedure & Usable & Fail & \shortstack{Mean\\shift} & MCSE & \shortstack{Central 95\%\\range} & \shortstack{Max. abs.\\shift} \\
\midrule
Marginal IPCW & 100/100 & \tablenum[table-format=1.0]{0} & \tablenum[table-format=-1.3]{0.000} & \tablenum[table-format=1.4]{0.0000} & [0.000, 0.000] & \tablenum[table-format=2.3]{0.000} \\
Period-stratified IPCW, labeled standardization & 100/100 & \tablenum[table-format=1.0]{0} & \tablenum[table-format=-1.3]{0.241} & \tablenum[table-format=1.4]{0.0114} & [0.019, 0.456] & \tablenum[table-format=2.3]{0.502} \\
Period-stratified IPCW, cohort standardization & 100/100 & \tablenum[table-format=1.0]{0} & \tablenum[table-format=-1.3]{0.256} & \tablenum[table-format=1.4]{0.0491} & [-0.824, 1.114] & \tablenum[table-format=2.3]{1.852} \\
Conditional IPCW, linear calendar-year effect & 100/100 & \tablenum[table-format=1.0]{0} & \tablenum[table-format=-1.3]{-0.556} & \tablenum[table-format=1.4]{0.0427} & [-1.450, 0.271] & \tablenum[table-format=2.3]{1.550} \\
Conditional IPCW, spline calendar-year effect & 97/100 & \tablenum[table-format=1.0]{3} & \tablenum[table-format=-1.3]{-0.550} & \tablenum[table-format=1.4]{0.0455} & [-1.407, 0.322] & \tablenum[table-format=2.3]{1.685} \\
Outcome-regression standardization, labeled covariates & 100/100 & \tablenum[table-format=1.0]{0} & \tablenum[table-format=-1.3]{-0.647} & \tablenum[table-format=1.4]{0.0753} & [-2.018, 0.632] & \tablenum[table-format=2.3]{2.168} \\
Outcome-regression standardization, cohort covariates & 100/100 & \tablenum[table-format=1.0]{0} & \tablenum[table-format=-1.3]{-0.205} & \tablenum[table-format=1.4]{0.3564} & [-6.953, 7.136] & \tablenum[table-format=2.3]{10.205} \\
Reverse-KM tie convention & 100/100 & \tablenum[table-format=1.0]{0} & \tablenum[table-format=-1.3]{-0.001} & \tablenum[table-format=1.4]{0.0001} & [-0.002, 0.000] & \tablenum[table-format=2.3]{0.003} \\
\bottomrule \multicolumn{7}{@{}p{\UniformWidth}@{}}{\textit{Note:} Shifts, MCSE, ranges and maxima are multiplied by $10^3$. The first 100 allocations use labeled and covariate-only pool sizes $(656,835)$ each. Ranges are allocation quantiles; spline summaries use the 97 successful fits.}\\
\end{tabular}
\end{minipage}\par\smallskip
\endgroup

\begin{figure}[H]\centering
\makebox[\linewidth][c]{\includegraphics[width=\linewidth]{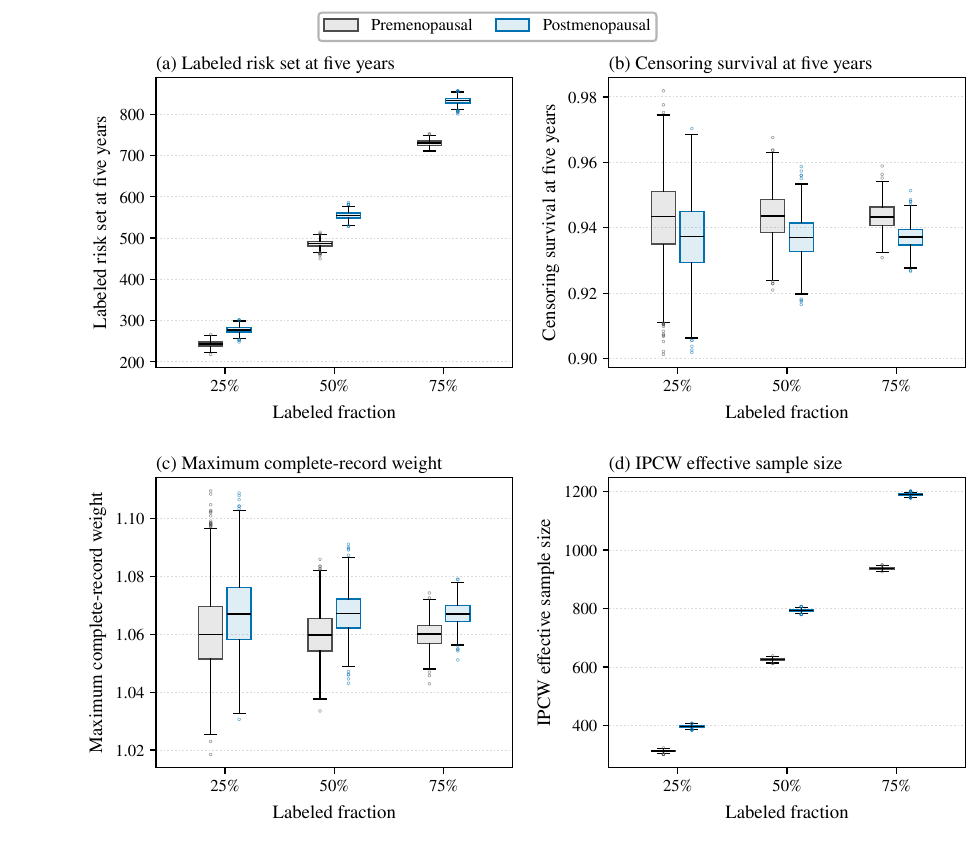}}
\caption{Rotterdam five-year labeled-sample support over all 1,000 label allocations at each budget, using the primary features. Panels give labeled risk-set size at day 1825, censoring survival $\widehat G(1825-)$, maximum inverse censoring weight among records with observed capped event time ($A=1$), and IPCW effective sample size. Boxes, medians, 1.5-IQR whiskers and all outliers describe allocation distributions within the fixed cohort; the repeated allocations condition on the observed cohort.}\label{fig:sSupport}
\end{figure}

Figure~\ref{fig:sSupport} shows how much five-year follow-up the labeled samples retain across allocations. The censoring survival, maximum weight and effective sample size indicate whether the retained information depends heavily on a small number of observations.

\FloatBarrier

\end{appendices}
\end{document}